
\input amstex
\documentstyle{amsppt}
\magnification = \magstephalf
\pageheight{9truein}
\pagewidth{6.5truein}
\baselineskip=10pt
\nologo
\NoRunningHeads
\NoBlackBoxes
\def\bhsp{{\widehat {\bold {sp}}}}
\def\hf{\frac{1}{2}}
\def\shf{{\tsize{\frac{1}{2}}}}
\def\sth{{\tsize{\frac{3}{2}}}}

\def\sfr{{\tsize{\frac{1}{4}}}}
\def\sei{{\tsize{\frac{l}{8}}}}
\def\slr{{\tsize{\frac{l}{4}}}}
\def\srd{{\tsize{\frac{1}{3}}}}
\def\sxt{{\tsize{\frac{1}{6}}}}
\def\std{{\tsize{\frac{2}{3}}}}
\def\sni{{\tsize{\frac{1}{9}}}}
\def\svx{{\tsize{\frac{7}{6}}}}
\def\fvx{{\tsize{\frac{5}{6}}}}

\def\one{\bold {\pmb 1}}
\def\pne{\bold {\pmb {1'}}}

\def\bV{\bold V}
\def\bW{\bold W}

\def\cI{{\Cal I}}
\def\oper{{(Y(w_1,z_1)Y({\bar w_2},z_2){\bar w_3},w_4)}}
\def\opre{{(Y({\bar w_2},z_2)Y(v_1,z_1){\bar w_3},w_4)}}
\def\opra{{(Y(Y(w_1,z_1-z_2){\bar w_2},z_2){\bar w_3},w_4)}}
\def\Dlt{{{z_1}^{\Delta(n_1,n_3)}{z_2}^{\Delta(n_2,n_3)}
{(z_1-z_2)}^{\Delta(n_1,n_2)}}}

\def\perm{{(Y(v_1,z_1)Y(v_2,z_2)v_3,v_4)}}
\def\prem{{(Y(v_2,z_2)Y(v_1,z_1)v_3,v_4)}}
\def\assoc{{(Y(Y(v_1,z_1-z_2)v_2,z_2)v_3,v_4)}}
\def\vsr{{x_r(-m_r) \dots x_1(-m_1)}}
\def\vst{{x_{r-1}(-m_{r-1}) \dots x_1(-m_1)}}
\def\vsq{{x_s(-m_s) \dots x_1(-m_1)}}
\def\Afr{\frac{1}{1+\delta_{1l}}}
\def\Afs{\frac{1+\delta_{il}}{1+\delta_{1l}}}
\def\Aft{\frac{1+\delta_{1i}}{1+\delta_{1l}}}
\def\Afw{\frac{2}{1+\delta_{1l}}}
\def\vva{\bold{vac}}
\def\vvb{a^+_1(-\shf)\vva}
\def\vvc{\bold{vac'}}
\def\vvd{a^-_l(0)\vvc}
\def\qqa{{\one}}
\def\qqb{a^+_1(-\shf)\qqa}
\def\qqc{{\pne}}
\def\qqd{a^-_l(0)\qqc}

\def\smj{\sum_{1 \leq j \leq l}}
\def\smk{\sum_{0 \leq P \in {\Bbb Z}}}
\def\smK{\sum_{0 \leq k \in {\Bbb Z}}}
\def\lmm{(\lambda^j_2-\delta_{1j}-\delta_{ij})}
\def\lmp{(\lambda^j_2-\delta_{1j}+\delta_{ij})}
\def\llmm{(\lambda^j_3-\delta_{1j}-\delta_{ij})}
\def\llmp{(\lambda^j_3-\delta_{1j}+\delta_{ij})}
\def\wwb{{{\bar w_2}}}
\def\wwc{{{\bar w_3}}}
\def\wwx{{{\bar w'_2}}}
\def\wwy{{{\bar w'_3}}}
\def\Bta{{\eta(n_1,n_2)}}
\def\bPhi{{{\Phi}}}
\def\bPsi{{{\Psi}}}
\def\bcV{{\bold{V(\shf{\Bbb Z})}}}

\def\bWM{\bold{WM}}

\def\BZ'{{\Bbb Z}+\shf}
\def\bvac'{\bold{vac'}}
\def\bvac{\bold{vac}}
\def\wt2{{wt({\bar w_2})}}

\def\ds{\displaystyle}

\def\sk1{\vskip 10pt}
\def\ni{\noindent}
\def\nl{\newline}

\def\ubr{\underbrace}

\let\pr\proclaim
\let\epr\endproclaim

\def\cl{\centerline}

\pageno=-1

\vbox to 0pt{}
\vskip 18pt

\cl {BOSONIC CONSTRUCTION OF VERTEX OPERATOR PARA-ALGEBRAS}

\cl {FROM SYMPLECTIC AFFINE KAC-MOODY ALGEBRAS}

\vskip 130pt

\cl {BY}
\sk1

\cl {MICHAEL DAVID WEINER}

\sk1

\cl {B.S., Rensselaer Polytechnic Institute, 1988}

\cl {M.A., State University of New York at Binghamton, 1990}
\vskip 195pt

\cl {DISSERTATION}
\sk1
\sk1

\cl {Submitted in partial fulfillment of the requirements for}
\cl {the degree of Doctor of Philosophy in Mathematics}
\cl {in The Graduate School of the}
\cl {State University of New York}
\cl {at Binghamton}
\cl {1994}

\newpage
\pageno = -2
\vbox to 4.2truein{}

\cl{\copyright\ \  Copyright by Michael David Weiner 1994}
\cl{All Rights Reserved}

\newpage








%
%
%
%
%
%
%
%
%
%
\pageno=-3

\cl {Table of Contents}
\sk1
\sk1
\sk1

\ni 1.  Introduction\dotfill 1
\sk1
\sk1
\ni 2. Bosonic Construction of Symplectic Affine Kac-Moody Algebras\dotfill 10

\sk1
\sk1
\ni 3. Bosonic Construction of Symplectic
Vertex Operator Algebras
and
Modules\dotfill 15
\sk1
\sk1

\ni 4. Bosonic Construction of Vertex Operator Para-Algebras\dotfill 40

\sk1

\newpage
\baselineskip=12pt

\def\bhsp{{\widehat{\bold{sp}}}}
\topmatter
\title\chapter{1} Introduction \endtitle \endtopmatter

\pageno = 1

The representation theory of affine Kac-Moody Lie algebras has grown
tremendously since their independent introduction by Robert V. Moody and
Victor G. Kac in 1968 \cite{M,K}. The most remarkable feature of the subject
has been its connectedness to other areas in mathematics and physics.
(See the introduction to \cite{FLM} and references therein.) The
Virasoro algebra is another infinite dimensional Lie algebra whose theory is
inextricably intertwined with the theory of affine Kac-Moody algebras
\cite{GO}. Inspired by mathematical structures found by theoretical physicists
\cite{BPZ}, and by the desire to understand the ``monstrous moonshine''
of the Monster group \cite{CN}, Borcherds \cite{B} introduced the concept of a
``vertex algebra'' into mathematics.
Frenkel, Lepowsky and Meurman \cite{FLM} further developed these ideas with
a slightly different axiom system, and succeeded in constructing the
moonshine module. A representation of the Virasoro algebra is a key point in
their definition of ``vertex operator algebras'' (VOA's).
In a very important class of examples, one also has a
representation of an affine algebra, so this new theory unifies and vastly
extends the representation theories of both kinds of infinite dimensional
Lie algebras.
With Borcherds axioms, one can even study representations of
non-affine Kac-Moody algebras with vertex algebra techniques. The
direction which this young field is currently taking is more towards the
study of modules for VOA's, and intertwining operators
between modules \cite{FHL}. Since the definitions of these structures are
rather complicated (details are given below), the actual
construction of specific examples can be a significant task. Several methods
of constructing VOA's and modules
are known \cite{F}. Among these are the bosonic and fermionic
constructions, based on representations of Weyl and Clifford algebras,
respectively. Explicit constructions of intertwining operators are difficult
to give in general, but their implicit definition is not so hard. The
problem is then to prove properties of these intertwiners. But the
main assumption about the intertwiners is their behavior with vertex
operators coming from the VOA. This assumption translates into reductive
formulas which can be used to give inductive proofs, using a grading on
the modules. Then properties of general intertwiners are based on properties
of intertwiners coming from the top graded piece of each module. In the
fermionic case studied in \cite{FFR} the top graded pieces are each finite
dimensional, but in the bosonic case studied here, some top graded pieces
are infinite dimensional. However, because of the action of the Cartan
subalgebra of the affine algebra, it is possible to reduce to a finite
number of cases.

Intertwining operators are of great importance because of their connection
with representations of braid groups and quantum groups.
Let $\bV$ be a VOA and let $\bW^i$, $0\leq i\leq n$, be $\bV$-modules with
$\bW^0 = \bV$.
Let $\cI(i,j,k) = \cI(W^i,W^j,W^k)$ be the vector space of intertwining
operators defined in detail below. The numbers $N_{ij}^k = dim(\cI(i,j,k))$
are called the fusion rules. A vertex operator para-algebra (VOPA)
includes a VOA, its modules, and its intertwiners in the special case when
the modules can be indexed by a finite abelian group and the fusion rules
are given by that group. In particular, $N_{ij}^k \leq 1$ if $i+j=k$, and
$N_{ij}^k = 0$ otherwise. The definition of VOPA was first given in
\cite{FFR}, but it is very similar to the concept of an abelian intertwining
algebra introduced in \cite{DL}. In \cite{FFR} a vertex
operator superalgebra and a twisted module for it were constructed from
the fermionic construction of
the four level 1 irreducible modules for the orthogonal affine
Kac-Moody Lie algebra of type $D_l^{(1)}$. But
their direct sum was made into a VOPA only in the case when the rank $l = 4$
using the classical triality of $D_4$. In this work, for any rank $l$,
a VOPA is constructed from the bosonic construction of the
four level -$\hf$ irreducible modules for the
symplectic affine Kac-Moody Lie algebra of type $C_l^{(1)}$ .
In \cite{FF} fermionic and bosonic constructions of these and other affine
algebras were given. In future work it should be possible to construct
VOA's, modules, intertwiners, and the unifying VOPA structure for all of
those classical affine algebras.

We begin with a brief review of the definitions of affine Kac-Moody Lie
algebras and of the Virasoro algebra.
Let $\bold g$ be a rank $l$ simple Lie algebra over the complex numbers, and
let $\bold h$ be a Cartan subalgebra of $\bold g$. Let $\langle\ ,\ \rangle$
be a nondegenerate symmetric invariant bilinear form on $\bold g$.
The homogeneously graded
affine Lie algebra ${\bold{\hat g}}$ associated with $\bold g$ is
defined to be
$${\bold{\hat g}} = {\bold g} \otimes {\Bbb C}[t,t^{-1}] \oplus {\Bbb C}c
\oplus {\Bbb C}d\eqno(1.1)$$
where $c$ is central and $d = t(d/dt)$.  For $x \in {\bold g}$, $n \in
{\Bbb Z}$, we write
$x(n) = x \otimes t^n$.  Then the brackets in ${\bold{\hat g}}$ are
$$\eqalign
{[x(m),y(n)] &= [x,y](m + n) + m \delta_{m,-n}\langle x,y\rangle c,\cr
[d,x(m)] &= m x(m).\cr}\eqno(1.2)$$
The value of $c$ is usually determined by normalizing the form
$\langle\ ,\ \rangle$ so that the induced form on the dual of the Cartan
subalgebra, $h^*$, gives $\langle \alpha,\alpha \rangle = 2$ for
$\alpha$ a long root.  In that case, the value of $c$ on an irreducible
${\bold{\hat g}}$-module is called the level of that module.
We identify ${\bold g} \subset {\bold{\hat g}}$ by
identifying $x$ with $x(0)$.  A Cartan subalgebra of ${\bold{\hat g}}$ is
$${\bold H} = {\bold h} \oplus {\Bbb C}c \oplus {\Bbb C}d,\eqno(1.3)$$
and the invariant form $\langle\ ,\ \rangle$ on $\bold g$ is extended
to an invariant form on ${\bold {\hat g}}$ by
$$\langle x(m) + r_1c + s_1d,\ y(n) + r_2c + s_2d\rangle = \langle
x,y\rangle\delta_{m,-n} + r_1s_2 + s_1r_2.\eqno(1.4)$$

The Virasoro algebra is the abstract Lie algebra $\bold{Vir}$ with basis
$\{ z,L_m \mid m \in {\Bbb Z} \}$ such that $z$ is central and
$$[L_m,L_n] = (m - n) L_{m + n} + \frac{1}{12} (m^3 - m) \delta_{m,-n} z.
\eqno(1.5)$$
We may form the semidirect product of $\bold{Vir}$ with the subalgebra of
${\bold{\hat g}}$,
$${\bold{\tilde g}} = {\bold g} \otimes {\Bbb C}[t,t^{-1}] \oplus {\Bbb C}c
\eqno(1.6)$$
where $c$ and $z$ are central and we have the brackets
$$[L_m,x(n)] = n x(m + n) . \eqno(1.7)$$
We identify $L_0 = d$ so that this semidirect product contains ${\bold{\hat
g}}$.

Following \cite{FFR}, we will now give the
definitions of vertex operator algebras (see also \cite{FLM} and
\cite{FHL}) and para-algebras.
Our definition only differs in two places, pointed out below.
These rather complex definitions can be motivated by a somewhat
nonstandard definition of a complex Lie algebra. Let
$\bV$ be
a vector space over the complex numbers $\Bbb C$, and let
$$ad : {\bV} \to \hbox{End}({\bV})\eqno(1.8)$$
be a linear transformation satisfying the Jacobi identity
$$ad(u)ad(v) - ad(v) ad(u) = ad(ad(u)v)\eqno(1.9)$$
for any $u,v \in {\bV}$. With the bracket $[u,v] = ad(u)v$,
$\bV$ becomes a Lie algebra
with trivial center if ker$(ad)$ is trivial. The Cauchy residue formula for a
rational function $f(z)$ with possible poles only at 0, $z_0$ and $\infty$
can be written as
$$-\hbox{Res}_{z=\infty}f(z) - \hbox{Res}_{z=0}f(z) =
\hbox{Res}_{z=z_0}f(z).\eqno(1.10)$$
The main axiom of vertex operator algebras combines these two formulas.

Let $\bV$ be a vector space and let ${\bV}[[z,z^{-1}]]$
denote the vector space of
formal Laurent series with coefficients in $\bV$. Let
$$ad_z : {\bV} \to \hbox{End}({\bV})[[z,z^{-1}]]\eqno(1.11)$$
be a linear transformation, which may be viewed
as a generating function
$$ad_z = \sum_{n\in{\Bbb Z}}ad_{(n)}z^{-n-1}\eqno(1.12)$$
for the linear transformations
$$ad_{(n)} : {\bV} \to \hbox{End}({\bV}),\eqno(1.13)$$
such that for $u,v \in {\bV}$,
we have $ad_{(n)}(u)v = 0$ if $n$ is sufficiently
large. For any $z_0 \in {\Bbb C}^*$ assume that $ad_z$
satisfies the {\it{Jacobi-Cauchy Identity}}
$$\eqalign
{&-\hbox{Res}_{z=\infty}(ad_z(u)ad_{z_0}(v)f(z)) -
\hbox{Res}_{z=0}(ad_{z_0}(v)ad_z(u)f(z))\cr
&= \hbox{Res}_{z=z_0}(ad_{z_0}(ad_{z-z_0}(u)v)f(z))\cr}\eqno(1.14)$$
for any $f(z) \in {\Bbb C}[z,z^{-1},(z-z_0)^{-1}]$.
Each residue in (1.14) is defined as the coefficient of the appropriate
singular term in the Laurent expansion around the singular point. In fact,
(1.14) is equivalent to infinitely many identities for the operators
$ad_{(n)}$. Taking $f(z) = z^m(z-z_0)^n$, $m,n\in{\Bbb Z}$, and equating
corresponding coefficients of $z_0$, one gets the identities
$$\eqalign
{&\sum_{0\leq i \in{\Bbb Z}} {r \choose i}
(-1)^i(ad_{(m+r-i)}(u)ad_{(n+i)}(v) - (-1)^rad_{(n+r-i)}(v)ad_{(m+i)}(u))\cr
&= \sum_{0\leq k \in {\Bbb Z}} {m \choose k}ad_{(m+n-k)}
(ad_{(r+k)}(u)v)\cr}\eqno(1.15)$$
for all $m,n,r\in {\Bbb Z}$. The finiteness of (1.15) when applied to any
particular vector in $\bV$ is guaranteed by our assumption that
$ad_{(n)}(u)v = 0$ for $n$ sufficiently large, so (1.14) is well-defined.
In all later chapters we use the notation
$Y(v,z)$ for $ad_z(v)$ and $\{v\}_n$ for $ad_{(n)}(v)$.
As before, we also require that
ker$(ad_z)$ be trivial. Additional axioms for vertex operator algebras
postulate the existence of two special vectors ${\pmb 1},\omega \in \bV$
such that
$$ad_z({\pmb 1}) = \hbox{Id}_{\bV},\eqno(1.16)$$
$$ad_z(\omega) = \sum_{n\in{\Bbb Z}} L(n) z^{-n-2}\eqno(1.17)$$
where $L(n) = ad_{(n+1)}(\omega)$,
$n \in {\Bbb Z}$, generate the Virasoro algebra
$$[L(m),L(n)] = (m-n) L(m+n) + \frac{1}{12} (m^3 - m)\delta_{m,-n}c
\hbox{ Id}_{\bV},\eqno(1.18)$$
the scalar $c$ is called the {\it rank} of $\bV$, and
$$[L(-1),ad_z(v)] = (d/dz)ad_z(v).\eqno(1.19)$$
Note that (1.15), with $m = 0$, $r = 0$ and $u = \omega$, gives
$$[L(-1),ad_z(v)] = ad_z(L(-1)v).\eqno(1.20)$$
Our final axioms require that $L(0)$ defines a $\shf\Bbb Z$-grading of $\bV$,
rather than the ${\Bbb Z}$-grading imposed in \cite{FFR},
$${\bV} = {\underset {n\in\shf{\Bbb Z}} \to {\bold \coprod}}
({\bV})_n\eqno(1.21)$$
where
$$({\bV})_n =\{v \in {\bV}\mid L(0)v = nv\}\eqno(1.22)$$
and
$({\bV})_n = \{0\}$ for $n$ sufficiently small.
We define a vertex operator algebra to be a quadruple
$(\bV,ad_z,{\pmb 1},\omega)$
satisfying the above axioms, but for brevity we sometimes just refer to
$\bV$ itself as a vertex operator algebra.
We define the {\it weight} of a homogeneous vector by the rule
$$wt(v) = n \ \hbox{ if }\ v \in ({\bV})_n,\eqno(1.23)$$
and note that (1.15), with $m = 0$, $r = 1$ and $u = \omega$, and (1.19) imply
$$wt(ad_{(n)}(v)w) = wt(v) + wt(w) - n - 1.\eqno(1.24)$$

Note that with $r = 0$ in (1.15) we get the commutator formula
$$[ad_{(m)}(u),ad_{(n)}(v)] = \sum_{0\leq k \in {\Bbb Z}} {m \choose k}
ad_{(m+n-k)}(ad_{(k)}(u)v)\eqno(1.25)$$
for all $m,n \in {\Bbb Z}$,
in which the summation is actually finite.

The definition of a module for a vertex operator algebra is analogous
to the definition of a Lie algebra module. Let $\bV$ be a Lie algebra with
bracket $[u,v] = ad(u)v$ as in (1.8)-(1.9), and let $\bW$
be a vector space. Then
$({\bW},\pi)$ is a $\bold V$-module if
$$\pi : {\bold V} \to \hbox{End}({\bW})\eqno(1.26)$$
is a linear transformation satisfying
$$\pi(u)\pi(v) - \pi(v)\pi(u) = \pi(ad(u)v)\eqno(1.27)$$
for any $u,v \in {\bold V}$. We say $\pi$ is {\it faithful} if ker$(\pi)$
is trivial. To define a vertex
operator algebra module we combine (1.27) with the Cauchy identity (1.10).
Let
$$\pi_z : {\bold V} \to \hbox{End}({\bW})[[z,z^{-1}]]\eqno(1.28)$$
be a linear transformation,
which may be viewed as a generating function
$$\pi_z = \sum_{n\in{\Bbb Z}}\pi_{(n)} z^{-n-1}\eqno(1.29)$$
for the linear transformations
$$\pi_{(n)} : {\bold V} \to \hbox{End}({\bW}),\eqno(1.30)$$
such that for $v \in {\bold V}$, $w \in {\bW}$,
we have $\pi_{(n)}(v)w = 0$ for $n$ sufficiently
large. For any $z_0 \in {\Bbb C}^*$ assume that $\pi_z$
satisfies the {\it {Jacobi-Cauchy Identity}}
$$\eqalign
{&-\hbox{Res}_{z=\infty}(\pi_z(u)\pi_{z_0}(v)f(z)) -
\hbox{Res}_{z=0}(\pi_{z_0}(v)\pi_z(u)f(z))\cr
&= \hbox{Res}_{z=z_0}(\pi_{z_0}(ad_{z-z_0}(u)v)f(z))\cr}\eqno(1.31)$$
for any $f(z) \in {\Bbb C}[z,z^{-1},(z-z_0)^{-1}]$.
As before, (1.31) is equivalent to the identities
$$\eqalign
{&\sum_{0\leq i \in {\Bbb Z}} {r\choose i}
(-1)^i(\pi_{(m+r-i)}(u)\pi_{(n+i)}(v) -
(-1)^r\pi_{(n+r-i)}(v)\pi_{(m+i)}(u))\cr
&= \sum_{0\leq k \in {\Bbb Z}} {m \choose k}
\pi_{(m+n-k)}(ad_{(r+k)}(u)v)\cr}\eqno(1.32)$$
for all $m,n,r\in {\Bbb Z}$, which imply
$$[\pi_{(m)}(u),\pi_{(n)}(v)] = \sum_{0\leq k \in {\Bbb Z}} {m \choose k}
\pi_{(m+n-k)}(ad_{(k)}(u)v).\eqno(1.33)$$
As additional axioms we require the analogs of (1.16)-(1.19) with $ad_z$
replaced by $\pi_z$ and Id$_{\bold V}$ replaced by Id$_{\bW}$.
The analog of (1.20) also follows.
The only new feature is that $\bW$ is
not necessarily ${\Bbb Z}$-graded, but it is ${\Bbb Q}$-graded.
In fact, if $\bW$ is irreducible
then it is graded by ${\Bbb Z}+\Delta_{\bW}$ where $\Delta_{\bW} \in
{\Bbb Q}$, defined mod $\Bbb Z$, is an important
characteristic of the module.
Note that the graded pieces of $\bW$ are not necessarily finite
dimensional.
Define a $\bold V$-module to be a pair
$({\bW},\pi_z)$
satisfying these axioms.

In constructing VOA's and their modules the main task is
the verification of the Jacobi Identity (1.14) and (1.31). To do this we
use correlation functions which can be defined from
a nondegenerate Hermitian form on $\bold V$ and $\bW$. It is
sufficient to establish three analytic properties, {\it rationality},
{\it permutability} and
{\it associativity}, of the correlation
functions made from products and compositions
of the operators $ad_z$ and $\pi_z$.
If $(\cdot\ ,\ \cdot)$ denotes a suitable Hermitian form on $\bold V$, then
by rationality we mean that for $u,v,w,w' \in {\bold V}$ the series
$(ad_z(u)ad_{z_0}(v)w,w')$ converges absolutely
in the domain $|z| > |z_0| > 0$ to a rational function
$f(z,z_0)$ which may only have poles at $z = 0$, $z_0 = 0$ or $z = z_0$.
By permutability we mean that the rational function, to which the series
$(ad_{z_0}(v)ad_z(u)w,w')$ converges absolutely
in the domain $|z_0| > |z| > 0$, is the same function $f(z,z_0)$.
By associativity we mean that the series
$(ad_{z_0}(ad_{z-z_0}(u)v)w,w')$ converges absolutely
in the domain $|z_0| > |z - z_0| > 0$ to the same function $f(z,z_0)$.
Analogous definitions of these properties can be made on a module $\bW$.
In fact, the Jacobi Identity implies these three properties.
See \cite{FHL} for further details on equivalent
axiomatic formulations of VOA's and modules.

In the theory of affine Lie algebras one also considers their twisted
constructions \cite{KP,L}. Analogous twisted constructions also apply to vertex
operator algebras \cite{FLM2} (see also Chapter 9 of \cite{FLM}, and
\cite{DL}). In this work we will only need a ${\Bbb Z}_2$-twisted module
for a VOA, so we just give that definition.
Let $({\bV},ad_z,{\pmb 1},\omega)$ be a VOA
and let $\vartheta$ be a vector space involution of $\bV$ such that
$$\vartheta ad_z(u)v = ad_z(\vartheta u)\vartheta v,\ \ \ \vartheta {\pmb
1} = {\pmb 1},\ \ \ \vartheta \omega = \omega.\eqno(1.34)$$
Let
$${\bV} = {\bV}^0 \oplus {\bV}^1 \eqno(1.35)$$
be the decomposition into the eigenspaces of $\vartheta$, where ${\bV}^j$
corresponds to
the eigenvalue $(-1)^j$. We say that $({\bW},\pi^\vartheta_z)$
is a $\vartheta$-twisted $\bV$-module (or a
${\bV}^\vartheta$-module) if
$$\pi^\vartheta_z : {\bV} \to
\hbox{End}({\bW})[[z^{1/2},z^{-1/2}]]\eqno(1.36)$$
is a linear transformation satisfying the axioms for a $\bV$-module modified
as follows. For $v \in {\bV}^j$, we have
$$\pi^\vartheta_z(v) = \sum_{n\in{\Bbb Z}+j/2}
ad_{(n)}(v)z^{-n-1}.\eqno(1.37)$$
On $\bW$ we have the twisted Jacobi-Cauchy Identity
$$\eqalign
{&-\hbox{Res}_{z=\infty} (\pi^\vartheta_z(u)\pi^\vartheta_{z_0}(v)f(z)) -
\hbox{Res}_{z=0}(\pi^\vartheta_{z_0}(v)\pi^\vartheta_z(u)f(z))\cr
&=
\hbox{Res}_{z=z_0}(\pi^\vartheta_{z_0}(ad_{z-z_0}(u)v)f(z))\cr}\eqno(1.38)$$
for $u \in {\bV}^i$, $v \in {\bV}^j$, $f(z) \in
z^{i/2}z_0^{j/2}{\Bbb C}[z,z^{-1},(z-z_0)^{-1}]$, which is
equivalent to the identities (1.32) with $r \in {\Bbb Z}$, $m \in {\Bbb
Z}+i/2$, $n \in {\Bbb Z}+j/2$.
$\bW$ is $\Bbb Q$-graded
by $L(0)$, and if it is irreducible then it is graded by $(1/2){\Bbb
Z}+\Delta_{\bW}$. Note that
the twisted Jacobi Identity with $r = 0$ implies (1.33) with $m \in {\Bbb
Z}+i/2$ and
$n \in {\Bbb Z}+j/2$ if $u \in {\bV}^i$ and $v \in {\bV}^j$.

We will now define intertwining operators between modules for a VOA.
For any vector space $\bW$, let $\bW\{z\}$ be the space of
$\bW$-valued formal series involving only rational powers of $z$.
For $1\leq i\leq 3$, let $(\bW^i,\pi^i_z)$ be
irreducible modules for a VOA $\bV$ and let $w_i\in \bW^i$. Let
$$\cI_z : \bW^1 \to \hbox{Hom}(\bW^2,\bW^3)\{z\}\eqno(1.39)$$
be a linear transformation, which may be viewed as a generating function
$$\cI_z = \sum_{n\in {\Bbb Q}} \cI_{(n)} z^{-n-1}\eqno(1.40)$$
for the linear transformations $\cI_{(n)} : \bW^1 \to \hbox{Hom}(\bW^2,\bW^3)$,
such that $\cI_{(n)}(w_1)w_2$ $= 0$ for n sufficiently large. For any $v\in\bV$
suppose that $\cI_z$ obeys the following version of the Jacobi-Cauchy
identity,
$$\eqalign
{&-\hbox{Res}_{z=\infty}(\pi^3_z(v)\cI_{z_0}(w_1)w_2 f(z)) -
\hbox{Res}_{z=0}(\cI_{z_0}(w_1)\pi^2_z(v)w_2 f(z))\cr
&= \hbox{Res}_{z=z_0}(\cI_{z_0}(\pi^1_{z-z_0}(v)w_1)w_2 f(z))\cr}\eqno(1.41)$$
for any $f(z) \in z_0^\alpha {\Bbb C}[z,z^{-1},(z-z_0)^{-1}]$ for
$\alpha$ an appropriate rational number. It is also assumed that
$$[L(-1),\cI_z(w_1)] = \frac{d}{dz} \cI_z(w_1) = \cI_z(L(-1)w_1).\eqno(1.42)$$
If it exists, such an operator $\cI_z$ is called an intertwining operator
of type $(\bW^1,\bW^2,$ $\bW^3)$, and the space of all such operators is
denoted by $\cI(\bW^1,\bW^2,$ $\bW^3)$. Note that
this space of intertwiners can have dimension greater than 1, and these
dimensions for all possible modules are called the fusion rules for $\bV$.
If $\bW$ is any $\bV$-module one also assumes that an intertwiner $\cI_z$ of
type $(\bW,\bV,\bW)$ satisfies the {\it creation property}
$$\cI_z(w){\pmb 1}\in \bW[[z]] \qquad \hbox{and}\qquad
\lim_{z\to 0} \cI_z(w){\pmb 1} = w \eqno(1.43)$$
for all $w\in \bW$.

The definition of a vertex operator algebra has a
further generalization when $\bV$ is graded by a finite abelian
group $\Gamma$.
Suppose that $\Gamma$ has exponent $g$. For each $\gamma \in \Gamma$,
suppose $\Delta_\gamma \in {\Bbb Q}$ and, for $\gamma_1,\gamma_2 \in
\Gamma$, let
$$\Delta(\gamma_1,\gamma_2) = \Delta_{\gamma_1} + \Delta_{\gamma_2}
- \Delta_{\gamma_1 + \gamma_2} \in (1/g){\Bbb Z}\eqno(1.44)$$
be bilinear $mod\  {\Bbb Z}$. Assume that
$${\bV} = \bigoplus_{\gamma\in\Gamma} {\bV}_\gamma\eqno(1.45)$$
is a vector space and
$$ad_z : {\bV} \to \hbox{End}(\bV)[[z^{1/g},z^{-1/g}]]\eqno(1.46)$$
is a linear transformation with trivial kernel such that for $u \in
\bV_{\gamma_1}$, and $v \in \bV_{\gamma_2}$, we have
$$ad_z(u)v = \sum_{n\in{\Bbb Z}+\Delta(\gamma_1,\gamma_2)}
ad_{(n)}(u)vz^{-n-1}\eqno(1.47)$$
and
$$ad_{(n)}(u)v \in {\bV}_{\gamma_1+\gamma_2}\eqno(1.48)$$
is zero for $n$ sufficiently large.
Suppose that there are distinguished vectors ${\pmb 1},\omega \in \bV_0$
satisfying the usual
axioms (1.16) - (1.19), that $L(0)$ defines a $\Bbb Q$-grading on $\bV$
such that ${\bV}_\gamma$ is graded by
${\Bbb Z}+\Delta_\gamma$,
and that
$({\bV}_\gamma)_n = \{0\}$
for $n$ sufficiently small.
Note that our definition differs from that in \cite{FFR} where each
graded piece $(\bV_{\gamma})_n$ is assumed to be finite dimensional.
The main axiom is the Jacobi-Cauchy Identity
$$\eqalign
{&-\hbox{Res}_{z=\infty}(ad_z(u)ad_{z_0}(v)wf(z,z_0)) -
 \eta\  \hbox{Res}_{z=0}(ad_{z_0}(v)ad_z(u)w f(z,z_0))\cr
&= \hbox{Res}_{z=z_0}(ad_{z_0}(ad_{z-z_0}(u)v)wf(z,z_0))\cr}\eqno(1.49)$$
for $u \in {\bV}_{\gamma_1}$, $v \in {\bV}_{\gamma_2}$, $w \in
{\bV}_{\gamma_3}$,
$$f(z,z_0) \in
z^{\Delta(\gamma_1,\gamma_3)}\ z_0^{\Delta(\gamma_2,\gamma_3)}\ (z-z_0)^{\Delta
(\gamma_1,\gamma_2)}
\ {\Bbb C}[z,z^{-1},z_0,z_0^{-1},(z-z_0)^{-1}]\eqno(1.50)$$
where
$\eta = \eta (\gamma_1,\gamma_2) \in {\Bbb C}^*$
is a bilinear function on $\Gamma \times \Gamma$.
Bilinearity implies that $\eta(\gamma_1,\gamma_2)$ is a root of
unity and that $\eta$ is a 2-cocycle representing a cohomology class in
$H^2(\Gamma,{\Bbb Z}_g)$. This class is restricted by the further assumption
that $\eta(\gamma_1,\gamma_2)\ e^{\pi i \Delta(\gamma_1,\gamma_2)}$
is skew-symmetric, that is,
$\eta(\gamma_1,\gamma_2)\ e^{\pi i \Delta(\gamma_1,\gamma_2)} =
\eta(\gamma_2,\gamma_1)^{-1}\ e^{-\pi i \Delta(\gamma_2,\gamma_1)}$.
We must be
precise about how the three residues in (1.49) are to be computed. Let
$p \in {\Bbb Z}+\Delta(\gamma_1,\gamma_2)$ and
$m \in {\Bbb Z}+\Delta(\gamma_1,\gamma_3)$.
In the first term of (1.49) we expand $(z - z_0)^p$ as
$$z^p \ (1 - z_0/z)^p = z^p \ \sum_{0\leq k \in {\Bbb Z}} {p \choose k}
(-1)^k (z_0/z)^k\eqno(1.51)$$
and find $-\hbox{Res}_{z=\infty}$ to be the coefficient of $z^{-1}$.
In the second term we expand $(z - z_0)^p$ as
$$e^{\pi i p} z_0^p \ (1 - z/z_0)^p = e^{\pi i p} z_0^p \
\sum_{0\leq k \in {\Bbb Z}} {p \choose k}
(-1)^k (z/z_0)^k\eqno(1.52)$$
and find $\hbox{Res}_{z=0}$ to be the coefficient of $z^{-1}$.
In the third term we let $\zeta = z - z_0$ and expand
$z^m = (z_0 + \zeta)^m$ as
$$z_0^m \ (1 + \zeta/z_0)^m =
z_0^m \ \sum_{0\leq k \in {\Bbb Z}} {m \choose k}
(\zeta/z_0)^k\eqno(1.53)$$
and find $\hbox{Res}_{z=z_0}$ to be the coefficient of $\zeta^{-1}$.
Note that the Jacobi-Cauchy Identity is equivalent to
the identities
$$\eqalign
{&\sum_{0\leq k \in {\Bbb Z}} {p \choose k} (-1)^k
(ad_{(m+p-k)}(u)ad_{(n+k)}(v) \cr
&\qquad - \eta(\gamma_1,\gamma_2) e^{\pi i p} ad_{(n+p-k)}(v)ad_{(m+k)}(u))w\cr
&= \sum_{0\leq k \in {\Bbb Z}} {m \choose k}
ad_{(m+n-k)}(ad_{(p+k)}(u)v)w\cr}\eqno(1.54)$$
for all $m \in {\Bbb Z}+\Delta(\gamma_1,\gamma_3)$, $n \in {\Bbb
Z}+\Delta(\gamma_2,\gamma_3)$, $p \in {\Bbb
Z}+\Delta(\gamma_1,\gamma_2)$. Under
these assumptions we say that $({\bold V},ad_z,{\pmb
1},\omega,\Gamma,\Delta,\eta)$ is a vertex operator para-algebra.

A vertex operator para-algebra is an algebraic structure including and
unifying a VOA, its modules, and its intertwiners in the special case when
the modules can be indexed by a finite abelian group and the fusion rules
are given by that group. In particular,
the
space of intertwiners of type $(V_{\gamma_1},V_{\gamma_2},V_{\gamma_3})$
is at most one-dimensional
if
$\gamma_1 + \gamma_2 = \gamma_3$ and zero otherwise.

The goal of this work is the construction of a vertex operator para-algebra
structure on the direct sum of the four level -$\hf$ irreducible
$\bhsp(2l)$-modules which are given by the bosonic construction of the type
$C_l^{(1)}$ affine Kac-Moody Lie algebra.

\noindent Finally we give a brief summary of the following chapters.

We begin Chapter II with the construction of four irreducible representations
of the symplectic Lie algebra {\bf sp}$(2l)$, the adjoint, the natural and
two oscillator representations.  These are constructed from the associative
Weyl algebra generated by a $2l$-dimensional complex vector space ${\bold A}$
equiped with a nondegenerate anti-symmetric bilinear form, and a polarization
of ${\bold A}$.  Our abstract description of the infinite dimensional
oscillator
representation can be given more concretely as multiplication and
differentiation
operators on the space of polynomials in $l$ commuting variables.  The main
goal of Chapter II is the construction of four level $-\hf$ irreducible
representations of the symplectic affine Kac-Moody Lie algebra $\bhsp(2l)$.
Explicit operators representing $\bhsp(2l)$ and the Virasoro algebra are given
in terms of two affinized Weyl algebras, acting on bosonic Fock spaces equiped
with a positive Hermitian form.
These two Fock spaces, called the Neveu-Schwarz (${\hbox {NS}}$) and Ramond
($R$) sectors
in the physics literature, are graded by the eigenvalues of the $L(0)$
Virasoro operator. If $L(0)v = nv$ then we write $wt(v) = n$ and say $v$ is
a vector of weight $n$. Each sector decomposes into two irreducible highest
weight $\bhsp(2l)$-modules, ${\hbox {NS}} = {\hbox {NS}}_0 \oplus {\hbox
{NS}}_1$ and $R = R_0 \oplus
R_1$, and the subspaces of minimal weight in these four modules have weights
$0$, $\hf$, $-\sei$ and $-\sei$, respectively. These ``top''  weight spaces are
{\bf sp}$(2l)$-modules, the trivial, the natural, and the two oscillator
modules. In fact, the adjoint module is just below the trivial
module in ${\hbox {NS}}_0$, consisting of the vectors of weight $1$.

In Chapter III we prove that ${\hbox {NS}}$ is a VOA and $R$ is a ${\Bbb
Z}_2$-twisted
${\hbox {NS}}$-module. The techniques for construction of the vertex operators
and for
the proofs of the theorems in this chapter, are taken from those in
\cite{FFR} with one fundamental modification, fermions are changed to
bosons. This means that in changing their underlying Clifford algebra to a
Weyl algebra, signs in various formulas had to be carefully modified. One
effect of this is that our ${\hbox {NS}}$ sector is a VOA while their ${\hbox
{NS}}$ sector is a
vertex operator superalgebra (a special case of VOPA with
$\Gamma = {\Bbb Z}_2$). It is built in to the construction that
for $v$ of weight $1$ in ${\hbox {NS}}_0$, the components of
the vertex operators $Y(v,z)$ represent $\bhsp(2l)$ on ${\hbox {NS}}$ and $R$.
Furthermore, there is a distinguished vector $\omega \in {\hbox {NS}}_0$ of
weight $2$
such that the components of $Y(\omega,z)$ represent the Virasoro algebra.

In Chapter IV we show that $V = {\hbox {NS}} \oplus R$ has the structure of a
VOPA
with group $\Gamma = {\Bbb Z}_4$. In fact, although it appears that
$\Gamma = {\Bbb Z}_2$ is more natural, we show that this choice of $\Gamma$
does not give a VOPA. To be precise about the $\Gamma$-grading of $V$,
let $\Gamma = \{0,1,2,3\}$ under addition mod 4, and let
$$V_0 = {\hbox {NS}}_0, \qquad V_2 = {\hbox {NS}}_1, \qquad V_1 = R_0, \qquad
V_3 = R_1.\eqno(1.55)$$
For $i\in\Gamma$, $\Delta_i$ is the minimal weight in $V_i$.
The main point is the proof of the Jacobi-Cauchy identity for intertwining
operators. The technique used to prove the Jacobi-Cauchy identity involves the
study of the three analytic properties of correlation functions, rationality,
permutability and associativity. In fact, one has to multiply the
correlation functions by the algebraic function
$$z_1^{\Delta(\gamma_1,\gamma_3)}\ z_2^{\Delta(\gamma_2,\gamma_3)}\
(z_1 - z_2)^{\Delta(\gamma_1,\gamma_2)}.\eqno(1.56)$$
The proof of these properties for general vectors in $V$ is reduced to
the case when the vectors $v_i\in V_{\gamma_i}$ are each of weight $\Delta_i$.
Note that the spaces of vectors of minimal weight in $V_1$ and $V_3$
are infinite dimensional.
But restrictions on the correlation functions coming from the Cartan
subalgebra of {\bf sp}$(2l)$ allow us to reduce to
a finite number of cases.
There are two equivalent methods of studying the correlation functions in
these cases. In one method, used in \cite{FFR}, recursive relations are
found among the coefficients of the correlation functions written as series.
In the other method, differential equations for the correlation functions
determine them. These equations are known in the physics literature as the
Knizhnik-Zamolodchikov (KZ) equations \cite{KZ}. Both methods are based on the
fact
that the Virasoro operators can be expressed in terms of certain
$\bhsp(2l)$ operators (the Sugawara construction). From the point of view of
the VOA ${\hbox {NS}}_0$, this fact comes from two expressions for $\omega$,
one in
terms of the fundamental bosons in the Weyl algebra, and the other in terms
of symplectic operators since ${\hbox {NS}}_0$ is an irreducible
$\bhsp(2l)$-module.
In the fermionic case studied in \cite{FFR}, $\omega$ has two
expressions, one in terms of the fundamental fermions in the Clifford
algebra, and the other in terms of operators from the orthogonal affine
algebra ${\bold {\widehat {so}}}(2l)$.  But, in fact, in the second expression,
only operators from the Heisenberg algebra are necessary, giving a
considerable simplification.  That simplification does not occur in our
symplectic case, making our work a bit harder.

A surprising point is that we found that the fusion rules $N_{ij}^{i+j} = 0$
for $i,j\in\{1,3\}$. Because of that, when the rank $l$ is even, the group
$\Gamma$ could be either ${\Bbb Z}_4$ or
${\Bbb Z}_2 \times {\Bbb Z}_2$. When $l$ is odd, (1.44) forces $\Gamma$ to
be ${\Bbb Z}_4$. When $l$ is even, it remains to be seen which of these
two groups is more fitting. If this VOPA can be embedded in a larger one,
then perhaps that will determine $\Gamma$.

\newpage
\def\eps{\epsilon}
\topmatter
\title\chapter{2} Bosonic Construction of Symplectic Affine
Kac-Moody Algebras\endtitle \endtopmatter

Begin with a nondegenerate anti-symmetric bilinear form
$\langle\ ,\ \rangle$ on a vector space ${\bold A} \cong {\Bbb
C}^{2l}$ such that ${\bold A} = {\bold A}^+ \oplus {\bold A}^-$ is a
polarization into maximally isotropic subspaces ${\bold A}^\pm \cong
{\Bbb C}^l$. Define an associative algebra {\bf Weyl} with unit element 1,
generators from $\bold A$ and relations
$$ab - ba = \langle a,b\rangle 1 \ \hbox{ for }\ a,b \in {\bold
A}.\eqno(2.1)$$
Letting
$$:ab:\ = {\tsize{\frac{1}{2}}}(ab + ba)\ \hbox{ for
}\ a,b \in {\bold A}, \eqno(2.2)$$
the span of all such elements in the Weyl algebra {\bf Weyl} is closed
under brackets, (see (2.9)),
providing the adjoint representation of {\bf sp}$(2l)$.  Three
more irreducible representations of {\bf sp}$(2l)$ are constructed as follows.
On $\bold A$, {\bf sp}$(2l)$ acts by brackets
$$[:ab:\ ,c] = \langle b,c\rangle a - \langle
a,c\rangle b,\eqno(2.3)$$
providing the natural representation.
Let ${\goth I}$ be the left ideal in {\bf Weyl} generated by ${\bold
A}^+$.  Then
$$\bold{WM} = \hbox{\bf Weyl}/{\goth I}\eqno(2.4)$$
is an irreducible left {\bf Weyl}-module having a ``vacuum vector''
$$\bold{vac} = 1 + {\goth I}\eqno(2.5)$$
satisfying ${\bold A}^+ \cdot \bold{vac} = 0$.
The subalgebra of {\bf Weyl} generated by
1 and ${\bold A}^-$ is the symmetric algebra of polynomials in the
commuting generators $a \in {\bold A}^-$, denoted by
$S({\bold A}^-)$, and
$$\bold{WM} = S({\bold A}^-) \cdot \bold{vac}.\eqno(2.6)$$
We have the decomposition
$$\bold{WM} = \bold{WM}^0 \oplus \bold{WM}^1\eqno(2.7)$$
into the even and odd parity subspaces,
$$\bold{WM}^0 = (S({\bold A}^-)^{even}) \cdot \bold
{vac},\ \bold{WM}^1 = (S({\bold A}^-)^{odd}) \cdot
\bold{vac}.\eqno(2.8)$$

If $x =:r_1r_2:$\ , $y =:
s_1s_2:\ \in ${\bf sp}$(2l)$ then
$$[x,y] = \langle r_1,s_2\rangle:r_2s_1:+ \langle
r_2,s_1\rangle:r_1s_2:+ \langle r_1,s_1\rangle
:r_2s_2:+ \langle r_2,s_2\rangle:
r_1s_1:\eqno(2.9)$$
and a nondegenerate invariant symmetric bilinear form on {\bf sp}$(2l)$ is
$$\langle x,y\rangle =
-\left[ \langle r_1,s_1\rangle\langle r_2,s_2\rangle
 + \langle r_1,s_2\rangle \langle r_2,s_1\rangle \right]
.\eqno(2.10)$$
Let $\{a^+_1,\hdots,a^+_l\}$ be a basis of ${\bold A}^+$ and let
$\{a^-_1,\hdots,a^-_l\}$ be a basis of ${\bold A}^-$ such that $\langle
a^+_i,a^-_j\rangle = \delta_{ij}$.
We call $\{a^+_1,\hdots,a^+_l,a^-_1,\hdots,a^-_l\}$ a ``canonical'' basis of
$\bold A$.
Then $\{ h_i =  -:
a^+_ia^-_i:\ \mid 1 \leq i \leq l\}$ is an orthonormal basis of
a Cartan subalgebra ${\bold h}$ of {\bf sp}$(2l)$.
Let $\{ \eps_1 \cdots \eps_l \}$ be a dual basis of the dual space
${\bold h}^*$, so that $\eps_i(h_j)=\delta_{ij}$.  Then, under the
form induced on ${\bold h}^*$, $\langle \eps_i,\eps_j \rangle = \delta_{ij}$.
The root system of {\bf sp}$(2l)$ is then
$$\{ \pm \eps_i \pm \eps_j , \pm 2 \eps_k | 1 \le i < j \le l, 1 \le k \le
l\}$$
and we choose simple roots
$$ \alpha_1=\eps_1-\eps_2, \alpha_2=\eps_2-\eps_3, \cdots,
\alpha_{l-1}=\eps_{l-1}
-\eps_l, \alpha_l=2\eps_l.$$
Note that the long simple root $\alpha_l$ has squared length $\langle
\alpha_l, \alpha_l \rangle = 4$, rather than the usual normalization of squared
length 2.

The bosonic construction of ${\bold{\hat g}}$ for $\bold g$ of type
$C_l$ begins as above with a nondegenerate symmetric bilinear form
$\langle\ ,\ \rangle$ on a vector space ${\bold A} \cong {\Bbb
C}^{2l}$ such that ${\bold A} = {\bold A}^+ \oplus {\bold A}^-$ is a
polarization into maximally isotropic subspaces ${\bold A}^\pm \cong
{\Bbb C}^l$.  Let
$${\bold A}({\Bbb Z}) = {\bold A} \otimes {\Bbb C}[t,t^{-1}]\ \ \hbox{ and
}\ \ {\bold A}({{\Bbb Z}+{\tsize{\frac{1}{2}}}}) = {\bold A}\otimes
t^{1/2}{\Bbb
C}[t,t^{-1}].\eqno(2.11)$$
For $Z = {\Bbb Z}$ or ${\Bbb Z} + {\tsize{\frac{1}{2}}}$ denote
$a \otimes t^n \in {\bold A}(Z)$ by $a(n)$, and let
$$\iota = \cases +1 &\text{if $Z = {\Bbb Z}$}\cr\\
-1 &\text{if $Z = {\Bbb Z}+{\tsize{\frac{1}{2}}}$.}\endcases \eqno(2.12)$$
Extend the form to ${\bold A}(Z)$ by setting
$$\langle a(m),b(n)\rangle = \langle a,b\rangle\delta_{m,-n}.\eqno(2.13)$$
Then ${\bold A}(Z) = {\bold A}(Z)^+ \oplus {\bold A}(Z)^-$ is a
polarization with respect to the extended form, where
$$\eqalign
{{\bold A}({\Bbb Z})^\pm &= {\bold A}^\pm \oplus ({\bold A} \otimes t^{\pm
1}{\Bbb C}[t^{\pm 1}]),\cr
{\bold A}({{\Bbb Z}+{\tsize{\frac{1}{2}}}})^\pm &= {\bold A} \otimes
t^{\pm 1/2} {\Bbb C}[t^{\pm 1}].\cr}\eqno(2.14)$$

Define an associative algebra {\bf Weyl}$(Z)$ with unit element 1,
generators from ${\bold A}(Z)$ and relations
$$a(m)b(n) - b(n)a(m) = \langle a(m),b(n)\rangle 1 = \langle
a,b\rangle\delta_{m,-n}1.\eqno(2.15)$$
Let ${\goth I}(Z)$ be the left ideal in {\bf Weyl}$(Z)$ generated by ${\bold
A}(Z)^+$.  Then
$$\bold{WM}(Z) = \hbox{\bf Weyl}(Z)/{\goth I}(Z)\eqno(2.16)$$
is an irreducible left {\bf Weyl}$(Z)$-module having a ``vacuum vector''
$$\bold{vac}(Z) = 1 + {\goth I}(Z)\eqno(2.17)$$
satisfying ${\bold A}(Z)^+ \cdot \bold{vac}(Z) = 0$.
The subalgebra of {\bf Weyl}$(Z)$ generated by
1 and ${\bold A}(Z)^-$ is the symmetric algebra of polynomials in the
commuting generators $a(m) \in {\bold A}(Z)^-$, denoted by
$S({\bold A}(Z)^-)$, and
$$\bold{WM}(Z) = S({\bold A}(Z)^-) \cdot \bold{vac}(Z).\eqno(2.18)$$
Therefore, {\bf WM}$(Z)$ is spanned by vectors of the form
$$a_1(-m_1) \hdots a_r(-m_r) \bold{vac}(Z) \eqno(2.19)$$
for $a_i(-m_i) \in {\bold A}(Z)^-$, $1 \leq i \leq r.$
We have the decomposition
$$\bold{WM}(Z) = \bold{WM}(Z)^0 \oplus \bold{WM}(Z)^1\eqno(2.20)$$
into the even and odd parity subspaces,
$$\bold{WM}(Z)^0 = (S({\bold A}(Z)^-)^{even}) \cdot \bold
{vac}(Z),\ \bold{WM}(Z)^1 = (S({\bold A}(Z)^-)^{odd}) \cdot
\bold{vac}(Z).\eqno(2.21)$$

Let $\zeta$ be a formal complex variable and for $a \in {\bold A}$ define
the generating function
$$a(\zeta) = \sum_{n \in Z} a(n)\zeta^{-n}.\eqno(2.22)$$
For $a(m),b(n) \in {\bold A}(Z)$ define the ``bosonic normal ordering''
$$:a(m)b(n):\ = \cases a(m)b(n) &\text{if $n
> m$}\cr\\ {\tsize{\frac{1}{2}}}(a(m)b(n) + b(n)a(m)) &\text{if $n = m$}\cr\\
b(n)a(m) &\text{if $n < m$.}\endcases \eqno(2.23)$$
Then
$$:a(\zeta)b(\zeta):\ = \sum_{k \in {\Bbb
Z}} \left(\sum_{n\in Z}:a(n)b(k - n):
\right)\zeta^{-k} \eqno(2.24)$$
is the generating function defining the homogeneous components
$$:a(\zeta)b(\zeta):_k = \sum_{n\in Z}
:a(n)b(k - n):\ \hbox{ for }\ k \in {\Bbb
Z}.\eqno(2.25)$$
These are well-defined operators on $\bold{WM}(Z)$.
For $1 \leq i \leq l$ define
the generating functions
$$L^Z_i(\zeta) = -\frac{1+\iota}{16} + \sum_{k \in {\Bbb Z}}
\ \ \sum_{n \in Z} (n-\shf k)\:
a^+_i(n)a^-_i(k - n):\zeta^{-k}\eqno(2.26)$$
and
$$L^Z(\zeta) = \sum_{1\leq i\leq l}L_i^Z(\zeta).\eqno(2.27)$$
Then for $k \in {\Bbb Z}$ we have their homogeneous components $L^Z_i(k)$
and $L^Z(k)$, so
$$L^Z(k) = \sum_{1\leq i\leq l}\ \ \sum_{n\in Z} (n - \shf k)\:
a^+_i(n)a^-_i(k - n):\ \hbox{ for }\ k \neq
0,\eqno(2.28)$$
and
$$L^Z(0) = -\frac{l(1 + \iota)}{16} + \sum_{1\leq i \leq l}\ \ \sum_{n \in Z}
n\:a^+_i(n)a^-_i(-n):\eqno(2.29)$$
These are well-defined operators on $\bold{WM}(Z)$, independent of the choice
of canonical basis $\{a^+_1,\hdots,a^+_l$,

\noindent $a^-_1,\hdots,a^-_l\}$.
Let $Z' = \frac{1}{2}{\Bbb Z} - Z$ and for $a \in {\bold A}$, $n \in Z$,
define $a(n)$ to act on $\bold{WM}(Z')$ as zero.  Then the operators
$$L(m) = L^{\Bbb Z}(m) + L^{{\Bbb Z}+\frac{1}{2}}(m),\ \ \ m \in {\Bbb
Z},\eqno(2.30)$$
are well-defined on $\bold{WM}({\Bbb Z}) \oplus \bold{WM}({\Bbb Z} +
\frac{1}{2})$ and $L(m) = L^Z(m)$ on $\bold{WM}(Z)$.
\sk1

\proclaim{Theorem 2.1} (\cite{FF}) For $Z = {\Bbb Z}$ or ${\Bbb Z} +
{\tsize{\frac{1}{2}}}$
the operators (2.25) along with the operator (2.29) and the identity
operator represent the affine algebra ${\bold {\widehat {sp}}}(2l)$ on
$\bold{WM}(Z)$.
$\bold{WM}(Z) = \bold{WM}(Z)^0 \oplus \bold{WM}(Z)^1$ is the decomposition of
$\bold{WM}(Z)$ into two irreducible ${\bold{\widehat {sp}}}(2l)$-modules.
The operators (2.30)
and the identity operator represent the Virasoro algebra on
$\bold{WM}({\Bbb Z}) \oplus \bold{WM}({\Bbb Z}+\frac{1}{2})$.

More precisely, we have the following. Let $a,a_1,a_2,a_3,a_4 \in {\bold A}$,
$x =\ :a_1a_2:\ , \ y =\ :a_3a_4: \
\in ${\bf sp}$(2l)$, $m,n \in {\Bbb Z}$, $k \in Z$, and let $a(k) \in
\hbox{\bf Weyl}(Z)$. Then on {\bf WM}$(Z)$ the operators
$$x(m) =\ :a_1(\zeta)a_2(\zeta):_m,
\ y(n) =\ :a_3(\zeta)a_4(\zeta):_n,
\ c = -1, \ d = L(0) \eqno(2.31)$$
represent ${\bold{\widehat {sp}}}(2l)$ and the operators
$L(m)$ and $z = -l$ represent ${\bold
{Vir}}$.
Note that the level of these representations is $-\shf$ because of our
choice of bilinear form given in (2.10).
In particular,  we have the brackets
$$[x(m),a(k)] = (x\cdot a)(m + k), \eqno(2.32)$$
$$[x(m),y(n)] = [x,y](m + n) - m\delta_{m,-n}\langle x,y\rangle
1,\eqno(2.33)$$
$$[L(n),a(k)] = -(k + {\tsize{\frac{1}{2}}} n)a(k+n),\eqno(2.34)$$
$$[L(n),x(m)] = -mx(n + m),\eqno(2.35)$$
$$[L(m),L(n)] = (m - n)L(m + n) - {\tsize{\frac{1}{12}}}
(m^3 - m)\delta_{m,-n} l,\eqno(2.36)$$
where $[x,y]$ and $\langle x,y\rangle$ are
given by (2.9) and (2.10).\endproclaim
\sk1

We can define a positive Hermitian form on the representation space
$\bold{WM}(Z)$ of ${\bold {\widehat {sp}}}(2l)$, and we can determine the
adjoints of
the operators acting on $\bold{WM}(Z)$. We may choose a ${\Bbb C}$-antilinear
involution $\vartheta:\bold{A} \to \bold{A}$ such that for
$a,b \in \bold{A}$ we have
$$\langle\vartheta a,\vartheta b\rangle = \overline{\langle
a,b\rangle},\eqno(2.37)$$
$$0 < \langle a,\vartheta a\rangle \in {\Bbb R} \ \hbox { if }\ a \neq
0,\eqno(2.38)$$
$$\bold{A}^- = \vartheta(\bold{A}^+).\eqno(2.39)$$
Note that given any ${\Bbb C}$-antilinear involution $\vartheta:\bold{A}\to
\bold{A}$ satisfying (2.37) and (2.38), for any choice of maximal isotropic
subspace $\bold{A}^+$, $\vartheta(\bold{A}^+)$
is a maximal isotropic subspace and $\bold{A}
= \bold{A}^+ \oplus \vartheta(\bold{A}^+)$
is a polarization of $\bold{A}$. For $a,b \in \bold{A}$
we define a form
$$(a,b) = \langle a,\vartheta b\rangle\eqno(2.40)$$
and note that it is positive Hermitian because of (2.37) and (2.38). We
extend this form to $\bold{A}(Z)^-$ by
$$(a(m),b(n)) = \delta_{m,n}(a,b),\eqno(2.41)$$
and we further extend it to $\bold{WM}(Z)$ (see (2.18)-(2.19)) by
$$\eqalign
{&(a_1(m_1)\hdots a_r(m_r)\bold{vac}(Z),\ b_1(n_1)\hdots
b_s(n_s)\bold{vac}(Z)) \cr
&\ \ \ = \delta_{r,s}\ \hbox{perm}[(a_i(m_i),b_j(n_j))] \cr
&\ \ \ = \delta_{r,s} \sum_{\sigma \in S_r} \prod_{1\leq k \leq
r}\big(a_k(m_k),b_{\sigma(k)}(n_{\sigma(k)})\big),\cr}\eqno(2.42)$$
giving a positive Hermitian form.   If $T$ is any linear operator on $\bold
{WM}(Z)$, denote by $T^*$ the adjoint of $T$ with respect to this form.

\pr{Proposition 2.2}  Let $(\ ,\ )$ denote the positive Hermitian form on
$\bold{WM}(Z)$ defined by (2.40)-(2.42).  Then we have
\roster
\item"(a)"  $a(m)^* = (\vartheta a)(-m)$ for $a(m) \in {\bold A}(Z)$,
\item"(b)"  $(\ :a(\zeta)b(\zeta):_k)^* =\
:(\vartheta b)(\zeta)(\vartheta a)(\zeta):_{-k}$ for
$a,b \in {\bold A}$, $k \in {\Bbb Z}$,
\item"(c)"  $L^Z(m)^* = L^Z(-m)$ for $m \in {\Bbb Z}$. \endroster \epr

\demo{Proof}  (a) follows from the definitions (2.40)-(2.42).  (b) From
(2.23) we have
$(\ :a(m)b(n):\ )^* =$ \ $:(\vartheta
b)(-n)(\vartheta a)(-m):$\ , giving (b).  Finally, since $\vartheta$
takes one canonical basis to another canonical basis, and the operators
$L^Z(k)$, $k \in {\Bbb Z}$, are independent of the choice of such a basis,
we get (c).  $\hfill\blacksquare$
\enddemo

The eigenspaces of $L(0)$ provide $\bold{WM}(Z)$ with a grading as follows.
Let $$(\bold{WM}(Z))_n = \{ u \in \bold{WM}(Z) \mid L(0)u = nu
\}\eqno(2.43)$$
be the subspace of vectors of weight $n$, and write $wt(u)$ for the
weight of $u$.
{}From (2.29) it is clear that
$$L(0)\bold{vac}(Z) = -\frac{l(1+\iota)}{16}\bold{vac}(Z),\eqno(2.44)$$
so using (2.34), for $u = a^+_1(-m_1) \hdots a^+_r(-m_r)\bold{vac}(Z)$ as in
(2.19), we have
$$L(0)u = \left(m_1+\hdots+m_r - \frac{l(1+\iota)}{16}\right)u.\eqno(2.45)$$
Then {\bf WM}$(Z)$ is the direct sum of the eigenspaces (2.43), and the
$\frac{1}{2}{\Bbb Z}$-grading of $\bold{WM}({\Bbb Z}+\frac{1}{2})$
begins at 0 and the $\Bbb Z$-grading of $\bold{WM}({\Bbb Z})$ begins at
$-\sei$. For ${\bold W} = \bold{WM}(Z)^\alpha$, $\alpha = 0,1$, we define
$\Delta_{\bold W}$ to be to be the minimal value of $wt(w)$ for $w \in
{\bold W}$, so that ${\bold W}$ is ${\Bbb Z}+\Delta_{\bold W}$-graded.
Since $L(0)$ is self-adjoint, the eigenspaces for distinct
eigenvalues are orthogonal with respect to the Hermitian form on {\bf
WM}$(Z)$. If $A(m)$ is any of the operators $a(m) \in \bold{Weyl}(Z)$,
$x(m) \in {\bold{\widehat {sp}}}(2l)$ or $L(m) \in \bold{Vir}$, then from
(2.34)-
(2.36) we see that
$$A(m):(\bold{WM}(Z))_n \to (\bold{WM}(Z))_{n-m}. \eqno(2.46)$$

\newpage
\topmatter
\title\chapter{3} Bosonic Construction of Symplectic Vertex
Operator Algebras and Modules \endtitle \endtopmatter
\TagsOnRight

We will now describe vertex operator algebras constructed from certain
representation spaces in the bosonic construction.
This chapter is almost a copy of Chapter 3 of [FFR], but certain sign
changes were necessary to change their Clifford algebra calculations into
Weyl algebra calculations.

First recall from
Chapter II the bosonic construction of the {\bf Weyl}$(Z)$-module
$$\bold{WM}(Z) = \bold{Weyl}(Z) \cdot \bold{vac}(Z) = S( \bold
{A}(Z)^-) \cdot \bold{vac}(Z),\eqno(3.1)$$
from a nondegenerate anti-symmetric bilinear form $\langle\ ,\ \rangle$
on ${\bold A} \approx {\Bbb C}^{2l}$, with
$Z = {\Bbb Z}$ or ${\Bbb Z} + \frac{1}{2}$,
and ${\bold A}(Z)^-$ defined in (2.14). $\bold{WM}(Z)$ has the decomposition
(2.20) into even and odd parity subspaces, each of which is preserved
by the operators $x(k)$, $k \in {\Bbb Z}$, defined in (2.25),
 representing
${\bold{\widehat {sp}}}(2l)$
and
by the Virasoro operators $L(m)$, $m \in {\Bbb Z}$,
defined in (2.28)-(2.30) using a
canonical basis of $\bold A$. Using a ${\Bbb C}$-antilinear
involution $\vartheta$ on $\bold A$ satisfying
(2.37)-(2.39), $\bold{WM}(Z)$ has a positive Hermitian form $(\ ,\ )$
defined by
(2.40)-(2.42). Proposition 2.2 gives the adjoints of the operators
$a(m) \in {\bold A}(Z)$, $x(k) \in {\bold{\widehat {sp}}}(2l)$,
and $L(m)$ with respect to $(\ ,\ )$. With the
notation as in Theorem 2.1, we have the formulas (2.32)-(2.36).

Let
$$\eqalign{
{\bold V}_0 = &\bold{WM}({\Bbb Z}+\shf)^0,\ \ {\bold V}_2 =
\bold{WM}({\Bbb Z}+\shf)^1,\cr
&{\bold V}_1 = \bold{WM}({\Bbb Z})^0,\ \
{\bold V}_3 = \bold{WM}({\Bbb Z})^1,\cr}\eqno(3.2)$$
$${\bold V}(\shf{\Bbb Z}) = \bold{WM}({\Bbb Z}+\shf) \oplus \bold{WM}({\Bbb
Z}) = {\bold V}_0 \oplus {\bold V}_2 \oplus {\bold V}_1 \oplus {\bold
V}_3,\eqno(3.3)$$
$$\bold{V} = \bold{V}_0 \oplus \bold{V}_2,\eqno(3.4)$$
and extend the Hermitian form $(\ ,\ )$ to $\bold{V}(\shf{\Bbb Z})$
so that $\bold{WM}({\Bbb Z}+\shf)$ and
$\bold{WM}({\Bbb Z})$ are orthogonal. For
$a \in \bold{A}$, $n \in \shf{\Bbb Z}$, let $a(n)$ act on $\bold{V}(\shf{\Bbb
Z})$ as usual
on $\bold{WM}(Z)$ if $n \in Z$, but as zero if $n \in Z' = \shf{\Bbb Z} - Z$.
{}From (2.34) and (2.43) we see that $L(0)$ provides a grading
of ${\bold V}_i$, $0\leq i\leq 3$, into mutually orthogonal eigenspaces
$${\bold V}_i = \bigoplus_{0 \leq k \in {\Bbb Z}+\Delta_i} ({\bold
 V}_i)_k\eqno(3.5)$$
where we define
$$\Delta_i = \Delta_{{\bold V}_i} = \cases 0 &\text{if $i = 0$}\cr\\
\shf &\text{if $i = 2$}\cr\\ -l/8 &\text{if $i = 1,3.$}\endcases\eqno(3.6)$$

Our objective is to construct a vertex operator algebra $(\bWM({\Bbb
Z}+\hf),Y(\ ,\zeta),\break \bvac({\Bbb Z}+\hf), L(-2)\bvac({\Bbb Z}+\hf))$
and the canonically ${\Bbb Z}_2$-twisted representation of it on $\bWM({\Bbb
Z})$.
The first step is to define generalized vertex operators $Y(v,\zeta)$ on
$\bold{WM}({\Bbb Z}+\shf)$ and on $\bold{WM}({\Bbb Z})$
for any $v \in {\bold V}$. The definition of $Y(v,\zeta)$ on
$\bold{WM}({\Bbb Z})$
is more complicated than it is on $\bold{WM}({\Bbb Z}+\shf)$,
but there is a common part
which we denote by ${\bar Y}(v,\zeta)$. With
$\bold{vac}
= \bold{vac}({\Bbb Z} + \hf)$ define
${\bar Y}(\bold{vac},\zeta) = 1$ (the identity operator) on
$\bold{V}(\shf{\Bbb Z})$, and for $a \in \bold{A}$, $0 \leq n \in {\Bbb Z}$,
on $\bold{WM}(Z)$ let
$${\bar Y}(a(-n-\shf)\bold{vac},\zeta) =
\frac{1}{n!}\left(\frac{d}{d\zeta}\right)^n
(\zeta^{-1/2}
a(\zeta))\eqno(3.9)$$
where
$$a(\zeta) = \sum_{m \in Z}a(m)\zeta^{-m}.\eqno(3.10)$$
Define the notations $a^{(n)}(\zeta)$, $a^{(n)}(\zeta)^\pm$ and
$a^{(n)}(\zeta)^0$ by
$$a^{(n)}(\zeta) = {\bar Y}(a(-n-\shf)\bold{vac},\zeta) = \ds{\sum_{m\in Z}}
{{-m-\shf} \choose n} a(m)\zeta^{-m-n-\hf},\eqno(3.11)$$
$$a^{(n)}(\zeta)^+ = \sum_{0 < m \in Z} {{-m-\hf} \choose n}
a(m)\zeta^{-m-n-\hf},\eqno(3.12)$$
$$a^{(n)}(\zeta)^- = \sum_{0>m \in Z} {{-m-\hf} \choose
n}a(m)\zeta^{-m-n-\hf},\eqno(3.13)$$
$$a^{(n)}(\zeta)^0 = {{-\hf} \choose n} a(0)\zeta^{-n-\hf}\ \hbox{ if }\ Z =
{\Bbb Z},\eqno(3.14)$$
so that
$$a^{(n)}(\zeta) = \cases a^{(n)}(\zeta)^+ + a^{(n)}(\zeta)^- &\text{on
$\bold{WM}({\Bbb Z}+\shf)$}\cr\\
a^{(n)}(\zeta)^+ + a^{(n)}(\zeta)^0 + a^{(n)}(\zeta)^- &\text{on
$\bold{WM}({\Bbb Z})$.}\endcases\eqno(3.15)$$

For $a_1,\hdots,a_r \in\bold{A}$, $n_1,\hdots,n_r \in {\Bbb Z}+\hf$
extend the definition of bosonic
normal ordering (2.23) to any product of Weyl generators
$$:a_1(n_1) \hdots a_r(n_r):\ =
a_{\sigma 1}(n_{\sigma 1}) \hdots a_{\sigma r}(n_{\sigma r}),\eqno(3.16)$$
where $\sigma \in S_r$ is any permutation such that
$n_{\sigma 1} \leq \hdots \leq n_{\sigma r}$. The Weyl
relations (2.15) show that this is well-defined and agrees with (2.23)
for $r = 2$. For $n_1,\hdots,n_r \in {\Bbb Z}$,
we have to modify the definition of the
normally ordered product $:a_1(n_1) \hdots
a_r(n_r):$ in (3.16) because of the
possibility that some $n_i = 0$. We wish to make the definition so that for
any permutation $\sigma \in S_r$ we have
$$:a_1(n_1) \hdots a_r(n_r):\ =
\ :a_{\sigma 1}(n_{\sigma 1}) \hdots a_{\sigma r}(n_{\sigma
r}):.\eqno(3.17)$$
It is therefore enough to make the definition when $n_1 \leq \hdots \leq
n_r$ and then define the other cases by (3.17). If $n_1\leq\hdots\leq
n_r$ and each $0 \neq n_i \in {\Bbb Z}$, let
$$:a_1(n_1) \hdots a_r(n_r):\ = a_1(n_1) \hdots
a_r(n_r),\eqno(3.18)$$
but if $n_{i-1} < 0 = n_i = n_{i+1} = \hdots = n_{i+s-1} < n_{i+s}$
for some $s \geq 1$, then define
$$\eqalign
{:&a_1(n_1) \hdots a_r(n_r):\cr
&= a_1(n_1) \hdots
a_{i-1}(n_{i-1})(:a_i(0) \hdots
a_{i+s-1}(0):)a_{i+s}(n_{i+s})\hdots a_r(n_r),\cr}\eqno(3.19)$$
where for any $a_1,\hdots,a_k \in \bold{A}$,
$$:a_1(0) \hdots a_k(0):\ = \frac{1}{k!} \sum_{\sigma
\in S_k} a_{\sigma 1}(0) \hdots a_{\sigma k}(0).\eqno(3.20)$$
For $0 \leq n_1,\hdots,n_r \in {\Bbb Z}$, and for elements of the form
$$v = a_1(-n_1-\shf) \hdots a_r(-n_r-\shf)\bold{vac} \in(\bold{V})_n,
\eqno(3.21)$$
$n = \hbox{wt}(v) = n_1 + \hdots + n_r + \shf r$, on $\bold{WM}(Z)$ we define
$$\eqalign
{{\bar Y}(v,\zeta) &=\ :{\bar Y}(a_1(-n_1-\shf)\bold{vac},\zeta)
\hdots
{\bar Y}(a_r(-n_r-\shf)\bold{vac},\zeta):\cr
&=\ :a_1^{(n_1)}(\zeta) \hdots
a_r^{(n_r)}(\zeta):. \cr}\eqno(3.22)$$
Then on $\bold{WM}({\Bbb Z}+\shf)$ we have
$${\bar Y}(v,\zeta) = \sum a_{\sigma 1}^{(n_{\sigma
1})}(\zeta)^\pm \hdots a_{\sigma r}^{(n_{\sigma
r})}(\zeta)^\pm\eqno(3.23)$$
where the summation is over the $2^r$ terms obtained by splitting each
factor as in the first line of
(3.15) and in each term permuting the resulting factors by
some $\sigma \in S_r$ so that all $^-$ factors are to the left of all
$^+$ factors. Note
that any two $^-$ factors commute, and any two $^+$ factors commute.
On $\bold{WM}({\Bbb Z})$ we have
$${\bar Y}(v,\zeta) = \sum a_{\sigma 1}^{(n_{\sigma 1})}(\zeta)^-
\hdots (:a_{\sigma i}^{(n_{\sigma i})}(\zeta)^0 \hdots
a_{\sigma j}^{(n_{\sigma j})}(\zeta)^0:) \hdots a_{\sigma
r}^{(n_{\sigma r})}(\zeta)^+\eqno(3.24)$$
where the summation is over the $3^r$ terms obtained by splitting each
factor as in the second line of (3.15) and in each term permuting the
resulting factors by some $\sigma \in S_r$ so that all $^-$
factors are to the left
and all $^+$ factors are to the right. Note again that any two $^-$ factors
commute, and any two $^+$ factors commute. Extend the definitions
of ${\bar Y}(v,\zeta)$ to any $v \in \bold{WM}({\Bbb Z}+\shf)$
by linearity. For $v \in ({\bold V})_k$ define the
homogeneous components ${\bar Y}_m(v)$ of ${\bar Y}(v,\zeta)$
and the notation ${\bar {\{v\}}}_n$ by
$${\bar Y}(v,\zeta) = \sum_{n\in\hf{\Bbb Z}}{\bar Y}_{n+1-{\text{wt}}(v)}
(v)\zeta^{-n-1} = \sum_{n\in\hf{\Bbb Z}}{\bar
{\{v\}}}_n\zeta^{-n-1}.\eqno(3.25)$$
On $\bold{WM}({\Bbb Z}+\shf)$ the coefficient of $\zeta^{-n-1}$
is zero unless $n \in {\Bbb Z}$. On $\bold{WM}({\Bbb Z})$ the
coefficient of $\zeta^{-n-1}$ is zero unless $n \in {\Bbb Z}$
for $v \in \bold{V}_0$, but for $v \in\bold{V}_2$, it is
zero unless $n \in {\Bbb Z}+\hf$. It is easy to see that the components
${\bar Y}_m(v)$ are
well-defined operators on $\bold{V}(\shf{\Bbb Z})$ such that
$$\hbox{wt}({\bar Y}_m(v)w) = \hbox{wt}(w) - m.\eqno(3.26)$$

\pr{Proposition 3.1} For $v \in {\bold V}$ we have
\roster
\item"(a)" $\ds{\lim_{\zeta\to 0}}\ {\bar Y}(v,\zeta)\bold{vac} = v$,
\item"(b)" ${\bar Y}(L(-1)v,\zeta) = (d/d\zeta){\bar Y}(v,\zeta)$ on
$\bold{V}(\shf{\Bbb Z})$,
\item"(c)" $[L(0),{\bar Y}(v,\zeta)] = \zeta(d/d\zeta)
{\bar Y}(v,\zeta) + {\bar Y}(L(0)v,\zeta)$ on $\bold{V}(\shf{\Bbb Z})$.
\endroster \epr

\demo{Proof} It is enough to check these for $v = a_1(-n_1-\hf)
\hdots a_r(-n_r-\hf)\bold{vac} \in \bold{V}$.
\roster
\item"(a)" From (3.23) we have
$${\bar Y}(v,\zeta)\bold{vac} = a_1^{(n_1)}(\zeta)^- \hdots
a_r^{(n_r)}(\zeta)^-\bold{vac}\eqno(3.27)$$
in which only terms with nonnegative integral
powers of $\zeta$ occur. From (3.13)
with $Z = {\Bbb Z}+\shf$, the $\zeta^0$ term is $a_1(-n-\hf) \hdots
a_r(-n_r-\hf)\bold{vac} = v$.
\item"(b)" It is clear from (2.28) and (2.34) that
$$L(-1)\bold{vac} = 0\eqno(3.28)$$
and
$$[L(-1),a(-n-\shf)] = (n+1)a(-n- \sth)\eqno(3.29)$$
so that
$$\eqalign
{&L(-1)v = \sum_{1\leq i \leq r} a_1(-n_1-\shf) \hdots [L(-1),a_i(-n_i-\shf)]
\hdots a_r(-n_r - \shf)\bold{vac}\cr
&= \sum_{1\leq i \leq r}(n_i+1)a_1(-n_1-\shf)\hdots a_i(-n_i - \sth) \hdots
a_r(-n_r-\shf)\bold{vac}.\cr}\eqno(3.30)$$
Therefore,
$$\eqalign
{{\bar Y}(L(-1)v,\zeta) &= \sum_{1\leq i\leq r}(n_i+1)
:a_1^{(n_1)}(\zeta) \hdots a_i^{(n_i+1)}(\zeta) \hdots
a_r^{(n_r)}(\zeta):\cr
&= (d/d\zeta):a_1^{(n_1)}(\zeta)\hdots
a_r^{(n_r)}(\zeta):\cr
&= (d/d\zeta){\bar Y}(v,\zeta).\cr}\eqno(3.31)$$
\item"(c)" With $v$ as above, we have
$$L(0)v = \sum_{1\leq i \leq r}(n_i+\shf)v,\eqno(3.32)$$
$$\eqalign
{[L(0),a^{(n)}(\zeta)] &= - \sum_{m\in Z} {{-m-\hf} \choose n}
ma(m)\zeta^{-m-n-\hf}\cr
&= \zeta(d/d\zeta)a^{(n)}(\zeta) + (n+\shf)a^{(n)}(\zeta)\cr}\eqno(3.33)$$
and so
$$\eqalign
{[L(0),&{\bar Y}(v,\zeta)] = \sum_{1\leq i \leq r}:
a_1^{(n_1)}(\zeta) \hdots [L(0),a_i^{(n_i)}(\zeta)] \hdots
a_r^{(n_r)}(\zeta):\cr
&= \sum_{1\leq i \leq r}:a_1^{(n_1)}(\zeta) \hdots
((\zeta(d/d\zeta) + (n_i+\shf))a_i^{(n_i)}(\zeta)) \hdots
a_r^{(n_r)}(\zeta):\cr
&= \zeta(d/d\zeta){\bar Y}(v,\zeta) + {\bar
Y}(L(0)v,\zeta).\cr}\eqno(3.34)$$
Note that we have used the fact that
$$\eqalign
{[L(0),\,&:a_1(m_1)\hdots a_r(m_r):\ ]\cr
&= \sum_{1\leq i \leq r}:a_1(m_1) \hdots [L(0),a_i(m_i)]
\hdots a_r(m_r):,\cr}\eqno(3.35)$$
which would not be true with $L(0)$ replaced by $L(n)$ for $n \neq 0$.
$\hfill\blacksquare$
\endroster \enddemo

We can write the generating function of the Virasoro operators on $\bold V$
as follows. Letting
$$\eqalign
{\omega &= L(-2)\bold{vac}\cr
&= {\shf} \sum_{1\leq i\leq l}:a^+_i(-\shf)a^-_i(-\sth) -
a^+_i(-\sth)a^-_i(-\shf):\bold{vac}\cr
&= {\shf} \sum_{1\leq i \leq l} (a^-_i(-\sth)a^+_i(-\shf) -
a^+_i(-\sth)a^-_i(-\shf))\bold{vac}\cr}\eqno(3.36)$$
we find that
$$\eqalign
{&{\bar Y}(\omega,\zeta) = {\shf} \sum_{1\leq i \leq l}
{\bar Y}(a^-_i(-\sth)a^+_i(-\shf)\bold{vac},\zeta) -
{\bar Y}(a^+_i(-\sth)a^-_i(-\shf) \bold{vac},\zeta)\cr
&= {\shf} \sum_{1\leq i\leq l}:
\left( \frac{d}{d\zeta} \sum_{m\in{\Bbb Z}+\hf}
a^-_i(m)\zeta^{-m-\hf}\right) \left(\sum_{n\in{\Bbb Z}+\hf}
a^+_i(n)\zeta^{-n-\hf}\right) \cr
&\qquad\qquad\quad - \left(\frac{d}{d\zeta} \sum_{n\in{\Bbb Z}+\hf}
a^+_i(n)\zeta^{-n-\hf}\right)
\left(\sum_{m\in{\Bbb Z}+\hf}a^-_i(m)\zeta^{-m-\hf}\right):\cr
&= -{\shf} \sum_{1\leq i \leq l}\ \sum_{m,n \in {\Bbb Z}+\hf} \{ (m+\shf)\,
:a^-_i(m)a^+_i(n):\cr
&\qquad\qquad\qquad\qquad\qquad - (n+\shf)\,:
a^+_i(n)a^-_i(m):\}\zeta^{-m-n-2}\cr
&= {\shf} \sum_{1\leq i\leq l}\ \sum_{m,n\in {\Bbb Z}+\hf}(n-m)\,:
a^+_i(n)a^-_i(m):\zeta^{-m-n-2}\cr
&= \sum_{1\leq i\leq l}\ \sum_{k\in{\Bbb Z}}\ \sum_{n\in {\Bbb Z}+\hf}
(n - {\shf} k)\,:a^+_i(n)a^-_i(k-n):\zeta^{-k-2}\cr
&= \zeta^{-2}L(\zeta)\cr}\eqno(3.37)$$
on $\bold V$, where $\{a^+_1,\hdots,a^+_l,a^-_1,\hdots,a^-_l\}$
is any canonical basis of $\bold A$.

Computing ${\bar Y}(\omega,\zeta)$ on $\bold{WM}({\Bbb Z})$
we find that the homogeneous components
agree with the definitions of the Virasoro operators in (2.28)-(2.29),
except that the scalar term $-l/8$ is missing from the operator which
should be giving $L(0)$. The effect of leaving off that scalar term, that
is, just using ${\bar Y}(\omega,\zeta) = \zeta^{-2}L(\zeta)$
to define the Virasoro operators on
$\bold{WM}({\Bbb Z})$, is to modify the scalar term in the
bracket formula (2.36) so
that $m^3-m$ is changed to $m^3+2m$. Since we want to have (2.36) on
$\bold{WM}({\Bbb Z})$, we must add to ${\bar Y}(\omega,\zeta)$
a term which acts as $-\zeta^{-2}l/8$, so
$${\bar Y}(\omega,\zeta) -
\frac{l}{8}
\zeta^{-2}
=
\zeta^{-2}L(\zeta)\eqno(3.38)$$
is the generating function of the Virasoro operators on $\bold{WM}({\Bbb
Z})$. The
vertex operators $Y(v,\zeta)$ which we will define later incorporates this
scalar term on $\bold{WM}({\Bbb Z})$.

There are two approaches one may use to deal with generating
functions of operators on $\bold{V}(\hf{\Bbb Z})$,
the formal variables approach and the
complex analytic approach. We deal formally with series whose
variables have fractional powers in $\frac{1}{n} {\Bbb Z}$ by treating
$\zeta^{1/n}$ as a formal
variable. When we deal with such series analytically, we treat any
expression $z^r$, $r \in {\Bbb Q}$, as $\exp(r\, \log\, z)$
using the principal branch of $\log\, z$
which is defined and analytic for
$$z \in {\Bbb C}^- = {\Bbb C} - \{x \in {\Bbb R} \mid x \leq
0\}.\eqno(3.39)$$
We will be dealing with series $f(z^{1/n})$, $g(z^{1/n})$ which converge to
algebraic functions in some domain with $z \in {\Bbb C}^-$,
such that the product
$f(z^{1/n})g(z^{1/n})$ converges in that domain to a rational function of
$z$. But then the convergence of the product is guaranteed in the domain
without the condition $z \in {\Bbb C}^-$. Note that the power law
$$z^rz^s = z^{r+s},\ \ \ r,s \in {\Bbb Q},\ \ z \in {\Bbb
C}^-,\eqno(3.40)$$
is valid, but $z^rw^r$ need not equal $(zw)^r$, even for $z,w,zw \in {\Bbb
C}^-$. Let
$$A(\zeta_1,\hdots,\zeta_r) = \sum_{m_1,\hdots,m_r\in \frac{1}{n} {\Bbb Z}}
A_{m_1,\hdots,m_r} \zeta_1^{-m_1}\hdots \zeta_r^{-m_r}\eqno(3.41)$$
be the generating function of the operators $A_{m_1,\hdots,m_r}$ on
$\bold{V}(\hf {\Bbb Z})$. The
identity $A(\zeta_1,\hdots,\zeta_r) = B(\zeta_1,\hdots,\zeta_r)$
means the equality of the coefficients
$A_{m_1,\hdots,m_r} = B_{m_1,\hdots,m_r}$ as operators on $\bold{V}(\hf
{\Bbb Z})$, for each $m_1,\hdots,m_r \in \frac{1}{n} {\Bbb Z}$.
In the formal approach, such an equality might be checked by a formal
manipulation of the series, and the series might involve formal expressions
such as the delta function $\delta (\zeta) = \sum _{m \in {\Bbb Z}} \zeta^m$
and its derivatives. In the complex analytic approach, such an
equality would be checked by showing that for any $w,w' \in \bold{V}(\hf
{\Bbb Z})$, the ``matrix coefficients'' $(A(\zeta_1,\hdots,\zeta_r)w,w')$
and $(B(\zeta_1,\hdots,\zeta_r)w,w')$ converge to
equal holomorphic functions for $(\zeta_1,\hdots,\zeta_r)$
in some open neighborhood in
${\Bbb C}^n$. This is sufficient because the Hermitian form on
$\bold{V}(\hf {\Bbb Z})$ is nondegenerate.
Most of our results follow because, after sometimes multiplying by an
appropriate algebraic function, the
matrix coefficients we need converge in some open domain in ${\Bbb C}^n$ to
polynomials in the ring ${\Bbb C}[\zeta_k,\zeta^{-1}_k$,$(\zeta_i -
\zeta_j)^{-1}$; $1 \leq k \leq n$, $1 \leq i < j \leq n]$.

\pr{Definition} Let $F_1(\zeta_1,\hdots,\zeta_n)$ and
$F_2(\zeta_1,\hdots,\zeta_n)$ be generating functions with
complex coefficients which represent rational functions of
$\zeta_1,\hdots,\zeta_n$ on
some domains $D_1$ and $D_2$ in ${\Bbb C}^n$.  We write $F_1 \sim F_2$
if $F_1$ and $F_2$ represent
the same rational function on ${\Bbb C}^n$. Note that if $F_1 \sim F_2$
and $P$ is any
linear differential operator with coefficients in ${\Bbb
C}(\zeta_1,\hdots,\zeta_n)$ then $PF_1 \sim PF_2$.  \epr

\pr{Definition} For $a_1,a_2 \in {\bold A}$,
$0 \leq n_1,n_2 \in {\Bbb Z}$ define the bosonic contraction
$$\eqalign
{&{\ubr{a_1^{(n_1)}(\zeta_1)a_2^{(n_2)}(\zeta_2)}} =
a_1^{(n_1)}(\zeta_1)a_2^{(n_2)}(\zeta_2) -:
a_1^{(n_1)}(\zeta_1)a_2^{(n_2)}(\zeta_2):\cr
&\ \ \ \ \ \ \ \ \ = \cases [a_1^{(n_1)}(\zeta_1),
a_2^{(n_2)}(\zeta_2)^-] &\text{on $\bold{WM}({\Bbb Z}+\hf)$} \cr\\
[a^{(n_1)}_1(\zeta_1),a_2^{(n_2)}(\zeta_2)^- + \hf a_2^{(n_2)}(\zeta_2)^0]
&\text{on $\bold{WM}({\Bbb Z})$.}\endcases\cr} \eqno(3.42)$$ \epr

\pr{Lemma 3.2}  For $a_1,a_2 \in \bold{A}$, $0\leq n_1,n_2 \in {\Bbb Z}$, as
formal series in $\zeta_1$ and $\zeta_2$, on $\bold{WM}({\Bbb Z}+\hf)$ we
have
$$\eqalign
{&{\ubr{a_1^{(n_1)}(\zeta_1)a_2^{(n_2)}(\zeta_2)}} \cr
&= (-1)^{n_1}\langle a_1,a_2\rangle \zeta_1^{-n_1-n_2-1}
 {{n_1+n_2} \choose
{n_1}} \sum_{0\leq k \in {\Bbb Z}} {{n_1+n_2+k} \choose k}
\left(\frac{\zeta_2}{\zeta_1}\right)^k\cr
&=n_1!^{-1}n_2!^{-1}(\partial/\partial \zeta_1)^{n_1}(\partial/\partial
\zeta_2)^{n_2} \langle a_1,a_2\rangle \zeta^{-1}_1 \sum_{0\leq m \in {\Bbb
Z}} \left(\frac{\zeta_2}{\zeta_1}\right)^m\cr}$$ \epr

\demo{Proof} This follows from (3.12), (3.13) and the commutation relations
(2.15) in $\bold{Weyl}({\Bbb Z}+\hf)$. $\hfill \blacksquare$
\enddemo

\pr{Corollary 3.3} For $a_1,a_2 \in \bold{A}$, $0 \leq n_1,n_2 \in {\Bbb
Z}$, $w,w' \in \bold{WM}({\Bbb Z}+\hf)$, when $|\zeta_1| > |\zeta_2|$
we have the absolute convergence of
$$\eqalign
{&({\ubr{a_1^{(n_1)}(\zeta_1)a_2^{(n_2)}(\zeta_2)}}w,w')\cr
&\ \ = \langle a_1,a_2\rangle
(w,w')n_1!^{-1}n_2!^{-1}(\partial/\partial\zeta_1)^{n_1}(\partial/
\partial\zeta_2)^{n_2}(\zeta_1 - \zeta_2)^{-1}\cr
&\ \ = \langle a_1,a_2\rangle(w,w') {{-n_2-1} \choose {n_1}} (\zeta_1 -
\zeta_2)^{-n_1-n_2-1}\cr}$$ \epr

\pr{Lemma 3.4} For $a_1,a_2 \in \bold{A}$, $0 \leq n_1,n_2 \in {\Bbb Z}$,
as formal series in $\zeta_1$ and $\zeta_2$, on
$\bold{WM}({\Bbb Z})$ we have
$$\eqalign
{&{\ubr{a_1^{(n_1)}(\zeta_1)a_2^{(n_2)}(\zeta_2)}} \cr
&= \langle a_1,a_2\rangle
\zeta_1^{-n_1-\hf}\zeta_2^{-n_2-\hf}\left({\shf} {{-\hf}
\choose {n_1}} {{-\hf} \choose {n_2}} +
\sum_{1\leq p \in{\Bbb Z}} {{-p-\hf} \choose {n_1}} {{p-\hf} \choose {n_2}}
\left(\frac{\zeta_2}{\zeta_1}\right)^p\right)\cr
&=n_1!^{-1}n_2!^{-1}(\partial/\partial \zeta_1)^{n_1}(\partial/\partial
\zeta_2)^{n_2} \langle a_1,a_2\rangle \zeta_1^{-\hf} \zeta_2^{-\hf}
\left(\shf + \sum_{1\leq p \in {\Bbb Z}}
\left(\frac{\zeta_2}{\zeta_1}\right)^p\right)\cr}$$ \epr

\demo{Proof} This follows from (3.12) - (3.14). $\hfill\blacksquare$
\enddemo

\pr{Corollary 3.5} For $a_1,a_2 \in\bold{A}$, $|\zeta_1| > |\zeta_2| > 0$,
$\zeta_1,\zeta_2 \in {\Bbb C}^-$, $w,w' \in \bold{WM}({\Bbb Z})$ and
$0 \leq n_1,n_2 \in{\Bbb Z}$, we have the absolute convergence of
$$({\ubr{a_1^{(0)}(\zeta_1)a_2^{(0)}(\zeta_2)}}w,w') = \langle
a_1,a_2\rangle(w,w')\zeta_1^{-\hf}\zeta_2^{-\hf} \shf(\zeta_1+
\zeta_2)(\zeta_1-\zeta_2)^{-1},$$
and
$$\eqalign
{&({\ubr{a_1^{(n_1)}(\zeta_1)a_2^{(n_2)}(\zeta_2)}}w,w') =
n_1!^{-1}n_2!^{-1}(\partial/\partial\zeta_1)^{n_1}(\partial/\partial
\zeta_2)^{n_2}({\ubr{a_1^{(0)}(\zeta_1)a_2^{(0)}(\zeta_2)}}w,w') \cr
&\ \ \ = \langle a_1,a_2\rangle (w,w')\zeta_1^{-n_1-\hf}\zeta_2^{-n_2-\hf}
(\zeta_1 - \zeta_2)^{-n_1-n_2-1} P_{n_1n_2}(\zeta_1,\zeta_2)\cr}$$
where $P_{n_1n_2}(\zeta_1,\zeta_2)$
is a homogeneous polynomial of degree $n_1+n_2+1$. \epr

For $v_1,\hdots,v_n \in \bold{V}$, $w,w' \in\bold{V}(\hf {\Bbb Z})$,
using products of vertex operators we define
the generating function of matrix coefficients
$$\eqalign
{&({\bar Y}(v_1,\zeta_1) \hdots {\bar Y}(v_n,\zeta_n)w,w')\cr
&= \sum_{m_1,\hdots,m_n \in \hf {\Bbb Z}} ({\bar Y}_{m_1}(v_1) \hdots
{\bar Y}_{m_n}(v_n)
w,w') \zeta_1^{-m_1-wt(v_1)} \hdots
\zeta_n^{-m_n-wt(v_n)}.\cr}\eqno(3.43)$$
Later we will find a domain in which this series converges absolutely.

We will now show how to express the vertex operators ${\bar Y}(v,\zeta)$ in
terms of normally ordered exponentials of various quadratic operators
on various
tensor products whose factors are
$\bold{WM}({\Bbb Z}+\hf)$ or
$\bold{WM}({\Bbb Z})$. Let ${\bold A}_1,\hdots,\bold{A}_r$
be copies of $\bold{A}$ with its polarization and let each of
$Z_1,\hdots,Z_r$ be either ${\Bbb Z}+\hf$ or $\Bbb Z$.
For $1 \leq i \leq r$ let $\bold{A}_i(Z_i)$ be spanned by $\{a_i(m)\mid a_i
\in \bold{A}_i,\ m \in Z_i\}$
and let $\bold{Weyl}(Z_1,\hdots,Z_r)$ be the Weyl algebra generated
by $\bold{A}_1(Z_1) \oplus \hdots \oplus \bold{A}_r(Z_r)$
and the form $\langle a_i(m),a'_j(n)\rangle = \langle a_i,a'_j\rangle
\delta_{ij}\delta_{m,-n}$.
Let ${\goth I}(Z_1,\hdots,Z_r)$ be the left ideal in
$\bold{Weyl}(Z_1,\hdots,Z_r)$ generated by
$\bold{A}_1(Z_1)^+ \oplus \hdots \oplus \bold{A}_r(Z_r)^+$
(see (2.11)-(2.21)). Then the
tensor
product $\bold{WM}(Z_1) \otimes \hdots \otimes \bold{WM}(Z_r)$
is isomorphic to the Weyl module
$\bold{WM}(Z_1,\hdots,Z_r) = \bold{Weyl}(Z_1,\hdots,Z_r)/{\goth
I}(Z_1,\hdots,Z_r)$.
Note that for $v_i \in\bold{WM}(Z_i)^{\alpha_i}$, $\alpha_i \in \{0,1\}$,
$1\leq i \leq r$, $a_i(n) \in \bold{A}_i(Z_i)$, we have
$$a_i(n) (v_1 \otimes \hdots \otimes v_r) = v_1 \otimes \hdots \otimes
 a_i(n) v_
i \otimes \hdots \otimes
v_r.\eqno(3.44)$$
Let $\{a^+_1,\hdots,a^+_l$,

\noindent $a_1^-,$ $\hdots,a_l^-\}$
be a canonical basis of $\bold{A}$ and for $1\leq j \leq r$
denote by $\{a_{j1}^+,\hdots,a_{jl}^+,a_{j1}^-,\hdots,a_{jl}^-\}$
a copy of that canonical basis in $\bold{A}_j$.
For
$$x = \sum_{1\leq i \leq l} (x_ia^+_i + x^*_ia_i^-) \in \bold{A},\ \ x_i,x^*_i
\in {\Bbb C},\eqno(3.45)$$
we sometimes abuse notation, denoting by $x$ the corresponding vector
$$\sum_{1\leq i \leq l}(x_ia^+_{ji} + x_i^*a^-_{ji}) \in
\bold{A}_j.\eqno(3.46)$$

Letting $1 \leq j \leq r$ and $u_j \in \bold{WM}(Z_j)$,
the Hermitian form $(\ ,\ )$ on $\bold{WM}(Z_j)$
provides a projection from $\bold{WM}(Z_1) \otimes \hdots \otimes
\bold{WM}(Z_r)$ onto $\bold{WM}(Z_1) \otimes \hdots \otimes
\bold{WM}(Z_{j-1}) \otimes \bold{WM}(Z_{j+1}) \otimes \hdots \otimes
\bold{WM}(Z_r)$ by
$$\pi_j(u_j)(v_1 \otimes \hdots \otimes v_r) = (v_j,u_j)(v_1 \otimes \hdots
v_{j-1}\otimes v_{j+1} \otimes \hdots \otimes v_r).\eqno(3.47)$$
The projection $\pi_j(\bold{vac}(Z_j))$
is denoted by $\pi_j$.

Fix $1\leq p \neq q\leq r$ with $Z_p = {\Bbb Z}+\hf$.
Define the generating function of
operators on $\bold{WM}(Z_1) \otimes \hdots \otimes \bold{WM}(Z_r)$,
$$\eqalign
{J_{pq}(\zeta) &= \sum \Sb {1\leq i\leq l}\\ 0\leq m \in {\Bbb Z},n\in Z_q
\endSb {{-n-\hf} \choose {m}} (
a_{qi}^-(n)a^+_{pi}(m+\shf)
-a^+_{qi}(n)a_{pi}^-(m+\shf)
)\zeta^{-n-m-\hf}\cr
&= \sum \Sb {1\leq i \leq l}\\ 0\leq m \in {\Bbb Z}\endSb (
a_{qi}^{-(m)}(\zeta)a^+_{pi}(m+\shf)
-a_{qi}^{+(m)}(\zeta)
a^-_{pi}(m+\shf)
)\cr}\eqno(3.48)$$
and write
$$J_{pq}(\zeta) = J_{pq}^-(\zeta) + J_{pq}^0(\zeta) +
J_{pq}^+(\zeta)\eqno(3.49)$$
where the superscript corresponds to taking the summation in the first
line of (3.48) over $n < 0$, $n = 0$ (if $0 \in Z_q$) or
$n > 0$, respectively, and in the second line to
replacing the generating functions by sums as in (3.12)-(3.14).

Define the normally ordered exponential of $J_{pq}(\zeta)$ by
$$\eqalign
{\colon \exp(J_{pq}(\zeta))\colon &= \exp(J_{pq}^-(\zeta))
\exp(J_{pq}^0(\zeta)) \exp(J_{pq}^+(\zeta))\cr
&= \exp(J_{pq}^-(\zeta)+\shf J_{pq}^0(\zeta)) \exp(J_{pq}^+(\zeta) + \shf
J_{pq}^0(\zeta))\cr}\eqno(3.50)$$
since $J_{pq}^0(\zeta)$ commutes with $J_{pq}^-(\zeta)$ and
$J_{pq}^+(\zeta)$.
It is easy to see that this is a generating function of well-defined
operators on $\bold{WM}(Z_1) \otimes \hdots \otimes \bold{WM}(Z_r)$.

\pr{Theorem 3.6} For $v_1 \otimes v_2 \in \bold{WM}({\Bbb Z}+\hf) \otimes
\bold{WM}(Z_2)$ we have
$$\pi_1(\colon \exp(J_{12}(\zeta))\colon v_1 \otimes v_2) = {\bar
Y}(v_1,\zeta)v_2.$$ \epr

\demo{Proof} Let $v_1 = x_1(-m_1-\hf) \hdots
x_k(-m_k-\hf)\bold{vac}\in\bold{WM}({\Bbb Z}+\hf)^{\alpha_1}$,
so $\alpha_1 \equiv k$ (mod 2),
and for $1\leq i_1,\hdots,i_n\leq k$ distinct but not necessarily in order,
let $v_1(i_1,\hdots,i_n)$
be the vector obtained from $v_1$ by removing the
$i_1^{\text{th}},\hdots,i_n^{\text{th}}$ factors.
Writing $J = J_{12}$, for $^\# =\ ^-,\,^0$ (if $Z_2 = {\Bbb Z}$), or $^+$
and $1 \leq n \in {\Bbb Z}$, from the definitions we have
$$\eqalign
{&(J^\#(\zeta))^nv_1 \otimes v_2\cr
&= \sum_{1\leq i_1\neq\hdots\neq i_n \leq
k}v_1(i_1,\hdots,i_n) \otimes
x_{i_1}^{(m_{i_1})}(\zeta)^\# \hdots x_{i_n}^{(m_{i_n})}(\zeta)^\#v_2\cr
&= n!\!\!\sum_{1\leq i_1 < \hdots < i_n\leq k}\!
v_1(i_1,\hdots,i_n) \otimes:x_{i_1}^{(m_{i_1})}(\zeta)^\#
\hdots x_{i_n}^{(m_{i_n})}(\zeta)^\#:v_2.\cr}\eqno(3.53)$$
Then, since $J^-(\zeta)^aJ^0(\zeta)^bJ^+(\zeta)^cv_1 \otimes v_2 = 0$
if $a+b+c > k$, and \hfill\break
$\pi_1(J^-(\zeta)^aJ^0(\zeta)^bJ^+(\zeta)^cv_1 \otimes v_2) = 0$
if $a+b+c < k$, we have
$$\eqalign
{&\pi_1(\exp(J^-(\zeta))\exp(J^0(\zeta))\exp(J^+(\zeta))v_1 \otimes v_2)\cr
&\ \ = \pi_1 \left(\sum \Sb a+b+c=k\\ 0\leq a,b,c\in {\Bbb Z} \endSb
\frac{J^-(\zeta)^a\ J^0(\zeta)^b\ J^+(\zeta)^c}{a!b!c!}\ v_1 \otimes
v_2\right).\cr}\eqno(3.54)$$
This matches the definition of ${\bar Y}(v_1,\zeta)v_2$
as in (3.24) if $Z_2 = {\Bbb Z}$ by using
(3.53) with $^\# =\ ^-,\,^0$, and $^+$,
and matches the definition of ${\bar Y}(v_1,\zeta)v_2$ as in
(3.23) if $Z_2 = {\Bbb Z}+\hf$ by using (3.53) with $^\# =\ ^-$ and $^+$.
$\hfill\blacksquare$
\enddemo

\pr{Corollary 3.7} For $v_i \in\bold{WM}({\Bbb Z}+\hf)$,
$1 \leq i \leq r$, $w \in \bold{WM}(Z)$ we have
$$\eqalign
{\pi_1\hdots \pi_r&(\colon \exp(J_{1(r+1)}(\zeta_1))\colon \hdots
\colon\exp(J_{r(r+1)}(\zeta_r))\colon v_1 \otimes \hdots \otimes v_r \otimes
w)\cr
&= {\bar Y}(v_1,\zeta_1) \hdots {\bar Y}(v_r,\zeta_r)w.\cr}$$ \epr

\pr{Corollary 3.8} For $v_i \in\bold{WM}(Z_i)^{\alpha_i}$, $\alpha_i \in
\{0,1\}$, $1 \leq i \leq r$, $1\leq p \neq q \leq r$,  $Z_p = {\Bbb Z}+\hf$,
we have
$$\eqalign
{&\pi_p(\colon\exp(J_{pq}(\zeta))\colon v_1 \otimes \hdots \otimes v_r)\cr
&= \cases v_1 \otimes
\hdots \otimes {\hat v}_p\otimes \hdots \otimes {\bar Y}(v_p,\zeta)v_q
\otimes \hdots \otimes v_r &\text{if $p < q$}\cr\\ v_1 \otimes \hdots \otimes
{\bar Y}(v_p,\zeta) v_q \otimes \hdots \otimes {\hat v}_p \otimes \hdots
\otimes v_r &\text{if $q < p$}\endcases\cr}$$
where $\hat{\ }$ indicates removal. \epr

\demo{Proof} This follows from the proof of Theorem 3.6.
$\hfill\blacksquare$
\enddemo

\pr{Definition} For $0 \leq m,n \in{\Bbb Z}$, let
$$C_{mn} = \shf \frac{m - n}{m + n + 1} {{-\hf} \choose m} {{-\hf} \choose
n},\eqno(3.55)$$
and note that
$$(m + 1)C_{(m+1)n} + (n + 1)C_{m(n+1)} = -(m+n+1)C_{mn}.\eqno(3.56)$$ \epr

\pr{Lemma 3.9} For $0\leq k,r,t \in{\Bbb Z}$
the coefficients defined in (3.55) satisfy
$$\sum_{0\leq m\leq k} {{m+r} \choose r} {{k-m+t} \choose t}
C_{(m+r)(k-m+t)} = {{-r-t-1} \choose k} C_{rt}.$$ \epr

\demo{Proof} The proof is by induction on $k$, using (3.56).
$\hfill\blacksquare$
\enddemo

We now define a generating function $\Delta(\zeta)$ of quadratic operators on
$\bold{WM}({\Bbb Z}+\hf)$
which allows us to define the operators $Y(v,\zeta)$ we want.

\pr{Definition} For $\{a^+_1,\hdots,a^+_l,a^-_1,\hdots,a^-_l\}$
a canonical basis of $\bold{A}$, define
$$\Delta(\zeta) = - \sum_{1\leq i \leq l}\ \  \sum_{0\leq m,n\in{\Bbb
Z}}C_{mn}a^+_i(m+\shf) a^-_i (n+\shf)\zeta^{-m-n-1}\eqno(3.57)$$
on $\bold{WM}({\Bbb Z}+\hf)$. Since $\Delta(\zeta)$
is quadratic, it preserves $\bold{V}_0$ and $\bold{V}_2$. \epr

\pr{Definition} For $v \in\bold{WM}({\Bbb Z}+\hf)$, let
$$Y(v,\zeta) = \cases {\bar Y}(v,\zeta) &\text{on $\bold{WM}({\Bbb
Z}+\hf)$}\cr\\
{\bar Y}(\exp(\Delta(\zeta))v,\zeta) &\text{on $\bold{WM}({\Bbb
Z})$.}\endcases\eqno(3.58)$$ \epr

For $v \in (\bold{V})_k$, define the homogeneous components $Y_m(v)$ of
$Y(v,\zeta)$ and the notation $\{v\}_n$ by
$$Y(v,\zeta) = \sum_{n\in\hf{\Bbb Z}}Y_{n+1-{\text{wt}}(v)}(v)\zeta^{-n-1}
= \sum_{n \in \hf{\Bbb Z}}\{v\}_n \zeta^{-n-1}.\eqno(3.59)$$
Since $\Delta(\zeta)$ preserves $\bold{V}_0$ and $\bold{V}_2$,
the discussion after (3.25) of what
coefficients of (3.25) are zero remains valid for (3.59). Let us use the
notation
$${\bar Y}'(v,\zeta) = (d/d\zeta){\bar Y}(v,\zeta).\eqno(3.60)$$

\pr{Lemma 3.10} For $a \in\bold{A}$, $0 \leq n \in{\Bbb Z}$, on
$\bold{WM}({\Bbb Z}+\hf)$, we have
\roster
\item"(a)" $[\Delta(\zeta),a(-n-\hf)] = \ds{\sum_{0\leq m\in{\Bbb Z}}}
C_{mn}a(m+\shf)\zeta^{-m-n-1}$,
\item"(b)" If $v = a_1(-n_1-\hf) \hdots a_k(-n_k-\hf)\bold{vac} \in
\bold{WM}({\Bbb Z}+\hf)$, and $v_{rs}$, $1\leq r < s \leq k$, is the
vector obtained from $v$ by removing $a_r(-n_r-\hf)$ and $a_s(-n_s-\hf)$, then
$$\Delta(\zeta)v = -\sum_{1\leq r < s \leq k} \langle
a_r,a_s\rangle \zeta^{-n_r-n_s-1} v_{rs},$$
\item"(c)" $[L(-1),\Delta(\zeta)] = - \Delta'(\zeta)$ on
$\bold{WM}({\Bbb Z}+\hf)$, and $\Delta'(\zeta) = (d/d\zeta)\Delta(\zeta)$
commutes with $\Delta(\zeta)$,
\item"(d)" $[L(-1),\exp(\Delta(\zeta))] =
-\Delta'(\zeta)\exp(\Delta(\zeta)) = -(d/d\zeta)\exp(\Delta(\zeta))$,
\item"(e)" $[L(0),\Delta(\zeta)] = \zeta\Delta'(\zeta)$,
\item"(f)" $[L(0),\exp(\Delta(\zeta))] =
\zeta\Delta'(\zeta)\exp(\Delta(\zeta))$. \endroster \epr

\demo{Proof} Part (a) follows from the commutation relations in
$\bold{Weyl}({\Bbb Z}+\hf)$, and (b) follows from (a) by induction.
Parts (c) and (e)
follow from the definitions using (3.56), and (d) and (f) follow from (c)
and (e).  $\hfill\blacksquare$ \enddemo

Note that
$$Y(L(-2)\bold{vac},\zeta) = \zeta^{-2}L(\zeta)\eqno(3.61)$$
includes the correction term $-l/8$ to the operator $L(0)$ on
$\bold{WM}({\Bbb Z})$, since from (3.36) we get
$$\Delta(\zeta)L(-2)\bold{vac} = -\tsize{\frac{l}{8}}
\bold{vac}\eqno(3.62)$$
and higher powers of $\Delta(\zeta)$ kill $L(-2)\bold{vac}$.

For $1\leq j \leq r$, if $Z_j = {\Bbb Z}+\hf$ let
$$\Delta_j(\zeta) = -\sum \Sb {1\leq i\leq l}\\ 0\leq m,n \in {\Bbb Z}\endSb
C_{mn}a^+_{ji}(m+\shf)a^-_{ji}(n+\shf)\zeta^{-m-n-1}\eqno(3.63)$$
on $\bold{WM}(Z_1) \otimes \hdots \otimes \bold{WM}(Z_r)$. It is clear that
$$[\Delta_j(\zeta),J_{pq}^\#(z)] = 0\ \hbox{ for }\ q \neq j \ \hbox{ and
}\ ^\# =\ ^-,\,^0\hbox{ or }\ ^+\eqno(3.64)$$
and
$$[\Delta_q(\zeta),J^+_{pq}(z)] = 0.\eqno(3.65)$$

\pr{Proposition 3.11} For $1\leq p\neq q\leq r$, $Z_p = Z_q = {\Bbb Z}+\hf$,
on $\bold{WM}(Z_1) \otimes \hdots \otimes \bold{WM}(Z_r)$ we
have
$$\eqalign{\hbox{\rm{(a)\ }}&[\Delta_q(\zeta),J^-_{pq}(z)] \cr
&= \ds{\sum\Sb 1\leq i\leq l\\ 0\leq m,t\in{\Bbb Z}\endSb}
(-a^+_{qi}(m+\shf)a^-_{pi}(t+\shf) +
a^-_{qi}(m+\shf)a^+_{pi}(t+\shf))\zeta^{-m-t-1}
 \cr
&\hbox to 2.5 truein{} \ds{\cdot\sum_{0\leq n\in{\Bbb Z}}}
{n \choose t} C_{mn} \left(\frac{z}{\zeta}\right)^{n-t},\cr}$$
$$\eqalign{\hbox{\rm{(b)\ }}&[[\Delta_q(\zeta),J^-_{pq}(z)],J^-_{pq}(z)] \cr
&= -\ds{\sum \Sb 1\leq i \leq l\\ 0\leq r,t\in{\Bbb Z}\endSb}
a^+_{pi}(r+\shf)a^-_{pi}(t+\shf)\zeta^{-r-t-1}2 C_{rt}\ds{\sum_{0\leq k
\in{\Bbb
Z}} {{-r-t-1} \choose k}} \left(\frac{z}{\zeta}\right)^k \cr
&= 2\Delta_p(\zeta+z)\quad \hbox{ expanded in nonnegative powers of z.}
\cr}$$ \epr

\demo{Proof} These are straightforward calculations using the definitions and
(3.44). Lemma 3.9 is required for part (b). $\hfill\blacksquare$
\enddemo

\pr{Corollary 3.12} For $1\leq p \neq q\leq r$, $Z_p = Z_q = {\Bbb Z}+\hf$,
on $\bold{WM}(Z_1)\otimes \hdots \otimes \bold{WM}(Z_r)$ we have
$$\eqalign
{&\exp(\Delta_q(\zeta))\colon \exp(J_{pq}(z))\colon \cr
&= \ \colon\exp(J_{pq}(z))\colon \exp(\Delta_q(\zeta) +
[\Delta_q(\zeta),J^-_{pq}(z)] + \Delta_p(\zeta+z))\cr}$$
with $\Delta_p(\zeta+z)$ expanded in nonnegative powers of $z$. \epr

\demo{Proof} For operators $A$, $B$ on a vector space such that $B$ is
locally nilpotent and $(ad_B)^nA = 0$ for some $n$, one has the
well-known formula
$$\exp(B)A\exp(-B) = \exp(ad_B)A\eqno(3.66)$$
where $ad_B(X) = [B,X]$. Letting $A = \Delta_q(\zeta)$ and
$B = -J_{pq}^-(z)$ we get
$$\eqalign
{&\exp(-J_{pq}^-(z))\Delta_q(\zeta)\exp(J^-_{pq}(z)) = \cr
&\Delta_q(\zeta) + [\Delta_q(\zeta),J^-_{pq}(z)] +
\shf[[\Delta_q(\zeta),J^-_{pq}(z)],J^-_{pq}(z)]\cr}\eqno(3.67)$$
and exponentiating both sides gives
$$\eqalign
{&\exp(-J_{pq}^-(z))\exp(\Delta_q(\zeta))\exp(J^-_{pq}(z)) = \cr
&\exp(\Delta_q(\zeta) + [\Delta_q(\zeta),J^-_{pq}(z)] +
\shf[[\Delta_q(\zeta),J^-_{pq}(z)],J^-_{pq}(z)]).\cr}\eqno(3.68)$$
{}From the last proposition, $[\Delta_q(\zeta),J^-_{pq}(z)]$ commutes with
$\Delta_q(\zeta)$,
$[[\Delta_q(\zeta),J^-_{pq}(z)],\break J^-_{pq}(z)] =\Delta_p(\zeta+z)$
commutes with both $\Delta_q(\zeta)$ and
$J^-_{pq}(z)$, and $J^+_{pq}(z)$
commutes with $\Delta_q(\zeta)$, $[\Delta_q(\zeta),J^-_{pq}(z)]$, and
$\Delta_p(\zeta+z)$.  Under the assumptions we have
$$\colon\exp(J_{pq}(z))\colon
=\exp(J^-_{pq}(z))\exp(J^+_{pq}(z))\eqno(3.69)$$
so the result follows. $\hfill\blacksquare$
\enddemo

For $1\leq p\leq r$, define the Virasoro operators $L_p(m)$, $m\in{\Bbb Z}$,
on the $p^{\text{th}}$
tensor factor of $\bold{WM}(Z_1)\otimes \hdots \otimes \bold{WM}(Z_r)$
by the usual formulas (2.23)-(2.25)
using the canonical basis in $\bold{A}_p$. For $1\leq p\neq q \leq r$,
let $L_{pq}(m) = L_p(m) + L_q(m)$.
{}From Lemma 3.10 (c) - (f) we have
$$[L_p(-1),\Delta_p(z)] = -\Delta'_p(z),\eqno(3.70)$$
$\Delta'_p(z)$ commutes with $\Delta_p(z)$, and
$$[L_p(-1),\exp(\Delta_p(z))] = -\Delta'_p(z)\exp(\Delta_p(z)) =
-(d/dz)\exp(\Delta_p(z)),\eqno(3.71)$$
where $\Delta'_p(z) = (d/dz)\Delta_p(z)$.

\pr{Proposition 3.13} For $1\leq p\neq q \leq r$, on
$\bold{WM}(Z_1)\otimes \hdots \otimes \bold{WM}(Z_r)$ we have
\roster
\item"(a)" $[L_p(-1),J^\#_{pq}(z)] = -(d/dz)J^\#_{pq}(z)$ for
$^\# =\ ^-,\,^0$ (if $Z_q = {\Bbb Z}$), or $^+$,
\item"(b)" $[L_q(-1),J^\#_{pq}(z)] = (d/dz)J^\#_{pq}(z)$ for
$^\# =\ ^-$ or $^+$ if $Z_q = {\Bbb Z}+\hf$,
\item"(c)"  $[L_q(-1),J^-_{pq}(z)] = (d/dz)J^-_{pq}(z)\ -$
\hfill if $Z_q = {\Bbb Z}$
\nl $\hf \ds{\sum \Sb {1\leq i\leq l}\\ 0\leq m\in{\Bbb Z}\endSb}
{-\hf \choose m}
 (-a^+_{qi}(-1)a^-_{pi}(m+\shf)+a^-_{qi}(-1)a^+_{pi}(m+\shf))z^{-m-\hf}$,
\item"(d)" $[L_q(-1),J^0_{pq}(z)]$
\nl $= \hf \ds{\sum \Sb {1\leq i\leq l}\\ 0\leq m \in{\Bbb Z}\endSb}
{-\hf \choose m}
(-a^+_{qi}(-1)a^-_{pi}(m+\shf) + a^-_{qi}(-1)a^+_{pi}(m+\shf))z^{-m-\hf}$
\nl if $Z_q = {\Bbb Z}$,
\item"(e)" $[L_q(-1),J^+_{pq}(z)] = (d/dz)J^+_{pq}(z) +
(d/dz)J^0_{pq}(z)$ if $Z_q = {\Bbb Z}$.\endroster \epr

\demo{Proof} Part (a) is immediate from the definition of $J_{pq}(z)$ using
\hfill\break $[L(-1),a(m+\hf)] = -ma(m-\hf)$. This also gives
$$\eqalign
{&[L_q(-1),J^\#_{pq}(z)] = \cr
&\sum \Sb {1\leq i\leq l}\\ 0\leq m\in {\Bbb Z}\\ n\in Z^\#_q\endSb
\! \! \! {{-n-\hf} \choose m} \!
(n-\shf)(-a^+_{qi}(n- \! 1)a^-_{pi}(m+\shf) +
a^-_{qi}(n- \! 1)a^+_{pi}(m+\shf))z^{-n-m-\hf}.\cr}\eqno(3.72)$$
Using
$${{-n-\hf} \choose m}(n-\shf) = -{{-n+\hf} \choose m}(-n-m+\shf)$$
and replacing $n$ by $n+1$, we get (b)-(e).
$\hfill\blacksquare$
\enddemo

\pr{Corollary 3.14} For $1\leq p\neq q \leq r$ and $Z_p = {\Bbb Z}+\hf$
on $\bold{WM}(Z_1) \otimes \hdots \otimes \bold{WM}(Z_r)$ we have
\roster
\item"(a)" $[L_{pq}(-1),J^\#_{pq}(z)] = 0$ for $^\# =\ ^-$ or $^+$
if $Z_q = {\Bbb Z}+\hf$,
\item"(b)" $[L_{pq}(-1),J^+_{pq}(z)] = (d/dz)J^0_{pq}(z)$ if
$Z_q = {\Bbb Z}$, which commutes with $J^+_{pq}(z)$ and $J^-_{pq}(z)$,
\item"(c)" $[L_{pq}(-1),J^-_{pq}(z)] = $
\nl $-\hf \ds{\sum \Sb {1\leq i\leq l}\\ 0\leq m\in{\Bbb Z}\endSb} {-\hf
\choose
m}(-a^+_{qi}(-1)a^-_{pi}(m+\shf) + a^-_{qi}(-1)a^+_{pi}(m+\shf))z^{-m-\hf}$
\nl if $Z_q = {\Bbb Z}$, which commutes with $J^-_{pq}(z)$ and $J^0_{pq}(z)$,
\item"(d)" $[L_{pq}(-1),J^0_{pq}(z)] = -(d/dz)J^0_{pq}(z)\ +$
\nl $\hf \ds{\sum \Sb {1\leq i\leq l}\\ 0\leq m\in{\Bbb Z}\endSb}
{-\hf \choose m}
(-a^+_{qi}(-1)a^-_{pi}(m+\shf) + a^-_{qi}(-1)a^+_{pi}(m+\shf))z^{-m-\hf}$
\nl if $Z_q = {\Bbb Z}$, which commutes with $J^-_{pq}(z)$,
\item"(e)" $[[L_{pq}(-1),J^0_{pq}(z)],J^0_{pq}(z)] = 2\Delta'_p(z)$
if $Z_q = {\Bbb Z}$, which commutes
with $J^\#_{pq}(z)$ for $^\# =\ ^-,\,^0$ or $^+$, and with $\Delta_p(z)$,
\item"(f)" $[L_{pq}(-1),\exp(J^0_{pq}(z))]
=\exp(J^0_{pq}(z))([L_{pq}(-1),J^0_{pq}(z)] + \Delta'_p(z))$
\nl if $Z_q = {\Bbb Z}$. \endroster \epr

\demo{Proof} Part (a) follows from parts (a) and (b) of the last proposition.
Parts (b)-(d) follow from parts (a) and (c)-(e) of the last proposition.
Part (e) follows from part (d) and the definitions. Part (f) follows from
part (e) and the formula (3.66) with $A = L_{pq}(-1)$ and $B =
-J_{pq}^0(z)$. $\hfill\blacksquare$
\enddemo

\pr{Theorem 3.15} For $1\leq p\neq q\leq r$ and $Z_p = {\Bbb Z}+\hf$
on $\bold{WM}(Z_1) \otimes \hdots \otimes \bold{WM}(Z_r)$ we have
$$[L_{pq}(-1),\colon\exp(J_{pq}(z))\colon] = \cases 0 &\text{if $Z_q = {\Bbb
Z}+\hf$}\cr\\
\Delta'_p(z)\colon\exp(J_{pq}(z))\colon &\text{if $Z_q = {\Bbb
Z}$}\endcases\eqno(3.73)$$
so
$$[L_{pq}(-1),\colon\exp(J_{pq}(z))\colon \exp(\Delta_p(z))] = 0 \ \hbox{ if
}\ Z_q = {\Bbb Z}.\eqno(3.74)$$ \epr

\demo{Proof} The result follows from the last corollary and (3.71).
$\hfill\blacksquare$
\enddemo

\pr{Corollary 3.16} For $1\leq p\neq q\leq r$ and $Z_p = {\Bbb Z}+\hf$
on $\bold{WM}(Z_1) \otimes \hdots \otimes \bold{WM}(Z_r)$ we have
$$\exp(\zeta L_{pq}(-1)) \colon\exp(J_{pq}(z))\colon\exp(-\zeta L_{pq}(-1))
= \colon \exp(J_{pq}(z))\colon\eqno(3.75)$$
if $Z_q = {\Bbb Z}+\hf$, which gives
$$\exp(\zeta L(-1))Y(v,z)\exp(-\zeta L(-1))w = Y(\exp(\zeta
L(-1))v,z)w\eqno(3.76)$$
for $v,w \in \bold{WM}({\Bbb Z}+\hf)$, and
$$\eqalign
{&\exp(\zeta L_{pq}(-1))\colon\exp(J_{pq}(z))\colon
\exp(\Delta_p(z))\exp(-\zeta L_{pq}(-1)) =\cr
&\colon\exp(J_{pq}(z))\colon \exp(\Delta_p(z))\cr}\eqno(3.77)$$
if $Z_q = {\Bbb Z}$, which gives
$$\exp(\zeta L(-1))Y(v,z) \exp(-\zeta L(-1))w = Y(\exp(\zeta
L(-1))v,z)w\eqno(3.78)$$
for $v \in \bold{WM}({\Bbb Z}+\hf)$ and $w \in \bold{WM}({\Bbb Z})$. \epr

\pr{Proposition 3.17} For $v \in \bold{V}$, on $\bold{V}(\hf {\Bbb Z})$ we have
\roster
\item"(a)" $[L(-1),Y(v,z)] = Y(L(-1)v,z) = (d/dz)Y(v,z)$,
\item"(b)" $[L(0),Y(v,z)] = z(d/dz)Y(v,z) + Y(L(0)v,z)$. \endroster \epr

\demo{Proof} (a) The first equality comes from (3.76) and (3.78) by taking the
derivative with respect to $\zeta$ and then setting $\zeta = 0$.
Proposition 3.1 (b)
gives the second equality on $\bold{WM}({\Bbb Z}+\hf)$, and with (3.71)
on $\bold{WM}({\Bbb Z})$ gives
$$\eqalign
{(d/dz)Y(v,z) &= (d/dz){\bar Y}(\exp(\Delta(z))v,z)\cr
&= {\bar Y}'(\exp(\Delta(z))v,z) + {\bar Y}((d/dz)\exp(\Delta(z))v,z)\cr
&= {\bar Y}'(\exp(\Delta(z))v,z) + {\bar
Y}(\Delta'(z)\exp(\Delta(z))v,z)\cr
&= {\bar Y}(L(-1)\exp(\Delta(z))v,z) - {\bar
Y}([L(-1),\exp(\Delta(z))]v,z)\cr
&= {\bar Y}(\exp(\Delta(z))L(-1)v,z)\cr
&= Y(L(-1)v,z).\cr}$$

(b) From Proposition 3.1 (c) and Lemma 3.10 (f) we have
$$\alignat2
&z(d/dz)Y(v,z) + Y(L(0)v,z)& \\
= &(d/dz){\bar Y}(\exp(\Delta(z))v,z) + {\bar
Y}(\exp(\Delta(z))L(0)v,z)& \\
= &z{\bar Y}'(\exp(\Delta(z))v,z) + z {\bar
Y}(\Delta'(z)\exp(\Delta(z))v,z)& \\
&+ {\bar Y}(L(0)\exp(\Delta(z))v,z) - {\bar Y}([L(0),\exp(\Delta(z))]v,z)&
\\
= &\ [L(0),{\bar Y}(\exp(\Delta(z))v,z)]& \\
= &\ [L(0),Y(v,z)].&\tag"$\blacksquare$"\endalignat$$
\enddemo

\pr{Proposition 3.18} For $1\leq p_1,p_2,q\leq r$ distinct and
$Z_{p_1} = Z_{p_2} = {\Bbb Z}+\hf$, on
$\bold{WM}(Z_1) \otimes \hdots \otimes \bold{WM}(Z_r)$ we have
$$\eqalign
{&[J_{p_1q}^{\#1}(\zeta_1),J_{p_2q}^{\#2}(\zeta_2)] =
\sum \Sb {1\leq i\leq l}\\ 0\leq m,n\in{\Bbb Z}\endSb
(-a^+_{p_2i}(n+\shf)a_{p_1i}^-(m+\shf)
+ a_{p_2i}^-(n+\shf)a^+_{p_1i}(m+\shf))  \cr
&\hbox to 3 truein{} \quad \cdot
[a_{qi}^{+(m)}(\zeta_1)^{\#1},a_{qi}^{-(n)}(\zeta_2)^{\#2}]\cr}$$
and the commutator
$[a_{qi}^{+(m)}(\zeta_1)^{\#1},a_{qi}^{-(n)}(\zeta_2)^{\#2}]$
is a formal series
independent of $i$ and $q$, depending only on $m,n$ and whether $Z_q = {\Bbb
Z}$ or ${\Bbb Z}+\hf$. \epr

\pr{Corollary 3.19} For $1\leq p_1,p_2,q\leq r$ distinct,
$Z_{p_1} = Z_{p_2} = {\Bbb Z}+\hf$ and $Z_q = {\Bbb Z}$, on
$\bold{WM}(Z_1) \otimes \hdots \otimes \bold{WM}(Z_r)$ we have
$$\eqalign
{&[J_{p_1q}^+(\zeta_1),J_{p_2q}^-(\zeta_2)] +
\shf[J_{p_1q}^0(\zeta_1),J_{p_2q}^0(\zeta_2)] =\cr
&\sum \Sb  {1\leq i\leq l}\\ 0\leq m,n\in{\Bbb Z}\endSb
(-a^+_{p_2i}(n+\shf)a_{p_1i}^-(m+\shf)
+ a_{p_2i}^-(n+\shf)a^+_{p_1i}(m+\shf))({\ubr{a_{qi}^{+(m)}(\zeta_1)
a_{qi}^{-(n)}(\zeta_2)}})\cr}$$
and the contraction is a formal series independent of $i$ and $q$, depending
only on $m,n$, given in Lemma 3.4. If $Z_q = {\Bbb Z}+\hf$ then we have the
same
formula with only the first bracket on the left side, and the
contraction given in Lemma 3.2. \epr

\pr{Definition}  For $1\leq p_1,p_2,q\leq r$ distinct, $Z_{p_1} = Z_{p_2} =
{\Bbb Z}+\hf$, let \break $K_{p_1p_2}(\zeta_1,\zeta_2;Z_q)$
denote the generating function of operators on $\bold{WM}(Z_1) \otimes
\hdots \otimes \bold{WM}(Z_r)$ given in Corollary 3.19. \epr

\pr{Proposition 3.20} For $1\leq p_1,p_2,q\leq r$ distinct,
$Z_{p_1} = Z_{p_2} = {\Bbb Z}+\hf$ and $Z_q = {\Bbb Z}$ or
$Z_q = {\Bbb Z}+\hf$, on $\bold{WM}(Z_1) \otimes \hdots \otimes
\bold{WM}(Z_r)$ we have
$$\eqalign
{&[J_{p_1q}^\#(\zeta_1),J_{p_2p_1}^-(\zeta_2)] =\cr
&\sum \Sb 1\leq i\leq l\\ 0\leq m \in{\Bbb Z},n\in Z^\#_q\endSb
(a^+_{qi}(n)a_{p_2i}^-(m+\shf) + a_{qi}^-(n)a^+_{p_2i}(m+\shf)) {{-n-\hf}
\choose
m} \zeta_1^{-m-n-\hf}\cr
&\hbox to 2.5 truein{} \cdot \sum_{0\leq t\in{\Bbb Z}} {{-m-n-\hf} \choose t}
\left(\frac{\zeta_2}{\zeta_1}\right)^t\cr
&\qquad\qquad\qquad\qquad
= J_{p_2q}^\#(\zeta_1 + \zeta_2)\cr}\eqno(3.79)$$
with all powers of $\zeta_1 + \zeta_2$
formally expanded in nonnegative powers of $\zeta_2$.
This operator commutes with $J_{p_2p_1}^\pm(\zeta_2)$,
so on $\bold{WM}(Z_1) \otimes \hdots \otimes \bold{WM}(Z_r)$ we have
$$\eqalign
{&\exp(J_{p_1q}^\#(\zeta_1))\exp(J_{p_2p_1}^-(\zeta_2)) = \cr
&\exp(J_{p_2p_1}^-(\zeta_2))\exp(J_{p_1q}^\#(\zeta_1) + J_{p_2q}^\#(\zeta_1
+ \zeta_2)).\cr}\eqno(3.80)$$ \epr

\demo{Proof} The definitions give
$$\eqalign
{&[J_{p_1q}^\#(\zeta_1),J_{p_2p_1}^-(\zeta_2)] \cr
&= \sum \Sb 1\leq i \leq l\\ 0\leq m,k\in{\Bbb Z}\endSb
(a_{qi}^{+(k)}(\zeta_1)^{\#} a_{p_2i}^-(m+\shf) +
a_{qi}^{-(k)}(\zeta_1)^{\#}a^+_{p_2i}(m+\shf)) {k \choose m} \zeta_2^{k-m}\cr
&= \sum \Sb 1\leq i \leq l\\ 0\leq m\in{\Bbb Z},n\in Z_q^\#\endSb
(a^+_{qi}(n)a_{p_2i}^-(m+\shf) + a^-_{qi}(n)a^+_{p_2i}(m+\shf))\cr
&\hbox to 1.7 truein{} \cdot \sum_{m\leq k\in {\Bbb Z}}
 {k \choose m} {{-n-\hf}
 \choose k}
\zeta_2^{k-m}\zeta_1^{-n-k-\hf}.\cr}\eqno(3.81)$$
Then using
$$\eqalign
{&\sum_{m\leq k\in{\Bbb Z}} {k \choose m} {{-n-\hf} \choose k}
\zeta_2^{k-m}\zeta_1^{-n-k-\hf}\cr
&= \zeta_1^{-m-n-\hf} \sum_{0\leq t\in{\Bbb Z}} {{m+t} \choose m} {{-n-\hf}
\choose {m+t}} \left(\frac{\zeta_2}{\zeta_1}\right)^t\cr
&= {{-n-\hf} \choose m} \zeta_1^{-m-n-\hf} \sum_{0\leq t\in{\Bbb Z}}
{{-m-n-\hf} \choose t} \left(\frac{\zeta_2}{\zeta_1}\right)^t\cr
&= {{-n-\hf} \choose m} \zeta_1^{-m-n-\hf} \left(1 +
\frac{\zeta_2}{\zeta_1}\right)^{-m-n-\hf}\cr}\eqno(3.82)$$
we get (3.79). The rest is clear.  $\hfill\blacksquare$
\enddemo

\pr{Definition} For $1\leq p_1,\hdots,p_t,q\leq r$ distinct,
$Z_{p_i} = {\Bbb Z}+\hf$, $1 \leq i \leq t$, let $J_i = J_{p_iq}(\zeta_i)$,
$J^\#_i = J_{p_iq}^\#(\zeta_i)$ for $^\# =\ ^-,\,^0$ or $^+$, and
$K_{ij} = K_{p_ip_j}(\zeta_i,\zeta_j;Z_q)$. Define
$$\eqalign
{&\colon \exp(J_1) \hdots \exp(J_t)\colon =\cr
&\exp(J^-_1 +\hdots + J^-_t)\exp(J^0_1 + \hdots + J^0_t)\exp(J^+_1 + \hdots
+ J^+_t).\cr}\eqno(3.83)$$ \epr

Note that Proposition 3.18 gives
$$[J^-_i,J^-_j] = 0 = [J^+_i,J^+_j]\ \hbox{ for }\ 1 \leq i,j \leq
t\eqno(3.84)$$
and
$$[[J^{\#1}_i,J^{\#2}_j],J^{\#3}_k] = 0\ \hbox{ for }\ 1\leq i,j,k \leq
t.\eqno(3.85)$$
Also note that for any permutation $\sigma \in S_t$, we have
$$\colon\exp(J_{\sigma 1}) \hdots \exp(J_{\sigma t})\colon = \colon
\exp(J_1) \hdots \exp(J_t)\colon.\eqno(3.86)$$

\pr{Lemma 3.21} With notation as above, on $\bold{WM}(Z_1) \otimes \hdots
\otimes \bold{WM}(Z_r)$ we have
$$\exp(J^0_1 + \hdots + J^0_t) \exp\left({\shf} \sum_{1\leq i<j\leq t}
[J^0_i,J^0_j]\right) = \exp(J^0_1)\hdots \exp(J^0_t).$$ \epr

\pr{Corollary 3.22} With notation as above, on $\bold{WM}(Z_1) \otimes
\hdots \otimes \bold{WM}(Z_r)$ we have
$$\colon \exp(J_1)\colon \hdots \colon \exp(J_t)\colon = \colon \exp(J_1)
\hdots \exp(J_t) \colon \exp\left(\sum_{1\leq i<j \leq t}K_{ij}\right).$$
\epr

\pr{Theorem 3.23} With notation as above, let $v_i \in
\bold{WM}(Z_i)^{\alpha_i}$, $1 \leq i \leq r$, $w' \in \bold{WM}(Z_q)$,
let $\bold{s}(v_{p_1},\hdots,v_{p_t};v_q) = (s_1,\hdots,s_t)$
be defined by
$$s_i = \cases \hf &\text{if $\alpha_{p_i} = 1$ and $Z_q = {\Bbb Z}$,}\cr\\
0 &\text{if $\alpha_{p_i} = 0$ or $Z_q = {\Bbb Z}+\hf$,}\endcases$$
and let $\zeta^{\bold s} = \ds{\prod_{1\leq i\leq t}}\zeta_i^{s_i}$.
Then we have the matrix coefficient
$$\zeta^{\bold s}\pi_1\hdots \pi_q(w') \hdots \pi_r(\colon \exp(J_1) \hdots
\exp(J_t)\colon v_1 \otimes \hdots \otimes v_r) \in {\Bbb
C}[\zeta_i,\zeta^{-1}_i,\ 1\leq i \leq t].$$ \epr

\demo{Proof} From (3.48) the powers of $\zeta_i$ in $J_i$
are in $\Bbb Z$ if $Z_q = {\Bbb Z}+\hf$, but they
are in ${\Bbb Z}+\hf$ if $Z_q = {\Bbb Z}$. We may assume $v_{p_i}$
is a vector with $k_i$ creation
operators applied to $\bold{vac}(Z_{p_i})$, so $k_i \equiv \alpha_{p_i}$
(mod 2). Let $J^\# = J^\#_1 + \hdots + J^\#_t$,
$^\# =\ ^-,\,^0$ or $^+$, and let $k = k_1 + \hdots + k_t$. Then, as in (3.54),
we have
$$\eqalign
{&\pi_1 \hdots \pi_q(w') \hdots\pi_r(\colon \exp(J_1) \hdots \exp(J_t)\colon
v_1
\otimes \hdots \otimes v_r) = \cr
&\pi_1\hdots \pi_q(w') \hdots \pi_r\left(\sum \Sb a+b+c=k\\ 0\leq a,b,c \in
{\Bbb Z}\endSb \frac{(J^-)^a (J^0)^b (J^+)^c}{a!b!c!} v_1 \otimes \hdots
\otimes
v_r\right)\cr}\eqno(3.87)$$
where, expanding each term of the sum, the only contributing terms
have total degree $k_i$ in $J^-_i$, $J^0_i$ and $J^+_i$,
$1\leq i \leq t$. It is clear from (3.53)
that $(J^0)^b(J^+)^c v_1 \otimes \hdots \otimes v_r$ is a finite sum.
Although $(J^-)^a$ applied to
that finite sum gives an infinite sum of terms, all but finitely many
are killed by $\pi_q(w')$. If $Z_q = {\Bbb Z}$, the total power of $\zeta_i$
in (3.87) is in $\Bbb Z$ if
$k_i$ is even $(\alpha_{p_i} = 0)$, but it is in ${\Bbb Z}+\hf$
if $k_i$ is odd $(\alpha_{p_i} = 1)$. $\hfill\blacksquare$
\enddemo

\pr{Lemma 3.24} With notation as in Theorem 3.23,
$$\zeta^{\bold s}\pi_1\hdots \pi_q(w') \hdots \pi_r(\colon \exp(J_1)\colon
\hdots \colon\exp(J_t)\colon v_1 \otimes \hdots \otimes v_r)$$
converges absolutely when $|\zeta_1| > \hdots > |\zeta_t| > 0$
to a rational function in
$${\Bbb C}[\zeta_i,\zeta^{-1}_i,(\zeta_j - \zeta_k)^{-1};\ 1\leq i \leq t,\
1\leq j < k \leq t].$$ \epr

\demo{Proof} The operators $K_{ij}$, $1\leq i < j \leq t$,
are given in Corollary 3.19. It is clear
that they commute with each other, and for any vector in
$\bold{WM}(Z_1) \otimes \hdots \otimes \bold{WM}(Z_r)$,
a sufficiently large power of $K_{ij}$ kills the vector.
If $Z_q = {\Bbb Z}+\hf$, then Corollary 3.22, Theorem 3.23, and Corollary 3.3
combine to give the result. If $Z_q = {\Bbb Z}$ then we can similarly get the
result using Corollary 3.5 in place of Corollary 3.3. The factor
$\zeta^{\bold s}$ occurs when $Z_q = {\Bbb Z}$ because $J_i$, $1\leq i\leq t$,
and $K_{ij}$, $1\leq i < j \leq t$, are series whose
variables occur with powers in ${\Bbb Z}+\hf$. If $v_{p_i}$ has
$k_i$ creation operators
applied to $\bold{vac}(Z_{p_i})$, then the only terms
contributing to the projection
$\pi_i$ have a total of $k_i$ operators $J_i$, $K_{ij}$
and $K_{li}$ applied, so the power of
$\zeta_i$ is in $\Bbb Z$ if $k_i$ is even $(\alpha_{p_i} = 0)$,
but is in ${\Bbb Z}+\hf$ if $k_i$ is odd $(\alpha_p = 1)$.
$\hfill\blacksquare$
\enddemo

\pr{Theorem 3.25} (Rationality) Let $v_1,\hdots,v_n \in \bold{V}_0 \cup
\bold{V}_2$, $w,w' \in \bold{WM}(Z)$, $\bold{s} = \bold{s}(v_1,\hdots,v_n;w)
= (s_1,\hdots,s_n)$.
Then the series $\zeta^{\bold s}(Y(v_1,\zeta_1) \hdots Y(v_n,\zeta_n)w,w')$
converges absolutely when $|\zeta_1| > \hdots > |\zeta_n| > 0$
to a rational function in ${\Bbb
C}[\zeta_i,\zeta_i^{-1}$,\newline
$(\zeta_j-\zeta_k)^{-1}$; $1 \leq i \leq n$, $1 \leq
j < k \leq n]$. \epr

\demo{Proof} For $Z = {\Bbb Z}+\hf$
this follows from Corollary 3.7, (3.58) and Lemma
3.24. For $Z = {\Bbb Z}$, one needs in addition the fact that for
$1\leq i \leq n$, $\exp(\Delta(\zeta_i))v_i$
is a finite sum of vectors (see Lemma 3.10 (b)) from $\bold{V}_0$
if $v_i \in\bold{V}_0$, or
from $\bold{V}_2$ if $v_i \in \bold{V}_2$.
So each $Y(v_i,\zeta_i)$ is a finite sum of $\bar Y$ terms.
$\hfill\blacksquare$
\enddemo

\pr{Theorem 3.26} (Permutability) Let $v_i \in \bold{WM}({\Bbb
Z}+\hf)^{\alpha_i}$, $1\leq i \leq r$, $w,w' \in \bold{WM}(Z)$, let
$\bold{s} = \bold{s}(v_1,\hdots,v_r;w) = (s_1,\hdots,s_r)$
and $\zeta^{\bold s} = \zeta_1^{s_1} \hdots \zeta_r^{s_r}$
be defined as in Theorem 3.23. Then for any $\sigma \in S_r$ we have
$$\zeta^{\bold s}(Y(v_1,\zeta_1) \hdots Y(v_r,\zeta_r)w,w') \sim
 \zeta^{\bold s}(Y(v_{\sigma 1},\zeta_{\sigma 1})
\hdots Y(v_{\sigma r},\zeta_{\sigma r})w,w').$$  \epr

\demo{Proof} Let $J_i = J_{i(r+1)}(\zeta_i)$ and $\Delta_i =
\Delta_i(\zeta_i)$ for $1 \leq i \leq r$, and $K_{ij} =
K_{ij}(\zeta_i,\zeta_j;Z)$
for $1\leq i\neq j \leq r$.
It is clear that it suffices to establish the
result for an adjacent transposition $\sigma = (t,t+1)$, $1\leq t < r$.
{}From Corollary 3.7 we see that when
$Z = {\Bbb Z}+\hf$ we have
$$\eqalign
{\zeta^{\bold s}&\pi_1\hdots \pi_r\pi_{r+1}(w')(\colon \exp(J_1)\colon
\hdots \colon\exp(J_r)\colon v_1\otimes \hdots \otimes v_r\otimes w)\cr
&= \zeta^{\bold s}(Y(v_1,\zeta_1) \hdots
Y(v_r,\zeta_r)w,w')\cr}\eqno(3.88)$$
but
$$\eqalign
{\zeta^{\bold s}&\pi_1\hdots \pi_r\pi_{r+1}(w')(\colon\exp(J_{\sigma
1})\colon \hdots \colon\exp(J_{\sigma r})\colon v_1\otimes \hdots \otimes
v_r\otimes w)\cr
&= \zeta^{\bold s}(Y(v_{\sigma 1},\zeta_{\sigma
1}) \hdots Y(v_{\sigma r},\zeta_{\sigma r})w,w').\cr}\eqno(3.89)$$
When $Z = {\Bbb Z}$, Corollary 3.7 gives the same result with
$\colon\exp(J_i)\colon$ replaced
by $\colon\exp(J_i)\colon\exp(\Delta_i)$ for $1\leq i \leq r$.
{}From Corollary 3.22, on $\bold{WM} = \bold{WM}({\Bbb Z}+\hf) \otimes \hdots
\otimes \bold{WM}({\Bbb Z}+\hf)\otimes \bold{WM}(Z)$ we have
$$\eqalign
{&\colon\exp(J_1)\colon \hdots \colon\exp(J_r)\colon\cr
&= \colon\exp(J_1)\hdots \exp(J_r)\colon \exp\left({\underset {1\leq i < j\leq
r}\to {\sum\nolimits{'}}}K_{ij}\right)\exp(K_{t(t+1)})\cr}\eqno(3.90)$$
and, using (3.82),
$$\eqalign
{&\colon \exp(J_{\sigma 1})\colon \hdots \colon \exp(J_{\sigma r})\colon\cr
&= \colon\exp(J_1) \hdots \exp(J_r)\colon \exp\left({\underset {1\leq i < j
\leq
r}\to {\sum\nolimits{'}}}K_{ij}\right)\exp(K_{(t+1)t})\cr}\eqno(3.91)$$
where $\sum'$ indicates that the summation excludes $(i,j) = (t,t+1)$.
Note that
the operators $\Delta_i$, $1\leq i \leq r$, commute with each other,
and with the $J$ and $K$
operators, so they are unaffected by $\sigma$. From (3.90) and (3.91) the
only issue is the relationship between $\exp(K_{t(t+1)})$ and
$\exp(K_{(t+1)t})$.
When they are applied to any vector in $\bold{WM}$, they each give a finite sum
of vectors with coefficients which are series in $\zeta_t$ and $\zeta_{t+1}$.
{}From
Corollary 3.19 we see that the operator part of $K_{t(t+1)}$ is the negative
of the operator part of $K_{(t+1)t}$. Thus, it suffices to show that the
function to which the contraction part of $K_{t(t+1)}$ converges (in the
appropriate domain) is the negative of the function to which the
contraction part of $K_{(t+1)t}$ converges (in the appropriate domain). When
$Z = {\Bbb Z}+\hf$ this is evident from Corollary 3.3, and when
$Z = {\Bbb Z}$ it follows from Corollary 3.5. $\hfill\blacksquare$
\enddemo

For $v_1,\hdots,v_n,w,w' \in \bold{V}$,
using compositions of vertex operators we
define the generating function of matrix coefficients
$$\eqalign
{&(Y(\hdots Y(Y(v_1,\zeta_1)v_2,\zeta_2) \hdots v_n,\zeta_n)w,w')\cr
&= \sum_{m_1,\hdots,m_n \in\hf {\Bbb Z}}(Y_{m_n}(\hdots
Y_{m_2}(Y_{m_1}(v_1)v_2)v_3 \hdots v_n)w,w')\  \cr
&\qquad\qquad\cdot\zeta_1^{-m_1-wt(v_1)} \zeta_2^{-m_2+m_1-wt(v_2)} \hdots
\zeta_n^{-m_n+m_{n-1}-wt(v_n)}.\cr}\eqno(3.92)$$

\pr{Theorem 3.27} (Associativity) Let $v_i \in\bold{WM}({\Bbb
Z}+\hf)^{\alpha_i}$, $i = 1,2$, $w,w' \in \bold{WM}(Z)$ and let
$(s_1,s_2) = \bold{s}(v_1,v_2;w)$ be as defined in Theorem 3.23.
Then we have
$$\eqalign
{&\zeta_1^{s_1}\ \zeta_2^{s_2}\ (Y(v_1,\zeta_1)Y(v_2,\zeta_2)w,w')
\sim\cr
&\zeta_2^{s_1+s_2}\ (1 + (\zeta_1 -
\zeta_2)/\zeta_2)^{s_1} (Y(Y(v_1,\zeta_1-\zeta_2)v_2,\zeta_2)w,w')\cr}$$
where the series on the right side converges absolutely when
$0 < |\zeta_1 - \zeta_2| < |\zeta_2|$. \epr

\demo{Proof} There are two cases:(a) $Z = {\Bbb Z}+\shf$, in which case
$s_1 = s_2 = 0$; (b) $Z = {\Bbb Z}$, in which case $s_1 = \shf \alpha_1$ and
$s_2 = \shf \alpha_2$.

(a) From Theorem 3.25 we know that the series in
$\zeta_1$ and $\zeta_2$,
$$(Y(v_1,\zeta_1)Y(v_2,\zeta_2)w,w') =
\pi_1\pi_2\pi_3(w')(\colon\exp(J_{13}(\zeta_1))\colon
\colon\exp(J_{23}(\zeta_2))\colon v_1 \otimes v_2 \otimes w)$$
converges absolutely when $|\zeta_1| > |\zeta_2| > 0$ to a rational function in
${\Bbb C}[\zeta_1,\zeta_1^{-1},\zeta_2,\break
\zeta_2^{-1}, (\zeta_1-\zeta_2)^{-1}]$.
We will show that the series in $\zeta_1 - \zeta_2$ and $\zeta_2$,
$$\eqalign
{&(Y(Y(v_1,\zeta_1-\zeta_2)v_2,\zeta_2)w,w')\cr
&= \pi_1\pi_2\pi_3(w')(\colon \exp(J_{23}(\zeta_2))\colon
\colon\exp(J_{12}(\zeta_1-\zeta_2))\colon v_1\otimes v_2\otimes w)\cr}$$
converges absolutely when $0 < |\zeta_1 - \zeta_2| < |\zeta_2|$
to the same rational function.

We begin in the middle by applying the ``normally ordered'' expression
$$\exp(J^-_{12}(\zeta))\exp(J^-_{13}(\zeta_2 + \zeta)
+ J^-_{23}(\zeta_2))\exp(J^+_{13}(\zeta_2 + \zeta)
+ J^+_{23}(\zeta_2) + J^+_{12}(\zeta))\eqno(3.93)$$
to $v_1 \otimes v_2 \otimes w$ and then applying the projections
$\pi_1(u_1)\pi_2(u_2)\pi_3(u_3)$ for
arbitrary vectors $u_1,u_2,u_3 \in \bold{WM}({\Bbb Z}+\hf)$.
The result is a rational function
$f(\zeta,\zeta_2)$ in
${\Bbb C}[\zeta,\zeta^{-1},\zeta_2,\zeta^{-1}_2,(\zeta_2+\zeta)^{-1}]$.
Note that $J^+_{13}(\zeta_2 + \zeta)$, $J^+_{23}(\zeta_2)$ and
$J^+_{12}(\zeta)$ commute, and $J^-_{13}(\zeta_2+\zeta)$ commutes with
$J^-_{12}(\zeta)$ and $J^-_{23}(\zeta_2)$, but
$J^-_{12}(\zeta)$ does not commute with $J^-_{23}(\zeta_2)$.
If in (3.93) we expand $J^-_{13}(\zeta_2+\zeta)$
and $J^+_{13}(\zeta_2+\zeta)$ in nonnegative powers of $\zeta$,
then the resulting series in $\zeta$
and $\zeta_2$ converges absolutely to $f(\zeta,\zeta_2)$ when
$0 < |\zeta| < |\zeta_2|$. From Propostion 3.20 we have
$$\exp(J^-_{12}(\zeta))\exp(J^-_{13}(\zeta_2+\zeta) + J^-_{23}(\zeta_2)) =
\exp(J^-_{23}(\zeta_2))\exp(J^-_{12}(\zeta))\eqno(3.94)$$
and then
$$\exp(J^-_{12}(\zeta))\exp(J^+_{13}(\zeta_2+\zeta)+J^+_{23}(\zeta_2))
= \exp(J^+_{23}(\zeta_2))
\exp(J^-_{12}(\zeta))\eqno(3.95)$$
so that
$$\pi_1(u_1)\pi_2(u_2)\pi_3(u_3)(\colon\exp(J_{23}(\zeta_2))\colon
\colon\exp(J_{12}(\zeta))\colon v_1 \otimes v_2 \otimes w)\eqno(3.96)$$
converges absolutely to $f(\zeta,\zeta_2)$ when $0 < |\zeta| < |\zeta_2|$.

Returning to expression (3.93), substituting
$\zeta_1 = \zeta_2 + \zeta$ gives the
rational function $f(\zeta_1-\zeta_2,\zeta_2)$.
{}From Corollary 3.3, the contraction part of
each term in $K_{12}(\zeta_1,\zeta_2;{\Bbb Z}+\hf)$
given in Corollary 3.19 converges absolutely when
$|\zeta_1| > |\zeta_2|$ to ${{-n-1} \choose m}(\zeta_1 - \zeta_2)^{-m-n-1}$,
so we get $K_{12}(\zeta_1,\zeta_2;{\Bbb Z}+\hf) = J^+_{12}(\zeta_1 - \zeta_2)$
when expanded in nonnegative powers of $\zeta_2$. Then Corollary
3.22 says that
$$\pi_1(u_1)\pi_2(u_2)\pi_3(u_3)(\exp(J^-_{12}(\zeta_1 - \zeta_2))
\colon\exp(J_{13}(\zeta_1))\colon \colon
\exp(J_{23}(\zeta_2))\colon v_1 \otimes v_2 \otimes w)\eqno(3.97)$$
converges absolutely to $f(\zeta_1 - \zeta_2,\zeta_2)$
when $|\zeta_1| > |\zeta_2| > 0$. For any
$v_1,v_2,w,w' \in \bold{WM}({\Bbb Z}+\hf)$
it is easy to see that with $u_1 = u_2 = \bold{vac}({\Bbb Z}+\hf)$, we
have
$$\pi_1\pi_2\pi_3(w')(J^-_{12}(\zeta_1 - \zeta_2)v_1 \otimes v_2 \otimes w) =
0.\eqno(3.98)$$
Therefore,
$$\pi_1\pi_2\pi_3(w')(\colon\exp(J_{13}(\zeta_1))\colon
\colon\exp(J_{23}(\zeta_2))\colon v_1 \otimes v_2 \otimes w)\eqno(3.99)$$
converges absolutely to $f(\zeta_1 - \zeta_2,\zeta_2)$ when
$|\zeta_1| > |\zeta_2| > 0$, and from (3.96),
$$\pi_1\pi_2\pi_3(w')(\colon\exp(J_{23}(\zeta_2))\colon \colon
\exp(J_{12}(\zeta_1-\zeta_2))\colon v_1 \otimes v_2 \otimes w)\eqno(3.100)$$
converges absolutely to $f(\zeta_1-\zeta_2,\zeta_2)$
when $0 < |\zeta_1-\zeta_2| < |\zeta_2|$.

(b) From Theorem 3.25 we know that the series in $\zeta_1$ and
$\zeta_2$,
$$\eqalign
{\zeta_1^{s_1}\zeta_2^{s_2}(Y(v_1,\zeta_1)Y(v_2,&\zeta_2)w,w')
\cr
= \zeta_1^{s_1}\zeta_2^{s_2}\pi_1\pi_2\pi_3(w')(&\colon\exp
(J_{12}(\zeta_1))\colon \exp(\Delta_1(\zeta_1))\cr
\cdot\ &\colon\exp(J_{23}(\zeta_2))\colon \exp(\Delta_2(\zeta_2) v_1\otimes v_2
\otimes w)\cr}\eqno(3.101)$$
converges absolutely when $|\zeta_1| > |\zeta_2| > 0$
to a rational function $f(\zeta_1,\zeta_2)$
in ${\Bbb C}[\zeta_1,\zeta_1^{-1},\zeta_2,\zeta_2^{-1},(\zeta_1-
\zeta_2)^{-1}]$.
We will show that the series in $\zeta = \zeta_1 - \zeta_2$ and $\zeta_2$,
$$\eqalign
{\zeta_2^{s_1+s_2}\
(1 + \zeta/\zeta_2)^{s_1}\ (Y(Y(v_1,\zeta) v_2,&\zeta_2)w,w') \cr
= \zeta_2^{s_1+s_2}\ (1 + \zeta/\zeta_2)^{s_1}\
\pi_1\pi_2\pi_3(w')(&\colon\exp(J_{23}(\zeta_2))\colon
\exp(\Delta_2(\zeta_2))\cr
\cdot\ &\colon\exp(J_{12}(\zeta))\colon v_1 \otimes v_2 \otimes
w)\cr}\eqno(3.102)$$
converges absolutely when $0 < |\zeta| < |\zeta_2|$ to
$f(\zeta_2+\zeta,\zeta_2)$.

{}From Corollary 3.5, the contraction part of each term in
$K_{12}(\zeta_1,\zeta_2;{\Bbb Z})$
converges absolutely when $|\zeta_1| > |\zeta_2| > 0$, $\zeta_1,\zeta_2 \in
{\Bbb C}^-$ to
$$\zeta_1^{-m-\hf}\zeta_2^{-n-\hf}(\zeta_1-\zeta_2)^{-m-n-1}P_{mn}
(\zeta_1,\zeta_2)\eqno(3.103)$$
for some homogeneous polynomial $P_{mn}(\zeta_1,\zeta_2)$ of degree $m+n+1$.
Let
\break ${\bar K}_{12}(\zeta_1,\zeta_2;{\Bbb Z})$
be the expression for $K_{12}(\zeta_1,\zeta_2;{\Bbb Z})$
with each contraction part
replaced by the corresponding algebraic function (3.103). Then
$$\exp({\bar K}_{12}(\zeta_1,\zeta_2;{\Bbb
Z}))\exp(\Delta_2(\zeta_2))\exp(\Delta_1(\zeta_1))v_1 \otimes v_2 \otimes
w\eqno(3.104)$$
is a finite sum of vectors with algebraic function coefficients analytic
for $\zeta_1,\zeta_2\in{\Bbb C}^-$, $\zeta_1\neq \zeta_2$,
so for any $u_1,u_2 \in\bold{WM}({\Bbb Z}+\hf)$, $u_3\in\bold{WM}({\Bbb
Z})$,
$$\eqalign
{\pi_1(u_1)\pi_2&(u_2)\pi_3(u_3)(\exp(J^-_{12}(\zeta_1-\zeta_2))
\colon\exp(J_{13}(\zeta_1))\exp(J_{23}(\zeta_2))\colon\cr
\cdot\ &\exp({\bar K}_{12}(\zeta_1,\zeta_2;{\Bbb
Z}))\exp(\Delta_2(\zeta_2))\exp(\Delta_1(\zeta_1))v_1 \otimes v_2 \otimes
w)\cr}\eqno(3.105)$$
is a polynomial in ${\Bbb
C}[\zeta_1^{1/2},\zeta_1^{-1/2},\zeta_2^{1/2},\zeta_2^{-1/2},(\zeta_1 -
\zeta_2)^{-1}]$ which is analytic
for $\zeta_1,\zeta_2 \in {\Bbb C}^-$, $\zeta_1\neq \zeta_2$.
By Corollary 3.22, when $|\zeta_1| > |\zeta_2| > 0$, $\zeta_1,\zeta_2\in{\Bbb
C}^-$,
$$\eqalign
{\pi_1(u_1)\pi_2(u_2)\pi_3(u_3)(&\exp(J^-_{12}(\zeta_1-\zeta_2))
\colon\exp(J_{13}(\zeta_1))\colon \colon\exp(J_{23}(\zeta_2))\colon \cr
\cdot&\ \exp(\Delta_2(\zeta_2)) \exp(\Delta_1(\zeta_1)) v_1 \otimes v_2 \otimes
w)\cr}\eqno(3.106)$$
converges absolutely to the algebraic function (3.105).

Letting $\zeta_1 = \zeta_2 + \zeta$ in (3.105), we get an algebraic function
$$\eqalign
{\pi_1&(u_1)\pi_2(u_2)\pi_3(u_3)\exp(J^-_{12}(\zeta))\exp(J^-_{13}(\zeta_2 +
\zeta) + J^-_{23}(\zeta_2)) \cr
\cdot\  &\exp(J^0_{13}(\zeta_2 +\zeta)+J^0_{23}(\zeta_2))\exp(J^+_{13}(\zeta_2
+
\zeta) + J^+_{23}(\zeta_2)) \cr
\cdot\ &\exp({\bar K}_{12}(\zeta_2 + \zeta,\zeta_2;{\Bbb Z}))
\exp(\Delta_2(\zeta_2)) \exp(\Delta_1(\zeta_2+\zeta)) v_1 \otimes v_2
\otimes w)\cr}\eqno(3.107)$$
analytic for $\zeta_2 + \zeta$, $\zeta_2 \in {\Bbb C}^-$, $\zeta \neq 0$.
If we further restrict to the domain
$$\alignat3
\zeta_2 \in {\Bbb C}^-, &\quad 0 < |\zeta| < |\zeta_2|
&\quad \hbox{if }Re(\zeta_2) \geq
0, & \\
&\quad 0 < |\zeta| < |Im(\zeta_2)| &\quad \hbox{ if }Re(\zeta_2) <
0,&\tag3.108\endalignat$$
then we have
$$(\zeta_2 + \zeta)^{1/2} = \zeta_2^{1/2}(1 +
\zeta/\zeta_2)^{1/2}\eqno(3.109)$$
and in (3.107) we can expand all powers of $\zeta_2 + \zeta$
in nonnegative powers of $\zeta$. Then Proposition 3.20 gives
$$\exp(J^-_{12}(\zeta))\exp(J^-_{13}(\zeta_2 + \zeta) + J^-_{23}(\zeta_2)) =
\exp(J^-_{23}(\zeta_2)) \exp(J^-_{12}(\zeta)),\eqno(3.110)$$
$$\exp(J^-_{12}(\zeta))\exp(J^0_{13}(\zeta_2 + \zeta) + J^0_{23}(\zeta_2)) =
\exp(J^0_{23}(\zeta_2)) \exp(J^-_{12}(\zeta)),\eqno(3.111)$$
and
$$\exp(J^-_{12}(\zeta))\exp(J^+_{13}(\zeta_2 + \zeta) + J^+_{23}(\zeta_2)) =
\exp(J^+_{23}(\zeta_2)) \exp(J^-_{12}(\zeta)).\eqno(3.112)$$
So
$$\eqalign
{\pi_1(u_1)\pi(u_2)\pi_3(u_3)(\colon&\exp(J_{23}(\zeta_2))\colon
\exp(J^-_{12}(\zeta)) \exp({\bar K}_{12}(\zeta_2 + \zeta,\zeta_2;{\Bbb Z}))
 \cr
\cdot\ &\exp(\Delta_2(\zeta_2)) \exp(\Delta_1(\zeta_2+\zeta)) v_1 \otimes v_2
\otimes w)\cr}\eqno(3.113)$$
converges absolutely in (3.108) to the algebraic function in (3.107). From
Lemma 3.28 (proven below) and Proposition 3.11 (a), we have
$${\bar K}_{12}(\zeta_2+\zeta,\zeta_2;{\Bbb Z}) =
[\Delta_2(\zeta_2),J^-_{12}(\zeta)] + J^+_{12}(\zeta)\eqno(3.114)$$
in the domain (3.108). From (3.114) and Corollary 3.12 we get that (3.113)
equals
$$\eqalign
{&\pi_1(u_1)\pi_2(u_2)\pi_3(u_3)(\colon \exp(J_{23}(\zeta_2))\colon
\colon\exp(J_{12}(\zeta))\colon \cr
&\qquad\qquad\cdot\ \exp(\Delta_2(\zeta_2) + [\Delta_2(\zeta_2),
 J^-_{12}(\zeta)] +
\Delta_1(\zeta_2 + \zeta)) v_1 \otimes v_2 \otimes w)\cr
&= \pi_1(u_1)\pi_2(u_2)\pi_3(u_3) (\colon \exp(J_{23}(\zeta_2))\colon
\exp(\Delta_2(\zeta_2)) \colon\exp(J_{12}(\zeta))\colon v_1 \otimes v_2
\otimes w)\cr}\eqno(3.115)$$
in the domain (3.108). Note that this converges absolutely to a
polynomial in ${\Bbb C}[(1 + \zeta/\zeta_2)^{1/2},(1 +
\zeta/\zeta_2)^{-1/2},\zeta^{1/2}_2,\zeta^{-1/2}_2,\zeta^{-1}]$, and
therefore has the larger domain
$0 < |\zeta| < |\zeta_2|$, $\zeta_2 \in {\Bbb C}^-$.

Letting $u_1 = u_2 = \bold{vac}({\Bbb Z}+\hf)$ and $u_3 = w' \in
\bold{WM}({\Bbb Z})$ in (3.106), only the
constant term of the factor $\exp(J^-_{12}(\zeta_1 - \zeta_2))$
contributes, so after
multiplying by $\zeta_1^{s_1}\zeta_2^{s_2}$,
we get the rational function $f(\zeta_1,\zeta_2)$.
Therefore, with the same values for $u_1$, $u_2$ and $u_3$ in (3.113),
and with $\zeta_1 = \zeta_2 + \zeta$,
multiplying by the algebraic function
$(\zeta_2+\zeta)^{s_1}\zeta_2^{s_2}$, yields
the rational function $f(\zeta_2 + \zeta,\zeta_2)$
in the domain (3.108). Using (3.109),
which is valid in that domain, and expanding
$(1 + \zeta/\zeta_2)^{s_1}$ in
nonnegative powers of $\zeta$, we get that (3.102) converges absolutely to
$f(\zeta_2+\zeta,\zeta_2)$.
Since this function is in ${\Bbb
C}[\zeta_1,\zeta^{-1}_1,\zeta_2,\zeta^{-1}_2,(\zeta_1 - \zeta_2)^{-1}]$, that
series actually converges absolutely in the larger domain $0 < |\zeta| <
|\zeta_2|$.  $\hfill\blacksquare$
\enddemo

\pr{Lemma 3.28} For $0 \leq m,n \in{\Bbb Z}$, the series
$${{-n-1} \choose m}\zeta^{-m-n-1} +\zeta_2^{-m-n-1} \sum_{0\leq k \in {\Bbb
Z}}{{m+k}\choose m}
C_{n(m+k)}\left(\frac{\zeta}{\zeta_2}\right)^k\eqno(3.116)$$
converges absolutely to the algebraic function
$$f_{mn}(\zeta,\zeta_2) = (\zeta_2 +\zeta)^{-m-\hf}\zeta_2^{-n-\hf}
\zeta^{-m-n-1}P_{mn}(\zeta_2 + \zeta,\zeta_2)\eqno(3.117)$$
defined in Corollary 3.5, in the domain
$$\alignat3
\zeta_2 \in{\Bbb C}^-, &\quad 0 < |\zeta| < |\zeta_2|
&\quad \hbox{ if }Re(\zeta_2) \geq
0,& \\
&\quad 0 < |\zeta| < |Im(\zeta_2)| &\quad \hbox{if }Re(\zeta_2) <
0.&\tag3.118\endalignat$$ \epr

\demo{Proof} From Corollary 3.5, the contraction
${\ubr{a_i^{+(m)}(\zeta_2+\zeta) a^{-(n)}_i(\zeta_2)}}$
converges absolutely when $|\zeta_2 +\zeta| > |\zeta_2|$
to the algebraic function
$f_{mn}(\zeta,\zeta_2)$. Therefore, ${\ubr{a_i^{+(m+1)}(\zeta_2 +
\zeta)a^{-(n)}_i(\zeta_2)}}$ converges absolutely in the
same domain to \nl $(m+1)^{-1}(\partial/\partial\zeta)f_{mn}(\zeta,\zeta_2)$
and ${\ubr{a_i^{+(m)}(\zeta_2 + \zeta)a^{-(n+1)}_i(\zeta_2)}}$
converges absolutely in that domain to $(n+1)^{-1}
(\partial/\partial\zeta_2 - \partial/\partial \zeta)f_{mn}(\zeta,\zeta_2)$.
So
$$(m+1)^{-1}(\partial/\partial\zeta)f_{mn}(\zeta,\zeta_2) =
f_{(m+1)n}(\zeta,\zeta_2)\eqno(3.119)$$
and
$$(n+1)^{-1} (\partial/\partial\zeta_2 -
\partial/\partial\zeta)f_{mn}(\zeta,\zeta_2) =
f_{m(n+1)}(\zeta,\zeta_2).\eqno(3.120)$$
It is easy to see that applying $(m+1)^{-1}(\partial/\partial\zeta)$
to the series (3.116)
yields the same series with $m$ replaced by $m+1$. So it suffices to check
the assertion for $m = 0$. Using (3.56) one finds that applying
$(n+1)^{-1}(\partial/\partial \zeta_2 - \partial/\partial\zeta)$
to (3.116) with $m = 0$ yields that series with $m = 0$
and $n$ replaced by $n+1$.
So it suffices to check the assertion for $m=n=0$.

We have
$$f_{00}(\zeta,\zeta_2)
= \frac{\zeta_2 + \shf\zeta} {(\zeta_2+\zeta)^{1/2}\zeta_2^{1/2}\zeta}
= \frac{\zeta_2 + \shf\zeta} {(1 + \zeta/\zeta_2)^{1/2}\zeta_2\zeta}
\eqno(3.121)$$
in the domain (3.118). For $|w| < 1$,
$$(1 + w)^{-1/2} = \sum_{0\leq k \in{\Bbb Z}} {{-\hf} \choose k}
w^k\eqno(3.122)$$
converges absolutely. Integrating from $0$ to $z$, $|z| < 1$, and dividing by
$2z$, we get
$$\frac{(1 + z)^{1/2} - 1}{z} = \sum_{0\leq k \in{\Bbb Z}}{-\hf \choose k}
\frac{z^k}{2(k+1)}.\eqno(3.123)$$
Differentiating and then multiplying by $-z$ yields
$$\frac{z+2}{2z(1 + z)^{1/2}} - \frac{1}{z} = - \sum_{0\leq k \in{\Bbb Z}}
{-\shf \choose k} \frac{k z^k}{2(k+1)}\eqno(3.124)$$
so that
$$\frac{z+2}{2z(1 + z)^{1/2}} = \frac{1}{z} + \sum_{0\leq k \in {\Bbb Z}}
C_{0k}z^k.\eqno(3.125)$$
Finally, substituting $z = \zeta/\zeta_2$ gives the result.
$\hfill\blacksquare$
\enddemo

\pr{Theorem 3.29} (The Jacobi-Cauchy Identity on $\bold{WM}(Z)$)
Let $v_i \in \bold{WM}({\Bbb Z}+\hf)^{\alpha_i} = \bold{V}_{2\alpha_i}$
for $i = 1,2$, $w,w' \in \bold{WM}(Z)$, and let
$f(\zeta_1,\zeta_2) \in {\Bbb
C}[\zeta_1,\zeta^{-1}_1,\zeta_2,\zeta^{-1}_2$,$(\zeta_1 - \zeta_2)^{-1}]$.
Let $(s_1,s_2) = \bold{s}(v_1,v_2;w)$ be as defined in Theorem 3.23.
We have
$$\eqalign{
&-Res_{\zeta_1=\infty}\zeta_1^{s_1}\zeta_2^{s_2}
(Y(v_1,\zeta_1)Y(v_2,\zeta_2)w,w')f(\zeta_1,\zeta_2)\cr
&-Res_{\zeta_1=0}\zeta_1^{s_1}\zeta_2^{s_2}
(Y(v_2,\zeta_2)Y(v_1,\zeta_1)w,w')f(\zeta_1,\zeta_2)\cr
&=Res_{\zeta_1=\zeta_2}
\zeta_2^{s_1+s_2} (1 + (\zeta_1 -
\zeta_2)/\zeta_2)^{s_1}
(Y(Y(v_1,\zeta_1 -
\zeta_2)v_2,\zeta_2)w,w')f(\zeta_1,\zeta_2)\cr}$$
where the function $f(\zeta_1,\zeta_2)$ is to be expanded in the three
residues as in (1.51)-(1.53),
 respectively, with $z = \zeta_1$ and $z_0 =
\zeta_2$. \epr

\demo{Proof} From the assumption on $f(\zeta_1,\zeta_2)$,
Theorems 3.26 and 3.27 say
that all three expressions represent the same rational function of $\zeta_1$
and
$\zeta_2$ having possible poles only at $\zeta_1 = 0$, $\zeta_2 = 0$
and $\zeta_1 = \zeta_2$.
Cauchy's residue theorem
gives the result. $\hfill\blacksquare$
\enddemo

\pr{Definition} For operators $L_1$ and $L_2$ on $\bold{V}(\hf{\Bbb Z})$
let $[L_1,L_2] = L_1L_2 - L_2L_1$. \epr

\pr{Corollary 3.30} Let $v_i \in\bold{WM}({\Bbb Z}+\hf)^{\alpha_i} =
\bold{V}_{2\alpha_i}$ for $i = 1,2$. Then for
any $r \in {\Bbb Z}$, for $m,n \in {\Bbb Z}$ on $\bold{WM}({\Bbb Z}+\hf)$,
and for $m \in {\Bbb Z}+\hf\alpha_1$, $n \in {\Bbb Z}+\hf\alpha_2$ on
$\bold{WM}({\Bbb Z})$, we have
$$\eqalign
{&\sum_{0 \leq i \in{\Bbb Z}}(-1)^i {r\choose i}
(\{v_1\}_{m+r-i}\{v_2\}_{n+i} - (-1)^r
\{v_2\}_{n+r-i}\{v_1\}_{m+i})\cr
&= \sum_{0\leq k\in{\Bbb Z}} {m \choose k}
\{\{v_1\}_{r+k}v_2\}_{m+n-k}\ .\cr}\eqno(3.126)$$
The sum over $k$ is finite, and if
$k_i = \hbox{wt}(v_i)$ then $0 \leq k \leq k_1 + k_2 - r - 1$.  \epr

\demo{Proof} Let
$v_i \in \bold{WM}({\Bbb Z}+\hf)^{\alpha_i}$, $w,w' \in \bold{WM}(Z)$,
$n \in{\Bbb Z} + s_i$, where $(s_1,s_2) = \bold{s}(v_1,v_2;w)$, that is,
$n \in{\Bbb Z}$ if $Z = {\Bbb Z}+\hf$, $n \in {\Bbb Z}+\hf\alpha_i$ if
$Z = {\Bbb Z}$.
{}From (3.59) and the discussion after
(3.25), Cauchy's theorem gives
$$Res_{\zeta_i=0}(Y(v_i,\zeta_i)w,w')\zeta^n_i=
(\{v_i\}_nw,w')\eqno(3.127)$$
Use Theorem 3.29 with
$f(\zeta_1,\zeta_2) = \zeta_1^{m-s_1}\zeta_2^{n-s_2}(\zeta_1 - \zeta_2)^r$.
In the first residue on the left side we have
$|\zeta_1| > |\zeta_2|$, so we should expand $(\zeta_1 - \zeta_2)^r$
as $\zeta^r_1(1 - \zeta_2/\zeta_1)^r$, but in the
second residue we have $|\zeta_2| > |\zeta_1|$, so we should expand
$(\zeta_1 - \zeta_2)^r$ as
$(-\zeta_2)^r(1 - \zeta_1/\zeta_2)^r$.
Then (3.127) gives the left side of (3.126)
applied to $w$ and paired with $w'$.
With the substitution $\zeta = \zeta_1 - \zeta_2$ where
$|\zeta_2|>|\zeta|>0$, we expand $\zeta_1^m$ as $\zeta_2^m\left(1+
\frac{\zeta}{\zeta_2}\right)^m$ in the residue on the right
side of Theorem 3.29.
This gives
$$\eqalign{
&Res_{\zeta_2=0}Res_{\zeta=0}
\sum \Sb {p \in {\Bbb Z}}\\
{q \in {\Bbb Z}+s_1+s_2} \endSb (\{\{v_1\}_pv_2\}_qw,w')
\zeta^{r-p-1}\ \zeta_2^{m+n-q-1}\ (1 + \zeta/\zeta_2)^m \cr
&=
Res_{\zeta_2=0}Res_{\zeta=0}
\sum \Sb {p \in {\Bbb Z}}\\
{q \in {\Bbb Z}+s_1+s_2} \endSb \!\! \sum_{0\leq k\in {\Bbb Z}}
\! {m \choose k} (\{\{v_1\}_pv_2\}_qw,w')
\zeta^{r-p-1+k}\zeta_2^{m+n-q-1-k}\cr
&= \sum_{0\leq k\in{\Bbb Z}} {m \choose k}
(\{\{v_1\}_{r+k}v_2\}_{m+n-k}w,w').\cr}\eqno(3.128)$$
For $k_i = \hbox{wt}(v_i)$, wt$(\{v_1\}_pv_2) = k_1
+ k_2 - p -1$,
so $\{v_1\}_pv_2 =0$ if $p \geq k_1 + k_2$. Since any $v_1$, $v_2$ are finite
sums of such homogeneous vectors, the sum is finite. Since the
Hermitian form on $\bold{WM}(Z)$ is nondegenerate, we have the assertion of the
corollary. $\hfill\blacksquare$
\enddemo

\pr{Corollary 3.31} Under the assumptions of Corollary 3.30, for $m,n\in{\Bbb
Z}$ on $\bold{WM}({\Bbb Z}+\hf)$,
and for $m \in {\Bbb Z}+\hf\alpha_1$, $n \in{\Bbb Z}+\hf\alpha_2$
on $\bold{WM}({\Bbb Z})$, we have
$$[\{v_1\}_m,\{v_2\}_n] = \sum_{0\leq k\in{\Bbb Z}} {m \choose k}
\{\{v_1\}_kv_2\}_{m+n-k}.\eqno(3.129)$$
This sum is
finite, and if $k_i = \hbox{wt}(v_i)$ then $0 \leq k \leq k_1 + k_2 - 1$.
\epr

\demo{Proof} Take $r = 0$ in Corollary 3.30. $\hfill\blacksquare$
\enddemo

Define the isomorphisms
$$\Upsilon:{\bold A} \to ({\bold V}_2)_{1/2},\ \ \ \ \Upsilon':{\bold
sp}(2l) \to ({\bold V}_0)_1\eqno(3.130)$$
by
$$\Upsilon(a_1) = a_1(-\shf)\bold{vac}({\Bbb Z}+\shf),\ \ \ \
\Upsilon'(:a_1a_2:) = a_1(-\shf)a_2(-\shf)
\bold{vac}({\Bbb Z}+\shf)\eqno(3.131)$$
for $a_1,a_2 \in {\bold A}$.  Recall from (3.9) and (3.59) that
$$a(n) = Y_n(\Upsilon(a)) = \{\Upsilon(a)\}_{n-\hf}\eqno(3.132)$$
for $a \in {\bold A}$, $n \in \hf{\Bbb Z}$.  For $x =\ :a_1a_2
:\ \in ${\bf sp}$(2l)$, $m \in {\Bbb Z}$, define the notation
$$x(m) = Y_m(\Upsilon'(x)) = \{\Upsilon'(x)\}_m\eqno(3.133)$$
and note that
$$x(m) = \sum_{k\in\hf{\Bbb Z}}:a_1(k)a_2(m-k):.
\eqno(3.134)$$
As was stated in Theorem 2.1, these are the operators on ${\bold
V}(\shf{\Bbb Z})$ representing ${\bold{\widehat {sp}}}(2l)$.  In Corollary 3.33
we
will see how this result follows from Corollary 3.31 with $v_1,v_2 \in
({\bold V}_0)_1$.  Note that we have
$$x(0)\Upsilon(a) = \Upsilon(x \cdot a)\eqno(3.135)$$
$$x(0)\Upsilon'(y) = \Upsilon'([x,y]),\eqno(3.136)$$
$$x(1)\Upsilon'(y) = \langle x,y\rangle \bold{vac}({\Bbb
Z}+\shf)\eqno(3.137)$$
for $x,y \in ${\bf sp}$(2l)$, $a \in {\bold A}$, which follow from
(2.3), (2.9)-(2.10) and (3.134).

\pr{Corollary 3.32} For $v\in\bold{V}$, $m\in{\Bbb Z}$, on $\bold{V}(\hf{\Bbb
Z})$ we have
\roster
\item"(a)" $[L(m),Y(v,z)] = \ds{\sum_{0\leq k\in{\Bbb Z}}} {{m+1}\choose k}
z^{m+1-k}Y(L(k-1)v,z)$ \hfill\break
and the sum is finite,
\item"(b)" If $L(n)v = 0$ for all $n > 0$, then we have
$$[L(m),Y(v,z)] = z^{m+1} \frac{d}{dz} Y(v,z) + (m+1)z^mY(L(0)v,z).$$
\endroster \epr

\demo{Proof} Part (a) follows from Corollary 3.31 using (3.61). The sum is
finite since wt$(L(k-1)v) = \hbox{wt}(v) - k + 1$
for homogeneous vectors $v$. Part (b) follows from (a). $\hfill\blacksquare$
\enddemo

Note that (2.27) - (2.31) follow from Corollary 3.31.

\pr{Corollary 3.33}  For $x \in ${\bf sp}$(2l)$, $m \in {\Bbb Z}$,
the operators $x(m) =
Y_m(\Upsilon'(x))$ represent ${\bold{\widehat {sp}}}(2l)$ on
$\bold{V}(\hf{\Bbb Z})$.  For any $v \in \bold{V}$ we have
$$[x(m),Y(v,\zeta)] = \zeta^m \sum_{0\leq k\in {\Bbb Z}} {m \choose k}
\zeta^{-k}Y(x(k)v,\zeta)\eqno(3.138)$$
and for $v \in (\bold{V}_0)_0$ or $v \in ({\bold V}_2)_{1/2}$, we have
$$[x(m),Y(v,\zeta)] = \zeta^m Y(x(0)v,\zeta).\eqno(3.139)$$ \epr

\demo{Proof} For $x,y \in ${\bf sp}$(2l)$, $m,n \in {\Bbb Z}$, from
Corollary 3.31 with $v = \Upsilon'(y)$, (3.136) and (3.137), we get
$$\eqalign
{[x(m),y(n)] &= \{x(0)\Upsilon'(y)\}_{m+n} - m\{x(1)\Upsilon'(y)\}_{m+n-1}\cr
&= \{\Upsilon'([x,y])\}_{m+n} - m\{\langle x,y\rangle \bold{vac}({\Bbb
Z}+\shf)\}_{m+n-1}\cr
&= [x,y](m+n) - m\langle x,y\rangle \delta_{m,-n}\cr}$$
so these operators represent ${\bold{\widehat {sp}}}(2l)$ on ${\bold
V}(\hf{\Bbb
Z})$.  The rest also follows from Corollary 3.31.
$\hbox{ } \hfill\blacksquare$
\enddemo

\pr{Corollary 3.34}  For $x \in ${\bf sp}$(2l)$, $m \in {\Bbb Z}$, $v \in
{\bold V}$, on ${\bold V}(\hf {\Bbb Z})$ we have
$$\eqalign
{&Y(x(-m)v,\zeta) =\cr
&\sum_{0\leq i \in {\Bbb Z}} {{m+i-1} \choose i} [\zeta^i
x(-m-i)Y(v,\zeta) - (-1)^m\zeta^{-m-i} Y(v,\zeta)x(i)]\cr}\eqno(3.140)$$
and
$$\eqalign
{&Y(L(-m-1)v,\zeta) =\cr
&\sum_{0\leq i \in {\Bbb Z}} {{m+i-1} \choose i}
[\zeta^iL(-m-i-1) Y(v,\zeta) -
(-1)^m\zeta^{-m-i}Y(v,\zeta)L(i-1)].\cr}\eqno(3.141)$$
\epr

\demo{Proof} The first formula follows from Corollary 3.30 with $m =
0$, $r$ replaced by $-m$, $v_1$ replaced by $x \in ({\bold V}_0)_1$, $v_2$
replaced by $v$, after multiplying both sides by $\zeta^{-n-1}$ and summing
over $n \in {\Bbb Z}+\hf \alpha_2$.  The second formula similarly follows
with $v_1$ replaced by $L(-2)\bold{vac}$.  $\hfill\blacksquare$
\enddemo

\pr{Theorem 3.35} With ${\bold V} = \bWM({\Bbb Z}+\shf)$, $\bvac =
\bvac({\Bbb Z}+\shf)$ and $\omega = L(-2)\bvac$, $({\bold V}, Y(\ ,z),
\bvac, \omega)$ is a vertex operator algebra as defined in
Chapter I. With ${\bold W} = \bWM({\Bbb Z})$, $({\bold W}, Y(\ ,z))$ is a
canonically ${\Bbb Z}_2$-twisted ${\bold V}$-module. \epr

\newpage
\def\pt{\partial}
\def\cRR{{\Bbb C}[z_1,z_2,{z_1}^{-1},{z_2}^{-1},{(z_1-z_2)}^{-1}]}
\def\cR{{\Cal R}}

\topmatter
\title\chapter{4} Bosonic Construction of Vertex Operator
Para-Algebras \endtitle
\endtopmatter

In Chapter III we gave the bosonic construction of a vertex operator algebra
from a construction of the affine
Kac-Moody algebra of type $C_l^{(1)}$.
In general, one wishes to understand a vertex
operator algebra, all of its modules, and the spaces of all intertwining
operators between modules.
Here we will construct a para-algebra from four irreducible level $-\shf$
modules of $C_l^{(1)}$. First we explicitly describe the vertex
operators which represent the symplectic affine Kac-Moody algebra
$\bhsp(2l)$ of type $C_l^{(1)}$ on each $\bV_i$, ${0 \leq i \leq 3}$.

\pr{Definition} For $1 \leq i,j \leq l$, $m \in {\Bbb Z}$, let
$$e_{ij}=
\shf a_i^+(-\shf)a_j^+(-\shf)\bvac  = e_{ji} \in \bV_0,\eqno(4.1)$$
$$f_{ij}=
 -\shf a_i^-(-\shf)a_j^-(-\shf)\bvac  = f_{ji} \in \bV_0,\eqno(4.2)$$
$$h_{ij}=
-a_i^+(-\shf)a_j^-(-\shf)\bvac \in \bV_0.\eqno(4.3)$$
For these vectors $v \in (\bV_0)_1$ we have $Y(v,z)=\sum_{n \in {\Bbb Z}}
\{ v \}_n z^{-n-1}$ and we write:
$$e_{ij}(m)=
\{e_{ij}
\}_m = e_{ji}(m),
f_{ij}(m)=
\{ f_{ij}
\}_m = f_{ji}(m),
h_{ij}(m)=
\{
h_{ij} \}_m.\eqno(4.4)$$
\epr

\pr{Proposition 4.1} For $1 \leq i,j,s,t \leq l$, $m,n \in {\Bbb Z}$,
and {$k \in \shf {\Bbb Z}$} we have
$$[e_{ij}(m),e_{st}(n)] = 0 = [f_{ij}(m),f_{st}(n)],\eqno(4.5)$$
$$[h_{ij}(m),e_{st}(n)] = \delta_{jt}e_{si}(m+n) +
\delta_{js}e_{ti}(m+n),\eqno(4.6)$$
$$\eqalign
{[e_{ij}(m),f_{st}(n)] = & \sfr (\delta_{is}h_{jt}(m+n) +
\delta_{it}h_{js}(m+n) + \delta_{js}h_{it}(m+n) \cr
&+ \delta_{jt}h_{is}(m+n))
+
{\tsize{\frac {n} {4}}}(\delta_{it}\delta_{js} +
\delta_{is}\delta_{jt})\delta_{m,-n},}\eqno(4.7)$$
$$[e_{ij}(m),a_s^+(k)] = 0 = [f_{ij}(m), a_s^-(k)],\eqno(4.8)$$
$$[e_{ij}(m),a_s^-(k)] = \shf(\delta_{is}a_j^+(m+k) +
\delta_{js}a_i^+(m+k)),\eqno(4.9)$$
$$[f_{ij}(m),a_s^+(k)] = \shf(\delta_{is}a_j^-(m+k) +
\delta_{js}a_i^-(m+k)),\eqno(4.10)$$
$$[h_{ij}(m),a_s^+(k)] = \delta_{js}a_i^+(m+k),\eqno(4.11)$$
$$[h_{ij}(m),a_s^-(k)] = -\delta_{is}a_j^-(m+k),\eqno(4.12)$$
$$[h_{ij}(m),h_{st}(n)] = \delta_{js}h_{it}(m+n) -
\delta_{it}h_{sj}(m+n) - m\delta_{js}\delta_{it}\delta_{m,-n},\eqno(4.13)$$
$$[L(m),a_s^\pm(k)] = -(k + \shf m)\ a_s^\pm(m+k),\eqno(4.14)$$
$$[L(m),h_{ij}(n)] = -n\ h_{ij}(n + m),\eqno(4.15)$$
$$[L(m),L(n)] = (m - n)L(m + n) - {\tsize{\frac{l}{12}}}
(m^3 - m)\delta_{m,-n} 1.\eqno(4.16)$$
\endproclaim
\demo{Proof} This follows from
 Corollary 3.31 and is consistent with Theorem 2.1.
$\hfill\blacksquare$

\sk1
\pr{Remark} We can see that
$e_{ij}(m)^* = -f_{ji}(-m), \
f_{ij}(m)^* = -e_{ji}(-m)$,
and $h_{ij}(m)^*$ $=$ $h_{ji}(-m)$.
For simplicity, define for ${1 \leq i \leq l}$,
$e_i(m)=e_{ii}(m)$,
$f_i(m)=f_{ii}(m)$,
and $h_i(m)=h_{ii}(m)$.  These are long root vectors and represent
${\widehat {\bold sl}}(2)^l$, a subalgebra of $\bhsp(2l)$.
\epr
\sk1

Let
$$\bV_0 = \bWM({\Bbb Z}+\shf)^0,\ \ \bV_2 = \bWM(\BZ')^1,\ \
\bV_1 = \bWM({\Bbb Z})^0,\ \ \bV_3 = \bWM({\Bbb Z})^1,$$
$$\bcV = \bV_0 \oplus \bV_2 \oplus \bV_1 \oplus \bV_3, \ \
\quad \bV = \bV_0 \oplus \bV_2,$$
and extend the Hermitian form $(\ ,\ )$ (see 2.40-2.42) to $\bcV$
so that $\bold{WM}({\Bbb Z}+\shf)$ and
$\bold{WM}({\Bbb Z})$ are orthogonal.
Let $\bvac = \bvac(\BZ')$ and $\bvac' = \bvac({\Bbb Z})$.

We will try to construct a vertex operator para-algebra on $\bcV$.
We must define vertex operators $Y(v,z)$ for $v \in \bWM({\Bbb Z})$
acting on all of $\bcV$.  Our basic assumptions are that such operators
are intertwining operators as defined in Chapter I (see (1.39)-(1.43)).
These operators obey the Jacobi-Cauchy Identity for VOPA's (1.48)
with one of the vectors $u$ and $v$ from $\bV_0$ and the other from any
$\bV_\gamma$. They also satisfy the creation property (1.43) and
$$[L(-1),Y(v,z)] = \frac{d}{dz} Y(v,z) \eqno(4.17)$$
for any $v\in \bcV$, (see (1.42)), which means that
$$[L(-1),Y_{n+1}(v)] = -(n+wt(v)) Y_n(v). \eqno(4.18)$$

Each sector $\bV_i$ is an irreducible representation of
${\bold {\widehat {sp}}}(2l)$,  and is preserved by the Virasoro
operators.
So it can be shown that it is enough
to define $(Y(v_1,z)v_2,v_3)$ for three vectors of minimal weight in
their sectors. The main problem is to prove analytic
properties of the two-point
functions
$$(Y(v_1,z_1)Y(v_2,z_2)v_3,v_4) \qquad\hbox{ and }\qquad
(Y(Y(v_1,z_1)v_2,z_2)v_3,v_4) \eqno(4.19)$$
from
these definitions. The three analytic properties, rationality,
permutability and associativity, then combine to give the Jacobi-Cauchy
Identity, the main axiom of vertex operator para-algebras.  Our main
technique is to use the operators $L(-1)$ and $h_j(-1),e_j(-1),f_j(-1)$,
$h_j(0),e_j(0),f_j(0)$, ${1 \leq j \leq
l,}$ to prove recursive
relations for the coefficients of the two-point functions.
This is equivalent to finding and solving the KZ equations which we
illustrate below.
Base cases are determined by the definitions of the one-point functions.
For all the following calculations and proofs it is assumed that Corollaries
3.32-3.34 are true for $v$ in $\bcV$, not just for $v$
in $\bV$. These corollaries are just special cases of (1.41).
Under these assumptions we will prove that $\bcV$ is a VOPA.

We give the set $\Gamma = \{0,1,2,3\}$ the additive group structure of ${\Bbb
Z}_4$.  With $\Delta_i = \Delta_{\bV_i}$, we have the following table of
basis vectors for $(\bV_i)_{\Delta_i}$, their weights and their $h_j(0)$
eigenvalues for ${1 \leq j \leq l}$.

\sk1
\centerline{Table A}
%
%
\message{cellular.TeX version 0.}%
\def\centertable{\leftskip=0pt plus1fill\rightskip=0pt plus1fill}

\def\begincellular#1#2\endcellular{\relax
   \begingroup
      \input cell1                                
      #1\relax
      \input cell2                                
      \ignorespaces
      #2\relax                                    
      \input cell3                                
      \input cell4                                
      \offinterlineskip
      \parskip=\zeropt
      \ignorespaces
      #2\relax                                    
      \par
      \endgroup
   }%

\vskip 15pt
\begincellular{\centertable}
\row{}\cell{$\bV_0$}\cell{$\bvac$}
\cell{$\Delta_0 = 0$}\cell{$0$}
\row{}\cell{$\bV_1$}\cell{$\bvac'$}
\cell{$\Delta_1 = -\sei$}\cell{$-\shf$}
\row{}\cell{$\bV_2$}\cell{$a_1^+(-\shf)\bvac$}\cell{$\Delta_2 = \shf$}
\cell{$\delta_{1j}$}
\row{}\cell{$\bV_3$}\cell{$a_l^-(0)\bvac'$}\cell{$\Delta_3 = -\sei$}
\cell{$-\shf-\delta_{jl}$}
\endcellular
\vskip 15pt

For $0 \leq i,j \leq 3$ define
$$\Delta(i,j) = \Delta_i + \Delta_j - \Delta_{i+j}\eqno(4.20)$$
and note that ${\Delta(i,j) \in \sfr{\Bbb Z}}$ is symmetric and bilinear
mod ${\Bbb Z}$.
We have
$$\Delta(0,j) = 0, \ \ \Delta(1,1) = \Delta(3,3) = -\slr-\shf,\eqno(4.21)$$
$$\Delta(1,2)= \Delta(2,3)= \shf, \ \ \Delta(2,2)= 1,
\ \ \Delta(1,3)= -\slr.\eqno(4.22)$$

\pr{Theorem 4.2} If $v_i \in \bV_{n_i}$ for $1\leq i \leq 4$, and $h_j(0)v_i =
\lambda^j_i v_i$ for $1\leq j \leq l$, $\lambda^j_i \in {\Bbb Q}$ then we have
$$(\lambda^j_1 + \lambda^j_2 - \lambda^j_3)\ (Y(v_1,z)v_2,v_3) =
0,\eqno(4.23)$$
$$(\lambda^j_1 + \lambda^j_2 + \lambda^j_3 - \lambda^j_4)\
(Y(v_1,z_1)Y(v_2,z_2)v_3,v_4) = 0,\eqno(4.24)$$
and
$$(\lambda^j_1 + \lambda^j_2 + \lambda^j_3 - \lambda^j_4)\
(Y(Y(v_1,z)v_2,z_2)v_3,v_4) = 0,\eqno(4.25)$$
so the 1-point function $(Y(v_1,z)v_2,v_3)$ is zero unless
$\lambda^j_1 + \lambda^j_2 = \lambda^j_3$, and the 2-point functions
$(Y(v_1,z_1)Y(v_2,z_2)v_3,v_4)$ and
$(Y(Y(v_1,z)v_2,z_2)v_3,v_4)$
are zero unless
$\lambda^j_1 + \lambda^j_2 + \lambda^j_3 = \lambda^j_4$. \epr

\pr{Remark} For homogeneous vectors, $(Y_m(v_1)v_2,v_3) = 0$ unless $m =
wt(v_2) - wt(v_3)$,
$(Y_m(v_1)Y_n(v_2)v_3 ,$
$v_4) = 0$ unless $m+n = wt(v_3)
- wt(v_4)$, and
$(Y_m(Y_n(v_1)$ $v_2)$ $v_3$,$v_4)$  $ = 0$ unless $m = wt(v_3) -
wt(v_4)$.
\epr

\demo{Proof} The proof of the first equation is a straight forward
calculation.  The other two follow exactly the same way.
{}From Corollary 3.33, (3.139), we have
$$\eqalign{
\lambda^j_3(Y(v_1,z)v_2,v_3)
&= (Y(v_1,z)v_2,\lambda^j_3v_3) \cr
&= (Y(v_1,z)v_2,h_j(0)v_3) \cr
&= (h_j(0)Y(v_1,z)v_2,v_3) \cr
&= (Y(v_1,z)h_j(0)v_2,v_3) + ([h_j(0), Y(v_1,z)]v_2,v_3) \cr
&= (Y(v_1,z)\lambda^j_2v_2,v_3) + (Y(h_j(0)v_1,z)v_2,v_3) \cr
&= \lambda^j_2(Y(v_1,z)v_2,v_3) + \lambda^j_1(y(v_1,z)v_2,v_3) \cr
&= (\lambda^j_2 +
\lambda^j_1)(Y(v_1,z)v_2,v_3).\qquad\qquad\qquad\blacksquare}$$
\sk1

Let

\     \ $\Omega={\{ \bvac, \bvac', a_1^+(-\shf)\bvac, a_l^-(0)\bvac' \}},$

\     \ $B_0 = {\{ h_{s1}(0), h_{lt}(0), f_{ij}(0) \ \ | \ \ {1 \leq i,j \leq
l} , {1 < s \leq l}, {1 \leq t < l}\}},$

\     \ $B = B_0 \ \ \cup \ \ {\{ e_{ij}(m), f_{ij}(m), h_{ij}(m) \ \ | \ \
{1 \leq i,j \leq l},{m < 0}\}}.$
\sk1
It is clear from Proposition 4.1 that the $\Bbb C$-span of $B$ is a Lie
subalgebra
so we have $\bcV = \Cal U (B) \cdot \Omega,$
where $\Cal U (B)$ is the universal enveloping algebra of the $\Bbb
C$-span of $B$.
Let ${\bar \Omega}$ = $\Cal U (B_0) \cdot \Omega.$
For the next proposition we will use the following:
for any ${T \in B},$
\roster
\item "1)" $T^*w$ = 0, for any $w \in \Omega,$
\item "2)" $T^*{\bar w} \in  {\bar \Omega}$ for any ${\bar w} \in {\bar
\Omega}.$
\endroster
\noindent (See the remark after Proposition 4.1.)

\pr{Proposition 4.3} Let ${v_i \in \bV_{\gamma_i}}$
for ${1 \leq i \leq 4}$, $\gamma_i\in\Gamma$, then
\roster
\item "i)" the
series $(Y(v_1,z)v_2,v_3)$ can be written as a linear combination of
series of the form ${z^n(Y(w_1,z){\bar w_2},w_3)},$
\item "ii)" the series $(Y(v_1,z_1)Y(v_2,z_2)v_3,v_4)$ can be written as
a linear combination of series of the form ${z_1}^{n_1}{z_2}^{n_2}
(Y(w_1,z_1)Y({\bar
w_2},z_2){\bar w_3},w_4)$,
\item "iii)" the series $(Y(Y(v_1,z)v_2,z_2)v_3,v_4)$ can be written as
a linear combination of series of the form
$z^{n_1}{z_2}^{n_2}(Y(Y(w_1,z){\bar w_2},z_2){\bar w_3},w_4),$
\endroster
where $w_i\in
\Omega \cap \bV_{\gamma_i}$, $ {\bar w_i} \in {\bar \Omega}\cap
\bV_{\gamma_i}$,
$1\leq i \leq 4$ and ${n,n_1,n_2 \in {\Bbb Z}}.$
\epr

\demo{Proof} We only give the proof for i) since the other two follow
in exactly the same way.
WLOG, suppose $v_2=x_r(-m_r) \cdots x_1(-m_1) w_2$ where
$x_i(-m_i) \in B,$ and $w_2 \in \Omega\cap\bV_{\gamma_2}$.  We will show by
induction
 on r that the series
$(Y(v_1,z)v_2,v_3)$ can be written as a linear combination of series of
the form $z^n(Y(v'_1,z)w_2,v'_3),$ where $wt(v'_j) \leq wt(v_j),$ $j \in
\{ 1,3 \}.$
For r=0 we have $v_2=w_2$ and we are finished.  Suppose the claim is
true for all $0 \leq k < r,$ i.e. suppose that for any $u_1,u_2,u_3$ with
$wt(u_i) \leq wt(v_i)$ for $1 \leq i \leq 3$ and $u_2=x'_k(-m'_k) \cdots
x'_1(-m'_1) w_2,0\leq k < r,$ we have that $(Y(u_1,z)u_2,u_3)$ can be written
as
above,
then
$$\eqalign{
&(Y(v_1,z)v_2,v_3)\cr
&= (Y(v_1,z)x_r(-m_r) \cdots x_1(-m_1) w_2,v_3) \cr
&= (x_r(-m_r)Y(v_1,z)x_{r-1}(-m_{r-1}) \cdots x_1(-m_1) w_2,v_3) \cr
&\quad  -
z^{-m_r} \sum_{0 \leq t \in {\Bbb Z}} {-m_r\choose t}
z^{-t}(Y(x_r(t)v_1,z)x_{r-1}(-m_{r-1})
 \cdots x_1(-m_1) w_2,v_3) \cr
&= (Y(v_1,z)x_{r-1}(-m_{r-1}) \cdots x_1(-m_1) w_2,x_r(-m_r)^*v_3) \cr
&\quad - z^{-m_r} \sum_{0 \leq t \in {\Bbb Z}} {-m_r\choose t}
z^{-t}(Y(x_r(t)v_1,z)x_{r-1}(-m_{r-1}) \cdots x_1(-m_1) w_2,v_3)
\cr}$$
so by induction, $(Y(v_1,z)v_2,v_3)$ can be written as claimed
 above noting that the sum
on the right is finite since $x_r(t)v_1 = 0$ unless $wt(v_1) - t
\ge -\sei,$ and that $wt(x_r(-m_r)^*v_3) = wt(v_3) - m_r.$

Now we claim that the series $(Y(v_1,z)w_2,v_3)$ can be written as a
linear combination of series of the form $z^n(Y(v'_1,z)w_2,w_3),$
where $w_i\in \Omega\cap\bV_{\gamma_i},i\in\{2,3\}, n \in {\Bbb Z}$  and
$wt(v'_1) \leq
wt(v_1)$.
WLOG,
suppose $v_3=x_r(-m_r) \cdots x_1(-m_1) w_3$ where $x_i(-m_i) \in B,$
and $w_3 \in \Omega\cap\bV_{\gamma_3}$.
For r=0 we have $v_3=w_3$ and we are finished.  Suppose the claim is
true for all $0 \leq k < r,$ then
$$\eqalign{
&(Y(v_1,z)w_2,v_3)\cr
&= (Y(v_1,z)w_2,x_r(-m_r)
\cdots x_1(-m_1)
w_3) \cr
&= (x_r(-m_r)^* Y(v_1,z)w_2,x_{r-1}(-m_{r-1})
\cdots x_1(-m_1)
w_3) \cr
&= z^{m_r} \sum_{0 \leq t \in {\Bbb Z}} {m_r\choose t} z^{-t}
(Y(x_r(-t)^*v_1,z)w_2,x_{r-1}(-m_{r-1})
\cdots x_1(-m_1)
w_3) \cr}$$
which is a finite sum since $m_r \ge 0$.  So by induction,
$(Y(v_1,z)w_2,v_3)$ can be written as claimed.

Finally, I claim that the series $(Y(v_1,z)w_2,w_3)$ can be written as
a linear combination of series of the form $z^n(Y(w_1,z){\bar w_2},w_3)$
where $w_i\in \Omega\cap\bV_{\gamma_i},i\in\{1,3\}, {\bar w_2} \in {\bar
\Omega}\cap\bV_{\gamma_2}$ and $n \in
{\Bbb Z}$.  WLOG, suppose $v_1=x_r(-m_r) \cdots x_1(-m_1) w_1$ where
$x_i(-m_i) \in B,$ and $w_1 \in \Omega\cap\bV_{\gamma_1}$.
For r=0 we have $v_1=w_1$ and we are finished.  Suppose the claim is
true for all $0 \leq k < r,$ then
$$\eqalign{
&(Y(v_1,z)w_2,w_3)\cr
&= (Y(x_r(-m_r) \cdots x_1(-m_1) w_1,z)w_2,w_3) \cr
&= \sum_{0 \leq t \in {\Bbb Z}} {m_r+t-1\choose
t}([z^tx_r(-m_r-t)Y(x_{r-1}(-m_{r-1}) \cdots x_1(-m_1) w_1,z) \cr
&\qquad - (-1)^{m_r}z^{-m_r-t}Y(x_{r-1}(-m_{r-1}) \cdots x_1(-m_1)
w_1,z)x_r(t)]w_2,w_3) \cr
&= \sum_{0 \leq t \in {\Bbb Z}} {m_r+t-1\choose
t}[z^t(Y(x_{r-1}(-m_{r-1}) \cdots x_1(-m_1) w_1,z)w_2,x_r(-m_r-t)^*w_3)
\cr
&\qquad - (-1)^{m_r}z^{-m_r-t}(Y(x_{r-1}(-m_{r-1}) \cdots x_1(-m_1)
w_1,z)x_r(t)w_2,w_3)]. \cr}$$
It is clear that the first part of the sum is zero since
$x_r(-m_r-t) \in B.$ Now notice that $x_r(t)w_2$=0 for any $t > 0,$ so the only
term that
survives is
$$-(-1)^{m_r}z^{-m_r}(Y(x_{r-1}(-m_{r-1}) \cdots x_1(-m_1)
w_1,z)x_r(0)w_2,w_3).$$
Clearly, $x_r(0)w_2 \in {\bar \Omega}\cap\bV_{\gamma_2}$ and therefore,
 by induction we see that the series $(Y(v_1,z)w_2,$ $w_3)$ can
be written as claimed.

Putting the above three steps together gives us that the series
$(Y(v_1,z)v_2,v_3)$ can be written as a linear combination of series of
the form $z^n(Y(w_1,z){\bar w_2},w_3)$ where $n \in {\Bbb Z},$ $w_i\in \Omega
\cap\bV_{\gamma_i},i\in\{1,3\}$ and
${\bar w_2} \in {\bar \Omega}\cap\bV_{\gamma_2}. \hfill\blacksquare$

\sk1

\pr{Corollary 4.4} Let $v_i \in \bV_{\gamma_i}$,$1 \leq i\leq 3$.  We have
$(Y(v_1,z_1)v_2,v_3)=0$ unless $\gamma_1+\gamma_2=\gamma_3$ in ${\Bbb Z}_4$.
\epr

\demo{Proof} Let us examine the possibilities for non-zero one-point functions
$(Y(v_1,z)$ $v_2$ $,$ $v_3)$ in order to understand the fusion rules.
Proposition 4.3 says we only need deal with the cases where the
first and third vector are from $\Omega$ and the second vector is from
${\bar \Omega}.$ Also, we have the restriction that the $h_j(0)$
eigenvalues must satisfy
Theorem 4.2 (4.23).
Since we have already seen that $\bV$ is a vertex operator algebra
(Theorem 3.35) we may
restrict our investigation to $v_1 \in \bWM({\Bbb Z}).$ There are
four
possible choices for $w_1,{\bar w_2},w_3$ satisfying
(4.23), and only two of those need to be determined.
The four series are:

\roster
\item "1)" $(Y(\bvac',z)\bvac,\bvac')$
\item "2)" $(Y(\bvac',z)f_{l1}(0)a^+_1(-\shf)\bvac,a^-_l(0)\bvac')$
\item "3)" $(Y(a^-_l(0)\bvac',z)\bvac,a^-_l(0)\bvac')$
\item "4)" $(Y(a^-_l(0)\bvac',z)h_{l1}(0)a^+_1(-\shf)\bvac,\bvac')$
\endroster
{}From the remark after Theorem 4.2 we have
\roster
\item "1)" $(Y_k(\bvac')\bvac,\bvac') = 0$ unless $k = \sei$
\item "2)" $(Y_k(\bvac')f_{l1}(0)a^+_1(-\shf)\bvac,a^-_l(0)\bvac') = 0$
unless $k = \shf + \sei$
\item "3)" $(Y_k(a^-_l(0)\bvac')\bvac,a^-_l(0)\bvac') = 0$ unless $k =
\sei$
\item "4)" $(Y_k(a^-_l(0)\bvac')h_{l1}(0)a^+_1(-\shf)\bvac,\bvac') = 0$
unless $k = \shf + \sei$
\endroster
and from the creation property we have
$(Y_{\sei}(\bvac')\bvac,\bvac')=1$ and
$(Y_{\sei}(a^-_l(0) $ $ \bvac') $ $ \bvac$,$a^-_l(0)\bvac')$ $=$ $1$.
$\hfill\blacksquare$

Let us define
$$
A_l
=\Afw(Y_{\shf+\sei}(\bvac')f_{l1}(0)a^+_1(-\shf)\bvac,a^-_l(0)\bvac'),
\eqno(4.26)$$
and
$$B_l
=(Y_{\shf+\sei}(a^-_l(0)\bvac')h_{l1}(0)a^+_1(-\shf)\bvac,\bvac').\eqno(4.27)$$

\pr{Remark} Noting the fact that there are no vectors $v_1,v_2 \in \bWM({\Bbb
Z})$,
$v_3 \in \bcV$ such that $(Y(v_1,z)v_2,v_3) \ne 0$, the group ${\Bbb Z}_4$ in
the
above corollary could also be ${\Bbb Z}_2 \times {\Bbb Z}_2$.
\epr

\ni In order to get a VOPA structure on $\bcV$ it is necessary to find
constants $A_l,B_l$ consistent with the Jacobi-Cauchy Identity.
To find the restrictions on $A_l,B_l$ we must
examine the two kinds of two-point functions,
$$(Y(v_1,z_1)Y(v_2,z_2)v_3,v_4) \qquad \hbox{ and } \qquad
(Y(Y(v_1,z)v_2,z_2)v_3,v_4),$$
product and composition, respectively.
In view of the proposition above, and the $h_j(0)$ eigenvalue
restrictions, the possibilities for
non-zero two-point functions with $v_i$, $1\leq i\leq 4$, of minimal
weight narrows down to twenty six cases.
Here are twenty two of them:


\centerline{Table B1}
\sk1
\begincellular{\centertable}
\row{}\cell{$w_1$}\cell{${\bar w_2}$}
\cell{${\bar w_3}$}\cell{$w_4$}
\row{}\cell{$\bvac$}\cell{${h_{li}(0)a^-_l(0)\bvac'}$}
\cell{$h_{i1}(0)a^+_1(-\shf)\bvac$}\cell{$\bvac'$}
\row{}\cell{$\bvac$}\cell{$\bvac'$}
\cell{$\bvac$}\cell{$\bvac'$}
\row{}\cell{$\bvac$}\cell{$f_{li}(0)\bvac'$}
\cell{$h_{i1}(0)a^+_1(-\shf)\bvac$}\cell{$a^-_l(0)\bvac'$}
\row{}\cell{$\bvac$}\cell{$a^-_l(0)\bvac'$}
\cell{$\bvac$}\cell{$a^-_l(0)\bvac'$}
\row{}\cell{$\bvac$}\cell{$\bvac'$}
\cell{$f_{l1}(0)a^+_1(-\shf)\bvac$}\cell{$a^-_l(0)\bvac'$}
\row{}\cell{$a^+_1(-\shf)\bvac$}\cell{$f_{i1}(0)\bvac'$}
\cell{$h_{i1}(0)a^+_1(-\shf)\bvac$}\cell{$\bvac'$}
\row{}\cell{$a^+_1(-\shf)\bvac$}\cell{$h_{l1}(0)a^-_l(0)\bvac'$}
\cell{$\bvac$}\cell{$\bvac'$}
\row{}\cell{$a^+_1(-\shf)\bvac$}\cell{$h_{11}(0)\bvac'$}
\cell{$f_{11}(0)a^+_1(-\shf)\bvac$}\cell{$\bvac'$}
\row{}\cell{$a^+_1(-\shf)\bvac$}\cell{$f_{i1}(0)a^-_l(0)\bvac'$}
\cell{$h_{i1}(0)a^+_1(-\shf)\bvac$}\cell{$a^-_l(0)\bvac'$}
\row{}\cell{$a^+_1(-\shf)\bvac$}\cell{$f_{l1}(0)\bvac'$}
\cell{$\bvac$}\cell{$a^-_l(0)\bvac'$}
\row{}\cell{$a^+_1(-\shf)\bvac$}\cell{$h_{11}(0)a^-_l(0)\bvac'$}
\cell{$f_{11}(0)a^+_1(-\shf)\bvac$}\cell{$a^-_l(0)\bvac'$}
\row{}\cell{$a^+_1(-\shf)\bvac$}\cell{$h_{l1}(0)a^-_l(0)\bvac'$}
\cell{$f_{l1}(0)a^+_1(-\shf)\bvac$}\cell{$a^-_l(0)\bvac'$}
\row{}\cell{$\bvac'$}\cell{$h_{i1}(0)a^+_1(-\shf)\bvac$}
\cell{$f_{i1}(0)a^+_1(-\shf)\bvac$}\cell{$\bvac'$}
\row{}\cell{$\bvac'$}\cell{$\bvac$}
\cell{$\bvac$}\cell{$\bvac'$}
\row{}\cell{$\bvac'$}\cell{$f_{i1}(0)a^+_1(-\shf)\bvac$}
\cell{$h_{i1}(0)a^+_1(-\shf)\bvac$}\cell{$\bvac'$}
\row{}\cell{$\bvac'$}\cell{$f_{l1}(0)a^+_1(-\shf)\bvac$}
\cell{$\bvac$}\cell{$a^-_l(0)\bvac'$}
\row{}\cell{$\bvac'$}\cell{$\bvac$}
\cell{$f_{l1}(0)a^+_1(-\shf)\bvac$}\cell{$a^-_l(0)\bvac'$}
\row{}\cell{$a^-_l(0)\bvac'$}\cell{$h_{l1}(0)a^+_1(-\shf)\bvac$}
\cell{$\bvac$}\cell{$\bvac'$}
\row{}\cell{$a^-_l(0)\bvac'$}\cell{$\bvac$}
\cell{$h_{l1}(0)a^+_1(-\shf)\bvac$}\cell{$\bvac'$}
\row{}\cell{$a^-_l(0)\bvac'$}\cell{$h_{i1}(0)a^+_1(-\shf)\bvac$}
\cell{$f_{i1}(0)a^+_1(-\shf)\bvac$}\cell{$a^-_l(0)\bvac'$}
\row{}\cell{$a^-_l(0)\bvac'$}\cell{$\bvac$}
\cell{$\bvac$}\cell{$a^-_l(0)\bvac'$}
\row{}\cell{$a^-_l(0)\bvac'$}\cell{$f_{i1}(0)a^+_1(-\shf)\bvac$}
\cell{$h_{i1}(0)a^+_1(-\shf)\bvac$}\cell{$a^-_l(0)\bvac'$}
\endcellular
\vskip 15pt

where $1 \leq i \leq l.$

 The following four possibilities occur only
when
$l=1.$

\newpage

\centerline{Table B2}
\sk1
\begincellular{\centertable}
\row{}\cell{$w_1$}\cell{${\bar w_2}$}
\cell{${\bar w_3}$}\cell{$w_4$}
\row{}\cell{$\bvac'$}\cell{$a^+_1(-\shf)\bvac$}
\cell{$\bvac'$}\cell{$\bvac$}
\row{}\cell{$\bvac'$}\cell{$\bvac'$}
\cell{$a^+_1(-\shf)\bvac$}\cell{$\bvac$}
\row{}\cell{$\bvac'$}\cell{$\bvac'$}
\cell{$\bvac'$}\cell{$a^-_1(0)\bvac'$}
\row{}\cell{$a^+_1(-\shf)\bvac$}\cell{$\bvac'$}
\cell{$\bvac'$}\cell{$\bvac$}
\endcellular
\vskip 15pt

The above four possibilities actually all turn out to be zero since all
the one point functions of the form $(Y(v_1,z_1)v_2,v_3)$ with
$v_1,v_2 \in \bWM({\Bbb Z})$ are zero.  Alternately, it can also be
shown that the first two functions and the last one are zero by
substituting in $e_1(0)f_1(0)\vvb$ for $\vvb$ and then bringing out the
$e_1(0)$.
For example,
$$\eqalign{
(Y&(\vvc,z_1)Y(\vvb,z_2)\vvc,\vva) \cr
&= (Y(\vvc,z_1)Y(e_1(0)f_1(0)\vvb,z_2)\vvc,\vva) \cr
&=(Y(\vvc,z_1)e_1(0)Y(f_1(0)\vvb,z_2)\vvc,\vva) \cr
&\quad -(Y(\vvc,z_1)Y(f_1(0)\vvb,z_2)e_1(0)\vvc,\vva) \cr
&=(e_1(0)Y(\vvc,z_1)Y(f_1(0)\vvb,z_2)\vvc,\vva) \cr
&\quad -(Y(e_1(0)\vvc,z_1)Y(f_1(0)\vvb,z_2)\vvc,\vva) \cr
&=-(Y(\vvc,z_1)Y(f_1(0)\vvb,z_2)\vvc,f_1(0)\vva) \cr
&=0.}$$

Here we will calculate the twenty six two-point correlation functions
corresponding to the rows of the preceeding table.  Later we will prove
the Jacobi-Cauchy identity for vertex operator para-algebras again by
proving the three analytic properties of correlation functions,
 Rationality, Permutability and Associativity.
We will need the following calculations for the recursion formulas
forthcoming.  We have

$$\eqalign{
\omega &= \shf \sum_{1 \leq i \leq l} (a^-_i(-\sth)a^+_i(-\shf) -
a^+_i(-\sth)a^-_i(-\shf))\bvac \cr
&= \{ \sxt \ \sum_{1 \leq i \leq l} [h_i(-1)^2 +
4e_i(-1)f_i(-1)]
- \srd \ \sum_{1 \leq i \leq l} h_i(-2) \} \bvac}\eqno(4.28)$$

so

$$\eqalign{
Y((\omega,z) &= \sum_{n \in {\Bbb Z}} L(n) z^{-n-2} \cr
&= \sxt \sum_{1 \leq i \leq l,m \in {\Bbb Z},0 \leq k}
[h_i(-k-1)h_i(m)z^{-m-1+k} + h_i(m)h_i(k)z^{-m-k-2} \cr
& \qquad \qquad \qquad \qquad +4e_i(-k-1)f_i(m)z^{-m-1+k} +
4f_i(m)e_i(k)z^{-m-k-2}] \cr
& \quad -\srd \sum_{1 \leq i \leq l, 0 \leq k}
(k+1)[h_i(-k-2)z^k - h_i(k)z^{-k-2}]}$$

and therefore we have that

$$\eqalign{
L(n)&=\sxt \sum_{1 \leq i \leq l,0 \leq k} [h_i(-k-1)h_i(k+n+1) +
h_i(n-k)h_i(k) \cr
& \qquad \qquad \qquad +4e_i(-k-1)f_i(k+n+1) +4f_i(n-k)e_i(k)] \cr
& \quad + \srd \sum_{1 \leq i \leq l} (n+1)h_i(n).}\eqno(4.29)$$

For $w \in \Omega$ and ${\bar w} \in {\bar \Omega},$

$$
L(0)w = \sxt \sum_{1 \leq i \leq l} [h_i(0)^2 + 2h_i(0)]w,$$
$$L(0){\bar w} = \sxt \sum_{1 \leq i \leq l} [h_i(0)^2 +
4f_i(0)e_i(0)+2h_i(0)]{\bar w},$$
$$L(-1){\bar w} = \srd \sum_{1 \leq i \leq l}
[h_i(-1)h_i(0)+2e_i(-1)f_i(0)+2f_i(-1)e_i(0)]{\bar w}.
\eqno(4.30)
$$

Suppose $v_1,v_2,v_3$ and $v_4$ are vectors of minimal weight with
$v_i \in \bV_{n_i}$.
Let $h_j(0)v_i=\lambda^j_iv_i$,
for
$1 \leq i \leq 4$,
$1 \leq j \leq l$.
Let $k=
wt(v_1)+wt(v_2)+wt(v_3)-wt(v_4)$ and $m = \Delta(n_2,n_3)$.
We may express the following product type of two-point function as a
series in $\left(\frac{z_2}{z_1}\right)$.
$$\perm = \sum_{p \in \shf {\Bbb Z}} \sum_{q \in \shf {\Bbb Z}} (Y_{p-wt(v_1)}
(v_1) Y_{q-wt(v_2)}(v_2) v_3,v_4){z_1}^{-p}{z_2}^{-q} \eqno(4.33)$$
with $p=-q + wt(v_1)+wt(v_2)+wt(v_3)-wt(v_4)=-q + k$
(see the remark after Theorem 4.2) gives
$$\eqalign{
\perm &=
\sum_{q \in \shf {\Bbb Z}} (Y_{-q+k-wt(v_1)}(v_1)Y_{q-wt(v_2)}
(v_2) v_3, v_4) {z_1}^{-k+q}{z_2}^{-q}\cr
&= {z_1}^{-k}
\sum_{q \in \shf {\Bbb Z}} (Y_{-q+k-wt(v_1)}(v_1)Y_{q-wt(v_2)}
(v_2) v_3, v_4) {\left(\frac{z_1}{z_2}\right)}^q.\cr
}$$
{}From Thereom 4.2, we know that the coefficients of $Y(v_2,z_2)v_3$ are in
$\bV_{n_2+n_3}$ so
$$wt(Y_{-q-wt(v_2)}(v_2)v_3) = wt(v_3)+q+wt(v_2) \ge \Delta_{n_2+n_3} =
wt(v_2) + wt(v_3) - m$$
therefore $0 \le q+m \in {\Bbb Z}$.  Let $z=\frac{z_2}{z_1}$ and we have
$$\eqalign{
&\perm =\cr
&{z_1}^{-k+m}{z_2}^{-m} \sum_{0 \le P \in {\Bbb Z}}(Y_{k+P-m-wt(v_1)} (v_1)
Y_{-P+m-wt(v_2)}
(v_2) v_3, v_4) {\left(\frac{z_2}{z_1}\right)}^P}\eqno(4.34)$$
or
$$\eqalign{
&\perm =\cr
&{z_1}^{-k}{z}^{-m} \sum_{0 \le P \in {\Bbb Z}}(Y_{k+P-m-wt(v_1)} (v_1)
Y_{-P+m-wt(v_2)}
(v_2) v_3, v_4) z^P.}\eqno(4.35)$$
We will also need
$$\eqalign{
[L(-1), Y(v,z_1 z)] &= -\sum_{p \in \shf{\Bbb Z}} (p-1) Y_{p-wt(v)-1}
(v) (z_1 z)^{-p} \cr
&=\frac{1}{z} \cdot \frac{\pt}{\pt z_1} Y(v,z_1 z)\cr
&=\frac{1}{z_1} \cdot \frac{\pt }{\pt z} Y(v,z_1 z),}\eqno(4.36)$$
And,
$$\eqalign{
\frac{\pt }{\pt z_1}
(Y(v_1,z_1)Y(v_2,z_1 z)v_3,v_4)
&=([\frac{\pt}{\pt z_1} Y(v_1,z_1)] Y(v_2, z_1 z) v_3,v_4) \cr
&\qquad + (Y(v_1,z_1)[\frac{\pt}{\pt z_1} Y(v_2,z_1 z)]v_3,v_4) \cr
&=([L(-1), Y(v_1,z_1)]Y(v_2,z_1 z)v_3, v_4) \cr
&\qquad + z(Y(v_1,z_1)[L(-1), Y(v_2,z_1 z)] v_3,v_4) \cr
&=-(Y(v_1,z_1)L(-1)Y(v_2,z_1 z)v_3,v_4) \cr
&\qquad + z(Y(v_1,z_1)[L(-1), Y(v_2,z_1 z)] v_3,v_4) \cr
&=-(Y(v_1,z_1)Y(v_2,z_1 z)L(-1)v_3,v_4) \cr
&\qquad + (z-1)(Y(v_1,z_1)[L(-1), Y(v_2,z_1 z)] v_3,v_4) \cr
&=-(Y(v_1,z_1)Y(v_2,z_1 z)L(-1)v_3,v_4) \cr
&\qquad + \frac{(z-1)}{z_1}\frac{\pt}{\pt z}(Y(v_1,z_1)Y(v_2,z_1 z) v_3,v_4).
\cr
}\eqno(4.37)$$

Then the general recursion for permutability is as follows:

$$\eqalign{
&-(P+k-m)\Phi_P(w_1,{\bar w_2},{\bar w_3},w_4) \cr
&=([L(-1),Y_{P+k-m-wt(w_1)+1}(w_1)]Y_{-P+m-\wt2}
({\bar w_2}){\bar w_3},w_4) \cr
&=-(Y_{P+k-m-wt(w_1)+1}(w_1)[L(-1),Y_{-P+m-\wt2}({\bar w_2})]{\bar
w_3},w_4) \cr
&\quad -(Y_{P+k-m-wt(w_1)+1}(w_1)Y_{-P+m-\wt2}
({\bar w_2})L(-1){\bar w_3},w_4) \cr
&=(-P+m-1)(Y_{P+k-m-wt(w_1)+1}(w_1)Y_{-P+m-\wt2-1}({\bar w_2}){\bar w_3},w_4)
\cr
&\quad -\srd \sum_{1 \leq j \leq l}
\lambda^j_3(Y_{P+k-m-wt(w_1)+1}(w_1)Y_{-P+m-\wt2}
({\bar
w_2})h_i(-1){\bar w_3},w_4) \cr
&\quad -\std \sum_{1 \leq j \leq l}(Y_{P+k-m-wt(w_1)+1}(w_1)Y_{-P+m-\wt2}({\bar
w_2})e_j(-1)f_j(0){\bar w_3},w_4) \cr
&\quad -\std \sum_{1 \leq j \leq l}(Y_{P+k-m-wt(w_1)+1}(w_1)Y_{-P+m-\wt2}({\bar
w_2})f_j(-1)e_j(0){\bar w_3},w_4) \cr
}$$
$$\eqalign{
&=(-P+m-1) \Phi_{P+1} + \srd \sum_{1 \leq j \leq l}
\lambda^j_2\lambda^j_3\Phi_{P+1} +\srd \sum_{1 \leq j \leq l}
\lambda^j_1\lambda^j_3\Phi_{P} \cr
&\quad +\std \sum_{1 \leq j \leq
l}(Y_{P+k-m-wt(w_1)+1}(w_1)Y_{-P+m-\wt2-1}(e_j(0){\bar
w_2})f_j(0){\bar w_3},w_4) \cr
&\quad +\std \sum_{1 \leq j \leq
l}(Y_{P+k-m-wt(w_1)+1}(w_1)Y_{-P+m-\wt2-1}(f_j(0){\bar
w_2})e_j(0){\bar w_3},w_4) \cr
&\quad +\std \sum_{1 \leq j \leq
l}(Y_{P+k-m-wt(w_1)}(f_j(0)w_1)Y_{-P+m-\wt2}({\bar
w_2})e_j(0){\bar w_3},w_4) \cr
}$$
$$\eqalign{
&=(q-1+wt({\bar w_2}) + \srd \sum_{1 \leq j \leq l}
\lambda^j_2\lambda^j_3)\Phi_{P+1} + \srd \sum_{1 \leq j \leq l}
\lambda^j_1\lambda^j_3\Phi_P \cr
&\quad + \std \sum_{1 \leq j \leq l} \Phi_{P+1}(w_1,e_j(0){\bar
w_2},f_j(0){\bar w_3},w_4) \cr
&\quad + \std \sum_{1 \leq j \leq l}
\Phi_{P+1}(w_1,f_j(0){\bar w_2},e_j(0){\bar w_3},w_4) \cr
&\quad -\std \sum_{1 \leq j \leq l} \Phi_P(w_1,{\bar
w_2},f_j(0)e_j(0){\bar w_3},w_4) - \std \sum_{1 \leq j \leq l}
\Phi_P(w_1,f_j(0){\bar w_2},e_j(0){\bar w_3}.w_4). }\eqno(4.38)$$

Now notice, out of the twenty six possibilities in the previous tables,
about half contain a vector of the form $f_{st}(0)w$ where $w \in
\Omega$. The ones that either don't have any `$f(0)$' term or are such
that ${\bar w_2}=\bvac$ or ${\bar w_3}=\bvac$ simplify immensely
since $e_j(0)[h_{st}(0)w] = 0$ for any $1 \leq j,s,t \leq l, \ w \in
\Omega$ and $e_{st}(0)\bvac=f_{st}(0)\bvac=0$.
The recursion for these reduces to
$$ -(P+k-m + \srd\sum_{1 \le j \le l} \lambda^j_1 \lambda^j_3) \Phi_P =
(-P +m -1 +\srd\sum_{1 \le j \le l} \lambda^j_2 \lambda^j_3) \Phi_{P+1}.
\eqno(4.39)$$
so if
$(-P +m -1 +\srd\sum_{1 \le j \le l} \lambda^j_2 \lambda^j_3) \ne 0$ we have
$$\Phi_{P+1} = \frac{
(P+k-m + \srd\sum_{1 \le j \le l} \lambda^j_1 \lambda^j_3)}
{(P -m +1 -\srd\sum_{1 \le j \le l} \lambda^j_2 \lambda^j_3)} \Phi_P.
\eqno(4.40)$$

The technique to solve for the functions with the `$f(0)$' terms will be
to pair up certain functions and solve the two simultaneously,
specifically pairing up functions of the form
$$(Y(w_1,z_1)Y(h_{s't'}(0)w_2,z_2)f_{st}(0)w_3,w_4)\eqno(4.41)$$
and
$$(Y(w_1,z_1)Y(f_{st}(0)w_2,z_2)h_{s't'}(0)w_3,w_4)\eqno(4.42)$$
where $w_i \in \Omega$ for $1 \leq i \leq 4$ and $1 \leq s,s',t,t' \leq l$.

An alternative method for solving for $\oper$ is using differential
equations as follows:

Let $x,y \in \{ 1, h_{i1}(0), h_{li}(0), f_{1i}(0) \}$.  Let
$\lambda^j_x, \lambda^j_y$ be defined by
$[h_j(0),x] = \lambda^j_x \cdot x$,
$[h_j(0),y] = \lambda^j_y \cdot y$,
and define $\gamma_{xy}=\gamma(w_2,w_3,x,y)$ and
$\beta_{xy}=\beta(w_2,w_3,x,y)$
by
$$e_j(0) \cdot x \cdot w_2 = \gamma^j_{xy} y \cdot w_2,
\qquad\sum_{1 \le j \le l} \gamma^j_{xy} f_j(0) \cdot y \cdot w_3 = \gamma_{xy}
x \cdot w_3\eqno(4.43)$$
and
$$e_j(0) \cdot y \cdot w_3 = \beta^j_{xy} x \cdot w_3,
\qquad\sum_{1 \le j \le l} \gamma^j_{xy} f_j(0) \cdot x \cdot w_2 = \beta_{xy}
y \cdot w_2.\eqno(4.44)$$
Define $\tau(v) \in {\Bbb Q}$ by
$$\sum_{1 \le j \le l} f_j(0)e_j(0) \cdot v = \tau(v) \cdot v.\eqno(4.45)$$
For $w_1,w_4 \in \Omega$,${\bar w_2},{\bar w_3} \in {\bar \Omega}$
$$\eqalign{
&\left[ \frac{(1-z)}{z_1} \frac{\pt}{\pt z} +\frac{\pt}{\pt z_1} \right]
(Y(w_1,z_1)Y({\bar w_2},z_1 z){\bar w_3},w_4) \cr
&= -
(Y(w_1,z_1)Y({\bar w_2},z_1 z)L(-1){\bar w_3},w_4) \cr
&= - \srd
\sum_{1 \le j \le l}
(Y(w_1,z_1)Y({\bar w_2},z_1 z)h_j(-1)h_j(0){\bar w_3},w_4) \cr
&\qquad - \std
\sum_{1 \le j \le l}
(Y(w_1,z_1)Y({\bar w_2},z_1 z)e_j(-1)f_j(0){\bar w_3},w_4) \cr
&\qquad - \std
\sum_{1 \le j \le l}
(Y(w_1,z_1)Y({\bar w_2},z_1 z)f_j(-1)e_j(0){\bar w_3},w_4) \cr
&= \frac{1}{z_1 z} \srd
\sum_{1 \le j \le l} \lambda^j_2\lambda^j_3
(Y(w_1,z_1)Y({\bar w_2},z_1 z){\bar w_3},w_4) \cr
&\qquad +\frac{1}{z_1} \srd
\sum_{1 \le j \le l} \lambda^j_1\lambda^j_3
(Y(w_1,z_1)Y({\bar w_2},z_1 z){\bar w_3},w_4) \cr
&\qquad - \std
\sum_{1 \le j \le l}
(Y(w_1,z_1)Y({\bar w_2},z_1 z)e_j(-1)f_j(0){\bar w_3},w_4) \cr
&\qquad - \std
\sum_{1 \le j \le l}
(Y(w_1,z_1)Y({\bar w_2},z_1 z)f_j(-1)e_j(0){\bar w_3},w_4) \cr
}$$
$$\eqalign{
&= \left(\frac{1}{z_1 z} \srd
\sum_{1 \le j \le l} \lambda^j_2\lambda^j_3
+\frac{1}{z_1} \srd
\sum_{1 \le j \le l} \lambda^j_1\lambda^j_3 \right)
(Y(w_1,z_1)Y({\bar w_2},z_1 z){\bar w_3},w_4) \cr
&\qquad + \frac{1}{z_1 z}\std
\sum_{1 \le j \le l}
(Y(w_1,z_1)Y(e_j(0){\bar w_2},z_1 z)f_j(0){\bar w_3},w_4) \cr
&\qquad + \frac{1}{z_1 z}\std
\sum_{1 \le j \le l}
(Y(w_1,z_1)Y(f_j(0){\bar w_2},z_1 z)e_j(0){\bar w_3},w_4) \cr
&\qquad + \frac{1}{z_1}\std
\sum_{1 \le j \le l}
(Y(f_j(0)w_1,z_1)Y({\bar w_2},z_1 z)e_j(0){\bar w_3},w_4) \cr
}$$
$$\eqalign{
&= \left(\frac{1}{z_1 z} \srd
\sum_{1 \le j \le l} \lambda^j_2\lambda^j_3
+\frac{1}{z_1} \srd
\sum_{1 \le j \le l} \lambda^j_1\lambda^j_3 \right)
(Y(w_1,z_1)Y({\bar w_2},z_1 z){\bar w_3},w_4) \cr
&\qquad + \frac{1}{z_1 z}\std
\sum_{1 \le j \le l}
(Y(w_1,z_1)Y(e_j(0){\bar w_2},z_1 z)f_j(0){\bar w_3},w_4) \cr
&\qquad + \frac{1}{z_1 z}\std
\sum_{1 \le j \le l}
(Y(w_1,z_1)Y(f_j(0){\bar w_2},z_1 z)e_j(0){\bar w_3},w_4) \cr
&\qquad - \frac{1}{z_1}\std
\sum_{1 \le j \le l}
(Y(w_1,z_1)f_j(0)Y({\bar w_2},z_1 z)e_j(0){\bar w_3},w_4)
,}$$
Therefore,
$$\eqalign{
&\left[ \frac{(1-z)}{z_1} \frac{\pt}{\pt z} +\frac{\pt}{\pt z_1} \right]
(Y(w_1,z_1)Y({\bar w_2},z_1 z){\bar w_3},w_4) \cr
&= \left(\frac{1}{z_1 z} \srd
\sum_{1 \le j \le l} \lambda^j_2\lambda^j_3
+\frac{1}{z_1} \srd
\sum_{1 \le j \le l} \lambda^j_1\lambda^j_3 \right)
(Y(w_1,z_1)Y({\bar w_2},z_1 z){\bar w_3},w_4) \cr
&\qquad + \frac{1}{z_1 z}\std
\sum_{1 \le j \le l}
(Y(w_1,z_1)Y(e_j(0){\bar w_2},z_1 z)f_j(0){\bar w_3},w_4) \cr
&\qquad + \frac{1}{z_1 z}\std
\sum_{1 \le j \le l}
(Y(w_1,z_1)Y(f_j(0){\bar w_2},z_1 z)e_j(0){\bar w_3},w_4) \cr
&\qquad - \frac{1}{z_1}\std
\sum_{1 \le j \le l}
(Y(w_1,z_1)Y(f_j(0){\bar w_2},z_1 z)e_j(0){\bar w_3},w_4) \cr
&\qquad - \frac{1}{z_1}\std
\sum_{1 \le j \le l}
(Y(w_1,z_1)Y({\bar w_2},z_1 z)f_j(0)e_j(0){\bar w_3},w_4)\cr
}$$
$$\eqalign{
&= \left(\frac{1}{z_1 z} \srd
\sum_{1 \le j \le l} \lambda^j_2\lambda^j_3
+\frac{1}{z_1} \srd
\sum_{1 \le j \le l} \lambda^j_1\lambda^j_3 \right)
(Y(w_1,z_1)Y({\bar w_2},z_1 z){\bar w_3},w_4) \cr
&\qquad + \frac{1}{z_1 z}\std
\sum_{1 \le j \le l}
(Y(w_1,z_1)Y(e_j(0){\bar w_2},z_1 z)f_j(0){\bar w_3},w_4) \cr
&\qquad + \frac{1-z}{z_1 z}\std
\sum_{1 \le j \le l}
(Y(w_1,z_1)Y(f_j(0){\bar w_2},z_1 z)e_j(0){\bar w_3},w_4) \cr
&\qquad - \frac{1}{z_1}\std
\sum_{1 \le j \le l}
(Y(w_1,z_1)Y({\bar w_2},z_1 z)f_j(0)e_j(0){\bar w_3},w_4) \cr
.}$$
Finally,
$$\eqalign{
&\left[ \frac{(1-z)}{z_1} \frac{\pt}{\pt z} +\frac{\pt}{\pt z_1}
- \frac{1}{z_1 z} \srd
\sum_{1 \le j \le l} \lambda^j_2\lambda^j_3
-\frac{1}{z_1} \srd
\sum_{1 \le j \le l} \lambda^j_1\lambda^j_3 \right] \cr
&\qquad\qquad\qquad \cdot
(Y(w_1,z_1)Y({\bar w_2},z_1 z){\bar w_3},w_4) \cr
&= \frac{1}{z_1 z}\std
\sum_{1 \le j \le l}
(Y(w_1,z_1)Y(e_j(0){\bar w_2},z_1 z)f_j(0){\bar w_3},w_4) \cr
&\quad+ \frac{1-z}{z_1 z}\std
\sum_{1 \le j \le l}
(Y(w_1,z_1)Y(f_j(0){\bar w_2},z_1 z)e_j(0){\bar w_3},w_4) \cr
&\quad - \frac{1}{z_1}\std
\sum_{1 \le j \le l}
(Y(w_1,z_1)Y({\bar w_2},z_1 z)f_j(0)e_j(0){\bar w_3},w_4) \cr
.}\eqno(4.46)$$
Let $z=\frac{z_2}{z_1}$, $r=\Delta (n_1,n_2)$ and let
$$\Phi_{xy}(z_1,z) = (Y(w_1,z_1) Y(x\cdot w_2,z_1 z)y \cdot
w_3,w_4).\eqno(4.47)$$
Define $f(z),g(z)$ by
$$\Phi_{xy}(z_1,z) = {z_1}^{-k}z^{-m}{(1-z)}^{-r} f(z),\qquad
\Phi_{yx}(z_1,z) = {z_1}^{-k}z^{-m}{(1-z)}^{-r} g(z).\eqno(4.48)$$
The general equation
reduces to the following two equations (one from $\Phi_{xy}$, the other
from $\Phi_{yx}$)
$$\eqalign{
z(1-z)f^{'}(z) =& [ (m+ \srd \sum_{1 \le j \le l} (\lambda^j_2 + \lambda^j_x)
(\lambda^j_3 + \lambda^j_y))\cr
&\qquad + ( k-m-r + \srd\sum_{1 \le j \le l} \lambda^j_1
(\lambda^j_3 + \lambda^j_y) -\std \tau(y \cdot w_3)) \cdot z ] f(z)\cr
&+(\std(\gamma_{xy}+\beta_{xy}) - \std\beta_{xy}\cdot z) g(z)
}\eqno(4.49)$$
and
$$\eqalign{
z(1-z)g^{'}(z) =& [ (m+ \srd \sum_{1 \le j \le l} (\lambda^j_2 + \lambda^j_y)
(\lambda^j_3 + \lambda^j_x)) \cr
&\qquad + ( k-m-r + \srd\sum_{1 \le j \le l} \lambda^j_1
(\lambda^j_3 + \lambda^j_x) -\std \tau(x \cdot w_3)) \cdot z ] g(z)\cr
&+(\std(\gamma_{yx}+\beta_{yx}) - \std\beta_{yx}\cdot z) f(z).
}\eqno(4.50)$$

Let
$$P=\bmatrix
m+ \srd \sum_{1 \le j \le l} (\lambda^j_2 + \lambda^j_x)
(\lambda^j_3 + \lambda^j_y)
&
\std(\gamma_{xy}+\beta_{xy})\\
\std(\gamma_{yx}+\beta_{yx})&
m+ \srd \sum_{1 \le j \le l} (\lambda^j_2 + \lambda^j_y)
(\lambda^j_3 + \lambda^j_x)
\endbmatrix$$
and for
$$F=
\srd \sum_{1 \le j \le l} (\lambda^j_1+ \lambda^j_2 + \lambda^j_x)
(\lambda^j_3 + \lambda^j_y) -\std \tau(y \cdot w_3)$$
and
$$G=
 \srd \sum_{1 \le j \le l} (\lambda^j_1+\lambda^j_2 + \lambda^j_y)
(\lambda^j_3 + \lambda^j_x)
-\std \tau(x \cdot w_3)$$
we have
$$Q=\bmatrix
k-r+F
&
\std\gamma_{xy}
\\
\std\gamma_{yx}
&
k-r+G
\endbmatrix.$$
Therefore
$$\bmatrix f^{'}(z) \\ g^{'}(z) \endmatrix \right]=
\frac{1}{1-z} Q
\bmatrix f(z)\\g(z)\endmatrix\right]
+ \frac{1}{z} P
\bmatrix f(z)\\g(z)\endmatrix\right]
.\eqno(4.51)$$

In all the cases, either $\beta_{xy}=0$ or $\beta_{yx}=0$ so let us assume
that $\beta_{xy}=0$ and combine the above two equations.  (Note: if an example
works out to have $\beta_{xy} \ne 0$ then switch $x$ and $y$.)
If $\beta_{xy}=0$ then $\tau(y\cdot w_3)=0$ and $\gamma_{yx}=0$ also.
$$\eqalign{
&z^2(1-z)^2 f^{''}(z)
\qquad
\qquad
\qquad
\qquad
\qquad
\qquad
\qquad
\qquad
\qquad
\qquad
\qquad
\cr
&+z(1-z)f^{'}(z) \cdot [
(1-2m-
\srd\sum_{1 \le j \le l} (\lambda^j_x + \lambda^j_y)(\lambda^j_2 + \lambda^j_3)
-
\std\sum_{1 \le j \le l} (\lambda^j_x  \lambda^j_y+\lambda^j_2  \lambda^j_3)
)\cr
&\quad\qquad\qquad-(2k-2r-2m+
\srd\sum_{1 \le j \le l} \lambda^j_1 (2\lambda^j_3 + \lambda^j_x + \lambda^j_y)
-\std\tau(x\cdot w_3))\cdot z]\cr
&+f(z)\{ (k-m-r+1-
\srd\sum_{1 \le j \le l} \lambda^j_1 (\lambda^j_3 + \lambda^j_x )
-\std\tau(x\cdot w_3))\cr
&\quad\qquad\qquad\quad \cdot (k-r-m+
\srd\sum_{1 \le j \le l} \lambda^j_1 (\lambda^j_3 + \lambda^j_y ))\cdot z^2
\cr
&\quad\qquad\qquad + \{ (m+
\srd\sum_{1 \le j \le l} (\lambda^j_2 + \lambda^j_y)(\lambda^j_3 + \lambda^j_x)
-1)\cr
&\qquad\qquad\qquad\qquad \cdot (k-r-m+
\srd\sum_{1 \le j \le l} \lambda^j_1 (\lambda^j_3 + \lambda^j_y ))\cr
&\quad\qquad\qquad\quad +  (m+
\srd\sum_{1 \le j \le l} (\lambda^j_2 + \lambda^j_x)(\lambda^j_3 + \lambda^j_y)
)\cr
&\quad\qquad\qquad\qquad\quad \cdot (k-r-m+
\srd\sum_{1 \le j \le l} \lambda^j_1 (\lambda^j_3 + \lambda^j_x )
-\std\tau(x\cdot w_3))\cr
&\quad\qquad\qquad\quad +  {\std}^2 \gamma_{xy}\beta_{yx} \} \cdot z \cr
&\quad\qquad\qquad + \{ (m+
\srd\sum_{1 \le j \le l} (\lambda^j_2 + \lambda^j_y)(\lambda^j_3 + \lambda^j_x)
)
(m+ \srd\sum_{1 \le j \le l} (\lambda^j_2 + \lambda^j_x)(\lambda^j_3 +
\lambda^j_y)
) \cr
&\quad\qquad\qquad\quad -
{\std}^2 \gamma_{xy}\beta_{yx} \} \}\cr
&\quad
\quad
\qquad
\qquad
\qquad
\qquad
\qquad
\qquad
\qquad
=0.}\eqno(4.52)$$
This second method is equivalent to solving KZ equations.  In this particular
case it was more convenient to solve the recursive equations from the first
method
than the differential equations in the second method and henceforth only the
recursive method is used.

Let us now proceed to solve for the functions of the first type, the
simpler cases.  The following table is a list of the vectors which will
be used in the recursions and their $h_j(0)$-eigenvalues, $\lambda^j$.
Let $1 \leq i \leq l$.

\newpage
\centerline{Table C}
\sk1
\begincellular{\centertable}
\row{}\cell{${\bar w}$}\cell{$\lambda^j,\ 1 \leq j \leq l$}
\row{}\cell{$\bvac$}\cell{0}
\row{}\cell{$h_{il}(0)a^+_1(-\shf)\bvac$}\cell{$\delta_{ij}$}
\row{}\cell{$f_{i1}(0)a^+_1(-\shf)\bvac$}\cell{$-\delta_{ij}$}
\row{}\cell{$\bvac'$}\cell{$-\shf$}
\row{}\cell{$f_{i1}(0)\bvac'$}\cell{$-\delta_{1j}-\delta_{ij} -\shf$}
\row{}\cell{$f_{li}(0)\bvac'$}\cell{$-\delta_{ij}-\delta_{lj}-\shf$}
\row{}\cell{$h_{li}(0)a^-_l(0)\bvac'$}\cell{$-\delta_{ij}-\shf$}
\row{}\cell{$f_{i1}(0)a^-_l(0)\bvac'$}\cell{$-\delta_{1j}-\delta_{ij}-
\delta_{lj}-\shf$}
\endcellular
\vskip 15pt

Here is an example of how to solve the simpler cases.
Let $w_1=\vvb,{\bar w_2}=h_{l1}(0)a^-_l(0)\bvac',$ ${\bar
w_3}=\bvac,w_4=\bvac'.$
We have $k=\shf$ and $m=0$ so
$$\oper={z_1}^{-\shf}\sum_P \Phi_P
{\left(\frac{z_2}{z_1}\right)}^P.$$
The recursion looks like
$$(-P-wt(w_1)-\srd \smj \lambda^j_1\lambda^j_3)\Phi_P = (-P +
\sei -1 +wt({\bar w_2}) + \srd \smj \lambda^j_2\lambda^j_3)\bPhi_{P+1}
$$
or
$(-P-\shf)\bPhi_P = (-P-1)\bPhi_{P+1}$
and so for all $P \ge 0$,
$$\bPhi_P = (-1)^P {-\shf \choose P} \bPhi_0.$$
The only question left is what is
the value of $\bPhi_0$.

$$\eqalign{
\bPhi_0&=(Y_0(\vvb)Y_{\sei}(h_{l1}(0)\vvd)\vva,\vvc) \cr
&=(a^+_1(0)Y_{\sei}(h_{l1}(0)\vvd)\vva,\vvc) \cr
&=(Y_{\sei}(h_{l1}(0)\vvd)\vva,a^-_1(0)\vvc) \cr
&=(Y_{\sei}(h_{l1}(0)\vvd)\vva,\frac{-1}{1+\shf\delta_{1l}} h_{l1}(0)\vvd) \cr
&=\frac{-1}{1+\shf\delta_{1l}}(h_{1l}(0)Y_{\sei}(h_{l1}(0)\vvd)\vva,\vvd) \cr
&=\frac{-1}{1+\shf\delta_{1l}}(Y_{\sei}(h_{1l}(0)h_{l1}(0)\vvd)\vva,\vvd) \cr
&=\frac{-1}{1+\shf\delta_{1l}}(Y_{\sei}(h_{1l}(0)\{-1-\shf\delta_{1l}\}
a^-_1(0)\vvc)\vva,\vvd) \cr
&=-(1+\shf\delta_{1l})(Y_{\sei}(\vvd)\vva,\vvd) \cr
&= -(1+\shf\delta_{1l}). }$$

Finally, we have that
$$\eqalign{
(Y(\vvb,z_1)&Y(h_{l1}(0)a^-_l(0)\bvac',z_2)\bvac, \bvac') \cr
&= {z_1}^{-\shf}\sum_{0 \leq P \in {\Bbb Z}} {-\shf \choose P} (-1)^P
\bPhi_0 {\left(\frac{z_2}{z_1}\right)}^k \cr
&=\bPhi_0 {z_1}^{-\shf} \sum_{0 \leq k \in {\Bbb Z}} {-\shf \choose k}
{\left(-\frac{z_2}{z_1}\right)}^k \cr
&=\bPhi_0 {z_1}^{-\shf} {\left(1 - \frac{z_2}{z_1}\right)}^{-\shf} \cr
&=-(1+\shf\delta_{1l}) \cdot {(z_1 - z_2)}^{-\shf}. }$$

Similarly, we may find the functions for each case in which either

\roster
\item "i)" has a $\bvac$ as one of the first three vectors,
\item "ii)" does not have an `$f(0)$' term.
\endroster

These cases are as follows:
$$
\aligned
&(Y(\bvac,z_1)Y(\bvac',z_2)\bvac,\bvac')=1, \\
&(Y(\bvac,z_1)Y(h_{li}(0)\vvd,z_2)h_{i1}(0)\vvb,\vvc)=-(1+\shf\delta_{il}
)B_l \cdot {z_2}^{-\shf}, \\
&(Y(\bvac,z_1)Y(f_{li}(0)\bvac',z_2)h_{i1}(0)a^+_1(-\shf)
\bvac,a^-_l(0)\bvac')= -\shf(1 + \delta_{il}) A_l \cdot {z_2}^{-\shf}, \\
&(Y(\vva,z_1)Y(\vvd,z_2)\vva,\vvd)=1, \\
&(Y(\vva,z_1)Y(\vvc,z_2)f_{l1}(0)\vvb,\vvd)= \shf(1 +\delta_{1l}) A_l
\cdot {z_2}^{-\shf}, \\
\endaligned$$
$$
\aligned
&(Y(\vvb,z_1)Y(f_{l1}(0)\vvc,z_2)\vva,\vvd)= -\shf(1+\delta_{1l}) \cdot
{(z_1 - z_2)}^{-\shf}, \\
&(Y(\vvc,z_1)Y(\vva,z_2)\vva,\vvc)=1, \\
&(Y(\vvc,z_1)Y(f_{l1}(0)\vvb,z_2)\vva,\vvd)= \shf(1+\delta_{1l})A_l \cdot
{(z_1-z_2)}^{-\shf}, \\
&(Y(\vvc,z_1)Y(\vva,z_2)f_{l1}(0)\vvb,\vvd)= \shf (1+\delta_{1l})
A_l \cdot {z_1}^{-\shf}, \\
&(Y(\vvd,z_1)Y(h_{l1}(0)\vvb,z_2)\vva,\vvc)= B_l \cdot {(z_1-z_2)}^
{-\shf}, \\
&(Y(\vvd,z_1)Y(\vva,z_2)h_{l1}(0)\vvb,\vvc)= B_l \cdot {z_1}^{-\shf}, \\
&(Y(\vvd,z_1)Y(\vva,z_2)\vva,\vvd)= 1. \\
\endaligned
$$

We are now left with nine unknown functions.  Looking at the nine
functions one will notice that each one is either of the form
$$(Y(w_1,z_1)Y(h_{i1}(0)w_2,z_2)f_{i1}(0)w_3,w_4) \ \hbox{ or }
\ (Y(w_1,z_1)Y(f_{i1}(0)w_2,z_2)h_{i1}(0)w_3,w_4)$$
where $1 \leq i \leq l$ and $w_1,w_2,w_3,w_4 \in \Omega$.

It will be convenient to define the functions $\alpha_i:\Omega
\to {\Bbb Q}$, such that
$$f_{ii}(0)h_{i1}(0)w = \alpha_i(w)f_{i1}(0)w.$$
So we have that
$$
\aligned
&\alpha_i(\vva)=1, \\
&\alpha_i(\vvb)=2-\delta_{1i}, \\
&\alpha_i(\vvc)=-\shf\delta_{1i}, \\
&\alpha_i(\vvd)=-(\delta_{il}+\shf\delta_{1i}). \\
\endaligned
\eqno(4.53)$$
We will also need that for any
$w \in \Omega$,
$$
\aligned
&h_{1i}(0)w=\delta_{1i}h_{i1}(0)w, \\
&e_j(0)f_{i1}(0)w=\shf\delta_{ij}(1+\delta_{1i})h_{i1}(0)w, \\
&
f_j(0)e_j(0)f_{i1}(0)w=\shf\delta_{ij}(1+\delta_{1i})\alpha_i(w)f_{i1}(0)
w. \\
\endaligned
\eqno(4.54)$$

Let us say that for any $w_1,w_2,w_3,w_4 \in \Omega,$
$$\Phi_P=(Y_{P+k-m-wt(w_1)}(w_1)Y_{-P+m-wt(w_2)}(h_{i1}(0)w_2)f_{i1}(0)w_3,w_4)
\eqno(4.55)$$
and
$$\Psi_P=(Y_{P+k-m-wt(w_1)}
(w_1)Y_{-P+m-wt(w_2)}(f_{i1}(0)w_2)h_{i1}(0)w_3,w_4).
\eqno(4.56)$$

Then the recursions are
$$\eqalign{
&\srd(1+\delta_{1i})\alpha(w_3)\Phi_{P+1} \cr
&=(P-m+1
-\srd\smj\lmm\llmp)\Psi_{p+1} \cr
&\quad -(P+k-m +\srd\smj\lambda^j_1\llmp)\Psi_P,}\eqno(4.57)$$
and
$$\eqalign{
&\srd(1+\delta_{1i})\alpha_i(w_2)[\Psi_{P+1}-\Psi_P] \cr
&=(P-m+1
-\srd\smj\lmp\llmm)\Phi_{P+1} \cr
&\quad
-(P+k-m+\srd\smj\lambda^j_1\llmm-\srd(1+\delta_{1i})
\alpha_i(w_3))\Phi_P.}\eqno(4.58)$$
Combining these two formulas gives us the following:
$$\eqalign{
&[\sni(1+3\delta_{1i})\alpha_i(w_2)\alpha_i(w_3)  \cr
&\quad - \{(P-m+1
-\srd\smj\lmp\llmm) \cr
&\quad \quad \cdot (P-m+1
-\srd\smj\lmm\llmp)\}] \Psi_{P+1}
\cr
& =[\sni(1+3\delta_{1i})\alpha_i(w_2)\alpha_i(w_3) \cr
&\quad - \{(P-m+1
-\srd\smj\lmp\llmm) \cr
& \quad \quad \cdot (P+k-m+\srd\smj\lambda^j_1\llmp)\} \cr
&\quad - \{(P+k-m + \srd\smj\lambda^j_1\llmm-\srd
(1+\delta_{1i})\alpha_i
(w_3)) \cr
& \quad \quad \cdot (P-m
-\srd\smj \lmm\llmp)\}] \Psi_P \cr
&\quad + [(P+k-m+ \srd\smj\lambda^j_1\llmm-\srd(1+\delta_{1i})
\alpha_i(w_3)) \cr
& \quad \quad \cdot (P+k-m-1 + \srd\smj\lambda^j_1\llmp)] \Psi_{P-1}.}
\eqno(4.59)$$

i) For the case where $w_1=\vvb,w_2=\vvc,w_3=\vvb$, and $w_4=\vvc$ we have that
$\lambda^j_1=\delta_{1j},\lambda^j_2=-\shf$, and $\lambda^j_3 =
\delta_{1j}$.  We find The recursion simplifies to
$$(P+1)(P+\std)\Psi_{P+1}= (2P^2+P+\srd)\Psi_P
-(P-\shf)(P-\sxt)\Psi_{P-1},$$
therefore
$$\bPsi_P = {(-1)}^P{ -\shf \choose P} \bPsi_0.$$
Thus we can solve for $\bPhi_P$, getting
$$\eqalign{
\bPhi_{P+1}&= \sth (P+1+\srd\delta_{1i}) \bPsi_{P+1} - \sth
(P+\shf+\srd\delta_{1i})\bPsi_P \cr
&= \sfr \delta_{i1} \frac{{(-1)}^{P+1}}{P+1} {-\shf \choose P}
\bPsi_0.}$$
We will also need
$$\eqalign{
\bPsi_0&=(Y_0(\vvb)Y_{\shf+\sei}(f_{i1}(0)\vvc)h_{i1
}(0)\vvb,\vvc) \cr
&=(a^+_1(0)Y_{\shf+\sei}(f_{i1}(0)\vvc)h_{i1}(0)\vvb,\vvc) \cr
&=(Y_{\shf+\sei}(f_{i1}(0)\vvc)h_{i1}(0)\vvb,a^-_1(0)\vvc) \cr
&=(Y_{\shf+\sei}(f_{i1}(0)\vvc)h_{i1}(0)\vvb,\frac{-1}{1+ \shf\delta_{1l}}
h_{l1}(0)\vvd) \cr
&=\frac{-1}{1+\shf\delta_{1l}}
(h_{1l}(0)Y_{\shf+\sei}(f_{i1}(0)\vvc)h_{i1}(0) \vvb,\vvd) \cr
&=\frac{-1}{1+\shf\delta_{1l}}[ (Y_{\shf+\sei}(h_{1l}(0)f_{i1}(0)\vvc)
h_{i1}(0) \vvb,\vvd) \cr
&= \qquad \qquad + (Y_{\shf+\sei}(f_{i1}(0)\vvc)h_{1l}(0)h_{i1}(0)
\vvb,\vvd)] \cr
&= \frac{-1}{1+\shf\delta_{1l}}[ -(1+\delta_{1i}+\shf\delta_{1l})
(Y_{\shf+\sei}(f_{li}(0)\vvc)h_{i1}(0)\vvb,\vvd) \cr
& \qquad \qquad + \delta_{il}(Y_{\shf+\sei}(f_{l1}(0)\vvc)\vvb,\vvd) ]
\cr
&= \frac{-1}{1+\shf\delta_{1l}} [(1+\delta_{1i}+\shf\delta_{1l})
(Y_{\shf+\sei}(\vvc)f_{li}(0)h_{i1}(0)\vvb,\vvd) \cr
&\qquad \qquad - \shf\delta_{il}
(1+\delta_{1l}) A_l] \cr
&= \frac{-1}{1+\shf\delta_{1l}}
[(1+\delta_{1i}+\shf\delta_{1l})\frac{1+\delta_{il}}{1+\delta_{1l}} -
\delta_{il}]\shf(1+\delta_{1l}) A_l \cr
&= -\shf(1+\delta_{1i})A_l,}$$
and
$$\eqalign{
\bPhi_0&=(Y_0(\vvb)Y_{\shf+\sei}(h_{i1}(0)\vvc)f_{i1}(0)\vvb,\vvc) \cr
&=\frac{-1}{1+\shf\delta_{1l}} (h_{1l}(0)Y_{\shf+\sei}(h_i1(0)\vvc)f_{i1}
(0)\vvb,\vvd) \cr
&=\frac{-1}{1+\shf\delta_{1l}} [(Y_{\shf+\sei}(h_{1l}(0) h_{i1}(0) \vvc)
f_{i1}(0)\vvb,\vvd) \cr
& \qquad \qquad + (Y_{\shf+\sei}(h_{i1}(0)\vvc) h_{1l}(0) f_{i1}(0)
\vvb,\vvd)] \cr
&= \frac{-1}{1+\shf\delta_{1l}} [\sfr\delta_{1l} \shf(1+\delta_{1l})
A_l \cr
&\qquad \qquad- \shf\delta_{1i}
(Y_{\shf+\sei}(\vvc)h_{1l}(0)f_{11}(0) \vvb,
\vvd)] \cr
&= \frac{-1}{1+\shf\delta_{1l}} [\sfr\delta_{1l} +
\frac{\shf\delta_{1i}}{\shf(1+\delta_{1l})}]\shf (1+\delta_{1l}) A_l \cr
}$$
$$\eqalign{
&= -\frac{\delta_{1i}+\shf\delta_{1l}}{1+2\delta_{1l}}\shf(1+\delta_{1l})
 A_l \cr
&= -\shf\delta_{1i}A_l.}$$
Therefore we have that
$$\eqalign{
&(Y(\vvb,z_1)Y(h_{i1}(0)\vvc,z_2)f_{i1}(0)\vvb,\vvc) \cr
&={z_1}^{-\shf}{z_2}^{-\shf}\smk\bPhi_P{\left(\frac{z_2}{z_1}\right)}^{P} \cr
&=\bPhi_0{z_1}^{-\shf}{z_2}^{-\shf} + {z_1}^{-\shf}{z_2}^{-\shf}\smk
\bPhi_{P+1} {\left(\frac{z_2}{z_1}\right)}^{P+1} \cr
&=\bPhi_0{z_1}^{-\shf}{z_2}^{-\shf} + \sfr\delta_{1i} \bPsi_0
{z_1}^{-\shf}{z_2}^{-\shf} \smk {-\shf \choose P}
\frac{(-1)^{P+1}}{P+1}{\left(\frac{z_2}{z_1}\right)}^{P+1} \cr
&=\bPhi_0{z_1}^{-\shf}{z_2}^{-\shf} + \sfr\delta_{1i} \bPsi_0 \{
2\left(1-\frac{z_2}{z_1}\right)^{\shf} - 2 \} \cr
&=-\delta_{1i}\shf A_l\cdot {z_1}^{-\shf}{z_2}^{-\shf} -\shf\delta_{1i}
\frac{1+\delta_{1i}}{2} A_l \cdot {z_1}^{-\shf}{z_2}^{-\shf}
\cr
&\qquad -\shf\delta_{1i}\shf(1+\delta_{1i})
 A_l\cdot {z_1}^{-1}{z_2}^{-\shf}{(z_1-z_2)}
^{\shf} \cr
&=-\shf\delta_{1i}\shf(1+\delta_{1i})
 A_l \cdot {z_1}^{-1}{z_2}^{-\shf}{(z_1-z_2)}
^{\shf} \cr
&=-\shf\delta_{1i}A_l \cdot {z_1}^{-1}{z_2}^{-\shf}{(z_1-z_2)}^{\shf},}$$
and
$$\eqalign{
&(Y(\vvb,z_1)Y(f_{i1}(0)\vvc,z_2)h_{i1}(0)\vvb,\vvc) \cr
&={z_1}^{-\shf}{z_2}^{-\shf}\smk\bPsi_P{\left(\frac{z_2}{z_1}\right)}^{P} \cr
}$$
$$\eqalign{
&={z_1}^{-\shf}{z_2}^{-\shf}\smk{-\shf \choose P} (-1)^P
{\left(\frac{z_2}{z_1}\right)}^P \bPsi_0 \cr
&=\bPsi_0{z_1}^{-\shf}{z_2}^{-\shf}{\left(1-\frac{z_2}{z_1}\right)}^{-\shf} \cr
&=-\shf(1+\delta_{1i}) A_l \cdot {z_2}^{-\shf}{(z_1-z_2)}^{-\shf}.
}$$

ii) For the case where $w_1=\vvb,w_2=\vvd,w_3=\vvb$, and $w_4=\vvd$ we have
that
$\lambda^j_1=\delta_{1j},\lambda^j_2=-\shf-\delta_{jl}$, and
$\lambda^j_3=\delta_{1j}$.
We find the recursion simplifies to
$$(P+1)(P+\std)\bPsi_{P+1}=(2P^2+P+\srd)\bPsi_P-(P-\sxt-\srd \delta_{1i})
(P-\shf+\srd\delta_{1i})\bPsi_{P-1},$$
which is the same as
$$(P+1)(P+\std)\bPsi_{P+1}=(2P^2+P+\srd)\bPsi_P-
(P-\sxt)(P-\shf)\bPsi_{P-1},$$
therefore,
$$\bPsi_P={-\shf \choose P}(-1)^P\bPsi_0.$$
Thus we can solve for $\bPhi_P$, getting
$$\eqalign{
\bPhi_{P+1}&= \sth(P+1+\srd\delta_{1i}+\srd\delta_{il})\bPsi_{P+1}
-\sth(P+\shf+\srd\delta_{1i})\Psi_P \cr
&=-\shf\frac{(-1)^{P+1}}{P+1}{-\shf \choose P}[\delta_{il}P+\shf
\delta_{il}-\shf\delta_{1i}].}$$
We find that
$$\eqalign{
\bPsi_0&=(Y_0(\vvb)Y_{\shf+\sei}(f_{i1}(0)\vvd)h_{i1}(0)\vvb,\vvd) \cr
&=(Y_{\shf+\sei}(f_{i1}(0)\vvd)h_{i1}(0)\vvb,a^-_1(0)\vvd) \cr
&=2(e_{l1}(0)Y_{\shf+\sei}(f_{i1}(0)\vvd)h_{i1}(0)\vvb,\vvc) \cr
&=2(Y_{\shf+\sei}(e_{l1}(0)f_{i1}(0)\vvd)h_{i1}(0)\vvb,\vvc) \cr
&=-\shf(1+3\delta_{1l}+\delta_{1i}+\delta_{il})(Y_{\shf+\sei}
(a^-_i(0)\vvc)h_{i1}(0)\vvb,\vvc) \cr
&=\frac{\shf(1+3\delta_{1l}+\delta_{1i}+\delta_{il})}{1+\shf\delta_{il}}
(Y_{\shf+\sei}(h_{li}(0)\vvd)h_{i1}(0)\vvb,\vvc) \cr
&=\frac{\shf(1+3\delta_{1l}+\delta_{1i}+\delta_{il})}{1+\shf\delta_{il}}
(-\shf\delta_{il}-1) B_l \cr
&=-\shf(1+3\delta_{1l}+\delta_{1i}+\delta_{il})B_l,
}$$
and
$$\eqalign{
\bPhi_0&=(Y_0(\vvb)Y_{\shf+\sei}(h_{i1}(0)\vvd)f_{i1}(0)\vvb,\vvd) \cr
&=2(e_{l1}(0)Y_{\shf+\sei}(h_{i1}(0)\vvd)f_{i1}(0)\vvb,\vvc) \cr
&=2(Y_{\shf+\sei}(h_{i1}(0)\vvd)e_{l1}(0)f_{i1}(0)\vvb,\vvc) \cr
&=\shf(1+\delta_{1i}+\delta_{1l}
)(Y_{\shf+\sei}(h_{i1}(0)\vvd)h_{li}(0)\vvb,\vvc) \cr
&\qquad +\shf\delta_{il}(Y_{\shf+\sei}(h_{i1}(0)\vvd)\vvb,\vvc) \cr
&=\shf\delta_{1i}(1+\delta_{1l}+\delta_{1i})
[-\shf
- \delta_{1l}]B_l
+\shf\delta_{il}[-\shf\delta_{1l}-1]B_l \cr
&=-\shf(\delta_{1i}+4\delta_{1l}+\delta_{il})B_l.
}$$
Therefore we have that
$$\eqalign{
&(Y(\vvb,z_1)Y(h_{i1}(0)\vvd,z_2)f_{i1}(0)\vvb,\vvd) \cr
&={z_1}^{-\shf}{z_2}^{-\shf}\smk \bPhi_P {\left(\frac{z_2}{z_1}\right)}^{P} \cr
&=\bPhi_0{z_1}^{-\shf}{z_2}^{-\shf} \cr
&\qquad -\shf\bPsi_0
{z_1}^{-\shf}{z_2}^{-\shf}\smk
{-\shf \choose P}
\frac{(-1)^{P+1}}{P+1}{\left(\frac{z_2}{z_1}\right)}^{P+1}[\delta_{il}P
+\shf\delta_{il}
-\shf\delta_{1i}] \cr
&=\bPhi_0{z_1}^{-\shf}{z_2}^{-\shf} + \shf\delta_{il}\bPsi_0
{z_1}^{-1}{z_2}^{\shf}{(z_1-z_2)}^{-\shf} \cr
&\qquad +\sfr(\delta_{1i}+\delta_{il})\bPsi_0{z_1}^{-\shf}{z_2}^{-\shf}
[2{\left(1-\frac{z_2}{z_1}\right)}^{\shf}-2] \cr
&=\shf\bPsi_0[\delta_{il}{z_1}^{-1}{z_2}^{\shf}{(z_1-z_2)}^{-\shf} +
(\delta_{1i}+\delta_{il}){z_1}^{-1}{z_2}^{-\shf}{(z_1-z_2)}^{\shf}] \cr
&=-\sfr(1+3\delta_{1l}+\delta_{1i}+\delta_{il})B_l
{z_1}^{-1}{z_2}^{-\shf}{(z_1-z_2)}^{-\shf}[\delta_{il}z_2+ (\delta_{1i}
+\delta_{il})(z_1-z_2)] \cr
&=-\sfr(1+3\delta_{1l}+\delta_{1i}+\delta_{il})B_l \cdot
[(\delta_{1i}+\delta_{il})z_1-\delta_{1i}z_2] {z_1}^{-1}{z_2}^{-\shf}
{(z_1-z_2)}^{-\shf},
}$$
and
$$\eqalign{
&(Y(\vvb,z_1)Y(f_{i1}(0)\vvd,z_2)h_{i1}(0)\vvb,\vvd) \cr
&={z_1}^{-\shf}{z_2}^{-\shf}\smk \bPsi_P {\left(\frac{z_2}{z_1}\right)}^P\cr
&=\bPsi_0{z_1}^{-\shf}{z_2}^{-\shf}\smk{-\shf \choose P}{(-1)}^P
{\left(\frac{z_2}{z_1}\right)}^P \cr
&=\bPsi_0{z_2}^{-\shf}{(z_1-z_2)}^{-\shf} \cr
&=-\shf(1+3\delta_{1l}+\delta_{1i}+\delta_{il})B_l \cdot {z_2}^{-\shf}
{(z_1-z_2)}^{-\shf}.}$$

iii) For the case where $w_1=\vvc,w_2=\vvb,w_3=\vvb$, and $w_4=\vvc$ we have
that $\lambda^j_1=-\shf,\lambda^j_2=\delta_{1j}$ and
$\lambda^j_3=\delta_{1j}$.
We find the recursion
simplifies to
$$(P+1)(P-\srd)\bPsi_{P+1}=(2P^2-P-\sxt)\bPsi_P-(P-\shf)(P-\svx)
\bPsi_{P-1},$$
therefore,
$$\bPsi_P={-\shf \choose P}(-1)^P\bPsi_0.$$
Thus we can solve for $\bPhi_P$, getting
$$\eqalign{
\bPhi_{P+1}&= \sth(P+\srd)\bPsi_{P+1}-\sth(P-\sxt)\bPsi_P \cr
&=-\shf\frac{(-1)^{P+1}}{P+1}{-\shf \choose P}\bPsi_0.}$$
We find that
$$\eqalign{
\bPsi_0&=(Y_{\sei}(\vvc)Y_{\shf}(f_{i1}(0)\vvb)h_{i1}(0)\vvb,\vvc) \cr
&=(Y_{\sei}(\vvc) [\shf(1+\delta_{1i})]a^-_i(\shf)a^+_i(-\shf)\vva,\vvc)
\cr
&=-\shf(1+\delta_{1i})(Y_{\sei}(\vvc)\vva,\vvc) \cr
&=-\shf(1+\delta_{1i}),}$$
and
$$\eqalign{
\bPhi_0&=(Y_{\sei}(\vvc)Y_{\shf}(h_{i1}(0)\vvb)f_{i1}(0)\vvb,\vvc) \cr
&=(Y_{\sei}(\vvc)a^+_i(\shf)[\shf(1+\delta_{1i})]a^-_i(-\shf)\vva,\vvc)
\cr
&=\shf(1+\delta_{1i})(Y_{\sei}(\vvc)\vva,\vvc) \cr
&=\shf(1+\delta_{1i}).}$$
Therefore we have that
$$\eqalign{
&(Y(\vvc,z_1)Y(h_{i1}(0)\vvb,z_2)f_{i1}(0)\vvb,\vvc) \cr
&={z_2}^{-1}\smk\bPhi_P {\left(\frac{z_2}{z_1}\right)}^{P} \cr
&=\bPhi_0{z_2}^{-1} -\shf\bPsi_0\smk{-\shf \choose P}
\frac{(-1)^{P+1}}{P+1} \left(\frac{z_2}{z_1}\right)^{P+1} \cr
&=\bPhi_0{z_2}^{-1} -\shf\bPsi_0 {z_2}^{-1}
(2\left(1-\frac{z_2}{z_1}\right)^{\shf}-2) \cr
&=-\bPsi_0{z_1}^{-\shf}{z_2}^{-1}{(z_1-z_2)}^{\shf} \cr
&=\shf(1+\delta_{1i})\cdot{z_1}^{-\shf} {z_2}^{-1}
{(z_1-z_2)}^{\shf},}$$
and
$$\eqalign{
&(Y(\vvc,z_1)Y(f_{i1}(0)\vvb,z_2)h_{i1}(0)\vvb,\vvc) \cr
&={z_2}^{-1}\smk\bPsi_P {\left(\frac{z_2}{z_1}\right)}{P} \cr
&=\bPsi_0{z_2}^{-1}{\left(1-\frac{z_2}{z_1}\right)}^{-\shf} \cr
&=-\shf(1+\delta_{1i})\cdot{z_1}^{\shf}{z_2}^{-1} {(z_1-z_2)}^{-\shf}.}$$

iv) For the case where $w_1=\vvd,w_2=\vvb,w_3=\vvb,$ and $w_4=\vvd$ we
have that $\lambda^j_1=-\shf-\delta_{jl},\lambda^j_2=\delta_{1j}$ and
$\lambda^j_3=\delta_{1j}$.
We find the recursion
simplifies to
$$(P+1)(P-\srd)\bPsi_{P+1} = (2P^2-P-\sxt-\srd\delta_{il})\bPsi_P -
(P-\shf+\srd\delta_{il})(P-\svx-\srd\delta_{il})\bPsi_{P-1},$$
therefore,
$$\bPsi_{P+1}=[(-1)^{P+1}{-\shf \choose P+1} + \delta_{il}(-1)^P{-\shf
\choose P}]\bPsi_0.$$
Thus we can solve for $\bPhi_P$, getting
$$\eqalign{
\bPhi_{P+1}&=\sth(P+\srd)\bPsi_{P+1} -\sth(P-\sxt-\srd\delta_{il})
\bPsi_P \cr
&=[\delta_{il}(-1)^P{-\shf \choose P}-\shf \frac{(-1)^{P+1}}{P+1}{-\shf
\choose P}]\bPsi_0.}$$
We find that
$$\eqalign{
\bPsi_0&=(Y_{\sei}(\vvd)Y_{\shf}(f_{i1}(0)\vvb)h_{i1}(0)\vvb,\vvd) \cr
&=(Y_{\sei}(\vvd)[\shf(1+\delta_{1i})]a^-_i(\shf)a^+_i(-\shf)\vva,\vvd)
\cr
&=-\shf(1+\delta_{1i})(Y_{\sei}(\vvd)\vva,\vvd) \cr
&=-\shf(1+\delta_{1i}), }$$
and
$$\eqalign{
\bPhi_0&=(Y_{\sei}(\vvd)Y_{\shf}(h_{i1}(0)\vvb)f_{i1}(0)\vvb,\vvd) \cr
&=(Y_{\sei}(\vvd)a^+_i(\shf)[\shf(1+\delta_{1i})]a^-_i(-\shf)\vva,\vvd
\cr
&=\shf(1+\delta_{1i})(Y(\vvd)\vva,\vvd) \cr
&=\shf(1+\delta_{1i}).}$$
Therefore we have that
$$\eqalign{
&(Y(\vvd,z_1)Y(h_{i1}(0)\vvb,z_2)f_{i1}(0)\vvb,\vvd) \cr
&={z_2}^{-1}\smk\bPhi_P {\left(\frac{z_2}{z_1}\right)}^{P} \cr
&=\bPhi_0{z_2}^{-1} +\delta_{il}\bPsi_0
{z_1}^{-1}\left(1-\frac{z_2}{z_1}\right)^{-\shf} -\shf\bPsi_0
{z_2}^{-1}[2\left(1-\frac{z_2}{z_1}\right)^{\shf}-2] \cr
&=-\shf\delta_{il}(1+\delta_{1i}){z_1}^{-\shf}{(z_1-z_2)}^{-\shf} +
\shf(1+\delta_{1i}){z_1}^{-\shf}{z_2}^{-1}{(z_1-z_2)}^{\shf} \cr
&=\shf(1+\delta_{1i})\cdot [z_1-(1+\delta_{il})z_2]
{z_1}^{-\shf}{z_2}^{-1}{(z_1-z_2)}^{-\shf},}$$
and
$$\eqalign{
&(Y(\vvd,z_1)Y(f_{i1}(0)\vvb,z_2)h_{i1}(0)\vvb,\vvd) \cr
&={z_2}^{-1}\smk\bPsi_P{\left(\frac{z_2}{z_1}\right)}^{P} \cr
&=\bPsi_0{z_2}^{-1} + \bPsi_0{z_2}^{-1}\smk(-1)^{P+1}{-\shf \choose P+1}
\left(\frac{z_2}{z_1}\right)^{P+1} \cr
&\qquad\qquad + \delta_{il}\bPsi_0{z_1}^{-1}\smk(-1)^P {-\shf
\choose P} \left(\frac{z_2}{z_1}\right)^P \cr
&=\bPsi_0{z_2}^{-1}\left(1-\frac{z_2}{z_1}\right)^{-\shf}+ \delta_{il}\bPsi_0
{z_1}^{-1} \left(1-\frac{z_2}{z_1}\right)^{-\shf} \cr
&=\bPsi_0[{z_1}^{\shf}{z_2}^{-1}(z_1-z_2)^{-\shf} +
\delta_{il}{z_1}^{-1}(z_1-z_2)^{-\shf}] \cr
&=-\shf(1+\delta_{1i})\cdot [z_1+\delta_{il}z_2]{z_1}^{-\shf}{z_2}^{-1}
{(z_1-z_2)}^{-\shf}.}$$

We can now summarize all the calculations we have done for these base cases
for Permutability
in Tables 1 and 2.

Let $\cR = \cRR$.  In the following tables we write
 $\qqa=\vva,\qqc=\vvc$.
Let $v_i \in \bV_{n_i}$ for $1 \leq i \leq 4$,
$s=\delta_{1i}+4\delta_{1l}+\delta_{il}$, $t=\shf(1+\delta_{1i})$ and
let $f(z_1,z_2)$ be the function to which the series
$${z_1}^{\Delta(n_1,n_3)}{z_2}^{\Delta(n_2,n_3)}(z_1-z_2
)^{\Delta(n_1,n_2)}(Y(v_1,z_1)Y(v_2,z_2)v_3,v_4)$$
converges in the domain $0<|z_2|<|z_1|$.
Let $\eta(n_1,n_2)$
be defined by
$$\eqalign{
&{z_1}^{\Delta(n_1,n_3)}{z_2}^{\Delta(n_2,n_3)}(z_1-z_2
)^{\Delta(n_1,n_2)}(Y(v_1,z_1)Y(v_2,z_2)v_3,v_4) \cr
&\sim \eta(n_1,n_2) {z_1}^{\Delta(n_1,n_3)}{z_2}^{\Delta(n_2,n_3)} (z_1-z_2
)^{\Delta(n_1,n_2)}(Y(v_2,z_2)Y(v_1,z_1)v_3,v_4),}$$
where $\sim$ means the two series converge in their respective domains
to the same rational function $f(z_1,z_2) \in \cR$.

\newpage
\centerline{Table 1}
\sk1
\begincellular{\centertable}
\row{}\cell{$v_1$}\cell{$v_2$}\cell{$v_3$}\cell{$v_4$}
\cell{$f(z_1,z_2)$}
\row{}\cell{$\qqa$}\cell{$\qqc$}\cell{$\qqa$}\cell{$\qqc$}
\cell{1}
\row{}\cell{$\qqa$}\cell{$h_{li}(0)\qqd$}\cell{$h_{i1}(0)\qqb$}
\cell{$\qqc$}
\cell{$-(1+\shf\delta_{il})B_l$}
\row{}\cell{$\qqa$}\cell{$f_{li}(0)\qqc$}\cell{$h_{i1}(0)\qqb$}
\cell{$\qqd$}
\cell{$-\shf(1+\delta_{il})A_l$}
\row{}\cell{$\qqa$}\cell{$\qqd$}\cell{$\qqa$}\cell{$\qqd$}
\cell{1}
\row{}\cell{$\qqa$}\cell{$\qqc$}\cell{$f_{l1}(0)\qqb$}\cell{$\qqd$}
\cell{$\shf(1+\delta_{1l})A_l$}
\row{}\cell{$\qqb$}\cell{$\qqc$}\cell{$\qqc$}\cell{$\qqa$}
\cell{0}
\row{}\cell{$\qqb$}\cell{$h_{l1}(0)\qqd$}\cell{$\qqa$}\cell{$\qqc$}
\cell{$-(1+\shf\delta_{1l})$}
\row{}\cell{$\qqb$}\cell{$f_{l1}(0)\qqc$}\cell{$\qqa$}\cell{$\qqd$}
\cell{$-\shf(1+\delta_{1l})$}
\row{}\cell{$\qqb$}\cell{$h_{i1}(0)\qqc$}\cell{$f_{i1}(0)\qqb$}
\cell{$\qqc$}
\cell{$-\shf\delta_{1i}A_l \cdot (z_1-z_2)$}
\row{}\cell{$\qqb$}\cell{$f_{i1}(0)\qqc$}\cell{$h_{i1}(0)\qqb$}
\cell{$\qqc$}
\cell{$-tA_l\cdot z_1$}
\row{}\cell{$\qqb$}\cell{$h_{i1}(0)\qqd$}\cell{$f_{i1}(0)\qqb$}\cell{$\qqd$}
\cell{$\shf B_l[(\delta_{1i}+2\delta_{1l})z_2-s
z_1]$}
\row{}\cell{$\qqb$}\cell{$f_{i1}(0)\qqd$}\cell{$h_{i1}(0)\qqb$}\cell{$\qqd$}
\cell{$-\shf(1+s-\delta_{1l})B_l\cdot z_1$}
\row{}\cell{$\qqc$}\cell{$\qqa$}\cell{$\qqa$}\cell{$\qqc$}
\cell{1}
\row{}\cell{$\qqc$}\cell{$\qqb$}\cell{$\qqc$}\cell{$\qqa$}
\cell{0}
\row{}\cell{$\qqc$}\cell{$f_{l1}(0)\qqb$}\cell{$\qqa$}\cell{$\qqd$}
\cell{$\shf(1+\delta_{1l})A_l$}
\row{}\cell{$\qqc$}\cell{$\qqa$}\cell{$f_{l1}(0)\qqb$}\cell{$\qqd$}
\cell{$\shf(1+\delta_{1l})A_l$}
\row{}\cell{$\qqc$}\cell{$h_{i1}(0)\qqb$}\cell{$f_{i1}(0)\qqb$}\cell{$\qqc$}
\cell{$t\cdot (z_1-z_2)$}
\row{}\cell{$\qqc$}\cell{$f_{i1}(0)\qqb$}\cell{$h_{i1}(0)\qqb$}\cell{$\qqc$}
\cell{$-t\cdot z_1$}
\row{}\cell{$\qqc$}\cell{$\qqc$}\cell{$\qqb$}\cell{$\qqa$}
\cell{0}
\row{}\cell{$\qqc$}\cell{$\qqc$}\cell{$\qqc$}\cell{$\qqd$}
\cell{0}
\row{}\cell{$\qqd$}\cell{$h_{l1}(0)\qqb$}\cell{$\qqa$}\cell{$\qqc$}
\cell{$B_l$}
\row{}\cell{$\qqd$}\cell{$\qqa$}\cell{$h_{l1}(0)\qqb$}\cell{$\qqc$}
\cell{$B_l$}
\row{}\cell{$\qqd$}\cell{$\qqa$}\cell{$\qqa$}\cell{$\qqd$}
\cell{1}
\row{}\cell{$\qqd$}\cell{$h_{i1}(0)\qqb$}\cell{$f_{i1}(0)\qqb$}\cell{$\qqd$}
\cell{$t\cdot[z_1-(1+\delta_{il})z_2]$}
\row{}\cell{$\qqd$}\cell{$f_{i1}(0)\qqb$}\cell{$h_{i1}(0)\qqb$}\cell{$\qqd$}
\cell{$-t\cdot[z_1+\delta_{il}z_2]$}
\endcellular


\newpage

\centerline{Table 2}
\sk1
\begincellular{\centertable}
\row{}\cell{$v_1$}\cell{$v_2$}\cell{$v_3$}\cell{$v_4$}
\cell{$\eta(n_1,n_2)$}
\row{}\cell{$\qqa$}\cell{$\qqc$}\cell{$\qqa$}\cell{$\qqc$}
\cell{1}
\row{}\cell{$\qqa$}\cell{$h_{li}(0)\qqd$}\cell{$h_{i1}(0)\qqb$}
\cell{$\qqc$}
\cell{1}
\row{}\cell{$\qqa$}\cell{$f_{li}(0)\qqc$}\cell{$h_{i1}(0)\qqb$}
\cell{$\qqd$}
\cell{1}
\row{}\cell{$\qqa$}\cell{$\qqd$}\cell{$\qqa$}\cell{$\qqd$}
\cell{1}
\row{}\cell{$\qqa$}\cell{$\qqc$}\cell{$f_{l1}(0)\qqb$}\cell{$\qqd$}
\cell{1}
\row{}\cell{$\qqb$}\cell{$\qqc$}\cell{$\qqc$}\cell{$\qqa$}
\cell{$-$}
\row{}\cell{$\qqb$}\cell{$h_{l1}(0)\qqd$}\cell{$\qqa$}\cell{$\qqc$}
\cell{$i{B_l}^{-1}$}
\row{}\cell{$\qqb$}\cell{$f_{l1}(0)\qqc$}\cell{$\qqa$}\cell{$\qqd$}
\cell{$i{A_l}^{-1}$}
\row{}\cell{$\qqb$}\cell{$h_{i1}(0)\qqc$}\cell{$f_{i1}(0)\qqb$}
\cell{$\qqc$}
\cell{$-iA_l$}
\row{}\cell{$\qqb$}\cell{$f_{i1}(0)\qqc$}\cell{$h_{i1}(0)\qqb$}
\cell{$\qqc$}
\cell{$-iA_l$}
\row{}\cell{$\qqb$}\cell{$h_{i1}(0)\qqd$}\cell{$f_{i1}(0)\qqb$}\cell{$\qqd$}
\cell{$-iB_l$}
\row{}\cell{$\qqb$}\cell{$f_{i1}(0)\qqd$}\cell{$h_{i1}(0)\qqb$}\cell{$\qqd$}
\cell{$-iB_l$}
\row{}\cell{$\qqc$}\cell{$\qqa$}\cell{$\qqa$}\cell{$\qqc$}
\cell{1}
\row{}\cell{$\qqc$}\cell{$\qqb$}\cell{$\qqc$}\cell{$\qqa$}
\cell{$-$}
\row{}\cell{$\qqc$}\cell{$f_{l1}(0)\qqb$}\cell{$\qqa$}\cell{$\qqd$}
\cell{$iA_l$}
\row{}\cell{$\qqc$}\cell{$\qqa$}\cell{$f_{l1}(0)\qqb$}\cell{$\qqd$}
\cell{1}
\row{}\cell{$\qqc$}\cell{$h_{i1}(0)\qqb$}\cell{$f_{i1}(0)\qqb$}\cell{$\qqc$}
\cell{$-i{A_l}^{-1}$}
\row{}\cell{$\qqc$}\cell{$f_{i1}(0)\qqb$}\cell{$h_{i1}(0)\qqb$}\cell{$\qqc$}
\cell{$-i{A_l}^{-1}$}
\row{}\cell{$\qqc$}\cell{$\qqc$}\cell{$\qqb$}\cell{$\qqa$}
\cell{$-$}
\row{}\cell{$\qqc$}\cell{$\qqc$}\cell{$\qqc$}\cell{$\qqd$}
\cell{$-$}
\row{}\cell{$\qqd$}\cell{$h_{l1}(0)\qqb$}\cell{$\qqa$}\cell{$\qqc$}
\cell{$iB_l$}
\row{}\cell{$\qqd$}\cell{$\qqa$}\cell{$h_{l1}(0)\qqb$}\cell{$\qqc$}
\cell{1}
\row{}\cell{$\qqd$}\cell{$\qqa$}\cell{$\qqa$}\cell{$\qqd$}
\cell{1}
\row{}\cell{$\qqd$}\cell{$h_{i1}(0)\qqb$}\cell{$f_{i1}(0)\qqb$}\cell{$\qqd$}
\cell{$-i{B_l}^{-1}$}
\row{}\cell{$\qqd$}\cell{$f_{i1}(0)\qqb$}\cell{$h_{i1}(0)\qqb$}\cell{$\qqd$}
\cell{$-i{B_l}^{-1}$}
\endcellular

\newpage

Let $w_i,{\bar w_i} \in \bV_{n_i}, 1 \leq i \leq 4$.
We may compare the functions
$$\oper \qquad\hbox{ and }\qquad (Y({\bar w_2},z_2)Y(w_1,z_1){\bar
w_3},w_4).$$
We wish to find a function $\eta(n_1,n_2)$ which
only depends on the sectors of $w_1$ and $w_2$ and which satisfies the
following:
\roster

\item "1)" In the domain $0 < |z_2| < |z_1|$, the series
$${z_1}^{\Delta(n_1,n_3)}{z_2}^{\Delta(n_2,n_3)}
{(z_1-z_2)}^{\Delta(n_1,n_2)}\oper
\eqno(4.60)$$
converges to a rational function $f(z_1,z_2)$,
\item "2)" In the domain $0 < |z_1| < |z_2|$, the series
$$\eta(n_1,n_2) {z_1}^{\Delta(n_1,n_3)}{z_2}^{\Delta(n_2,n_3)}
{(z_1-z_2)}^{\Delta(n_1,n_2)}(Y({\bar w_2},z_2)Y(w_1,z_1){\bar
w_3},w_4)
\eqno(4.61)$$
converges to the same rational function $f(z_1,z_2)$.
\endroster
For the cases where ${\bar w_2}=x(0)w_2$, $x(0) \in B_0$,$w_2 \in
\Omega$, we will use
$$\eqalign{
&(Y(x(0)w_2,z_2)Y(w_1,z_1){\bar w_3},w_4) \cr
&=(x(0)Y(w_2,z_2)Y(w_1,z_1){\bar w_3},w_4)-(Y(w_2,z_2)x(0)Y(w_1,z_1)
{\bar w_3},w_4) \cr
&=-(Y(w_2,z_2)Y(w_1,z_1)x(0){\bar w_3},w_4) -
(Y(w_2,z_2)Y(x(0)w_1,z_1){\bar w_3},w_4).}
\eqno(4.62)$$
Checking Table 2, the required conditions on $\eta$
are $-iA^{-1}_l=iA_l$, and
$iB^{-1}_l=-iB_l$,
that is, ${A_l}^2=-1={B_l}^2$.

Now we may proceed to the proof of Associativity.
{}From a previous theorem we have the same
restrictions on the $h_j(0)$ eigenvalues for associativity as we did for
permutability so the list of two-point functions of vectors of
minimal weight is the
same list of twenty six in Tables B1 and B2.
We need another recursion to solve for these functions.
For vectors $v_1,v_2,v_3,v_4$ of minimal weight with $v_i \in
 \bV_{n_i}$ let $k=
wt(v_1)+wt(v_2)+wt(v_3)-wt(v_4)$ and $m = \Delta(n_2,n_3)$.
Let $h_j(0)v_i=\lambda^j_iv_i$,
for $1 \leq i \leq 4,1 \leq j \leq l$.
Also, let $r=\Delta(n_1,n_2)$.
Let us
express the two-point functions in terms of series in
$\left(\frac{z_1-z_2}{z_2}
\right)$.
$$\eqalign{
&\assoc \cr
&= \sum_{p \in \shf {\Bbb Z}} \sum_{q \in \shf {\Bbb Z}}
(Y_{p+q-wt(v_1)-wt(v_2)}
(Y_{q-wt(v_1)}(v_1)v_2) v_3,v_4){(z_1-z_2)}^{-q}{z_2}^{-p}}$$
with $p+q=wt(v_1)+wt(v_2)+wt(v_3)-wt(v_4)=k$ gives
$$\eqalign{
&\assoc \cr
&=\sum_{p \in \shf {\Bbb Z}} (Y_{wt(v_3)-wt(v_4)}(Y_{-p+k-wt(v_1)}
(v_1)v_2) v_3, v_4) {(z_1-z_2)}^{p-k}{z_2}^{-p}.\cr
}$$
{}From Theorem 4.2,
we know that the coefficients of $Y(v_1,z_1-z_2)v_2$ are in $\bV_{n_1+n_2}$ so
$$wt(Y_{-p+k-wt(v_1)}(v_1)v_2) = wt(v_2)+p-k+wt(v_1) \ge \Delta_{n_1+n_2} =
wt(v_1) + wt(v_2) - r$$
therefore $0 \le p-(k-r) \in {\Bbb Z}$ and so
$$\eqalign{
&\assoc \cr
&=
{(z_1-z_2)}^{-r}{z_2}^{-k+r} \sum_{0 \le P \in {\Bbb Z}}(Y_{wt(v_3)-wt(v_4)}
(Y_{-P+r-wt(v_1)}
(v_1)v_2) v_3, v_4) {\left(\frac{z_1-z_2}{z_2}\right)}^P.}\eqno(4.63)$$

Let
$\Phi_P=\Phi_P(w_1,{\bar w_2},{\bar w_3},w_4) = (Y_{wt(v_3)-wt(v_4)}(
Y_{-P+r-wt(w_1)}(w_1){\bar w_2}){\bar w_3},w_4)$.
Using the Jacobi-Caucy identity, the general recursion for associativity is as
follows:
$$\eqalign{
&-(-P+r)\Phi_P(w_1,{\bar w_2},{\bar w_3},w_4) \cr
&=-(-P+r)(Y_{wt(v_3)-wt(w_4)}(Y_{-P+r-wt(w_1)}(w_1){\bar w_2}){\bar w_3},w_4)
\cr
&=(Y_{wt(w_3)-wt(w_4)}([L(-1),Y_{-P+r-wt(w_1)+1}(w_1)]{\bar w_2}){\bar
w_3},w_4) \cr
&=(Y_{wt(w_3)-wt(w_4)}(L(-1)Y_{-P+r-wt(w_1)+1}(w_1){\bar w_2}){\bar w_3},w_4)
\cr
&\qquad -(Y_{wt(w_3)-wt(w_4)}(Y_{-P+r-wt(w_1)+1}(w_1)L(-1){\bar w_2}){\bar
w_3},w_4) \cr
&=-(P+k-r-1)(Y_{wt(w_3)-wt(w_4)}(Y_{-P+r-wt(w_1)+1}(w_1){\bar w_2}){\bar
w_3},w_4)
\cr
&\qquad-\srd\smj\lambda^j_2
(Y_{wt(w_3)-wt(w_4)}(Y_{-P+r-wt(w_1)+1}(w_1)h_j(-1){\bar w_2}){\bar
w_3},w_4) \cr
&\qquad - \std\smj(Y_{wt(w_3)-wt(w_4)}(Y_{-P+r-wt(w_1)+1}(w_1)
e_j(-1)f_j(0){\bar w_2}) {\bar
w_3},w_4) \cr
&\qquad - \std\smj(Y_{wt(w_3)-wt(w_4)}(Y_{-P+r-wt(w_1)+1}(w_1)
f_j(-1)e_j(0){\bar w_2}) {\bar
w_3},w_4) \cr
&=-(P+k-r-1)\Phi_{P+1}+ \srd\smj\lambda^j_1\lambda^j_2
\Phi_P -\srd \smj\lambda^j_2\lambda^j_3 \Phi_{P+1} \cr
&\qquad -\std\smj(Y_{wt(w_3)-wt(w_4)}(Y_{-P+r-wt(w_1)+1}(w_1)f_j(0){\bar
w_2})e_j(0){\bar w_3},w_4)
\cr
&\qquad -\std\smj(Y_{wt(w_3)-wt(w_4)}(Y_{-P+r-wt(w_1)+1}(w_1)e_j(0){\bar
w_2})f_j(0){\bar w_3},w_4)
\cr
&\qquad -\std\smj(Y_{wt(w_3)-wt(w_4)}(Y_{-P+r-wt(w_1)}(w_1)e_j(0){\bar
w_2})f_j(0){\bar w_3},w_4)
\cr
&\qquad -\std\smj(Y_{wt(w_3)-wt(w_4)}(Y_{-P+r-wt(w_1)}(w_1)f_j(0)e_j(0){\bar
w_2}){\bar w_3},w_4).}$$

The recursion can be reduced to three specific cases based on the breakdown
used
in permutability.
Let $w_1,w_2,w_3,w_4 \in \Omega$.
Let
$$\Phi_P=(Y_{wt(w_3)-wt(w_4)}(Y_{-P+r-wt(w_1)}(w_1)f_{i1}(0)w_2)h_{i1}(0)w_3,w_4)\eqno(4.64)$$
and
$$\Psi_q=
(Y_{wt(w_3)-wt(w_4)}(Y_{-P+r-wt(w_1)}(w_1)h_{i1}(0)w_2)f_{i1}(0)w_3,w_4).\eqno(4.65)$$

$$\eqalign{
Case \ 1: \quad&-(-P+r+\srd\smj\lambda^j_1\lambda^j_2)(Y_{wt(w_3)-wt(w_4)}(Y_
{-P+r-wt(w_1)}(w_1)
w_2)w_3,w_4) \cr
&=-(P+k-r-1\cr
&\qquad+\srd\smj\lambda^j_2\lambda^j_3)(
Y_{wt(w_3)-wt(w_4)}(Y_{-P+r-wt(w_1)+1}(w_1)w_2)w_3,w_4),
}\eqno(4.66)$$

$$\eqalign{
\qquad Case \ 2: \quad&-(-P+r + \srd\smj\lambda^j_1\lmm)\Phi_P \cr
&=-(P+k-r-1+\srd\smj\lmm\llmp)\Phi_{P+1} \cr
&\qquad-\srd(1+\delta_{1i})\alpha_i(w_3) \Psi_{P+1} \cr
&\qquad-\srd(1+\delta_{1i})\alpha_i(w_3) \Psi_P \cr
&\qquad-\srd(1+\delta_{1i})\alpha_i(w_2))\Phi_P, }$$
and so
$$\eqalign{
\qquad\quad  \quad&-(-P+r+\srd\smj\lambda^j_1\lmm -
\srd(1+\delta_{1i})\alpha_i(w_2))\Phi_P \cr
&=-(P+k-r-1+ \srd\smj\lmm\llmp)\Phi_{P+1} \cr
&\qquad-\srd(1+\delta_{1i})\alpha(w_3)\{ \Psi_{P+1} + \Psi_P \}, }\eqno(4.67)$$

$$\eqalign{
Case \ 3: \quad&-(-P+r+\srd\smj\lambda^j_1\lmp)\Psi_P \cr
&=-(P+k-r-1+\srd\smj\lmp\llmm)\Psi_{P+1} \cr
&\qquad -\srd(1+\delta_{1i})\alpha_i(w_2)\Phi_{P+1}. }\eqno(4.68)$$

We will now proceed to solve for the functions that either occur in Case 1
or are such that $w_i=\vva$ for $i \in \{1,2,3\}$.
Here is a typical example, the rest will follow similarly.
We will use $z=z_1-z_2$.

Let $w_1=\vvd,w_2=\vva,w_3=h_{l1}(0)\vvb$, and $w_4=\vvc$.
Then let $\bPhi_P=(Y_{\shf+\sei}(Y_{-P+\sei}(w_1)w_2)w_3,w_4)$
so
$$\eqalign{
&(Y(Y(w_1,z)w_2,z_2)w_3,w_4) \cr
&={z_2}^{-\shf}\smk(Y_{\shf+\sei}(Y_{-P+\sei}(w_1)w_2)w_3,w_4){\left(
\frac{z}{z_2}\right)}^P \cr
&={z_2}^{-\shf}\smk\bPhi_P{\left(\frac{z}{z_2}\right)}^P }$$
and the recursion is
$$-(-P+\sei-\sei+0)\bPhi_P=-(\shf+\sei+0+P-\sei-1+0)\bPhi_{P+1}$$
or
$$\bPhi_{P+1}=-\frac{P+\shf}{P+1}\bPhi_P$$
and so
$$\bPhi_P={-\shf \choose P}\bPhi_0.$$
We also have that
$$\eqalign{
\bPhi_0&=(Y_{\shf+\sei}(Y_{\sei}(\vvd)\vva)h_{l1}(0)\vvb,\vvc) \cr
&=(Y_{\shf+\sei}(\vvd)h_{l1}(0)\vvb,\vvc) \cr
&=B_l.}$$
Therefore,
$$\eqalign{
&(Y(Y(\vvd,z)\vva,z_2)h_{l1}(0)\vvb,\vvc) \cr
&=B_l{z_2}^{-\shf}\smk{-\shf \choose P}{\left(\frac{z}{z_2}\right)}^P \cr
&=B_l \cdot {(z_2+z)}^{-\shf}.}$$

Similarly,
$$\aligned
&(Y(Y(\vva,z)\vvc,z_2)\vva,\vvc)=1, \\
&(Y(Y(\vva,z)\vvd,z_2)\vva,\vvd)=1, \\
&(Y(Y(\vva,z)f_{li}(0)\vvc,z_2)h_{i1}(0)\vvb,\vvd) =-\shf(1+\delta_{il})
A_l \cdot {z_2}^{-\shf}, \\
&(Y(Y(\vva,z)\vvc,z_2)f_{l1}(0)\vvb,\vvd) =\shf(1+\delta_{1l})A_l
\cdot {z_2}^{-\shf}, \\
&(Y(Y(\vva,z)h_{li}(0)\vvd,z_2)h_{i1}(0)\vvb,\vvc) =-(1+\shf\delta_{il})
B_l \cdot {z_2}^{-\shf}, \\
&(Y(Y(\vvb,z)f_{l1}(0)\vvc,z_2)\vva,\vvd) =-\shf(1+\delta_{1l})\cdot
z^{-\shf}, \\
&(Y(Y(\vvb,z)h_{l1}(0)\vvd,z_2)\vva,\vvc)= -(1+\shf\delta_{1l}) \cdot
z^{-\shf}, \\
&(Y(Y(\vvc,z)\vva,z_2)\vva,\vvc)=1, \\
&(Y(Y(\vvc,z)\vva,z_2)f_{l1}(0)\vvb,\vvd)= \shf(1+\delta_{1l}) A_l
\cdot {(z+z_2)}^{-\shf}, \\
&(Y(Y(\vvc,z)f_{l1}(0)\vvb,z_2)\vva,\vvd)= \shf(1+\delta_{1l}) A_l
\cdot z^{-\shf}, \\
&(Y(Y(\vvd,z)\vva,z_2)\vva,\vvd)=1, \\
&(Y(Y(\vvd,z)h_{l1}(0)\vvb,z_2)\vva,\vvc)= B_l \cdot z^{-\shf}. \\
\endaligned
$$

We are again left with nine unknown functions.  These will be solved
using recursions from $Case \ 2$ and $Case \ 3$.

i) For the case where $w_1=\vvb,w_2=\vvc,w_3=\vvb$, and $w_4=\vvc$.
The recursions are
$$ (P-\srd\delta_{1i})\bPsi_P = -(P-\srd\delta_{1i})\bPsi_{P-1} +
\srd\delta\bPhi_{P-1}$$
and
$$ P\bPhi_P= -(P-1-\srd\delta_{1i})\bPhi_{P-1}- \sxt(1+3\delta_{1i})
\{ \bPsi_P + \bPsi_{P-1} \}.$$
These combine to give
$$ \bPhi_P = -\frac{(P-1)(P-\std\delta_{1i})}{P(P-\srd\delta_{1i})}
\bPhi_{P-1}, \quad P>0$$
so
$$\bPhi_P=0, \quad P >0.$$
We need
$$\eqalign{
\bPhi_0&=(Y_{\shf+\sei}(Y_0(\vvb)f_{i1}(0)\vvc)h_{i1}(0)\vvb,\vvc) \cr
&=(Y_{\shf+\sei}(a^+_1(0)f_{i1}(0)\vvc)a^+_i(-\shf)\vva,\vvc) \cr
&=-\shf(1+\delta_{1i})(Y_{\shf+\sei}(a^-_i(0)\vvc)a^+_i(-\shf)\vva,
\vvc) \cr
&=\shf\frac{1+\delta_{1i}}{1+\shf\delta_{il}} (Y_{\shf+\sei}(h_{li}(0)
\vvd)a^+_i(-\shf)\vva,\vvc) \cr
&=-\shf(1+\delta_{1i})B_l.
}$$
Hence we have
$$\eqalign{
&(Y(Y(\vvb,z)f_{i1}(0)\vvc,z_2)h_{i1}(0)\vvb,\vvc) \cr
&={z}^{-\shf}{z_2}^{-\shf}\smk\bPhi_P {\left(\frac{z}{z_2}\right)}^{P} \cr
&=\bPhi_0 z^{-\shf}{z_2}^{-\shf} \cr
&=-\shf(1+\delta_{1i})B_l \cdot z^{-\shf} {z_2}^{-\shf}.}$$
On the other side the recursions give
$$(P-\srd\delta_{1i})\{\bPsi_P + \bPsi_{P-1}\} = \srd\delta_{1i} \bPhi_{P-1}
= 0,\quad P>1$$
so for $P>1$,
$$\bPsi_P=-\bPsi_{P-1}$$
or
$$\bPsi_P={(-1)}^{P-1}\bPsi_1.$$
We find that $\bPsi_0=0$ and so we need
$$\eqalign{
\bPsi_1&=-\shf\delta_{1i}(Y_{\shf+\sei}(a^+_1(-1)\vvc)a^-_1(-\shf)\vvb,
\vvc) \cr
&=-\delta_{1i}\Afr(Y_{\shf+\sei}(e_{l1}(-1)\vvd)a^-_1(-\shf)\vva,\vvc) \cr
&=-\delta_{1i}\Afr(Y_{\shf+\sei}(\vvd)e_{l1}(0)a^-_1(-\shf)\vva,\vvc) \cr
&=-\shf\delta_{1i}B_l.}$$
Therefore,
$$\eqalign{
&(Y(Y(\vvb,z)h_{i1}(0)\vvc,z_2)f_{i1}(0)\vvb,\vvc) \cr
&=z^{-\shf}{z_2}^{-\shf}\sum_{1 \leq P \in {\Bbb Z}} \bPsi_P
{\left(\frac{z}{z_2}\right)}^P \cr
&=-\bPsi_1 z^{-\shf}{z_2}^{-\shf}\sum_{1 \leq P \in {\Bbb Z}}
{\left(\frac{-z}{z_2}\right)}^P \cr
&=\bPsi_1 z^{\shf}{z_2}^{-\sth} \frac{1}{1+\frac{z}{z_2}} \cr
&=-\shf\delta_{1i}B_l \cdot z^{\shf}{z_2}^{-\shf}{(z+z_2)}^{-1}.
}$$

ii) For the case where $w_1=\vvb,w_2=\vvd,w_3=\vvb$, and $w_4=\vvd$.
The recursions are
$$-(P-\srd\delta_{1l}-\srd\delta_{il})\bPhi_P = (P-1-\srd\delta_{1i}
-\srd\delta_{il})\bPhi_{P-1} +\std\{\bPsi_P+\bPsi_{P-1}\}$$
and
$$(P-\srd\delta_{1i}+\srd\delta_{1l})\bPsi_P =-(P-\std+\srd\delta_{1i}
+\srd\delta_{il})\bPsi_{P-1} +\srd(\delta_{1i}+\delta_{1l}+\delta_{il})
\bPhi_{P-1}.$$
They reduce to
$$\eqalign{
&-(P-\srd\delta_{1l}-\srd\delta_{il})(P+1-\srd\delta_{1i}+\srd\delta_{1l})
\bPsi_{P+1} \cr
&=(P-1)(P-\std)\bPsi_{P-1}+(2P^2-(\std+\std\delta_{1i}
+\srd\delta_{il})P+\std\delta_{1i}-\std\delta_{1l})\bPsi_P.}$$
Let's solve this in two parts:
\roster
\item "1)" $i=1$
\item "2)" $i>1$
\endroster
When $i=1$ all of the terms will cancel giving
$\bPsi_P={(-1)}^{P-1}\bPsi_1$ for $P>0$.
When $i>1$ we see that $h_{i1}(0)\vvd=0$ unless $i=l$ so the recursion
gives that
$\bPsi_P=0$ for $P>0$.
In both cases we see that
$\bPsi_P={(-1)}^{P-1}\bPsi_1$ for $P>0$.
Let us now solve for $\bPsi_0$ and $\bPsi_1$.
$$\eqalign{
\bPsi_0&=(Y_{\shf+\sei}(a^+_1(0)h_{i1}(0)\vvd)f_{i1}(0)\vvb,\vvd) \cr
&=-(\shf\delta_{1l}+\delta_{il})(Y_{\shf+\sei}(\vvc)f_{i1}(0)\vvb,\vvd) \cr
&=-\shf(1+\delta_{1l})(\shf\delta_{1l}+\delta_{il})A_l \cr
&=-\shf(2\delta_{1l}+\delta_{il})A_l,}$$
$$\eqalign{
\bPsi_1&=(Y_{\shf+\sei}(a^+_1(-1)h_{i1}(0)\vvd)f_{i1}(0)\vvb,\vvd) \cr
&=\shf(1+\delta_{1i})(Y_{\shf+\sei}([h_{il}(-1)-h_{i1}(0)h_{1l}(0)]\vvc)
a^-_i(-\shf)\vva,\vvd) \cr
&= -\shf(1+\delta_{1i})(\shf\delta_{1i}+\delta_{1l})A_l \cr
&=-\shf(2\delta_{1l}+\delta_{1i})A_l.}$$
Therefore,
$$\eqalign{
&(Y(Y(\vvb,z)h_{i1}(0)\vvd,z_2)f_{i1}(0)\vvb,\vvd) \cr
&={z}^{-\shf}{z_2}^{-\shf}\smk {\left(\frac{z}{z_2}\right)}^P \cr
&=\bPsi_0 z^{-\shf}{z_2}^{-\shf} -\bPsi_1 z^{-\shf}{z_2}^{-\shf}
\sum_{1 \leq P \in {\Bbb Z}} {\left(\frac{-z}{z_2}\right)}^P \cr
&=\bPsi_0 z^{-\shf}{z_2}^{-\shf} +\bPsi_1 z^{\shf}{z_2}^{-\sth}
\frac{1}{1+\frac{z}{z_2}} \cr
&= [\bPsi_0(z+z_2)+\bPsi_1z]z^{-\shf}{z_2}^{-\shf}{(z+z_2)}^{-1} \cr
&=-\shf A_l[(\delta_{1i}+4\delta_{1l}+\delta_{il})z+(\delta_{1i}+2\delta_{1l})
z_2]z^{-\shf}{z_2}^{-\shf}{(z+z_2)}^{-1} \cr
&=-\sfr(1+\delta_{1i}+3\delta_{1l}+\delta_{il})A_l \cdot
[(\delta_{1i}+\delta_{il})z+\delta_{il}z_2]z^{-\shf}{z_2}^{-\shf}
{(z+z_2)}^{-1}.}$$
For the other half, the recursion reduces to
$$\srd(\delta_{1i}+\delta_{1l}+\delta_{il})\bPhi_{P-1} =
(P-\srd\delta_{1i}+\srd\delta_{1l})\bPsi_P+(P-\std+\srd\delta_{1i}
+\srd\delta_{il})\bPsi_{P-1}$$
so for $P>1$,
$$\srd(\delta_{1i}+\delta_{1l}+\delta_{il})\bPhi_{P-1}=
\srd(2-2\delta_{1i}+\delta_{1l}-\delta_{il}){(-1)}^{P-1}\bPsi_1.$$
We also need
$$\eqalign{
\bPhi_0&=(Y_{\shf+\sei}(a^+_1(0)f_{i1}(0)\vvd)a^+_i(-\shf)\vva,\vvd) \cr
&=\delta_{1l}(Y_{\shf+\sei}(f_{l1}(0)\vvc)\vvb,\vvd) \cr
&\quad-\shf(1+\delta_{1i})(Y_{\shf+\sei}(a^-_i(0)\vvd)a^+_i(-\shf)\vva,
\vvd) \cr
&=-\delta_{1l}A_l-(1+\delta_{1i})(Y_{\shf+\sei}(\vvc)f_{li}a^+_i(-\shf)
\vva,\vvd) \cr
&=-\delta_{1l}A_l-\shf(1+\delta_{1i})(1+\delta_{il})A_l \cr
&=-\shf(1+\delta_{1i})(1+\delta_{1l}+\delta_{il})A_l.}$$
Therefore, if $i \in \{1,l\}$ then
$$\eqalign{
&(Y(Y(\vvb,z)f_{i1}(0)\vvd,z_2)h_{i1}(0)\vvb,\vvd) \cr
&=z^{-\shf}{z_2}^{-\shf}\bPhi_0 +
z^{-\shf}{z_2}^{-\shf}
\sum_{1 \leq P \in {\Bbb Z}} \bPhi_P
{\left(\frac{z}{z_2}\right)}^P\cr
&=\bPhi_0z^{-\shf}{z_2}^{-\shf} - \Psi_1\frac{2-2\delta_{1i}+\delta_{1l}
-\delta_{il}}{\delta_{1i}+\delta_{1l}+\delta_{il}}z^{\shf}{z_2}^{-\sth}
\frac{1}{1+\frac{z}{z_2}} \cr
&=z^{-\shf}{z_2}^{-\shf}{(z+z_2)}^{-1}[\bPhi_0(z+z_2) -\bPsi_1(1-\delta_{1i})
z] \cr
&=-\shf(1+\delta_{1i})(1+\delta_{1l}+\delta_{il})A_l \cdot
z^{-\shf}{z_2}^{-\shf},}$$
and if $1 < i < l$ then $\bPsi_P=0$ so for $P>0$
$$ -P\bPhi_P=(P-1)\bPhi_{P-1}$$
and so $\bPhi_P=0$ for $P>0$.
Therefore, for $1<i<l$
$$\eqalign{
&(Y(Y(\vvb,z)f_{i1}(0)\vvd,z_2)h_{i1}(0)\vvb,\vvd) \cr
&=\bPhi_0 z^{-\shf} {z_2}^{-\shf} \cr
&=-\shf A_l \cdot z^{-\shf}{z_2}^{-\shf} \cr
&=-\shf(1+\delta_{1i})(1+\delta_{1l}+\delta_{il})A_l \cdot
z^{-\shf} {z_2}^{-\shf},}$$
which is consistent with the $i \in \{1,l\}$ case.

iii) For the case where $w_1=\vvc,w_2=\vvb,w_3=\vvb$ and $w_4=\vvc$.
The recursions are
$$P\Phi_P= -(P-\fvx)\bPhi_{P-1} + \std\{\bPsi_P+\bPsi_{P-1}\}$$
and
$$(P-\srd)\bPsi_P = -(P-\fvx)\Psi_{P-1} -\std\bPhi_{P-1}$$
so therefore
$$-P(P+\std)\bPsi_{P+1}= (P-\sxt)(P-\sth)\bPsi_{P-1} +
(2P^2-P-\sxt)\bPsi_P.$$
Thus we have that for $P \ge 0$,
$$\bPsi_{P+1}= {-\shf \choose P}\bPsi_1$$
and so
$$\eqalign{
\bPhi_P&= -\sth[(P+\std){-\shf \choose P} + (P+\sxt){-\shf \choose
P-1}] \bPsi_1 \cr
&=-\shf\bPsi_1{-\shf \choose P-1} \frac{1}{P}.}$$
To find $\bPhi_0$ we need
$$\eqalign{
&(Y_{\shf+\sei}(\vvc)f_{i1}(0)\vvb,h_{li}(0)\vvd) \cr
&=(Y_{\shf+\sei}(\vvc)h_{il}(0)f_{i1}(0)\vvb,\vvd) \cr
&\quad +(Y_{\shf+\sei}(h_{il}(0)\vvc)f_{i1}(0)\vvb,\vvd) \cr
&=-\Aft (Y_{\shf+\sei}(\vvc)f_{l1}(0)\vvb,\vvd) \cr
&\quad -\shf\delta_{il}(Y_{\shf+\sei}(\vvc)f_{l1}(0)\vvb,\vvd) \cr
&=-\shf(1+\delta_{1i}+\shf\delta_{1l}+\shf\delta_{il}) A_l,}$$
and
$$(h_{li}(0)\vvd,h_{li}(0)\vvd)=(1+\shf\delta_{il})^2.$$
Then
$$\eqalign{
\bPhi_0&=(Y_{\shf+\sei}(Y_{\shf+\sei}(\vvc)f_{i1}(0)\vvb) h_{i1}(0)\vvb
,\vvc) \cr
&=-\shf\frac{1+\delta_{1i}+\shf\delta_{1l}+\shf\delta_{il}}{(1+\shf
\delta_{il})^2} A_l (Y_{\shf+\sei}(h_{li}(0)\vvd)h_{i1}(0)\vvb,\vvc) \cr
&=-\shf \frac{1+\delta_{1i}+\shf\delta_{1l}+\shf\delta_{il}}{(1+\shf
\delta_{il})^2}[-\shf\delta_{il}-1]A_l B_l \cr
&=\shf(1+\delta_{1i}) A_l B_l.}$$
For $\bPsi$ we have
$$\bPsi_0=(Y_{\shf+\sei}(Y_{\shf+\sei}(\vvc)h_{i1}(0)\vvb)
f_{i1}(0)\vvb,\vvc)$$
but
$$(Y_{\shf+\sei}(\vvc)h_{i1}(0)\vvb,e_{li}(0)\vvd)=0$$
so
$$\bPsi_0=0.$$
To find $\bPsi_1$ we need
$$\eqalign{
&(Y_{-\shf+\sei}(\vvc)h_{i1}(0)\vvb,e_{li}(-1)\vvd) \cr
&=-(Y_{-\shf+\sei}(\vvc)f_{li}(1)h_{i1}(0)\vvb,\vvd) \cr
&\quad -(Y_{\shf+\sei}(f_{li}(0)\vvc)h_{i1}(0)\vvb,\vvd) \cr
&=(Y_{\shf+\sei}(\vvc)f_{li}(0)h_{i1}(0)\vvb,\vvd) \cr
&=\Afs (Y_{\shf+\sei}(\vvc)f_{l1}(0)\vvb,\vvd) \cr
&=\shf(1+\delta_{il}) A_l,}$$
and
$$(e_{li}(-1)\vvd,e_{li}(-1)\vvd)=-\sfr(1+\delta_{il})^2.
$$
Then
$$\eqalign{
\bPsi_1&=(Y_{\shf+\sei}(Y_{-\shf+\sei}(\vvc) h_{i1}(0)\vvb)
f_{i1}(0)\vvb,\vvc) \cr
&=-\frac{2}{1+\delta_{il}}
 (Y_{\shf+\sei}(e_{li}(-1)\vvd)f_{i1}(0)\vvb,\vvc) \cr
&=-\frac{2}{1+\delta_{il}}[\sfr(1+\delta_{1i}+\delta_{1l}+\delta_{il}]
(Y_{\shf+\sei}(\vvd)h_{l1}(0)\vvb,\vvc) \cr
&=-\shf(1+\delta_{1i}) A_l B_l.}$$
Therefore,
$$\eqalign{
&(Y(Y(\vvc,z)h_{i1}(0)\vvb,z_2)f_{i1}(0)\vvb,\vvc) \cr
&=z^{-\shf}{z_2}^{-\shf}\sum_{1 \leq P \in {\Bbb Z}} \bPsi_P
{\left(\frac{z}{z_2}\right)}^P \cr
&=\bPsi_1 z^{\shf}{z_2}^{-\sth}\smk {-\shf \choose P}
{\left(\frac{z}{z_2}\right)}^P \cr
&=-\shf(1+\delta_{1i})A_lB_l\cdot
z^{\shf}{z_2}^{-1}{(z+z_2)}^{-\shf},}$$
and
$$\eqalign{
&(Y(Y(\vvc,z)f_{i1}(0)\vvb,z_2)h_{i1}(0)\vvb,\vvc) \cr
&=z^{-\shf}{z_2}^{-\shf}\smk \bPhi_P {\left(\frac{z}{z_2}\right)}^P \cr
&=\bPhi_0 z^{-\shf} {z_2}^{-\shf} + z^{-\shf}{z_2}^{-\shf} \sum_{1 \leq
P \in {\Bbb Z}} \bPhi_P {\left(\frac{z}{z_2}\right)}^P \cr
&=\bPhi_0 z^{-\shf}{z_2}^{-\shf} - \shf\bPsi_1 z^{-\shf}{z_2}^{-\shf}
\sum_{1 \leq P \in {\Bbb Z}} {-\shf \choose P-1}\frac{1}{P}
{\left(\frac{z}{z_2}\right)}^P \cr
&=\bPhi_0 z^{-\shf}{z_2}^{-\shf} -\shf\bPsi_1 z^{-\shf}{z_2}^{-\shf}
[2\left(1+\frac{z}{z_2}\right)^{\shf}-2] \cr
&=-\bPsi_1 z^{-\shf}{z_2}^{-1}{(z+z_2)}^{\shf} \cr
&=\shf(1+\delta_{1i}) A_l B_l \cdot z^{-\shf} {z_2}^{-1}
{(z+z_2)}^{\shf}.}$$

iv) For the case where $w_1=\vvd,w_2=\vvb,w_3=\vvb$ and $w_4=\vvd$.
The recursions are
$$-(P-\srd\delta_{il})\bPhi_P = (P-\fvx)\bPhi_{P-1}+\std
[\bPsi_P+\bPsi_{P-1}]$$
and
$$-(P-\srd+\srd\delta_{il})\bPsi_P = (P-\fvx)\bPsi_{P-1} +
\std\bPhi_{P-1}$$
or
$$-(P-\srd\delta_{il})(P+\std+\srd\delta_{il})\bPsi_{P+1} =
(P-\sth)(P-\sxt)\bPsi_{P-1} + (2P^2-P-\sxt-\srd\delta_{il})\bPsi_P.$$
Therefore, for $P \ge 0$,
$$\bPsi_{P+1}=[(1-\srd\delta_{il}){-\shf \choose P} -
\std\delta_{il}{-\shf \choose P+1}]\bPsi_1,$$
and
$$\bPhi_{P+1}=[\std\delta_{il}{-\shf \choose P} - (\shf+\sxt\delta_{il})
{-\shf \choose P}\frac{1}{P+1}]\bPsi_1.$$
We need to solve for $\bPhi_0$ and $\bPsi_0$, and possibly $\bPsi_1$
but first we will need the following:
\roster
\item "1)" $$\eqalign{
&(Y_{\shf+\sei}(\vvd)f_{i1}(0)\vvb,f_{li}(0)\vvc) \cr
&=-(Y_{\shf+\sei}(\vvd) e_{li}(0)f_{i1}(0)\vvb,\vvc) \cr
&=-\sfr(1+\delta_{1i}+\delta_{1l}+\delta_{il})B_l,}$$
\item "2)" $$(f_{li}(0)\vvc,f_{li}(0)\vvc)=\sfr(1+\delta_{il}),$$
\item "3)" $$(Y_{\shf+\sei}(\vvd)h_{i1}(0)\vvb,h_{il}(0)\vvc)=-
\shf\delta_{il} B_l,$$
\item "4)" $$(h_{il}(0)\vvc,h_{il}(0)\vvc)=\sfr\delta_{il},$$
\item "5)" $$\eqalign{
&(Y_{-\shf+\sei}(\vvd)h_{i1}(0)\vvb,h_{il}(-1)\vvc) \cr
&=(Y_{\shf+\sei}(h_{li}(0)\vvd)h_{i1}(0)\vvb,\vvc) \cr
&=-\shf\delta_{il}B_l-(Y_{\shf+\sei}(\vvd)h_{li}(0)h_{i1}(0)\vvb,\vvc)
\cr
&=-(1+\shf\delta_{il}) B_l,}$$
\item "6)" $$(h_{il}(-1)\vvc,h_{il}(-1)\vvc)=-1.$$
\endroster
Now, using the above, we see that
$$\eqalign{
\bPhi_0&=(Y_{\shf+\sei}(Y_{\shf+\sei}(\vvd)f_{i1}(0)\vvb)h_{i1}(0)
\vvb,\vvd) \cr
&=\frac{1+\delta_{1i}+\delta_{1l}+\delta_{il}}{1+\delta_{il}}
B_l (Y_{\shf+\sei}(f_{li}(0)\vvc)h_{i1}(0)\vvb,\vvd) \cr
&=\frac{1+\delta_{1i}+\delta_{1l}+\delta_{il}}{1+\delta_{il}} B_l
(Y_{\shf+\sei}(\vvc)f_{li}(0)h_{i1}(0)\vvb,\vvd) \cr
&=\shf(1+\delta_{1i})(1+\delta_{il}) A_l B_l,}$$
$$\eqalign{
\bPsi_0&=(Y_{\shf+\sei}(Y_{\shf+\sei}(\vvd)h_{i1}(0)\vvb)
f_{i1}(0)\vvb,\vvd) \cr
&=-2\delta_{il}B_l (Y_{\shf+\sei}(h_{il}(0)\vvc)f_{i1}(0)\vvb, \vvd)
\cr
&=\shf\delta_{il}(1+\delta_{1i}) A_l B_l,}$$
and
$$\eqalign{
\bPsi_1&=(Y_{\shf+\sei}(Y_{-\shf+\sei}(\vvd)h_{i1}(0)\vvb)f_{i1}(0)
\vvb,\vvd) \cr
&=(1+\shf\delta_{il})B_l
(Y_{\shf+\sei}(h_{il}(-1)\vvc)f_{i1}(0)\vvb,\vvd) \cr
&=(1+\shf\delta_{il})B_l (Y_{\shf+\sei}(\vvc)h_{il}(0)f_{i1}(0)
\vvb,\vvd) \cr
&=(1+\shf\delta_{il})B_l [-\shf(1+\delta_{1i}) A_l] \cr
&=-\shf(1+\delta_{1i})(1+\shf\delta_{il}) A_l B_l.}$$

Finally we have that
$$\eqalign{
&(Y(Y(\vvd,z)h_{i1}(0)\vvb,z_2)f_{i1}(0)\vvb,\vvd) \cr
&=z^{-\shf}{z_2}^{-\shf}\smk \bPsi_P \left(\frac{z}{z_2}\right)^P \cr
&=\bPsi_0 z^{-\shf}{z_2}^{-\shf} + z^{-\shf}{z_2}^{-\shf} \smk \bPsi_{P+1}
\left(\frac{z}{z_2}\right)^{P+1} \cr
&=\bPsi_0z^{-\shf}{z_2}^{-\shf} + (1-\srd\delta_{il})\bPsi_1 z^{\shf}
{z_2}^{-\sth}\left(1+\frac{z}{z_2}\right)^{-\shf} \cr
&\qquad\quad -\std \delta_{il}\bPsi_1
z^{-\shf}{z_2}^{-\shf}[\left(1+\frac{z}{z_2}\right)^{-\shf}-1] \cr
&=\bPsi_0z^{-\shf}{z_2}^{-\shf} + (1-\srd\delta_{il})\bPsi_1
z^{\shf}{z_2}^{-1}(z+z_2)^{-\shf} \cr
&\qquad\quad -\std\delta_{il}\bPsi_1 z^{-\shf}
(z+z_2)^{-\shf} +\std\delta_{il}\bPsi_1 z^{-\shf}{z_2}^{-\shf} \cr
&=-\shf(1+\delta_{1i})A_l B_l \cdot [z-\delta_{il}z_2]z^{-\shf} {z_2}^{-1}
(z+z_2)^{-\shf},}$$
and
$$\eqalign{
&(Y(Y(\vvd,z)f_{i1}(0)\vvb,z_2)h_{i1}(0)\vvb,\vvd) \cr
&=z^{-\shf} {z_2}^{-\shf} \smk \bPhi_P \left(\frac{z}{z_2}\right)^P \cr
&=\bPhi_0 z^{-\shf}{z_2}^{-\shf} + z^{-\shf}{z_2}^{-\shf} \smk \bPhi_{P+1}
\left(\frac{z}{z_2}\right)^{P+1} \cr
&=\bPhi_0 z^{-\shf}{z_2}^{-\shf} + \bPsi_1 z^{-\shf}{z_2}^{-\shf}
\smk[ \std\delta_{il}-(\shf
+\sxt\delta_{il})\frac{1}{P+1}]{-\shf \choose
P}\left(\frac{z}{z_2}\right)^{P+1}
\cr
&=\shf(1+\delta_{1i})A_l B_l \cdot [z+(1+\delta_{il})z_2]z^{-\shf}
{z_2}^{-1} (z+z_2)^{-\shf}.}$$

We can now summarize all the calculations we have done for associativity in
Table 3.

Let $v_i \in \bV_{n_i}$ for $1 \leq i \leq 4$, $s=\delta_{1i} +
4\delta_{1l}+\delta_{il}$, $t=\shf(1+\delta_{1i})$
 and

\ni let $g(z,z_2)$ be the function to which the series
$$(z+z_2)^{\Delta(n_1,n_3)}
{z_2}^{\Delta(n_2,n_3)}z^{
\Delta(n_1,n_2)}(Y(Y(v_1,z)v_2,z_2)v_3,v_4)$$
converges in the domain $0<|z|<|z_2|$.


\centerline{Table 3}
\sk1
\begincellular{\centertable}
\row{}\cell{$v_1$}\cell{$v_2$}\cell{$v_3$}\cell{$v_4$}
\cell{$g(z,z_2)$}
\row{}\cell{$\qqa$}\cell{$\qqc$}\cell{$\qqa$}\cell{$\qqc$}
\cell{1}
\row{}\cell{$\qqa$}\cell{$h_{li}(0)\qqd$}\cell{$h_{i1}(0)\qqb$}
\cell{$\qqc$}
\cell{$-(1+\shf\delta_{il})B_l$}
\row{}\cell{$\qqa$}\cell{$f_{li}(0)\qqc$}\cell{$h_{i1}(0)\qqb$}
\cell{$\qqd$}
\cell{$-\shf(1+\delta_{il})A_l$}
\row{}\cell{$\qqa$}\cell{$\qqd$}\cell{$\qqa$}\cell{$\qqd$}
\cell{1}
\row{}\cell{$\qqa$}\cell{$\qqc$}\cell{$f_{l1}(0)\qqb$}\cell{$\qqd$}
\cell{$\shf(1+\delta_{1l})A_l$}
\row{}\cell{$\qqb$}\cell{$\qqc$}\cell{$\qqc$}\cell{$\qqa$}
\cell{0}
\row{}\cell{$\qqb$}\cell{$h_{l1}(0)\qqd$}\cell{$\qqa$}\cell{$\qqc$}
\cell{$-(1+\shf\delta_{1l})$}
\row{}\cell{$\qqb$}\cell{$f_{l1}(0)\qqc$}\cell{$\qqa$}\cell{$\qqd$}
\cell{$-\shf(1+\delta_{1l})$}
\row{}\cell{$\qqb$}\cell{$h_{i1}(0)\qqc$}\cell{$f_{i1}(0)\qqb$}
\cell{$-\qqc$}
\cell{$-\shf\delta_{1i}B_l \cdot z$}
\row{}\cell{$\qqb$}\cell{$f_{i1}(0)\qqc$}\cell{$h_{i1}(0)\qqb$}
\cell{$\qqc$}
\cell{$-tB_l\cdot (z+z_2)$}
\row{}\cell{$\qqb$}\cell{$h_{i1}(0)\qqd$}\cell{$f_{i1}(0)\qqb$}\cell{$\qqd$}
\cell{$-\shf A_l[(\delta_{il}+2\delta_{1l})z_2+s
z]$}
\row{}\cell{$\qqb$}\cell{$f_{i1}(0)\qqd$}\cell{$h_{i1}(0)\qqb$}\cell{$\qqd$}
\cell{$-\shf(1+s-\delta_{1l})A_l\cdot (z+z_2)$}
\row{}\cell{$\qqc$}\cell{$\qqa$}\cell{$\qqa$}\cell{$\qqc$}
\cell{1}
\row{}\cell{$\qqc$}\cell{$\qqb$}\cell{$\qqc$}\cell{$\qqa$}
\cell{0}
\row{}\cell{$\qqc$}\cell{$f_{l1}(0)\qqb$}\cell{$\qqa$}\cell{$\qqd$}
\cell{$\shf(1+\delta_{1l})A_l$}
\row{}\cell{$\qqc$}\cell{$\qqa$}\cell{$f_{l1}(0)\qqb$}\cell{$\qqd$}
\cell{$\shf(1+\delta_{1l})A_l$}
\row{}\cell{$\qqc$}\cell{$h_{i1}(0)\qqb$}\cell{$f_{i1}(0)\qqb$}\cell{$\qqc$}
\cell{$-tA_lB_l\cdot z$}
\row{}\cell{$\qqc$}\cell{$f_{i1}(0)\qqb$}\cell{$h_{i1}(0)\qqb$}\cell{$\qqc$}
\cell{$tA_lB_l\cdot (z+z_2)$}
\row{}\cell{$\qqc$}\cell{$\qqc$}\cell{$\qqb$}\cell{$\qqa$}
\cell{0}
\row{}\cell{$\qqc$}\cell{$\qqc$}\cell{$\qqc$}\cell{$\qqd$}
\cell{0}
\row{}\cell{$\qqd$}\cell{$h_{l1}(0)\qqb$}\cell{$\qqa$}\cell{$\qqc$}
\cell{$B_l$}
\row{}\cell{$\qqd$}\cell{$\qqa$}\cell{$h_{l1}(0)\qqb$}\cell{$\qqc$}
\cell{$B_l$}
\row{}\cell{$\qqd$}\cell{$\qqa$}\cell{$\qqa$}\cell{$\qqd$}
\cell{1}
\row{}\cell{$\qqd$}\cell{$h_{i1}(0)\qqb$}\cell{$f_{i1}(0)\qqb$}\cell{$\qqd$}
\cell{$-tA_lB_l\cdot[z-\delta_{il}z_2]$}
\row{}\cell{$\qqd$}\cell{$f_{i1}(0)\qqb$}\cell{$h_{i1}(0)\qqb$}\cell{$\qqd$}
\cell{$tA_lB_l\cdot[z+(1+\delta_{il})z_2]$}
\endcellular

\newpage
\ni In the domain, $0<|z_2|<|z_1|$, we would like the series
$${z_1}^{\Delta(n_1,n_3)}{z_2}^{\Delta(n_2,n_3)}(z_1-z_2)^{
\Delta(n_1,n_2)} (Y(v_1,z_1)Y(v_2,z_2)v_3,v_4)\eqno(4.69)$$
to converge to the same rational function as
$${z_1}^{\Delta(n_1,n_3)}{z_2}^{\Delta(n_2,n_3)}(z_1-z_2)^{\Delta(n_1,n_2)}
(Y(Y(v_1,z_1-z_2)v_2,z_2)v_3,v_4)\eqno(4.70)$$
converges in the domain $0<|z_1-z_2|<|z_2|$,
and comparing the results from the associativity part with the
permutability part (Tables 1 and 3), the above comparison will hold if
$$A_lB_l=-1\eqno(4.71)$$
and so
$$A_l=B_l=\pm i.\eqno(4.72)$$
This determines $\eta$.
\sk1
\centerline{Table D}
\sk1
\begincellular{\centertable}
\row{}\cell{$\eta(\gamma_1,\gamma_2)$}\cell{$\gamma_2=0$}\cell{$\gamma_2=1$}\cell{$\gamma_2=2$}
\cell{$\gamma_2=3$}
\row{}\cell{$\gamma_1=0$}\cell{$1$}\cell{$1$}\cell{$1$}
\cell{$1$}
\row{}\cell{$\gamma_1=1$}\cell{$1$}\cell{$-$}\cell{$iA_l$}
\cell{$-$}
\row{}\cell{$\gamma_1=2$}\cell{$1$}\cell{$-iA_l$}\cell{$1$}
\cell{$-iA_l$}
\row{}\cell{$\gamma_1=3$}\cell{$1$}\cell{$-$}\cell{$iA_l$}
\cell{$-$}
\endcellular

\sk1

Now we will prove the Jacobi-Cauchy Identity for Para-algebras.
We have calculated all the ``base case two-point functions'' and compared them.
We will do an inductive proof with the results from Tables 1,2 and 3 as the
base cases.
It should
be stated here that though $\bV_0$ is a VOA with modules $\bV_1,\bV_2,\bV_3$,
we also have that $\bV=\bV_0 \oplus \bV_2$ is a VOA with $\bW=\bV_1 \oplus
\bV_3$ a module.  It can be shown that $\bV$ and $\bW$ do not form a VOPA
since there are no constants $A_l,B_l$ that could define
$\eta(\gamma_1,\gamma_2)$.

We will need the following:
\pr{Lemma 4.5} For any $m_r \in {\Bbb Z},0 \leq p \in {\Bbb Z}$ we have that
\roster
\item "i)" for $0<|z_2|<|z_1|$,
$$\eqalign{
&{z_1}^{-m_r-p} \sum_{0 \le k \in {\Bbb Z}}
 {m_r+k-1 \choose k}{-m_r-k \choose p}
{\left(\frac{z_2}{z_1}\right)}^k \cr
&= (-1)^p(z_1-z_2)^{-m_r-p}{m_r+p-1 \choose p},}\eqno(4.73)$$
\item "ii)" for $0<|z_1|<|z_2|$,
$$\eqalign{
&{z_1}^{-p}{z_2}^{-m_r}\sum_{0 \le k \in {\Bbb Z}}
 {m_r+k-1 \choose k}{k \choose p}
{\left(\frac{z_1}{z_2}\right)}^k \cr
&= (z_2-z_1)^{-m_r-p}{m_r+p-1 \choose p}.}\eqno(4.74)$$
\endroster
\epr

\demo{Proof} Here we use induction on p.
$\hfill\blacksquare$

\pr{Corollary 4.6} For any $w_1,w_4 \in \Omega,{\bar w_2},{\bar w_3} \in
{\bar \Omega}$ with $w_i,{\bar w_i} \in \bV_{n_i}$ we have that, for
$0<|z_2|<|z_1|$, the series
$$\Dlt\oper\eqno(4.75)$$
converges to a function in ${\Bbb C}[z_i,{z_i}^{-1},(z_1-z_2)^{-1}]$.
\epr

\demo{Proof}  See Table 1.
$\hfill\blacksquare$

\pr{Theorem 4.7} (Rationality) For any $v_i \in \bV_{n_i}$,
$1 \leq i \leq 4$,
$0<|z_2|<|z_1|$, the series
$$\Dlt\perm\eqno(4.76)$$
converges to a function in ${\Bbb C}[z_i,{z_i}^{-1},(z_1-z_2)^{-1}]$.
\epr

\demo{Proof} We will prove this theorem in four steps.  Suppose
$0<|z_2|<|z_1|$.

\demo{Step 1} For any $v_1 \in \bV_{n_1},{\bar w_i} \in \bV_{n_i} \cap
{\bar \Omega}$,$i \in \{2,3\}$ and $w_4 \in \bV_{n_4} \cap \Omega$
we have that the series
$$\Dlt(Y(v_1,z_1)Y({\bar w_2},z_2){\bar w_3},w_4)\eqno(4.77)$$
converges to a function in
${\Bbb C}[z_i,{z_i}^{-1},(z_1-z_2)^{-1}]$.
\demo{Proof}(Step 1) WLOG, let $v_1=\vsr w_1$ where $x_j(-m_j) \in B$,
$1 \leq j \leq r$, and $w_1 \in \Omega$. We will use induction on $r$.
When $r=0$, then $v_1=w_1$ and by Corollary 4.6 we are done.
For $r>0$, suppose that for any $0 \leq s <r$, $v'_1=\vsq w_1$, and any
${\bar w'_i} \in \bV_{n_i} \cap {\bar \Omega}$,$i \in \{2,3\}$ we have that
the series
$$\Dlt (Y(v'_1,z_1)Y({\bar w'_2},z_2){\bar w'_3},w_4)$$
converges to a function in ${\Bbb C}[z_i,
{z_i}^{-1},(z_1-z_2)^{-1}]$,
then
$$\eqalign{
&\Dlt (Y(v_1,z_1)Y({\bar w_2},z_2){\bar w_3},w_4) \cr
&=\Dlt \smK {m_r+k-1 \choose k} \cdot \cr
&\quad [{z_1}^k (x_r(-m_r-k)Y(\vst w_1,z_1)
Y({\bar w_2},z_2){\bar w_3},w_4) \cr
&\quad -(-1)^{m_r}{z_1}^{-m_r-k}(Y(\vst w_1,z_1)x_r(k)
Y({\bar w_2},z_2){\bar w_3},w_4)] \cr
&=\Dlt \smK {m_r+k-1 \choose k} (-1)^{m_r+1}{z_1}^{-m_r-k} \cdot \cr
&\quad [(Y(\vst w_1,z_1)Y({\bar w_2},z_2)x_r(k){\bar w_3},w_4) \cr
&\quad \quad+ {z_2}^k
(Y(\vst w_1,z_1)Y(x_r(0){\bar w_2},z_2){\bar w_3},w_4)] \cr
&=\Dlt (-1)^{m_r+1}{z_1}^{-m_r} \cdot \cr
&\quad [\smK {m_r+k-1 \choose k} {z_1}^{-k} \cdot\cr
&\quad \quad \qquad(Y(\vst w_1,z_1)
Y(x_r(0){\bar w_2},z_2){\bar w_3},w_4) \cr
&\quad\quad + (Y(\vst w_1,z_1)Y({\bar w_2},z_2) x_r(0) \wwc,w_4)] \cr
&=\Dlt (-1)^{m_r+1} \cdot \cr
&\quad [(z_1-z_2)^{-m_r} (Y(\vst w_1,z_1)Y(x_r(0)\wwb,z_2)
\wwc,w_4) \cr
&\quad\quad +{z_1}^{-m_r}(Y(\vst w_1,z_1)Y(\wwb,z_2) x_r(0)
\wwc,w_4)]
}$$
but $x_r(0){\bar w_2},x_r(0){\bar w_3} \in {\bar \Omega}$ so the above
satisfies the induction hypothesis.  Therefore, since $m_r \in {\Bbb Z}$ we
have proved Step 1.

\demo{Step 2}  For any $v_i \in \bV_{n_i}$,$i \in \{1,4\}$,
${\bar w_t} \in \bV_{n_t} \cap {\bar \Omega}$,$t \in \{2,3\}$
we have that the series
$$\Dlt (Y(v_1,z_1)Y(\wwb,z_2)\wwc,v_4)\eqno(4.78)$$
converges to a function in ${\Bbb C}[z_i,{z_i}^{-1},
(z_1-z_2)^{-1}]$.
\demo{Proof}(Step 2) WLOG, let $v_4 = \vsr w_4$ where $x_j(-m_j) \in B$,
$1 \leq j \leq r$, and $w_4 \in \Omega$.  We will again use induction on
r.  When $r=0$ we have $v_4=w_4$ and by Step 1 we are done.
For $r>0$ suppose that for any $0 \leq s < r$, $v'_4=\vsq w_4$,
${\bar w'_i} \in
\bV_{n_i} \cap {\bar \Omega}$, and $v'_1 \in \bV_{n_1}, wt(v'_1) \leq
wt(v_1)$ we have that the series
$$\Dlt (Y(v'_1,z_1)Y(\wwx,z_2)\wwy,v'_4)$$
converges to a function in ${\Bbb C}[z_i,{z_i}^{-1},
(z_1-z_2)^{-1}]$,
then
$$\eqalign{
&\Dlt (Y(v_1,z_1)Y(\wwb,z_2)\wwc,v_4) \cr
&=\Dlt \cdot \cr
&\quad(x^*_r(m_r)Y(v_1,z_1)Y(\wwb,z_2)\wwc,\vst w_4) \cr
&=\Dlt \cdot \cr
&\quad [\smK {m_r \choose k} {z_1}^{-k} (Y(x^*_r(k)v_1,z_1)
Y(\wwb,z_2)\wwc, \vst w_4)  \cr
&\quad\quad+(Y(v_1,z_1)x^*_r(m_r)Y(\wwb,z_2)\wwc,\vst w_4)] \cr
&=\Dlt \cdot \cr
&\quad [\smK {m_r \choose k} {z_1}^{m_r-k} (Y(x^*_r(k)v_1,z_1)
Y(\wwb,z_2)\wwc,\vst w_4) \cr
&\quad \quad+(Y(v_1,z_1)Y(\wwb,z_2)x^*_r(m_r)\wwc,\vst w_4) \cr
&\quad\quad + {z_2}^{m_r} (Y(v_1,z_1)Y(x^*_r(0)
\wwb,z_2)\wwc,\vst w_4)],}$$
but since $wt(x^*_r(k) v_1) = wt(v_1)-k \leq wt(v_1)$ and
$x^*_r(m_r)\wwc,x^*_r(k)\wwb \in {\bar \Omega}$ the above satisfies
the induction hypothesis and we have proven Step 2.

\demo{Step 3} For any $v_i \in \bV_{n_i}$,$i \in \{1,2,4\}$,
$\wwc \in \bV_{n_3} \cap {\bar \Omega}$ we have that the series
$$\Dlt (Y(v_1,z_1)Y(v_2,z_2)\wwc,v_4)\eqno(4.79)$$
converges to a function in ${\Bbb C}[z_i,{z_i}^{-1},
(z_1-z_2)^{-1}]$.
\demo{Proof}(Step 3) WLOG, let $v_2= \vsr w_2$ where $x_j(-m_j) \in B$,
$1 \leq j \leq r$, and $w_2 \in \Omega$.  We will use induction on r.
When $r=0$ we have that $v_2 = w_2$ and by Step 2 we are done.
For $r>0$ suppose that for any $0 \leq s < r$, $v'_2= \vsq w_2$,
$\wwy \in
\bV_{n_3} \cap {\bar \Omega}$, and $v'_i \in \bV_{n_i}$,$wt(v'_i) \leq
wt(v_i)$,$i \in \{1,2,4\}$ we have that the series
$$\Dlt (Y(v'_1,z_1)Y(v'_2,z_2)\wwy,v'_4)$$
converges to a function in ${\Bbb C}[z_i,{z_i}^{-1},
(z_1-z_2)^{-1}]$,
then
$$\eqalign{
&\Dlt (Y(v_1,z_1)Y(v_2,z_2)\wwc,v_4) \cr
&=\Dlt \smK {m_r+k-1 \choose k} \cdot \cr
&\quad [{z_2}^k(Y(v_1,z_1)x_r(-m_r-k) Y(\vst w_2,
z_2)\wwc,v_4) \cr
&\quad\quad -(-1)^{m_r}{z_2}^{-m_r-k}(Y(v_1,z_1)Y(\vst w_2,z_2)
x_r(k) \wwc,v_4)] \cr
&=\Dlt \{ \smK {z_2}^k {m_r+k-1 \choose k} \cdot \cr
&\quad [(Y(v_1,z_1)Y(\vst w_2,z_2)\wwc,x^*_r(m_r+k) v_4) \cr
& -\sum_{0 \leq p \in {\Bbb Z}}
{-m_r-k \choose p}{z_1}^{-m_r-k-p}\cr
&\quad\quad\quad\quad(Y(x_r(p)v_1,z_1)Y(\vst w_2,z_2) \wwc,v_4)] \cr
& \quad-(-1)^{-m_r}{z_2}^{-m_r}(Y(v_1,z_1)Y(\vst w_2,z_2)x_r(0)\wwc,v_4)\}
\cr
&=\Dlt \{ \smK {m_r+k-1 \choose k} \cdot \cr
&\quad [{z_2}^k  (Y(v_1,z_1)Y(\vst w_2, z_2)
\wwc, x^*_r(m_r+k)v_4) \cr
&\quad \quad - (-1)^k(z_1-z_2)^{-m_r-k} \cr
&\quad\quad\quad\quad(Y(x_r(k)v_1,z_1) Y(\vst
w_2,z_2)\wwc,v_4)] \cr
&\quad -(-1)^{-m_r}{z_2}^{-m_r}(Y(v_1,z_1)Y(\vst w_2,z_2)x_r(0)\wwc,
v_4)\}
}$$
but since $x_r(0)\wwc \in {\bar \Omega}$, and $wt(x_r(p)v_1)=wt(v_1)-p \leq
wt(v_1)$, with the
$wt(x^*_r(m_r+k)v_4)=wt(v_4)-m_r-k \leq wt(v_4)$ and for
$k$ sufficiently large we have that $x_r(k)v_1=0$,$x^*_r(m_r+k)v_4=0$,
the above satisfies the induction hypothesis and
we have proven Step 3.

\demo{Step 4} For any $v_i \in \bV_{n_i},1 \leq i \leq 4$,
the series
$$\Dlt \perm\eqno(4.80)$$
converges to a function in ${\Bbb C}[z_i,{z_i}^{-1},(z_1-z_2)^{-1}]$.
\demo{Proof}(Step 4) WLOG, let $v_3=\vsr w_3$, where $x_j(-m_j) \in B$,
$1 \leq j \leq r$, and $w_3 \in \Omega$.  We will again use induction on r.
When $r=0$ we have that $v_3=w_3$ and by Step 3 we are done.
For $r>0$ suppose that for any $0 \leq s <r$, $v'_3= \vsq w_3$, and
$v'_i \in \bV_{n_i}$, $wt(v'_i) \leq wt(v_i)$,$i \in \{1,2,4\}$ we have that
the series
$$\Dlt (Y(v'_1,z_1)Y(v'_2,z_2)v'_3,v'_4)$$
converges to a function in ${\Bbb C}[z_i,{z_i}^{-1},
(z_1-z_2)^{-1}]$,
then
$$\eqalign{
&\Dlt \perm \cr
&=\Dlt \cdot \cr
&\quad [(Y(v_1,z_1)x_r(-m_r)Y(v_2,z_2)\vst w_3,v_4) \cr
&\quad\quad -{z_2}^{-m_r}\smK{-m_r \choose k} {z_2}^{-k}\cdot\cr
&\quad\quad\quad\quad
(Y(v_1,z_1)Y(x_r(k)v_2,z_2) \vst w_3,v_4)] \cr
&=\Dlt \cdot \cr
&\quad[(x_r(-m_r)Y(v_1,z_1)Y(v_2,z_2)\vst w_3,v_4) \cr
&\quad\quad -{z_1}^{-m_r}\smK{-m_r \choose k} {z_1}^{-k}\cdot\cr
&\quad\quad\quad\quad
(Y(x_r(k)v_1,z_1)Y(v_2,z_2)\vst w_3,v_4) \cr
&\quad\quad -{z_2}^{-m_r}\smK{-m_r \choose k} {z_2}^{-k}\cdot\cr
&\quad\quad\quad\quad
(Y(v_1,z_1)Y(x_r(k)v_2,z_2)\vst w_3,v_4)] \cr
&=\Dlt \cdot \cr
&\quad[(Y(v_1,z_1)Y(v_2,z_2)\vst w_3,x^*_r(m_r)v_4) \cr
&\quad\quad -{z_1}^{-m_r}\smK{-m_r \choose k} {z_1}^{-k}\cdot\cr
&\quad\quad\quad\quad
(Y(x_r(k)v_1,z_1)Y(v_2,z_2)\vst w_3,v_4) \cr
&\quad\quad -{z_2}^{-m_r}\smK{-m_r \choose k} {z_2}^{-k}\cdot\cr
&\quad\quad\quad\quad
(Y(v_1,z_1)Y(x_r(k)v_2,z_2)\vst w_3,v_4) ]}$$
but since $wt(x_r(k)v_i)=wt(v_i)-k \leq wt(v_i)$,$i \in \{1,2\}$,
$wt(x^*_r(m_r)v_4)=
wt(v_4)-m_r \leq wt(v_4)$ and for $k$ large enough, $x_r(k)v_i=0$,
$i \in \{1,2\}$, the
above satisfies the induction hypothesis and we have proven Step 4 and
therefore
we have proven the Rationality Theorem.
$\hfill\blacksquare$

\pr{Corollary 4.8} For any $w_1,w_4 \in \Omega,{\bar w_2},{\bar w_3} \in
{\bar \Omega}$ with $w_i,{\bar w_i} \in \bV_{n_i}$ we have that
the series
$$\Dlt \oper\eqno(4.81)$$
converges to some rational function $f(z_1,z_2)$ in the domain
$0<|z_2|<|z_1|$ and the series
$$\Bta \Dlt \opre\eqno(4.82)$$
converges to the same rational function $f(z_1,z_2)$ but in the
domain $0<|z_1|<|z_2|$.
\epr

\demo{Proof}  See Table 2 and Corollary 4.6.
$\hfill\blacksquare$

\pr{Theorem 4.9} (Permutability) For any $v_i \in \bV_{n_i}$,
 $1 \leq i \leq 4$, the series
$$\Dlt\perm \eqno(4.83)$$
converges to some rational function $f(z_1,z_2)$ in the domain
$0<|z_2|<|z_1|$ and the series
$$\Bta\Dlt\prem\eqno(4.84)$$
converges to the same rational function $f(z_1,z_2)$ in the
domain $0<|z_1|<|z_2|$.
\epr

\demo{Proof} We will prove this theorem in four steps.

\demo{Step 1} For any $v_1 \in \bV_{n_1},{\bar w_i} \in \bV_{n_i} \cap
{\bar \Omega}$,$i \in \{2,3\}$ and $w_4 \in \bV_{n_4} \cap \Omega$
we have that the series
$$\Dlt(Y(v_1,z_1)Y({\bar w_2},z_2){\bar w_3},w_4)\eqno(4.85)$$
converges in the domain $0 < |z_2|<|z_2|$ to the same rational function
as the series
$$\Bta\Dlt(Y(\wwb,z_2)Y(v_1,z_1)\wwc,w_4)\eqno(4.86)$$
does in the domain $0<|z_1|<|z_2|$.
\demo{Proof}(Step 1) WLOG, let $v_1=\vsr w_1$ where $x_j(-m_j) \in B$,
$1 \leq j \leq r$, and $w_1 \in \Omega$. We will use induction on $r$.
When $r=0$, then $v_1=w_1$ and by the above Corollary we are done.
For $r>0$, suppose that for any $0 \leq s <r$, $v'_1=\vsq w_1$, and any
${\bar w'_i} \in \bV_{n_i} \cap {\bar \Omega}$,$i \in \{2,3\}$ we have that
the series
$$\Dlt (Y(v'_1,z_1)Y({\bar w'_2},z_2){\bar w'_3},w_4)$$
converges in the domain $0<|z_2|<|z_1|$ to the same rational function as
the series
$$\Bta\Dlt (Y(\wwx,z_2)Y(v'_1,z_1)\wwy,w_4)$$
does in the domain $0<|z_1|<|z_2|$,
then for $0<|z_2|<|z_1|$,
$$\eqalign{
&\Dlt (Y(v_1,z_1)Y({\bar w_2},z_2){\bar w_3},w_4) \cr
&=\Dlt \smK {m_r+k-1 \choose k} \cdot \cr
&\quad [{z_1}^k (x_r(-m_r-k)Y(\vst w_1,z_1)
Y({\bar w_2},z_2){\bar w_3},w_4) \cr
&\quad \quad -(-1)^{m_r}{z_1}^{-m_r-k}\cdot \cr
&\quad\quad\quad\qquad (Y(\vst w_1,z_1)x_r(k)
Y({\bar w_2},z_2){\bar w_3},w_4)] \cr
&=\Dlt \smK {m_r+k-1 \choose k} (-1)^{m_r+1}{z_1}^{-m_r-k} \cdot \cr
&\quad [(Y(\vst w_1,z_1)Y({\bar w_2},z_2)x_r(k){\bar w_3},w_4) \cr
&\quad \quad+ {z_2}^k
(Y(\vst w_1,z_1)Y(x_r(0){\bar w_2},z_2){\bar w_3},w_4)] \cr
&=\Dlt (-1)^{m_r+1}{z_1}^{-m_r} \cdot \cr
&\quad[\smK {m_r+k-1 \choose k} {z_1}^{-k} {z_2}^k\cdot \cr
&\quad\quad\quad\qquad(Y(\vst w_1,z_1)
Y(x_r(0){\bar w_2},z_2){\bar w_3},w_4) \cr
&\quad\quad +(Y(\vst w_1,z_1)Y(\wwb,z_2) x_r(0) \wwc,w_4)] \cr
&=\Dlt (-1)^{m_r+1} \cdot \cr
&\quad[(z_1-z_2)^{-m_r} (Y(\vst w_1,z_1)Y(x_r(0)\wwb,z_2)
\wwc,w_4) \cr
&\quad\quad +{z_1}^{-m_r}(Y(\vst w_1,z_1)Y(\wwb,z_2) x_r(0) \wwc,w_4)] \cr
}$$
and for $0<|z_1|<|z_2|$,
$$\eqalign{
&\Dlt (Y(\wwb,z_2)Y(v_1,z_1)\wwc,w_4) \cr
&=\Dlt \smK {m_r+k-1 \choose k} \cdot \cr
&\quad [{z_1}^k(Y(\wwb,z_2)x_r(-m_r-k)Y(\vst w_1,z_1)\wwc,w_4) \cr
&\quad\quad -(-1)^{m_r}{z_1}^{-m_r-k} \cdot\cr
&\quad\quad\quad\qquad(Y(\wwb,z_2)Y(\vst w_1,z_1)
x_r(k)\wwc,w_4)]
\cr
&=-\Dlt \cdot \cr
&[\smK {m_r+k-1 \choose k} {z_1}^k{z_2}^{-m_r-k} \cdot\cr
&\quad\quad\quad\qquad(Y(x_r(0)\wwb,z_2)Y(\vst w_1,z_1)\wwc,w_4) \cr
&\quad\quad +(-1)^{m_r}{z_1}^{-m_r} \cdot\cr
&\quad\quad\quad\qquad(Y(\wwb,z_2)Y(\vst w_1,z_1)
x_r(0)w_3,w_4)] \cr
}$$
$$\eqalign{
&=-\Dlt \cdot \cr
&\quad[(z_2-z_1)^{-m_r} (Y(x_r(0)\wwb,z_2)Y(\vst w_1,z_1)
\wwc,w_4) \cr
&\quad\quad +(-1)^{m_r}{z_1}^{-m_r} \cdot\cr
&\quad\quad\quad\qquad(Y(\wwb,z_2)Y(\vst w_1,z_1)
x_r(0)w_3,w_4)] \cr
&=\Dlt (-1)^{m_r+1} \cdot \cr
&\quad [(z_1-z_2)^{-m_r} (Y(x_r(0)\wwb,z_2)Y(\vst w_1,z_1)
\wwc,w_4) \cr
&\quad\quad +{z_1}^{-m_r}(Y(\wwb,z_2)Y(\vst w_1,z_1) x_r(0)w_3,w_4)]
}$$
but $x_r(0){\bar w_2},x_r(0){\bar w_3} \in {\bar \Omega}$ so the above
satisfies the induction hypothesis.  Therefore, since $m_i \in {\Bbb Z}$ we
have proved Step 1.

\demo{Step 2}  For any $v_i \in \bV_{n_i}$,$i \in \{1,4\}$,
${\bar w_t} \in \bV_{n_t} \cap {\bar \Omega}$,$t \in \{2,3\}$
we have that the series
$$\Dlt (Y(v_1,z_1)Y(\wwb,z_2)\wwc,v_4) \eqno(4.87)$$
converges in the domain $0<|z_2|<|z_1|$ to the same rational function as
the series
$$\Bta\Dlt (Y\wwb,z_2)Y(v_1,z_1)\wwc,v_4)\eqno(4.88)$$
does in the domain $0<|z_1|<|z_2|$.
\demo{Proof}(Step 2) WLOG, let $v_4 = \vsr w_4$ where $x_j(-m_j) \in B$,
$1 \leq j \leq r$, and $w_4 \in \Omega$.  We will again use induction on
r.  When $r=0$ we have $v_4=w_4$ and by Step 1 we are done.
For $r>0$ suppose that for any $0 \leq s < r$, $v'_4=\vsq w_4$,
${\bar w'_i} \in
\bV_{n_i} \cap {\bar \Omega}$, and $v'_1 \in \bV_{n_1}, wt(v'_1) \leq
wt(v_1)$ we have that the series
$$\Dlt (Y(v'_1,z_1)Y(\wwx,z_2)\wwy,v'_4) $$
converges in the domain $0<|z_2|<|z_1|$ to the same rational function as
the series
$$\Bta\Dlt (Y(\wwx,z_2)Y(v'_1,z_1)\wwy,v'_4)$$
does in the domain $0<|z_1|<|z_2|$,
then for $0 <|z_2|<|z_1|$,
$$\eqalign{
&\Dlt (Y(v_1,z_1)Y(\wwb,z_2)\wwc,v_4) \cr
&=\Dlt \cdot \cr
&\quad(x^*_r(m_r)Y(v_1,z_1)Y(\wwb,z_2)\wwc,\vst w_4) \cr
}$$
$$\eqalign{
&=\Dlt \cdot \cr
&\quad[(Y(v_1,z_1)x^*_r(m_r)Y(\wwb,z_2)\wwc,\vst w_4) \cr
&\quad \quad +\smK {m_r \choose k} {z_1}^{m_r-k} \cdot\cr
&\quad\quad\quad\qquad(Y(x^*_r(k)v_1,z_1)
Y(\wwb,z_2)\wwc, \vst w_4) ] \cr
&=\Dlt \cdot \cr
&\quad [(Y(v_1,z_1)Y(\wwb,z_2)x^*_r(m_r)\wwc,\vst w_4) \cr
&\quad\quad +\smK {m_r \choose k} {z_1}^{m_r-k}\cdot\cr
&\quad\quad\quad\qquad (Y(x^*_r(k)v_1,z_1)
Y(\wwb,z_2)\wwc,\vst w_4) \cr
&\quad\quad + {z_2}^{m_r} (Y(v_1,z_1)Y(x^*_r(0)
\wwb,z_2)\wwc,\vst w_4)],}$$
and for $0 < |z_1|<|z_2|$,
$$\eqalign{
&\Dlt (Y(\wwb,z_2)Y(v_1,z_1)\wwc,v_4) \cr
&=\Dlt \cdot \cr
&\quad (x^*_r(m_r)Y(\wwb,z_2)Y(v_1,z_1)\wwc,\vst w_4) \cr
&=\Dlt \cdot \cr
&\quad [(Y(\wwb,z_2)x^*_r(m_r)Y(v_1,z_1)\wwc,\vst w_4) \cr
&\quad \quad + {z_2}^{m_r}(Y(x^*_r(0)\wwb,z_2)Y(v_1,z_1)\wwc,\vst w_4) \cr
&=\Dlt \cdot \cr
&\quad [(Y(\wwb,z_2)Y(v_1,z_1)x^*_r(m_r)\wwc,\vst w_4) \cr
&\quad \quad+\smK {m_r \choose k} {z_1}^{m_r-k}\cdot\cr
&\quad\quad\quad\qquad(Y(\wwb,z_1) Y(x^*_r(k)
v_1,z_1)\wwc, \vst w_4) \cr
&\quad \quad +{z_2}^{m_r}(Y(x^*_r(0)\wwb,z_2)Y(v_1,z_1) \wwc,\vst w_4)
] \cr
}$$
but since $wt(x^*_r(k) v_1) = wt(v_1)-k \leq wt(v_1)$ and
$x^*_r(m_r)\wwc,x^*_r(k)\wwb \in {\bar \Omega}$ the above satisfies
the induction hypothesis and we have proven Step 2.

\demo{Step 3} For any $v_i \in \bV_{n_i}$,$i \in \{1,2,4\}$,
$\wwc \in \bV_{n_3} \cap {\bar \Omega}$
we have that
the series
$$\Dlt (Y(v_1,z_1)Y(v_2,z_2)\wwc,v_4) \eqno(4.89)$$
converges in the domain $0<|z_2|<|z_1|$ to the same rational function as
the series
$$\Bta\Dlt (Y(v_2,z_2)Y(v_1,z_1)\wwc,v_4)\eqno(4.90)$$
does in the domain $0<|z_1|<|z_2|$.
\demo{Proof}(Step 3) WLOG, let $v_2= \vsr w_2$ where $x_j(-m_j) \in B$,
$1 \leq j \leq r$, and $w_2 \in \Omega$.  We will use induction on r.
When $r=0$ we have that $v_2 = w_2$ and by Step 2 we are done.
For $r>0$ suppose that for any $0 \leq s < r$, $v'_2= \vsq w_2$,
$\wwy \in
\bV_{n_3} \cap {\bar \Omega}$, and $v'_i \in \bV_{n_i}$,$wt(v'_i) \leq
wt(v_i)$,$i \in \{1,2,4\}$ we have that the series
$$\Dlt (Y(v'_1,z_1)Y(v'_2,z_2)\wwy,v'_4)$$
converges in the domain $0<|z_2|<|z_1|$ to the same rational function as
the series
$$\Bta\Dlt (Y(v'_2,z_2)Y(v'_1,z_1)\wwy,v'_4)$$
does in the domain $0<|z_1|<|z_2|$,
then for $0<|z_2|<|z_1|$,
$$\eqalign{
&\Dlt (Y(v_1,z_1)Y(v_2,z_2)\wwc,v_4) \cr
&=\Dlt \smK {m_r+k-1 \choose k} \cdot \cr
&\quad [{z_2}^k(Y(v_1,z_1)x_r(-m_r-k) Y(\vst w_2,
z_2)\wwc,v_4) \cr
&\quad\quad -(-1)^{m_r}{z_2}^{-m_r-k}\cdot\cr
&\quad\quad\quad\qquad(Y(v_1,z_1)Y(\vst w_2,z_2)
x_r(k) \wwc,v_4)] \cr
&=\Dlt \{ \smK {z_2}^k {m_r+k-1 \choose k} \cdot \cr
&\quad [(Y(v_1,z_1)Y(\vst w_2,z_2)\wwc,x^*_r(m_r+k) v_4) \cr
&-\sum_{0 \leq p \in {\Bbb Z}}
{-m_r-k \choose p}{z_1}^{-m_r-k-p}\cdot\cr
&\quad\quad\quad\qquad(Y(x_r(p)v_1,z_1)Y(\vst w_2,z_2) \wwc,v_4)] \cr
&\quad-(-1)^{-m_r}{z_2}^{-m_r}(Y(v_1,z_1)Y(\vst w_2,z_2)x_r(0)\wwc,v_4)\}
\cr
&=\Dlt \{ \smK {m_r+k-1 \choose k} \cdot \cr
&\quad [{z_2}^k(Y(v_1,z_1)Y(\vst w_2,z_2)\wwc,x^*_r(m_r+k)v_4) \cr
&\quad\quad -(-1)^k(z_1-z_2)^{-m_r-k}\cdot\cr
&\quad\quad\quad\qquad(Y(x_r(k)v_1,z_1) Y(\vst w_2,z_2)
\wwc,v_4)] \cr
&\quad -(-1)^{-m_r}{z_2}^{-m_r}(Y(v_1,z_1)Y(\vst
w_2,z_2)x_r(0)\wwc,v_4) \}
}$$
and for $0<|z_1|<|z_2|$,
$$\eqalign{
&\Dlt (Y(v_2,z_2)Y(v_1,z_1) \wwc,v_4) \cr
&=\Dlt \smK {m_r+k-1 \choose k} \cdot \cr
&\quad [{z_2}^k(x_r(-m_r-k) Y(\vst w_2,z_2)Y(v_1,z_1)\wwc,v_4) \cr
&\quad\quad -(-1)^{m_r}{z_2}^{-m_r-k}\cdot\cr
&\quad\quad\quad\qquad(Y(\vst w_2,z_2)x_r(k) Y(v_1,z_1)
\wwc,v_4)] \cr
&=\Dlt \smK {m_r+k-1 \choose k} \cdot \cr
&[{z_2}^k (Y(\vst w_2,z_2)Y(v_1,z_1)\wwc,x^*_r(m_r+k)v_4) \cr
&-(-1)^{m_r}{z_2}^{-m_r-k}(Y(\vst
w_2,z_2)Y(v_1,z_1)x_r(k)\wwc, v_4) \cr
&-(-1)^{m_r}{z_2}^{-m_r-k}
\sum_{0 \leq p \in {\Bbb Z}} {k \choose p}{z_1}^{k-p} \cdot\cr
&\quad\quad\quad\qquad(Y(\vst w_2,z_2)
Y(x_r(p)v_1,z_1)\wwc,v_4)] \cr
&=\Dlt \{ \smK {m_r+k-1 \choose k} \cdot \cr
&\quad[{z_2}^k (Y(\vst w_2,z_2)Y(v_1,z_1)\wwc,x^*_r(m_r+k)v_4) \cr
&\quad\quad -(-1)^{m_r} (z_2-z_1)^{-m_r-k} \cdot\cr
&\quad\quad\quad\qquad(Y(\vst w_2,z_2)Y(x_r(k)
v_1,z_1)\wwc,v_4) ] \cr
&\quad -(-1)^{m_r}{z_2}^{-m_r}(Y(\vst w_2,z_2)Y(v_1,z_1)x_r(0)\wwc,
v_4) \} \cr
&=\Dlt \{ \smK {m_r+k-1 \choose k} \cdot \cr
&\quad[{z_2}^k(Y(\vst w_2,z_2)Y(v_1,z_1)\wwc,x^*_r(m_r+k)v_4) \cr
&\quad\quad -(-1)^k(z_1-z_2)^{-m_r-k} \cdot\cr
&\quad\quad\quad\qquad(Y(\vst w_2,z_2) Y(x_r(k) v_1,z_1)
\wwc,v_4) ] \cr
&\quad -(-1)^{m_r}{z_2}^{-m_r}(Y(\vst w_2,z_2)Y(v_1,z_1)x_r(0)\wwc,v_4)
\}
}$$
but since $x_r(0)\wwc \in {\bar \Omega}$, and $wt(x_r(p)v_1)=wt(v_1)-p \leq
wt(v_1)$, with the
$wt(x^*_r(m_r+k)v_4)=wt(v_4)-m_r-k \leq wt(v_4)$ and for $k$ large
we have that $x_r(k)v_1=0$,$x^*_r(m_r+k)v_4=0$,
the above satisfies the induction hypothesis and
we have proven Step 3.

\demo{Step 4} For any $v_i \in \bV_{n_i},1 \leq i \leq 4$,
we have that
the series
$$\Dlt \perm \eqno(4.91)$$
converges in the domain $0<|z_2|<|z_1|$ to the same rational function as
the series
$$\Bta\Dlt\prem\eqno(4.92)$$
does in the domain $0<|z_1|<|z_2|$.
\demo{Proof}(Step 4) WLOG, let $v_3=\vsr w_3$, where $x_j(-m_j) \in B$,
$1 \leq j \leq r$, and $w_3 \in \Omega$.  We will again use induction on r.
When $r=0$ we have that $v_3=w_3$ and by Step 3 we are done.
For $r>0$ suppose that for any $0 \leq s <r$, $v'_3= \vsq w_3$, and
$v'_i \in \bV_{n_i}$, $wt(v'_i) \leq wt(v_i)$,$i \in \{1,2,4\}$ we have that
the series
$$\Dlt (Y(v'_1,z_1)Y(v'_2,z_2)v'_3,v'_4)$$
converges in the domain $0<|z_2|<|z_1|$ to the same rational function as
the series
$$\Bta\Dlt (Y(v'_2,z_2)Y(v'_1,z_1)v'_3,v'_4)$$
does in the domain $0<|z_1|<|z_2|$,
then for $0<|z_2|<|z_1|$,
$$\eqalign{
&\Dlt \perm \cr
&=\Dlt \cdot \cr
&\quad [(Y(v_1,z_1)x_r(-m_r)Y(v_2,z_2)\vst w_3,v_4) \cr
&\quad\quad -{z_2}^{-m_r}\smK{-m_r \choose k} {z_2}^{-k} \cdot\cr
&\quad\quad\quad\qquad
(Y(v_1,z_1)Y(x_r(k)v_2,z_2) \vst w_3,v_4)] \cr
&=\Dlt \cdot \cr
&\quad[(x_r(-m_r)Y(v_1,z_1)Y(v_2,z_2)\vst w_3,v_4) \cr
&\quad\quad -{z_1}^{-m_r}\smK{-m_r \choose k} {z_1}^{-k}
\cdot\cr
&\quad\quad\quad\qquad
(Y(x_r(k)v_1,z_1)Y(v_2,z_2)\vst w_3,v_4) \cr
&\quad\quad -{z_2}^{-m_r}\smK{-m_r \choose k} {z_2}^{-k}
\cdot\cr
&\quad\quad\quad\qquad
(Y(v_1,z_1)Y(x_r(k)v_2,z_2)\vst w_3,v_4)] \cr
}$$
$$\eqalign{
&=\Dlt \cdot \cr
&\quad[(Y(v_1,z_1)Y(v_2,z_2)\vst w_3,x^*_r(m_r)v_4) \cr
&\quad\quad -{z_1}^{-m_r}\smK{-m_r \choose k} {z_1}^{-k}
\cdot\cr
&\quad\quad\quad\qquad
(Y(x_r(k)v_1,z_1)Y(v_2,z_2)\vst w_3,v_4) \cr
&\quad\quad -{z_2}^{-m_r}\smK{-m_r \choose k} {z_2}^{-k}
\cdot\cr
&\quad\quad\quad\qquad
(Y(v_1,z_1)Y(x_r(k)v_2,z_2)\vst w_3,v_4) ]}$$
and for $0<|z_1|<|z_2|$,
$$\eqalign{
&\Dlt \prem \cr
&=\Dlt \cdot \cr
&\quad [(Y(v_2,z_2) x_r(-m_r)Y(v_1,z_1)\vst w_3,v_4) \cr
&\quad\quad -{z_1}^{-m_r}\smK{-m_r \choose k} {z_1}^{-k}
\cdot\cr
&\quad\quad\quad\qquad
(Y(v_2,z_2)Y(x_r(k)v_1,z_1) \vst w_3,v_4)] \cr
&=\Dlt \cdot \cr
&\quad[(Y(v_2,z_2)Y(v_1,z_1)\vst w_3,x^*_r(m_r)v_4) \cr
&\quad\quad -{z_1}^{-m_r}\smK {-m_r \choose k}{z_1}^{-k}
\cdot\cr
&\quad\quad\quad\qquad
(Y(v_2,z_2)Y(x_r(k)v_1,z_1)\vst w_3,v_4) \cr
&\quad\quad -{z_2}^{-m_r}\smK{-m_r \choose k} {z_2}^{-k}
\cdot\cr
&\quad\quad\quad\qquad
(Y(x_r(k)v_2,z_2)Y(v_1,z_1)\vst w_3,v_4)] }$$
but since $wt(x_r(k)v_i)=wt(v_i)-k \leq wt(v_i)$,$i \in \{1,2\}$,
$wt(x^*_r(m_r)v_4)=
wt(v_4)-m_r \leq wt(v_4)$ and for $k$ large enough, $x_r(k)v_i=0$,
$i \in \{1,2\}$, the
above satisfies the induction hypothesis and we have proven Step 4 and
therefore
we have proven the Permutability Theorem.
$\hfill\blacksquare$

\pr{Corollary 4.10} For any $w_1,w_4 \in \Omega,{\bar w_2},{\bar w_3} \in
{\bar \Omega}$ with $w_i,{\bar w_i} \in \bV_{n_i}$ we have that
the series
$$\Dlt \oper \eqno(4.93)$$
converges in the domain $0 < |z_2|<|z_1|$ to the same rational function
as the series
$$\Dlt \opra \eqno(4.94)$$
does in the domain $0<|z_1-z_2|<|z_2|$.
\epr

\demo{Proof}  See Table 3 and Corollary 4.6.
$\hfill\blacksquare$

\pr{Theorem 4.11} (Associativity) For any $v_i \in \bV_{n_i}$,
 $1 \leq i \leq 4$,
the series
$$\Dlt\perm \eqno(4.95)$$
converges in the domain $0 < |z_2|<|z_1|$ to the same rational function
as the series
$$\Dlt\assoc\eqno(4.96)$$
does in the domain $0<|z_1-z_2|<|z_2|$.
\epr

\demo{Proof} We will prove this theorem in four steps.

\demo{Step 1} For any $v_1 \in \bV_{n_1},{\bar w_i} \in \bV_{n_i} \cap
{\bar \Omega}$,$i \in \{2,3\}$ and $w_4 \in \bV_{n_4} \cap \Omega$
we have that
the series
$$\Dlt(Y(v_1,z_1)Y({\bar w_2},z_2){\bar w_3},w_4) \eqno(4.97)$$
converges in the domain $0 < |z_2|<|z_1|$ to the same rational function
as the series
$$\Dlt(Y(Y(v_1,z_1-z_2)\wwb,z_2)\wwc,w_4)\eqno(4.98)$$
does in the domain $0<|z_1-z_2|<|z_2|$.
\demo{Proof}(Step 1) WLOG, let $v_1=\vsr w_1$ where $x_j(-m_j) \in B$,
$1 \leq j \leq r$, and $w_1 \in \Omega$. We will use induction on $r$.
When $r=0$, then $v_1=w_1$ and by the above Corollary we are done.
For $r>0$, suppose that for any $0 \leq s <r$, $v'_1=\vsq w_1$, and any
${\bar w'_i} \in \bV_{n_i} \cap {\bar \Omega}$,$i \in \{2,3\}$ we have that
the series
$$\Dlt (Y(v'_1,z_1)Y({\bar w'_2},z_2){\bar w'_3},w_4)$$
converges in the domain $0 < |z_2|<|z_1|$ to the same rational function
as the series
$$\Dlt (Y(Y(v'_1,z_1-z_2)\wwx,z_2)\wwy,w_4)$$
does in the domain $0<|z_1-z_2|<|z_2|$,
then for $0<|z_2|<|z_1|$,
$$\eqalign{
&\Dlt (Y(v_1,z_1)Y({\bar w_2},z_2){\bar w_3},w_4) \cr
&=\Dlt \smK {m_r+k-1 \choose k} \cdot \cr
&\quad [{z_1}^k (x_r(-m_r-k)Y(\vst w_1,z_1)
Y({\bar w_2},z_2){\bar w_3},w_4) \cr
&\quad \quad -(-1)^{m_r}{z_1}^{-m_r-k}
\cdot\cr
&\quad\quad\quad\qquad
(Y(\vst w_1,z_1)x_r(k)
Y({\bar w_2},z_2){\bar w_3},w_4)] \cr
&=\Dlt \smK {m_r+k-1 \choose k} (-1)^{m_r+1}{z_1}^{-m_r-k} \cdot \cr
&\quad [(Y(\vst w_1,z_1)Y({\bar w_2},z_2)x_r(k){\bar w_3},w_4) \cr
&\quad \quad+ {z_2}^k
(Y(\vst w_1,z_1)Y(x_r(0){\bar w_2},z_2){\bar w_3},w_4)] \cr
&=\Dlt (-1)^{m_r+1}{z_1}^{-m_r} \cdot \cr
&\quad[\smK {m_r+k-1 \choose k} {z_1}^{-k} {z_2}^k
\cdot\cr
&\quad\quad\quad\qquad
(Y(\vst w_1,z_1)
Y(x_r(0){\bar w_2},z_2){\bar w_3},w_4) \cr
&\quad\quad +(Y(\vst w_1,z_1)Y(\wwb,z_2) x_r(0) \wwc,w_4)] \cr
&=\Dlt (-1)^{m_r+1} \cdot \cr
&\quad[(z_1-z_2)^{-m_r} (Y(\vst w_1,z_1)Y(x_r(0)\wwb,z_2)
\wwc,w_4) \cr
&\quad\quad +{z_1}^{-m_r}(Y(\vst w_1,z_1)Y(\wwb,z_2) x_r(0) \wwc,w_4)] \cr
}$$
and for $0<|z_1-z_2|<|z_2|$,
$$\eqalign{
&\Dlt (Y(Y(v_1,z_1-z_2)\wwb,z_2)\wwc,w_4) \cr
&=\Dlt \smK {m_r+k-1 \choose k} \cdot \cr
&\quad[{(z_1-z_2)}^k
\cdot\cr
&\qquad\quad
(Y(x_r(-m_r-k)Y(\vst w_1,z_1-z_2)\wwb,z_2)\wwc,w_4)
\cr
&-(-1)^{m_r}(z_1-z_2)^{-m_r-k}
\cdot\cr
&\qquad\quad
(Y(Y(\vst w_1,z_1-z_2)x_r(k)
\wwb,z_2)\wwc,w_4)] \cr
&=\Dlt \cdot \cr
&\{ \smK {m_r+k-1 \choose k}(z_1-z_2)^k \sum_{0 \leq p \in {\Bbb
Z}} {m_r+k+p-1 \choose p} \cdot \cr
&[{z_2}^p(x_r(-m_r-k-p)Y(Y(\vst w_1,z_1-z_2)\wwb,z_2) \wwc,w_4) \cr
&-(-1)^{m_r+k}{z_2}^{-m_r-k-p}
\cdot\cr
&\quad\qquad
(Y(Y(\vst w_1,z_1-z_2)\wwb,z_2)
x_r(p)\wwc,w_4)] \cr
}$$
$$\eqalign{
&\quad -(-1)^{m_r}(z_1-z_2)^{-m_r}
\cdot\cr
&\quad\qquad
(Y(Y(\vst
w_1,z_1-z_2)x_r(0)\wwb,z_2)\wwc,w_4) \} \cr
&=\Dlt (-1)^{m_r+1} \cdot \cr
&\quad [(z_1-z_2)^{-m_r}(Y(Y(\vst w_1,z_1-z_2)x_r(0)\wwb,z_2)\wwc,w_4)
\cr
&\quad\quad +\smK {m_r+k-1 \choose k}(z_1-z_2)^k(-1)^k{z_2}^{-m_r-k} \cdot \cr
&\quad\quad
(Y(Y(\vst w_1,z_1-z_2)\wwb,z_2)x_r(0)\wwc,w_4)] \cr
&=\Dlt (-1)^{m_r+1} \cdot \cr
&\quad[(z_1-z_2)^{-m_r}(Y(Y(\vst w_1,z_1-z_2)x_r(0)\wwb,z_2)\wwc,w_4)
\cr
&\quad\quad +{z_1}^{-m_r}(Y(Y(\vst w_1,z_1-z_2)\wwb,z_2)x_r(0)\wwc,w_4)]
}$$
but $x_r(0){\bar w_2},x_r(0){\bar w_3} \in {\bar \Omega}$ so the above
satisfies the induction hypothesis.  Therefore, since $m_i \in {\Bbb Z}$ we
have proved Step 1.

\demo{Step 2}  For any $v_i \in \bV_{n_i}$,$i \in \{1,4\}$,
${\bar w_t} \in \bV_{n_t} \cap {\bar \Omega}$,$t \in \{2,3\}$
we have that
the series
$$\Dlt (Y(v_1,z_1)Y(\wwb,z_2)\wwc,v_4)\eqno(4.99)$$
converges in the domain $0 < |z_2|<|z_1|$ to the same rational function
as the series
$$\Dlt (Y(Y(v_1,z_1-z_2)\wwb,z_2)\wwc,v_4)\eqno(4.100)$$
does in the domain $0<|z_1-z_2|<|z_2|$.
\demo{Proof}(Step 2) WLOG, let $v_4 = \vsr w_4$ where $x_j(-m_j) \in B$,
$1 \leq j \leq r$, and $w_4 \in \Omega$.  We will again use induction on
r.  When $r=0$ we have $v_4=w_4$ and by Step 1 we are done.
For $r>0$ suppose that for any $0 \leq s < r$, $v'_4=\vsq w_4$,
${\bar w'_i} \in
\bV_{n_i} \cap {\bar \Omega}$, and $v'_1 \in \bV_{n_1}, wt(v'_1) \leq
wt(v_1)$ we have that
the series
$$\Dlt (Y(v'_1,z_1)Y(\wwx,z_2)\wwy,v'_4)$$
converges in the domain $0 < |z_2|<|z_1|$ to the same rational function
as the series
$$\Dlt (Y(Y(v'_1,z_1-z_2)\wwx,z_2)\wwy,v'_4)$$
does in the domain $0<|z_1-z_2|<|z_2|$,
then for $0 <|z_2|<|z_1|$,
$$\eqalign{
&\Dlt (Y(v_1,z_1)Y(\wwb,z_2)\wwc,v_4) \cr
&=\Dlt \cdot \cr
&\quad(x^*_r(m_r)Y(v_1,z_1)Y(\wwb,z_2)\wwc,\vst w_4) \cr
&=\Dlt \cdot \cr
&\quad[(Y(v_1,z_1)x^*_r(m_r)Y(\wwb,z_2)\wwc,\vst w_4) \cr
&\quad \quad +\smK {m_r \choose k} {z_1}^{m_r-k}
\cdot\cr
&\quad\quad\quad\qquad
(Y(x^*_r(k)v_1,z_1)
Y(\wwb,z_2)\wwc, \vst w_4) ] \cr
&=\Dlt \cdot \cr
&\quad [(Y(v_1,z_1)Y(\wwb,z_2)x^*_r(m_r)\wwc,\vst w_4) \cr
&\quad\quad +\smK {m_r \choose k} {z_1}^{m_r-k}
\cdot\cr
&\quad\quad\quad\qquad
 (Y(x^*_r(k)v_1,z_1)
Y(\wwb,z_2)\wwc,\vst w_4) \cr
&\quad\quad + {z_2}^{m_r} (Y(v_1,z_1)Y(x^*_r(0)
\wwb,z_2)\wwc,\vst w_4)],}$$
and for $0 < |z_1-z_2|<|z_2|$,
$$\eqalign{
&\Dlt (Y(Y(v_1,z_1-z_2)\wwb,z_2)\wwc,v_4) \cr
&=\Dlt \cdot \cr
&\quad (x^*_r(m_r)Y(Y(v_1,z_1-z_2)\wwb,z_2)\wwc,\vst w_4) \cr
&=\Dlt \cdot \cr
&\quad [(Y(Y(v_1,z_1-z_2)\wwb,z_2)x^*_r(m_r)\wwc,\vst w_4) \cr
&\quad \quad +\smK{m_r \choose k}
{z_2}^{m_r-k}
\cdot\cr
&\quad\quad\quad\qquad
(Y(x^*_r(k)Y(v_1,z_1-z_2)\wwb,z_2)\wwc,\vst w_4) \cr
&=\Dlt \cdot \cr
&\quad [(Y(Y(v_1,z_1-z_2)\wwb,z_2)x^*_r(m_r)\wwc,\vst w_4) \cr
&\quad+\smK {m_r \choose k} {z_2}^{m_r-k}
 \sum_{0 \leq p \in {\Bbb Z}} {k \choose p} (z_1-z_2)^{k-p}
\cdot \cr
&\quad\quad\quad\qquad
(Y(Y(x^*_r(p)
v_1,z_1-z_2)\wwb,z_2)\wwc, \vst w_4) \cr
&\quad +{z_2}^{m_r}(Y(Y(v_1,z_1-z_2)x^*_r(0)\wwb,z_2)\wwc,\vst w_4)
] \cr
&=\Dlt \cdot \cr
&\quad[(Y(Y(v_1,z_1-z_2)\wwb,z_2)x^*_r(m_r)\wwc,\vst w_4) \cr
}$$
$$\eqalign{
&\quad\quad +\smK {m_r \choose k}{z_2}^{m_r-k}
(Y(Y(x^*_r(k)v_1,z_1-z_2)\wwb,z_2)\wwc,w_4) \cr
&\quad\quad +{z_2}^{m_r}(Y(Y(v_1,z_1-z_2)x^*_r(0)\wwb,z_2)\wwc,w_4) ]
}$$
but since $wt(x^*_r(k) v_1) = wt(v_1)-k \leq wt(v_1)$ and
$x^*_r(m_r)\wwc,x^*_r(k)\wwb \in {\bar \Omega}$ the above satisfies
the induction hypothesis and we have proven Step 2.

\demo{Step 3} For any $v_i \in \bV_{n_i}$,$i \in \{1,2,4\}$,
$\wwc \in \bV_{n_3} \cap {\bar \Omega}$ we have that
the series
$$\Dlt (Y(v_1,z_1)Y(v_2,z_2)\wwc,v_4) \eqno(4.101)$$
converges in the domain $0 < |z_2|<|z_1|$ to the same rational function
as the series
$$\Dlt (Y(Y(v_1,z_1-z_2)v_2,z_2)\wwc,v_4) \eqno(4.102)$$
does in the domain $0<|z_1-z_2|<|z_2|$.
\demo{Proof}(Step 3) WLOG, let $v_2= \vsr w_2$ where $x_j(-m_j) \in B$,
$1 \leq j \leq r$, and $w_2 \in \Omega$.  We will use induction on r.
When $r=0$ we have that $v_2 = w_2$ and by Step 2 we are done.
For $r>0$ suppose that for any $0 \leq s < r$, $v'_2= \vsq w_2$,
$\wwy \in
\bV_{n_3} \cap {\bar \Omega}$, and $v'_i \in \bV_{n_i}$,$wt(v'_i) \leq
wt(v_i)$,$i \in \{1,2,4\}$ we have that
the series
$$\Dlt (Y(v'_1,z_1)Y(v'_2,z_2)\wwy,v'_4)$$
converges in the domain $0 < |z_2|<|z_1|$ to the same rational function
as the series
$$\Dlt (Y(Y(v'_1,z_1-z_2)v_2',z_2)\wwy,v'_4)$$
does in the domain $0<|z_1-z_2|<|z_2|$,
then for $0<|z_2|<|z_1|$,
$$\eqalign{
&\Dlt (Y(v_1,z_1)Y(v_2,z_2)\wwc,v_4) \cr
&=\Dlt \smK {m_r+k-1 \choose k} \cdot \cr
&\quad [{z_2}^k(Y(v_1,z_1)x_r(-m_r-k) Y(\vst w_2,
z_2)\wwc,v_4) \cr
&\quad\quad -(-1)^{m_r}{z_2}^{-m_r-k}
\cdot \cr
&\quad\quad\quad\qquad
(Y(v_1,z_1)Y(\vst w_2,z_2)
x_r(k) \wwc,v_4)] \cr
&=\Dlt \{ \smK {z_2}^k {m_r+k-1 \choose k} \cdot \cr
&\quad [(Y(v_1,z_1)Y(\vst w_2,z_2)\wwc,x^*_r(m_r+k) v_4) \cr
&-\sum_{0 \leq p \in {\Bbb Z}}
{-m_r-k \choose p}{z_1}^{-m_r-k-p}
\cdot \cr
&\quad\quad\quad\qquad
(Y(x_r(p)v_1,z_1)Y(\vst w_2,z_2) \wwc,v_4)] \cr
&\quad-(-1)^{-m_r}{z_2}^{-m_r}(Y(v_1,z_1)Y(\vst w_2,z_2)x_r(0)\wwc,v_4)\}
\cr
&=\Dlt \{ \smK {m_r+k-1 \choose k} \cdot \cr
&\quad [{z_2}^k(Y(v_1,z_1)Y(\vst w_2,z_2)\wwc,x^*_r(m_r+k)v_4) \cr
&\quad\quad -(-1)^k(z_1-z_2)^{-m_r-k}
\cdot \cr
&\quad\quad\quad\qquad
(Y(x_r(k)v_1,z_1) Y(\vst w_2,z_2)
\wwc,v_4)] \cr
&\quad -(-1)^{-m_r}{z_2}^{-m_r}(Y(v_1,z_1)Y(\vst
w_2,z_2)x_r(0)\wwc,v_4) \}
}$$
and for $0<|z_1-z_2|<|z_2|$,
$$\eqalign{
&\Dlt (Y(Y(v_1,z_1-z_2)v_2,z_2)\wwc,v_4) \cr
&=\Dlt \cdot \cr
&\quad[(Y(x_r(-m_r)Y(v_1,z_1-z_2)\vst w_2,z_2)\wwc,v_4) \cr
&\quad\quad-\smK{-m_r \choose k}(z_1-z_2)^{-m_r-k} \cdot \cr
&\quad\qquad\quad\quad(Y(Y(x_r(k)v_1,z_1-z_2)\vst w_2,z_2)\wwc,v_4)] \cr
&=\Dlt \{ \sum_{0 \leq p \in {\Bbb Z}}{m_r+p-1 \choose p} \cdot \cr
&\quad[{z_2}^p(x_r(-m_r-p)Y(Y(v_1,z_1-z_2)\vst w_2,z_2)\wwc,v_4) \cr
}$$
$$\eqalign{
&\quad\quad-(-1)^{m_r}{z_2}^{-m_r-p}
\cdot \cr
&\quad\quad\quad\qquad
(Y(Y(v_1,z_1-z_2)\vst w_2,z_2)x_r(p)
\wwc,v_4)] \cr
&\quad-\smK{m_r+k-1 \choose
k}(-1)^k(z_1-z_2)^{-m_r-k} \cdot \cr
&\quad\qquad\quad\quad(Y(Y(x_r(k)v_1,z_1-z_1) \vst w_2,z_2)\wwc,v_4)
\} \cr
&=\Dlt \{ \smK {m_r+k-1 \choose k} \cdot \cr
&[{z_2}^k(Y(Y(v_1,z_1-z_2)\vst w_2,z_2)\wwc,x^*_r(m_r+k)v_4) \cr
&\quad -(-1)^k(z_1-z_2)^{-m_r-k}
\cdot \cr
&\quad\quad\quad\qquad
(Y(Y(x_r(k)v_1,z_1-z_2)\vst
w_2,z_2) \wwc,v_4)] \cr
&-(-1)^{m_r}{z_2}^{-m_r}(Y(Y(v_1,z_1-z_2)\vst w_2,z_2)x_r(0)
\wwc,v_4) \}
}$$
but since $x_r(0)\wwc \in {\bar \Omega}$, and $wt(x_r(p)v_1)=wt(v_1)-p \leq
wt(v_1)$,$wt(x^*_r(m_r+k)v_4)=wt(v_4)-m_r-k \leq wt(v_4)$ and for $k$ large
we have that $x_r(k)v_1=0$,$x^*_r(m_r+k)v_4=0$,
the above satisfies the induction hypothesis and
we have proven Step 3.

\demo{Step 4} For any $v_i \in \bV_{n_i},1 \leq i \leq 4$,
the series
$$\Dlt \perm \eqno(4.103)$$
converges in the domain $0 < |z_2|<|z_1|$ to the same rational function
as the series
$$\Dlt\assoc\eqno(4.104)$$
does in the domain $0<|z_1-z_2|<|z_2|$.
\demo{Proof}(Step 4) WLOG, let $v_3=\vsr w_3$, where $x_j(-m_j) \in B$,
$1 \leq j \leq r$, and $w_3 \in \Omega$.  We will again use induction on r.
When $r=0$ we have that $v_3=w_3$ and by Step 3 we are done.
For $r>0$ suppose that for any $0 \leq s <r$, $v'_3= \vsq w_3$, and
$v'_i \in \bV_{n_i}$, $wt(v'_i) \leq wt(v_i)$,$i \in \{1,2,4\}$ we have that
the series
$$\Dlt (Y(v'_1,z_1)Y(v'_2,z_2)v'_3,v'_4)$$
converges in the domain $0 < |z_2|<|z_1|$ to the same rational function
as the series
$$\Dlt (Y(Y(v'_1,z_1-z_2)v'_2,z_2)v'_3,v'_4)$$
does in the domain $0<|z_1-z_2|<|z_2|$,
then for $0<|z_2|<|z_1|$,
$$\eqalign{
&\Dlt \perm \cr
&=\Dlt \cdot \cr
&\quad [(Y(v_1,z_1)x_r(-m_r)Y(v_2,z_2)\vst w_3,v_4) \cr
&\quad\quad -{z_2}^{-m_r}\smK{-m_r \choose k} {z_2}^{-k}
\cdot \cr
&\quad\quad\quad\qquad
(Y(v_1,z_1)Y(x_r(k)v_2,z_2) \vst w_3,v_4)] \cr
&=\Dlt \cdot \cr
&\quad[(x_r(-m_r)Y(v_1,z_1)Y(v_2,z_2)\vst w_3,v_4) \cr
&\quad\quad -{z_1}^{-m_r}\smK{-m_r \choose k} {z_1}^{-k}
\cdot \cr
&\quad\quad\quad\qquad
(Y(x_r(k)v_1,z_1)Y(v_2,z_2)\vst w_3,v_4) \cr
&\quad\quad -{z_2}^{-m_r}\smK{-m_r \choose k} {z_2}^{-k}
\cdot \cr
&\quad\quad\quad\qquad
(Y(v_1,z_1)Y(x_r(k)v_2,z_2)\vst w_3,v_4)] \cr
&=\Dlt \cdot \cr
&\quad[(Y(v_1,z_1)Y(v_2,z_2)\vst w_3,x^*_r(m_r)v_4) \cr
&\quad\quad -{z_1}^{-m_r}\smK{-m_r \choose k} {z_1}^{-k}
\cdot \cr
&\quad\quad\quad\qquad
(Y(x_r(k)v_1,z_1)Y(v_2,z_2)\vst w_3,v_4) \cr
&\quad\quad -{z_2}^{-m_r}\smK{-m_r \choose k} {z_2}^{-k}
\cdot \cr
&\quad\quad\quad\qquad
(Y(v_1,z_1)Y(x_r(k)v_2,z_2)\vst w_3,v_4) ]}$$
and for $0<|z_1-z_2|<|z_2|$,
$$\eqalign{
&\Dlt \assoc \cr
&=\Dlt \cdot \cr
&\quad [(x_r(-m_r)Y(Y(v_1,z_1-z_2)v_2,z_2)\vst w_3,v_4) \cr
&-{z_2}^{-m_r}\smK{-m_r \choose k} {z_2}^{-k}
\cdot \cr
&\quad\quad\quad\qquad
(Y(x_r(k)Y(v_1,z_1-z_2)v_2,z_2) \vst w_3,v_4)] \cr
}$$
$$\eqalign{
&=\Dlt \cdot \cr
&\quad[(Y(Y(v_1,z_1-z_2)v_2,z_2)\vst w_3,x^*_r(m_r)v_4) \cr
&-{z_2}^{-m_r}\smK {-m_r \choose k}{z_2}^{-k}
\cdot \cr
&\quad\quad\quad\qquad
(Y(Y(v_1,z_1-z_2)x_r(k)v_2,z_2)\vst w_3,v_4) \cr
&-{z_2}^{-m_r}\smK{-m_r \choose k} {z_2}^{-k}
\sum_{0 \leq p \in {\Bbb Z}} {k \choose p} (z_1-z_2)^{k-p}
\cdot \cr
&\quad\qquad\quad\quad (Y(Y(x_r(p)v_1,z_1-z_2)v_2,z_2)\vst w_3,v_4)] \cr
&=\Dlt \cdot \cr
&\quad[(Y(Y(v_1,z_1-z_2)v_2,z_2)\vst w_3, x^*_r(m_r)v_4) \cr
&-{z_2}^{-m_r}\smK{-m_r \choose k}{z_2}^{-k}
\cdot \cr
&\quad\quad\quad\qquad
(Y(Y(v_1,z_1-z_2)x_r(k)v_2,z_2)\vst w_3,v_4) \cr
&-{z_1}^{-m_r}\smK{-m_r \choose k}{z_1}^{-k}
\cdot \cr
&\quad\quad\quad\qquad
(Y(Y(x_r(k)v_1,z_1-z_2)v_2,z_2)\vst w_3,v_4)]
}$$
but since $wt(x_r(k)v_i)=wt(v_i)-k \leq wt(v_i)$,$i \in \{1,2\}$,
$wt(x^*_r(m_r)v_4)=
wt(v_4)-m_r \leq wt(v_4)$ and for $k$ large enough, $x_r(k)v_i=0$,
$i \in \{1,2\}$, the
above satisfies the induction hypothesis and we have proven Step 4 and
therefore
we have proven the Associativity Theorem.
$\hfill\blacksquare$

\pr{Theorem 4.12} (The Jacobi-Cauchy Identity for VOPAs) Let
$v_i \in \bV_{n_i}$, for $i=1,2,3,4$, let $\eta=\eta(n_1,n_2)$ be defined
as in Table D,
$\Delta(n_1,n_2)$ be defined as in (4.20), and let
$$f(z_1,z_2) \in
{z_1}^{\Delta(n_1,n_3)}{z_2}^{\Delta(n_2,n_3)}{(z_1-z_2)}^{\Delta(n_1,n_2)}
{\Bbb C}[z_1,{z_1}^{-1},z_2,{z_2}^{-1},{(z_1-z_2)}^{-1}].$$
we have
$$\eqalign{
&-Res_{z_1=\infty}\perm f(z_1,z_2)\cr
&-\eta Res_{z_1=0}\prem f(z_1,z_2)\cr
&=Res_{z_1=z_2}\assoc f(z_1,z_2)\cr
}$$
where the function $f(z_1,z_2)$ is to be expanded in the three residues
as in (1.51)-(1.53), respectively with $z=z_1$ and $z_0=z_2$.
\epr

\demo{Proof} From the assumption on $f(z_1,z_2)$, Theorems 4.7,
4.9 and 4.11 say that all three expressions represent the same rational
function of $z_1$ and $z_2$ having possible poles only at
$z_1=0$,$z_2=0$ and $z_1=z_2$.  Cauchy's residue thereom gives the result.
$\hfill\blacksquare$

\pr{Theorem 4.13} With $\bV=$ $\bcV=$ $\bV_0 \oplus
\bV_1 \oplus \bV_2 \oplus \bV_3$, $\Gamma={\Bbb Z}_4$, $ad_z=Y(\cdot,z)$,
${\bold 1}=\bvac$, $\omega=L(-2)\bvac$, $\Delta$ defined in (4.20), and $\eta$
defined
in Table D, $(\bV, ad_z, {\bold 1}, \omega, \Gamma, \Delta, \eta)$ is a
vertex operator para-algebra.
\epr

\vskip 10pt
\pr{Remark} Here it should be noted that since for $w_1,w_2 \in \bV_1 \oplus
\bV_3$ we have $Y(w_1,z)w_2=0$ the group $\Gamma={\Bbb Z}_4$ is not forced.
In the case when $l$ is even, the group $\Gamma={\Bbb Z}_2 \times {\Bbb Z}_2$
works just as well.   It remains to be seen if there is a larger structure that
will determine which group is the best choice in this situation.
We may also look inside $\bcV$ to see if there is any substructures
which give insight into this choice.
If we look at the three structures, $\bV_0 \oplus \bV_1$,$\bV_0 \oplus
\bV_2$,  and $\bV_0 \oplus \bV_3$, one can see that these each form a
VOPA with group structure $\Gamma={\Bbb Z}_2$.  Therefore, since all
three spaces are ``sub-VOPA's'' of $\bcV$, perhaps $\Gamma={\Bbb Z}_2 \times
{\Bbb Z}_2$ is the best choice for the VOPA, $\bcV$, and the restriction
(1.44) should be relaxed. \epr

\newpage
\topmatter
\title References \endtitle \endtopmatter
\document
\widestnumber\key{ABCDE}

\ref\key{B}\by R.E. Borcherds
\paper Vertex algebras, Kac-Moody algebras, and the Monster
\jour Proc. Natl. Acad. Sci. USA
\vol 83 \yr 1986 \pages 3068--3071 \endref

\ref\key{BPZ} \by A. A. Belavin, A. M. Polyakov, A. B. Zamolodchikov
\paper Infinite conformal symmetry in two-dimensional quantum field theory
\jour Nuclear Physics \vol B241 \yr 1984 \pages 333-380 \endref

\ref\key{CN} \by J. H. Conway, S. P. Norton \paper Monstrous moonshine
\jour Bull. London Math. Soc. \vol 11 \yr 1979 \pages 308-339 \endref

\ref\key{DL} \by C.-Y. Dong, J. Lepowsky
\book Generalized vertex algebras and relative vertex operators
\bookinfo Progress in Mathematics, Vol. 112
\publ Birkhauser \publaddr Boston \yr 1993 \endref

\ref\key{F} \by A. J. Feingold
\paper Constructions of Vertex Operator Algebras
\inbook Algebraic Groups and Their Generalizations
\bookinfo Proceedings of Symposia in Pure Mathematics, Vol. 56, Part 2
\eds William J. Haboush and Brian J. Parshall
\publ American Mathematical Society
\publaddr Providence, RI \yr 1994 \endref

\ref\key{FF} \by A. J. Feingold, I. B. Frenkel
\paper Classical affine algebras
\jour Advances in Mathematics \vol 56 \yr 1985 \pages 117--172\endref

\ref\key{FFR} \by A. J. Feingold, I. B. Frenkel, J. F. X. Ries
\book Spinor Construction of Vertex Operator Algebras, Triality, and
$E_8^{(1)}$
\bookinfo Contemporary Mathematics, Vol. 121
\publ American Mathematical Society
\publaddr Providence, RI \yr 1991 \endref

\ref\key{FHL} \by I.B. Frenkel, Y.-Z. Huang, J. Lepowsky
\book On Axiomatic Approaches to Vertex Operator Algebras
and Modules
\bookinfo Memoirs of the Amer. Math. Soc.,
Vol. 104, No. 594
\publ American Mathematical Society
\publaddr Providence, RI \yr 1993 \endref

\ref\key{FLM} \by I.B. Frenkel, J. Lepowsky, A. Meurman
\book Vertex Operator Algebras and the Monster
\bookinfo Pure and Applied Mathematics, Vol. 134
\publ Academic Press \publaddr Boston \yr 1988 \endref

\ref\key{FLM2} \bysame
\paper Vertex operator calculus
\inbook Mathematical Aspects of String Theory, Proc. 1986
Conference, San Diego
\ed S. -T. Yau \publ World Scientific \publaddr Singapore
\yr 1987 \pages 150--188 \endref

\ref\key{GO} \by P. Goddard, D. Olive
\paper Kac-Moody and Virasoro Algebras in Relation to Quantum Physics
\jour Int. J. Mod. Physics A1 \yr 1986 \pages 303--414
\endref

\ref\key{K} \by V.G. Kac
\paper Simple irreducible graded Lie algebras of finite growth
\jour Izv. Akad. Nauk SSSR
\vol 32 \yr 1968 \pages 1323--1367
\transl English transl., Math. USSR Izv.
\vol 2 \yr 1968 \pages 1271--1311 \endref

\ref\key{KP} \by V.G. Kac, D.H. Peterson
\paper 112 Constructions of the basic
representations of the loop group of $E_8$
\inbook Anomalies, Geometry and Topology
\bookinfo Proc. of the Conf., Argonne 1985
\publ World Scientific \yr 1985 \pages 276--298 \endref

\ref\key{KZ} \by V. G. Knizhnik, A. B. Zamolodchikov
\paper Current algebra and Wess-Zumino model in two dimensions
\jour Nucl. Phys. \vol B247
\yr 1984 \pages 83--103 \endref

\ref\key{L} \by J. Lepowsky \paper Calculus of twisted vertex operators
\jour Proc. Natl. Acad. Sci. USA
\vol 82 \yr 1985 \pages 8295--8299 \endref

\ref\key{M} \by R. V. Moody
\paper Lie Algebras Associated with Generalized Cartan Matrices
\jour Bull. Amer. Math. Soc. \vol 73 \yr 1967
\pages 217--221 \endref

\bye